# Gravitational Collapse and Black Hole Formation in a Braneworld

by

Daoyan Wang

A THESIS SUBMITTED IN PARTIAL FULFILLMENT OF
THE REQUIREMENTS FOR THE DEGREE OF

DOCTOR OF PHILOSOPHY

in

The Faculty of Graduate and Postdoctoral Studies

(Physics)

THE UNIVERSITY OF BRITISH COLUMBIA

(Vancouver)

April 2015

# Abstract

In this thesis we present the first numerical study of gravitational collapse in braneworld within the framework of the single brane model by Randall-Sundrum (RSII). We directly show that the evolutions of sufficiently strong initial data configurations result in black holes (BHs) with finite extension into the bulk. The extension changes from sphere to pancake (or cigar, as seen from a different perspective) as the size of BH increases. We find preliminary evidences that BHs of the same size generated from distinct initial data profiles are geometrically indistinguishable. As such, a no-hair theorem of BH (uniqueness of BH solution) is suggested to hold in the RSII spacetimes studied in this thesis—these spacetimes are axisymmetric without angular momentum and non-gravitational charges. In particular, the BHs we obtained as the results of the dynamical system, are consistent with the ones previously obtained from a static vacuum system by Figueras and Wiseman. We also obtained some results in closed form without numerical computation such as the equality of ADM mass of the brane with the total mass of the braneworld.

The calculation within the braneworld requires advances in the formalism of numerical relativity (NR). The regularity problem in previous numerical calculations in axisymmetric (and spherically symmetric) spacetimes, is actually associated with neither coordinate systems nor the machine precision. The numerical calculation is regular in any coordinates, provided the fundamental variables (used in numerical calculations) are regular, and their asymptotic behaviours at the vicinity of the axis (or origin) are compatible with the finite difference scheme. The generalized harmonic (GH) formalism and the BSSN formalism for general relativity are developed to make them suitable for calculations in non-Cartesian coordinates under non-flat background. A conformal function of the metric is included into the GH formalism to simulate the braneworld.

# Preface

Chapter 1 is the introduction and nothing in the chapter is original work. The author rederived the formulae in the literature while keeping $d$ (the dimension of spacetime) and $\epsilon$ (the sign to characterize spacelike or timelike foliation of spacetime) general.

All the works in the rest chapters (Chap. 2, 3, 4 and 5), except for the works properly cited, are original. Within these:

— A significant part of the work regarding initial data (presented in Chap. 4) was conducted in the collaboration with Evgeny Sorkin. Specifically, Evgeny proposed "direct method" (defined in the chapter), taught the author to calculate the total energy in general relativity following [87], and proposed to study the mass-area relation.

— Eq. (5.34), the form of the relation among the circumstances of black holes, was proposed by Toby Wiseman.

— All the remaining works were carried out by the author, with the guidance of Matthew W. Choptuik and William G. Unruh.

# Table of Contents

## Appendices

# List of Tables

# List of Figures

# Notations and Abbreviations

Please pay particular attention to sign conventions: the sign of metric, the sign of the Christoffel symbol, and the sign of the Riemann tensor. The conventions we employ in this thesis, are the same as those in Baumgarte-Shapiro [34] (therefore the same as those in Wald [32]). The sign of extrinsic curvature is the same as Baumgarte-Shapiro [34] (therefore the opposite of Wald [32]).

| | |
|---|---|
| GR | General Relativity |
| BH | Black Hole |
| BS | Black String |
| PDE | Partial Differential Equation |
| FDA | Finite Difference Approximation |
| GH | Generalized Harmonic |
| RSII | Randall-Sundrum braneworld model II (the single brane model) |
| $g_{\mu\nu}$ | spacetime metric |
| $\gamma_{\mu\nu}$ | spatial metric |
| $h_{\mu\nu}$ | brane metric |
| $\nabla$ | the covariant derivative associated with $g$ |
| $D$ | the covariant derivative associated with $\gamma$ |
| $\mathcal{D}$ | the covariant derivative associated with $h$ |
| $d$ | the dimension of the whole spacetime |
| Metric signature | $(-,+,+,+,+)$ |
| $\alpha$ | lapse function |
| $\beta^{\mu}$ | shift function |
| Christoffel symbols ${\Gamma^{\alpha}}_{\beta\gamma}$ | $= \frac{1}{2} g^{\alpha\mu} \left( g_{\mu\beta,\gamma} + g_{\mu\gamma,\beta} - g_{\beta\gamma,\mu} \right)$, where $_{,\mu} \equiv \partial_\mu$ |
| Riemann Tensor ${R^{\gamma}}_{\mu\alpha\beta}$ | defined via $\nabla_\alpha \nabla_\beta v^\gamma - \nabla_\beta \nabla_\alpha v^\gamma = {R^{\gamma}}_{\mu\alpha\beta}\, v^\mu, \forall$ vector field $v^\mu$ |
| $n_\mu$ | unit normal vector of $t =$ const hypersurface |
| $K_{\mu\nu}$ | $\equiv -{\gamma_\mu}^\alpha {\gamma_\nu}^\beta \nabla_\alpha n_\beta$, extrinsic curvature of $t =$ const hypersurface |
| $\mathbb{n}_\nu$ | unit normal vector of the brane |
| $\mathcal{K}_{\mu\nu}$ | $\equiv -{h_\mu}^\alpha {h_\nu}^\beta \nabla_\alpha \mathbb{n}_\beta$, extrinsic curvature of the brane |
| tilde˜ | quantities and operations associated with conformal metric |
| bar¯ | quantities and operations associated with background metric |
| $\Lambda$ | cosmological constant in the bulk |
| $\lambda$ | tension of the brane |
| $k_n$ | $= 8\pi G_n$ where $G_n$ is Newton's constant in $n$ dimension |

# Acknowledgements

First and foremost I would like to thank my supervisor, Matthew Choptuik. He taught me how solid scientific work should be conducted. He encouraged me to pursue a challenging yet important project. Among other supports, I especially appreciate the freedom that he gave me during these years.

I would like to thank William Unruh for his inspirations and his guidance, as well as his involvement and support during my PhD studies.

I thank Evgeny Sorkin for the collaboration on the initial value problem and discussions on various other topics. I thank Frans Pretorius for the PAMR/AMRD package and his continuous advise, and his generous share of his code on binary black holes. The conversation with Frans, Evgeny and Matt at Matt's home in August 2011, about regularity and Generalized Harmonic formalism in non-Cartesian coordinate, was crucial to help me realize the confront of the research and motivated me to pursue further.

I thank Toby Wiseman and Pau Figueras for their generosity to send me their final data to compare, and the useful discussions.

I would like to thank the members of my committee: Colin Gay and Mark Van Raamsdonk for the questions and comments which helped me to improve my thesis.

It was a pleasure being a member of the numerical relativity group at UBC. I am happy to thank my friends and collegues, Arman, Ben, Bruno, Daniel, Graham, Jason and Silvestre. Special thanks go to Arman for his good questions which helped to improve the thesis. I also thank Arman, Graham and Silvestre for proofreading certain parts of my thesis.

I want to thank Junqi Guo for the collaboration on $f(R)$ study (which is not a part of this thesis).

I am grateful to my wife, Yuanyuan Gao, for her love, all kinds of support, understanding and companionship during these years, which enable me to do what I do.

I would like to thank my parents and my sister for their love, sacrifice and the freedom they gave me, and thank my parents in law for their support and understanding.

To my family.

# Chapter 1

# Introduction

## 1.1 Overview

Our observable universe is 3+1 dimensional. The exploration on the possibility of the existence of extra dimensions can trace back to the work by Kaluza and Klein in 1920s [1, 2], where they tried to unify electromagnetism with gravity by using the metric tensor in five dimensional (5D) spacetime in which our universe was a 3+1 dimensional hypersurface. The basic idea of braneworld scenarios is that our observable universe could be a 3+1 dimensional hypersurface (the *brane*) embedded in a higher dimensional spacetime (the *bulk*). Only gravity can propagate freely into the bulk while all the matters are confined onto the brane. Among all the braneworld models, the single brane model constructed by Randall and Sundrum (also known as RSII) [3], is remarkable for its simplicity. It assumes one extra dimension with infinite size, a negative cosmological constant $\Lambda$ in the bulk, tension $\lambda$ on brane which makes the brane a gravitating object, and $Z_2$ symmetry of the bulk with respect to the brane. $\lambda$ is set to a value such that general relativity (GR) is recovered on the brane at low energies [4]. However, the high energy behaviour, where the dynamics could be rather different from GR, is not understood as clearly.

Black holes (BHs), among others, are objects that can illustrate the high energy behaviour. BHs are predicted by GR, and there is strong observational evidence for their existence [5], therefore RSII needs to reproduce BHs in order to be a viable physical theory. In the absence of matter, there exist a particular class of solutions in RSII constructed from vacuum solutions of four dimensional (4D) GR using the construction method in [6]. These solutions are called black strings when the corresponding 4D solutions are black holes. However, the black strings are unstable due to the Gregrory-Laflamme instability [7, 8]. In fact, Lehner-Pretorius [69] showed that certain type of black string (which is different from the black string in braneworld) evolves into a series of 3D spherical BHs connected by black strings and the black strings continue to evolve in the same pattern, which yields a self-similar configuration. Within finite asymptotic time, a naked singularity is created, which violates the cosmic censorship hypothesis. Besides, the Kretschmann scalar $R_{\mu\nu\alpha\beta}R^{\mu\nu\alpha\beta}$ (where $R_{\mu\nu\alpha\beta}$ is the Riemann tensor) diverges at the AdS horizon [6], which means the black string (in RSII) solution itself has pathology. Therefore, black strings can not be formed via natural processes such as gravitational collapse.

What is the bulk counterpart of the 4D black hole? Noticing that the instability is most severe at the AdS horizon where black strings might "pinch off", Chamblin-Hawking-Reall [6] proposed that BH with finite extension into the bulk was a stable state and was actually the end state of gravitational collapse. The investigation on this proposal started from studying one aspect of this proposal: the existence of such black objects. Tanaka [9] and Emparan et al. [10] conjectured, via holography, that static black holes with radii much greater than the AdS length, can not exist in RSII. Fitzpatrick et al. [11] and Gregory et al. [12] then argued that such a conjecture may not be justified since the arguments in [9, 10] did not take into account the strongly coupled nature of the holographic theory.

Several other groups investigated this proposal using numerical techniques and tried to find specific solutions describing the BHs. As a first step, Shiromizu-Shibata [13] studied time symmetric initial datasets for the Einstein's equations, demanding that the data contained apparent horizons, so that subsequent evolution would presumably settle down to black holes. Kudoh-Tanaka-Nakamura [14] carried out calculations on static BHs in the braneworld, finding solutions with finite extension into the bulk, but only for radii that were small compared to the AdS radius. Subsequent work by Kudoh [15] showed that corresponding solutions could also be found in the 6D braneworld, and that the horizon sizes could be larger than the AdS radius in that case. Further numerical calculations in 5D by Tanahashi-Tanaka [16] found time symmetric initial data with large apparent horizons. In 2008, Yoshino [17] repeated the calculation done in [14], using a more accurate numerical scheme, and argued that due to certain systematic errors exhibited by the solutions—and hence presumably affecting the previous computations—there was no evidence that BHs with finite extension into the bulk could exist at all (i.e. not even small ones). In 2011, Kleihaus et al [19] also claimed the same.

Finally, the debate was settled by the remarkable work carried out by Figueras-Wiseman [20] in 2011. Via perturbing AdS/CFT solution (which itself was also numerically constructed [21]), they obtained static black holes with a range of sizes, including ones large compared to the AdS scale. Recently (2014), Figueras presented an improved calculation [22] which obtained black holes with much wider range, thus presumably black holes can be of any size. In 2012, large BHs were also obtained by Abdolrahimi et al [24] using a different method, and there was preliminary evidence that their solution was the same as the Figueras-Wiseman solution.

Without proving the stability and uniqueness of the solution however, it is not clear whether the Figueras-Wiseman BHs can be formed via natural processes such as gravitational collapse. *More generally*, despite the success on the statics side, the full dynamics of RSII—a more important and interesting aspect—remains poorly understood, especially in the high energy regime. The answers to fundamental questions, such as the end state of gravitational collapse, or the interaction between the brane and bulk, remain unknown or vague.

The main goal of this thesis work is to study the full dynamics in RSII via numerical calculation. Specifically, we will study the dynamical process of black hole formation as a result of gravitational collapse of massless scalar field (which lives only on the brane). To our knowledge, there is no previous calculations of the *full dynamics* of (any) braneworld scenario. The study is crucial to understand the dynamics of braneworld models. Technically, as it turns out, the study deepens our understanding and also extends the formalism of numerical relativity.

The calculation in the braneworld is significantly more difficult than that in GR. The confinement of matter to the brane and the interaction between the brane and the bulk due to brane content including matter and brane tension, are the roots of the difficulties [25]. Gravitational collapse inevitably produces energies high enough to make the interaction between the brane and the bulk significant. This rules out any attempts to solve the problem using brane content only. Due to this interaction, the brane equations are not closed, thus the dynamics of the whole spacetime (including both the brane and the bulk) needs to be studied.

Here we successfully performed a numerical study, and we directly show that the results of the gravitational collapse of sufficiently strong initial configurations are BHs with finite extension into the bulk. The extension changes from sphere to pancake (or cigar, seen from a different perspective) as the size increases. There are indications that BHs with the same size produced from different initial data, have the same shape, which means the details of initial data are lost in the resulting BHs. Therefore a no-hair theorem (uniqueness of BH) is suggested to hold also for BHs in the RSII spacetimes that are studied in this thesis. In particular, the BHs we obtained as the results of dynamical systems, are consistent with the ones previously obtained from a static problem by Figueras-Wiseman [20]. Please refer to Fig. 1.1 for a preview.

There are gaps between the existing knowledge in numerical relativity and the knowledge needed for the numerical calculation of braneworlds. We developed some of these to make our numerical calculation possible. With these developments, we solved the regularity problem in numerical relativity (the irregularity associated with simulations in axisymmetric or spherically symmetric spacetime which appeared in previous numerical relativity calculations) by studying it from a different perspective than previous works did. Our work deepens the understanding of the problem, and can potentially completely solve the problem. We generalized evolution formalisms (generalized harmonic and BSSN) of GR to make them suitable for simulations in non-Cartesian coordinate under non-flat background. A conformal function of the metric is included in the formalism to simulate the braneworld. The constraint violations found in the braneworld calculations, are cured by imposing the constraints at the brane.

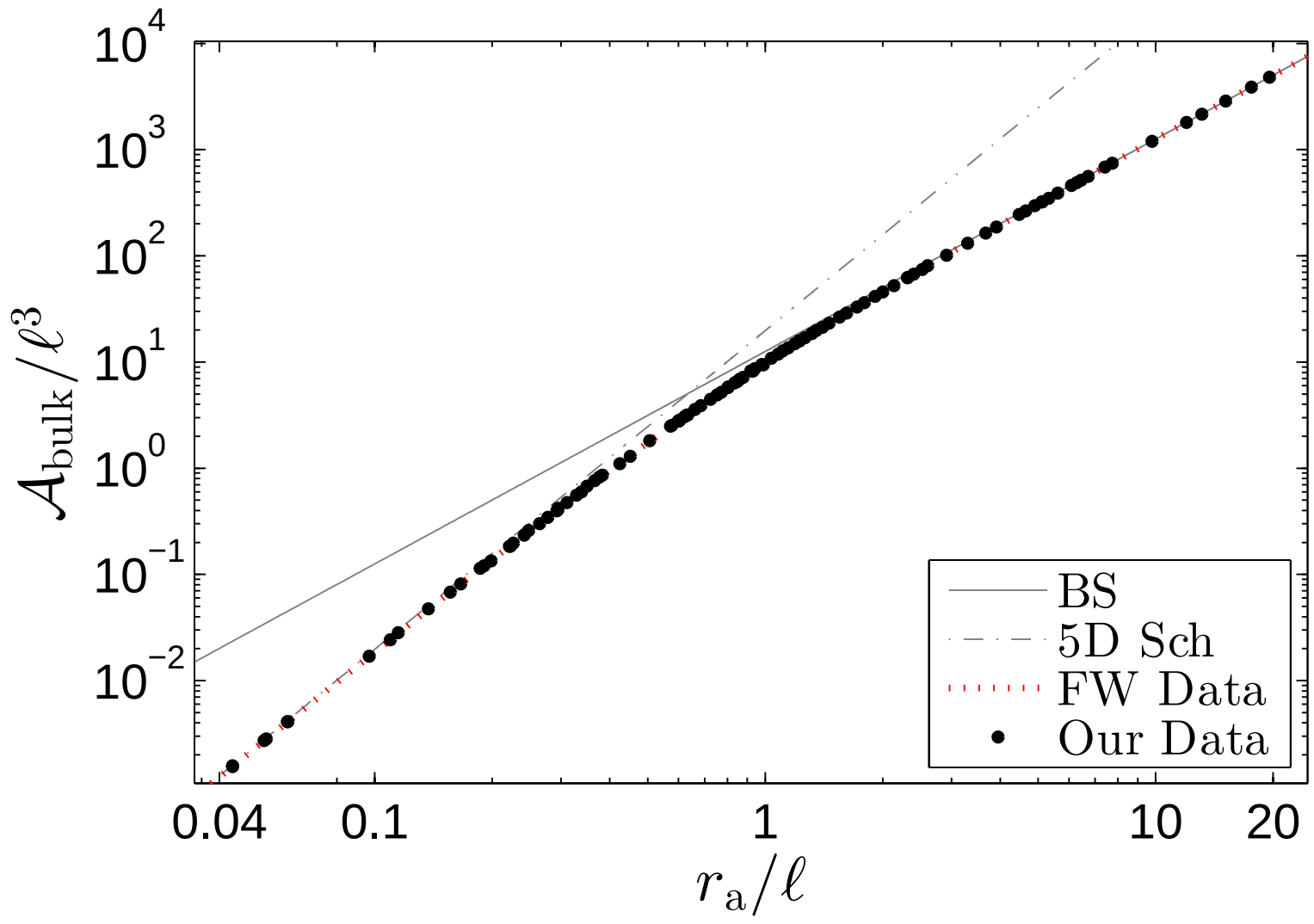

Figure 1.1: The area of bulk apparent horizon as a function of the areal radius of the horizon on the brane. The coupling strength of the brane tension is proportional to $1/\ell$ (where $\ell$ is the AdS length), which is invisible to the BHs whose sizes (the areal radii on the brane) are much smaller than $\ell$. Therefore small BHs behave as 5D Schwarzschild. When the size is much larger than $\ell$, the asymptotic relation is that of the corresponding black string. Our data obtained from the evolutionary system, is consistent with the data obtained by Figueras-Wiseman [20] from a static system.

### 1.1.1 Overview of Our Approach

The system is formulated as an initial value problem (IVP), with Einstein's equations in the bulk as the governing equation. The brane content, including brane tension and the matter, is imposed via Israel's junction conditions [26], which serve as parts of the boundary conditions of the IVP. The matter on the brane evolves as ordinary 4D matter (which only "feels" the 4D metric on the brane).

Given that calculations of this sort (or even published attempts) do not exist, and due to the prohibitive computational cost of performing the calculations in the fully 5D context, we restrict ourselves to spherical symmetry on the brane, which makes the system axisymmetric in the bulk (i.e. functions depend on two spatial dimensions plus time).

Even with this significant simplification, we are faced with a challenging task, not least since there are several features of the problem that have not been addressed, or fully resolved, in previous numerical relativity studies. Apriori, there are the following challenges

(1) The numerical treatment of *delta-function* (distributional) matter

(2) The numerical treatment of a brane with tension

(3) Regularity issues induced by axisymmetry

(4) Appropriate evolution schemes for use in non-Cartesian coordinate systems, and under non-flat backgrounds

(5) Properly incorporating the AdS boundary conditions

(6) Higher dimensional numerical relativity

Among the features mentioned above, some are more challenging and cause more severe problems than others, and could be made into good projects by themselves. In fact, Oliver Rinne's PhD thesis (at University of Cambridge, 2005) [71] was on regularity issues in axisymmetry. He studied a specific formalism (Z(2+1)+1 system) for axisymmetric system. On the other hand, what we are going to present in this thesis is a "direction change" of the study on this topic. We will conjecture that the regularity problems are caused by the the fundamental variables used in numerical simulations, rather than coordinates. We further developed a few general methods to produce formalisms that yield regular results. Hans Bantilan's PhD thesis (at Princeton University, 2013) [68] was on the AdS spacetime in 5D, where lightlike signals can propagate to the spatial infinity and come back to its departure location within a finite proper time as measured by a timelike observer. In this thesis, by employing the formalisms we developed to simulate braneworld spacetime (in Chap. 3), there are no issues associated with the higher dimensions. The spacetime of the braneworld is asymptotically a part of the Poincaré patch of the AdS spacetime whose causal structure is similar to that of Minkowski spacetime. Therefore Cauchy surfaces exist in the spacetime (so that Cauchy problem is well-defined), and the above mentioned problem associated with lightlike signals will not occour. There will be complications related to the Killing horizons of the Poincare patch, however, such as a timelike curve of finite length can reach the Killing horizons. Please refer to Sec. 1.4 for more details about AdS spacetime, the Poincaré patch and the background of the RSII braneworld.

During the investigation, we encountered peculiar properties and challenges which were not anticipated. In solving for time symmetric initial data, in most situations the problem can reduce to solving the Hamiltonian constraint for one unknown variable. In RSII, however, there have to be at least two unknown variables in the initial data to make Israel's boundary conditions consistent; normally the shift and lapse functions can be arbitrarily specified utilizing the gauge freedom in choosing coordinates, while they have to satisfy certain conditions in the braneworld due to Israel's boundary conditions. In most situations, the constraint violation problem in free evolution schemes, such as the generalized harmonic formalism and BSSN formalism, can be considered solved by constraint damping method [58, 61–63]. However, the same method does not control the severe violation at the brane of the braneworld. We found that we could solve the violation by properly imposing the constraints at the brane boundary.

Therefore we add the following two additinal problems to the feature list

(7) Peculiar properties in initial data problem

(8) Constraint violation in braneworld

In the following chapters, we will show how we solve these challenges and achieve the simulation of the dynamical process of black hole formation as the result of gravitational collapse on the brane. The thesis is divided into three parts. In the remaining sections of Chap. 1, we will introduce the braneworld in more detail plus some basic aspects of numerical computation. The second part is devoted to develop the machinery to perform simulations and to extract physical informations in the braneworld. This part consists of Chap. 2 and Chap. 3. Chap. 2 develops the conceptual aspects, such as the smoothness of apparent horizon across the brane, the relation between brane horizons and bulk horizons, the coordinate gauge at the brane, and energy aspects of braneworld. In Chap. 3, our approach to the regularity problem will be presented. We also show how evolution schemes of GR, such as the generalized harmonic formalism, can be developed and generalized in non-Cartesian coordinate under non-flat asymptotic spacetime background, to be made suitable for simulations in the braneworld. The final part is the numerical simulation. Initial data is obtained in Chap. 4, which also provides numerical results to support the discussion on energy aspects in Chap. 2. In Chap. 5, we show how the constraint violations are solved, and also show the simulation per se, and the physical results. Other than Chap. 1, all the works in this thesis, except for the ones properly cited, are original.

## 1.2 Spacetime Foliation and the Decomposition of Einstein's Equations

In this section, we will introduce the notations and conventions in the context of the $(d-1)+1$ decomposition formalism of general relativity (GR), and Israel's junction conditions, which are needed to introduce the braneworld. Here $d$ is the dimension of the whole spacetime. Metric signature is $(-,+,\ldots,+)$, which is the same as that in [29, 32, 34, 36].

The $(d-1)+1$ decomposition of Einstein's equations, is a generalization of the $3+1$ formalism of GR [34, 36]. We start with a $(d-1)$ dimensional hypersurface $\Sigma$ embedded in $d$ dimensional spacetime $M$. $\Sigma$ can be thought as the brane in braneworld, or a "time"= constant hypersurface in an evolutionary problem. I.e. the formalism introduced in this section applies to both cases. Especially, the parameter $t$ in this section, is not reserved for "time".

$\Sigma$ divides $M$ into two parts $M^{\pm}$. Please refer to Fig. 1.2.

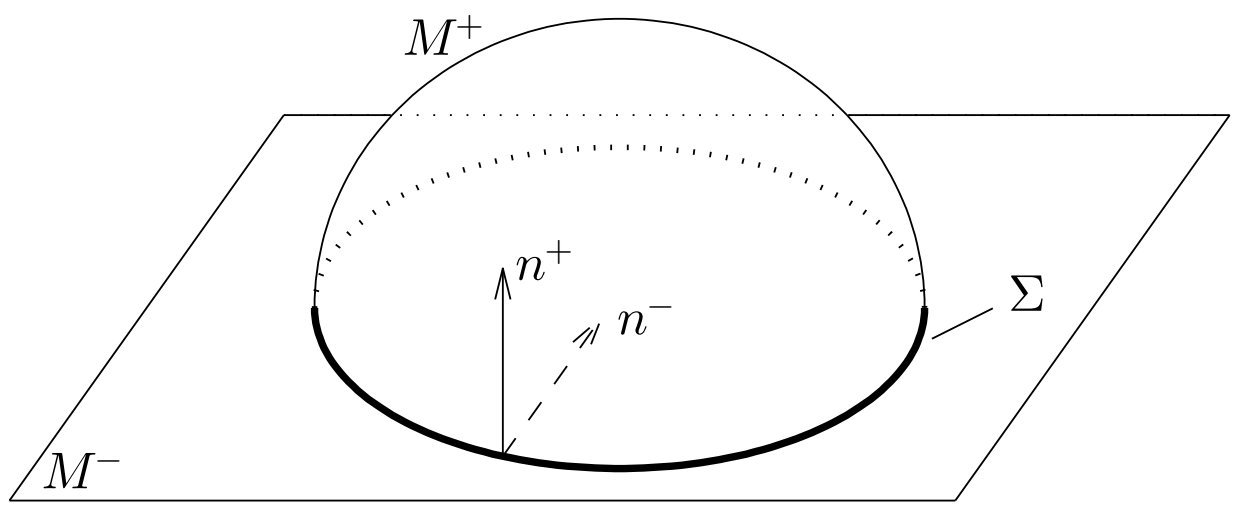

Figure 1.2: The embedding of a hypersurface in a higher dimensional spacetime. The whole spacetime $M$ is composed of $M^+$ (the half sphere) and $M^-$ (the plane with a big hole in the middle to fit $M^+$). The intersection of $M^+$ and $M^-$ is hypersurface $\Sigma$ (the circle). $n^{\pm}$ are unit normal vectors of $\Sigma$, as seen in $M^{\pm}$. The convention is that $n^+$ *points into* $M^+$, and $n^-$ *points out of* $M^-$. Note: (i) the extrinsic curvature of $\Sigma$ in $M^+$ can be different from that in $M^-$. For example, in this figure $\Sigma$ is a curved line in $M^-$, but a straight line in $M^+$; (ii) In general $n^+ \neq n^-$ (but a few authors incorrectly assumed $n^+ = n^-$ when deriving junction conditions). For example, in the figure $n^+$ is not defined in $M^-$ and $n^-$ is not defined in $M^+$. Another example is $n^- = -n^+$ *in Randall-Sundrum braneworlds.*

Generally $\Sigma$ can be a hypersurface within a hypersurface *family* that *locally* foliates the spacetime. Let us use parameter $t$ to characterize the foliation, where $t=$ constant is a specific hypersurface $(\Sigma_t)$ within the family. $t\,(M^+) > t\,(M^-)$ is the convention we adopt. Let $n^{\pm}$ be unit normal vectors of $\Sigma$, as seen in $M^{\pm}$. The convention is that the direction of $n^{\pm}$ is the same as the increasing direction of $t$. Therefore $n^+$ *points into* $M^+$, and $n^-$ *points out of* $M^-$. The super-index

$\pm$ is omitted whenever we express anything that is valid for both signs. $n_\alpha$ is normalized as

$$n^\alpha n_\alpha = \epsilon, \tag{1.1}$$

where $\epsilon$ is $+1$ when $\Sigma$ is the brane in the braneworld and $-1$ when $\Sigma$ is a "time"=constant hypersurface. In terms of $t$, the normal vector $n_\mu$ is

$$n_\mu = \epsilon\ \alpha \nabla_\mu t, \tag{1.2}$$

where $\alpha \equiv \left(\sqrt{\epsilon \nabla_\mu t \nabla^\mu t}\right)^{-1}$ is the lapse function. $\nabla$ is the covariant derivative associated with the metric $g_{\mu\nu}$. The greek index $\mu, \nu = 0, \dots (d-1)$, where 0 is the time coordinate. Define

$$m_\mu \equiv \alpha n_\mu, \tag{1.3}$$

which satisfies $m^\mu \nabla_\mu t = 1$.

The decomposition of Einstein's equations is expressed in terms of the following quantities.

The *induced metric* on $\Sigma$

$$\gamma_{\mu\nu} \equiv g_{\mu\nu} - \epsilon\ n_\mu n_\nu. \tag{1.4}$$

The *extrinsic curvature* is defined as

$$K_{\mu\nu} \equiv -\gamma_\mu{}^\alpha \gamma_\nu{}^\beta \nabla_\alpha n_\beta = -\gamma_\mu{}^\alpha \nabla_\alpha n_\nu. \tag{1.5}$$

Note the geometrical meaning: the extrinsic curvature is defined as the "change rate" of the normal vector $n_\alpha$ along the hypersurface, as seen in the bulk ($M^\pm$). Since the intrinsic observer on $\Sigma$ can not even "feel" $n_\alpha$, the extrinsic curvature characterizes the extrinsic nature of the embedding. [1]

The definition of Riemann tensor is the same as that in [29, 32, 34, 36], which is defined via an arbitrary vector field $v^\mu$

$$\nabla_\alpha \nabla_\beta v^\gamma - \nabla_\beta \nabla_\alpha v^\gamma = R^\gamma{}_{\mu\alpha\beta}\, v^\mu. \tag{1.6}$$

We also adopt the same sign convention of Christoffel symbols as that in [29, 32, 34, 36]

$$\Gamma^\alpha{}_{\beta\gamma} = \frac{1}{2} g^{\alpha\mu} \left( g_{\mu\beta,\gamma} + g_{\mu\gamma,\beta} - g_{\beta\gamma,\mu} \right), \tag{1.7}$$

[1] The extrinsic curvature can also be obtained as the measure of the deviation of the geodesics in $\Sigma$ and $M$, produced by the same vector lying within $\Sigma$. Please refer to section 2.3.3 and appendix B for the study from this perspective.

where $_{,\mu} \equiv \partial_\mu$.

### 1.2.1 The Decomposition of Einstein's Equations

Repeating the Gauss-Codazzi-Ricci decomposition [29, 32, 34, 36] while keeping $d$ and $\epsilon$ general, the Einstein tensor $G_{\mu\nu} \equiv R_{\mu\nu} - \frac{1}{2}g_{\mu\nu}R$ is decomposed as

$$^{(d)}G_{\mu\nu}n^\mu n^\nu = \frac{1}{2}\left(-\epsilon \cdot {}^{(d-1)}R + K^2 - K_{\mu\nu}K^{\mu\nu}\right); \tag{1.8}$$

$$^{(d)}G_{\mu\nu}n^\nu \gamma^\mu{}_\alpha = D_\alpha K - D_\mu K^\mu{}_\alpha; \tag{1.9}$$

$$\begin{aligned} ^{(d)}G_{\mu\nu}\gamma^\mu{}_\alpha \gamma^\nu{}_\beta = {} & \frac{\epsilon}{\alpha}\left(\mathcal{L}_\mathbf{m} K_{\alpha\beta} - \gamma_{\alpha\beta}\mathcal{L}_\mathbf{m} K\right) + {}^{(d-1)}G_{\alpha\beta} \\ & + \epsilon\left(2K_{\alpha\mu}K^\mu{}_\beta - KK_{\alpha\beta} + \frac{1}{2}\gamma_{\alpha\beta}\left(K^2 + K_{\mu\nu}K^{\mu\nu}\right)\right) \\ & + \frac{1}{\alpha}\left(\gamma_{\alpha\beta}D^\mu D_\mu \alpha - D_\alpha D_\beta \alpha\right), \end{aligned} \tag{1.10}$$

where $D$ is the covariant derivative associated with the hypersurace metric $\gamma_{\mu\nu}$. $\mathcal{L}_\mathbf{m}$ is the Lie derivative with respect to $m^\mu$. The super-indices $(d)$ and $(d-1)$ are here to characterize dimension. For example, $^{(d-1)}G_{\mu\nu}$ and $^{(d-1)}R$ are the intrinsic Einstein tensor and the intrinsic Ricci scalar of $\Sigma$.

Imposing Einstein's equations in the $d$ dimensional spacetime (which relates the geometry with matter via matter's stress tensor $T_{\mu\nu}$)

$$^{(d)}G_{\mu\nu} = k_d T_{\mu\nu}, \tag{1.11}$$

we obtain Hamiltonian constraint, momentum constraint and evolution equation

$$k_d\ \rho = \frac{1}{2}\left(-\epsilon^{(d-1)}R + K^2 - K_{\mu\nu}K^{\mu\nu}\right); \tag{1.12}$$

$$\epsilon\ k_d\ S_\alpha = D_\alpha K - D_\mu K^\mu{}_\alpha; \tag{1.13}$$

$$\begin{aligned} k_d\ S_{\alpha\beta} = {} & \frac{\epsilon}{\alpha}\left(\mathcal{L}_\mathbf{m} K_{\alpha\beta} - \gamma_{\alpha\beta}\mathcal{L}_\mathbf{m} K\right) + {}^{(d-1)}G_{\alpha\beta} \\ & + \epsilon\left(2K_{\alpha\mu}K^\mu{}_\beta - KK_{\alpha\beta} + \frac{1}{2}\gamma_{\alpha\beta}\left(K^2 + K_{\mu\nu}K^{\mu\nu}\right)\right) \\ & + \frac{1}{\alpha}\left(\left(\gamma_{\alpha\beta}D^\mu D_\mu \alpha - D_\alpha D_\beta \alpha\right), \end{aligned} \tag{1.14}$$

where $k_d = 8\pi G_d$ and $G_d$ is Newton's constant in $d$ dimension. We have defined $\rho \equiv T_{\mu\nu}n^\mu n^\nu$, $S_\alpha \equiv \epsilon\ T_{\mu\nu}n^\nu \gamma^\mu{}_\alpha$, $S_{\alpha\beta} \equiv T_{\mu\nu}\gamma^\mu{}_\alpha \gamma^\nu{}_\beta$, which yield the following decompostion: $T_{\mu\nu} = \rho n_\mu n_\nu + n_\mu S_\nu +$ $S_\mu n_\nu + S_{\mu\nu}$. Taking trace gives $T = S + \epsilon\rho$, where $S \equiv S_{\mu\nu}\gamma^{\mu\nu}$.

The evolution equation (1.14) can be equivalently expressed as [34, 36]

$$\mathcal{L}_{\mathbf{m}} K_{\alpha\beta} = \epsilon D_\alpha D_\beta \alpha + \alpha\left[-\epsilon\ {}^{(d-1)}R_{\alpha\beta} + KK_{\alpha\beta} - 2K_{\alpha\mu}K^\mu{}_\beta + \epsilon\ k_d\left(S_{\alpha\beta} - \gamma_{\alpha\beta}\frac{S+\epsilon\rho}{d-2}\right)\right]. \quad (1.15)$$

The definition equation of extrinsic curvature (1.5), can be equivalently expressed as

$$\mathcal{L}_{\mathbf{m}}\gamma_{\alpha\beta} = -2\alpha K_{\alpha\beta}. \quad (1.16)$$

(1.15) or (1.14), together with (1.16) form a complete set of evolution equations [34, 36] using $\gamma_{\mu\nu}$ and $K_{\mu\nu}$ as fundamental variables. This formalism is called ADM-York formalism of GR.

### 1.2.2 Israel's Junction Conditions

In classical electromagnetism, there are situations where electric charges are highly concentrated on two dimensional surfaces such as the interfaces of different materials. The distribution of the electric charges is then singular in the 3D space. The electric fields appear to be discontinuous across the singular layers, and the relation relating the discontinuities of the electric fields with the singular distributions of the electric charges, is called a junction condition.

Similarly, in GR, if the stress tensor is highly concentrated on $\Sigma$, it will cause discontinuities. The discontinuities are described by Israel's junction conditions.

*Israel's first junction condition* [26] states that the intrinsic geometry of the hypersurface is well defined. i.e. the induced metric of $\Sigma$ obtained from $M^+$ agrees with the induced metric obtained from $M^-$

$$\hat{[}\gamma_{\mu\nu}\hat{]} = 0, \quad (1.17)$$

where the notation $\hat{[}a\hat{]} \equiv a^+ - a^-$.

Integrating (1.14) over an infinitesimal layer across $\Sigma$, we get *Israel's second junction condition* [26]

$$k_d\ \mathscr{S}_{\mu\nu} = \epsilon\Big(\hat{[}K_{\mu\nu}\hat{]} - \gamma_{\mu\nu}\hat{[}K\hat{]}\Big) \quad \text{or} \quad \hat{[}K_{\mu\nu}\hat{]} = \epsilon k_d\left(\mathscr{S}_{\mu\nu} - \gamma_{\mu\nu}\frac{\mathscr{S}}{d-2}\right), \quad (1.18)$$

where $\mathscr{S}_{\mu\nu}$ is the singular part of the projected energy-momentum tensor on $\Sigma$ defined as

$$\mathscr{S}_{\mu\nu} \equiv \int_{0^-}^{0^+} S_{\mu\nu}\mathrm{d}l, \quad (1.19)$$

where $\mathrm{d}l \equiv \mathrm{d}t\,(\partial_t)^\mu\, n_\mu = \alpha \mathrm{d}t$ is the proper length/time across $\Sigma$, and $l$ is (arbitrarily) set to be

$l = 0$ at $\Sigma$.

Integrating (1.12) and (1.13) gives

$$\mathscr{S}_\mu \equiv \int_{0^-}^{0^+} S_\mu \mathrm{d}l = 0; \qquad \varrho \equiv \int_{0^-}^{0^+} \rho \mathrm{d}l = 0, \tag{1.20}$$

which is because the right hand sides (RHSs) of (1.12) and (1.13), as well as the non-$\mathcal{L}_\mathbf{m}$ terms on the RHS of (1.14), are all well-defined and finite on both sides of the hypersurface.

Eq. (1.13) and (1.18) give

$$D_\mu \mathscr{S}^\mu{}_\nu = -\hat{[} S_\nu \hat{]}, \tag{1.21}$$

which is the singular matter's conservation law on $\Sigma$.

Because of the identity $\hat{[} a^2 \hat{]} = \hat{[} a \hat{]} \hat{\{} a \hat{\}}$, where $\hat{\{} a \hat{\}} \equiv a^+ + a^-$, eq. (1.12) gives $k_d \hat{[} \rho \hat{]} = \frac{1}{2} \left( \hat{[} K \hat{]} \hat{\{} K \hat{\}} - \hat{[} K_{\mu\nu} \hat{]} \hat{\{} K^{\mu\nu} \hat{\}} \right)$. Combining with (1.18), we have

$$\hat{[} \rho \hat{]} = \frac{\epsilon}{2} \left( \frac{\mathscr{S}}{2-d} \hat{\{} K \hat{\}} - \left( \mathscr{S}_{\mu\nu} - \gamma_{\mu\nu} \frac{\mathscr{S}}{d-2} \right) \hat{\{} K^{\mu\nu} \hat{\}} \right) = -\frac{\epsilon}{2} \mathscr{S}_{\mu\nu} \hat{\{} K^{\mu\nu} \hat{\}}. \tag{1.22}$$

Eq. (1.21) and (1.22) are constraint/conservation of the singular matter.

### 1.2.3 Coordinate Description

A coordinate system is needed to perform numerical relativity. A coordinate system using $t$ as a coordinate, can be constructed as the following. On each $\Sigma_t$, a coordinate system $x^i$ is assigned and $x^i$ is set to be differentiable across $\Sigma_t$. Here the Latin index $i$ runs from 1 to $(d-1)$. $(t, x^i)$ is then a coordinate system.

Since $(\partial_t)^\mu \nabla_\mu t = 1 = m^\mu \nabla_\mu t$, the shift function $\beta^\mu \equiv (\partial_t)^\mu - m^\mu$ is perpendicular to $\nabla_\mu t$, therefore in the tangent space of $\Sigma_t$. The evolution equations can be written in coordinate system by using $\mathcal{L}_{m^\mu} = \mathcal{L}_{(\partial_t)^\mu} - \mathcal{L}_{\beta^\mu}$, where $\mathcal{L}_{(\partial_t)^\mu}$ is simply $\partial_t$ in the coordinate system in which $t$ serves as one of the coordinate [31]. In general $\boldsymbol{\beta}^\mu = \beta^i (\partial_i)^\mu$ and $\boldsymbol{\gamma}_{\mu\nu} = \gamma_{ij} \left(\mathrm{d}x^i\right)_\mu \left(\mathrm{d}x^j\right)_\nu$. They can be further reduced to $\beta^i$ and $\gamma_{ij}$ in this coordinate system by using $(\partial_i)^\mu = (0, 0, ..., 1, ..., 0)$ and $(\mathrm{d}x^i)_\mu = (0, 0, ..., 1, ..., 0)$, where only the $i$-th position is 1. An infinitesimal vector pointing from $(t, x^i)$ to $(t + \mathrm{d}t, x^i + \mathrm{d}x^i)$ is then $\mathrm{d}t\, (\partial_t)^\mu + \mathrm{d}x^i\, (\partial_i)^\mu$, and its length is derived as

$$\mathrm{d}s^2 = \left(\epsilon\ \alpha^2 + \beta_i \beta^i\right) \mathrm{d}t^2 + 2\beta_i \mathrm{d}t \mathrm{d}x^i + \gamma_{ij} \mathrm{d}x^i \mathrm{d}x^j, \tag{1.23}$$

where $\beta_i \equiv \gamma_{ij} \beta^j$.

## 1.3 Randall-Sundrum Braneworld II (RSII)

In this section we will introduce the single brane model by Randall-Sundrum (RSII). We will first introduce the results regarding the general ideas of braneworld (there is one extra dimension, and matter is confined on the brane), then introduce the setup that is unique in RSII.

From now on, both the brane and "time"=constant hypersurface will be discussed. Therefore we use different notations to distinguish between them. We will use $\mathcal{K}_{\mu\nu}$, $\mathbb{n}_\nu$, $h_{\mu\nu}$, $\mathcal{D}_\mu$, ${}^{(d-1)}\mathcal{G}_{\mu\nu}$, ${}^{(d-1)}\mathcal{R}_{\mu\nu}$ to denote the extrinsic curvature, unit normal vector, induced metric, covariant derivative, Einstein tensor, Ricci tensor of the brane (therefore $\mathbb{n}_\mu \mathbb{n}^\mu = 1$), and use $K_{\mu\nu}$, $n_\mu$, $\gamma_{\mu\nu}$, $D_\mu$, ${}^{(d-1)}G_{\mu\nu}$, ${}^{(d-1)}R_{\mu\nu}$ to denote the corresponding quantities of the "time"=const hypersurfaces (therefore $n_\mu n^\mu = -1$). Especially, the parameter $t$ is reserved for "time"=const hypersurfaces.

### 1.3.1 Formalism for General Braneworld

A class of braneworlds are defined as a five dimensional spacetime where the matter is trapped on a (3+1) dimensional brane $\Sigma$, but gravity can propagate freely into the bulk. Since the brane $\Sigma$ is our observed universe, we would expect that the induced equations on the brane become Einstein's equations at a certain limit. Using Gauss' equations and the $d$-dimensional Einstein's equations, one gets [4]

$$\begin{aligned} {}^{(d-1)}\mathcal{G}_{\alpha\beta} = &\frac{d-3}{d-2} k_d \left\{ T_{\mu\nu} h^\mu{}_\alpha h^\nu{}_\beta + h_{\alpha\beta} \left( T_{\mu\nu} \mathbb{n}^\mu \mathbb{n}^\nu + \frac{T}{1-d} \right) \right\} \\ &+ \left( \mathcal{K}\mathcal{K}_{\alpha\beta} - \mathcal{K}_{\alpha\gamma}\mathcal{K}^h{}_\beta \right) - \frac{1}{2} h_{\alpha\beta} \left( \mathcal{K}^2 - \mathcal{K}^{\mu\nu}\mathcal{K}_{\mu\nu} \right) - E_{\alpha\beta}, \end{aligned} \tag{1.24}$$

where $T_{\alpha\beta}$ is the stress tensor in $d$ dimensional spacetime, and $T$ is the trace. $E_{\mu\nu}$ is defined as

$$E_{\mu\nu} \equiv {}^{(d)}C_{\alpha\beta\gamma\delta}\, \mathbb{n}^\alpha \mathbb{n}^\gamma h^\beta{}_\mu h^\delta{}_\nu. \tag{1.25}$$

${}^{(d)}C_{\mu\nu\gamma\delta}$ is the Weyl tensor defined by the following relation [4]

$${}^{(d)}R_{\mu\nu\gamma\delta} = {}^{(d)}C_{\mu\nu\gamma\delta} + \frac{2}{d-2} \left( g_{\mu[\gamma}\, {}^{(d)}R_{\delta]\nu} - g_{\nu[\gamma}\, {}^{(d)}R_{\delta]\mu} \right) - \frac{2}{(d-1)(d-2)} {}^{(d)}R\; g_{\mu[\gamma} g_{\delta]\nu}, \tag{1.26}$$

where we used the notation $A_{[\alpha\beta]} \equiv \frac{1}{2}(A_{\alpha\beta} - A_{\beta\alpha})$.

From (1.24) one sees that the matter content $T_{\mu\nu}$ needs to be specified. The extrinsic curvature and the projection of the Weyl tensor need to be addressed.

### 1.3.2 Randall-Sundrum Braneworld II

In RSII, the size of the extra dimension is infinite. The bulk is empty except for a negative cosmological constant. Therefore the stress tensor is

$$T_{\alpha\beta} = -\Lambda g_{\alpha\beta} + \mathscr{S}_{\alpha\beta}\delta(l), \tag{1.27}$$

where $\Lambda$ is the cosmological constant in the bulk, $l$ is the proper length discussed in section 1.2.2. The brane is located at $l = 0$. In general, $l$ is only defined in a neighbourhood of the brane. The brane content is

$$\mathscr{S}_{\alpha\beta} = -\lambda h_{\alpha\beta} + \tau_{\alpha\beta}, \tag{1.28}$$

where $\tau_{\alpha\beta}$ is the stress energy tensor of matter on brane, and $\lambda$ is the tension of the brane, which is required for the consistency of the theory (see below).

There is a $Z_2$ symmetry with respect to the brane [3] in Randall-Sundrum braneworlds, so that $\mathtt{n}^- = -\mathtt{n}^+$. $Z_2$ symmetry is mirror symmetry followed by an identification operation: a spacetime $M$ with $Z_2$ symmetry can be obtained by taking a piece of spacetime with a boundary, then generating its image by parity transformation with respect to the boundary, and then gluing these two pieces together by identifying the piece with its image (i.e. identifying point $p$ with its image as the same point, for all $p \in M$). In another word, Randall-Sundrum braneworlds are "one-sided" spacetimes. This point will be made clearer in Sec. 1.3.4. As a consequence of $\mathtt{n}^- = -\mathtt{n}^+$, one can obtain $\mathcal{K}^+_{\alpha\beta} = -\mathcal{K}^-_{\alpha\beta}$. Israel's second junction condition (1.18) is then reduced to

$$\mathcal{K}^+_{\alpha\beta} = -\mathcal{K}^-_{\alpha\beta} = \frac{1}{2}k_d\left(\mathscr{S}_{\alpha\beta} - h_{\alpha\beta}\frac{\mathscr{S}}{d-2}\right) = \frac{1}{2}k_d\left(\lambda\frac{h_{\alpha\beta}}{d-2} + \tau_{\alpha\beta} - h_{\alpha\beta}\frac{\tau}{d-2}\right). \tag{1.29}$$

This equation relates $\mathcal{K}_{\alpha\beta}$ with the matter distribution on the brane, which can eliminate the extrinsic curvatures in (1.24). The $E_{\alpha\beta}$ term in (1.24) is related to the geometry of the bulk and can not be eliminated easily. However, as analysed in [4], this term is only important at high energy regime.

Now, (1.24) can be evaluated on either side of the brane, which may be realized by performing the evaluation at $l \neq 0$ and then taking $l \to 0$, because it does not make sense to do calculation exactly on the brane due to the delta-functions. One bonus relation obtained from this procedure is $\hat{[}E_{\alpha\beta}\hat{]} = 0$, which is a consequence of the $Z_2$ symmetry. The Einstein tensor on the brane is therefore [4]

$$^{(d-1)}\mathcal{G}_{\alpha\beta} = -\Lambda_{d-1}h_{\alpha\beta} + k_{d-1}\tau_{\alpha\beta} + k_d^2\pi_{\alpha\beta} - E^{\pm}_{\alpha\beta}, \tag{1.30}$$

where

$$
\begin{aligned}
\Lambda_{d-1} =&\ (d-3)k_d\left(\frac{1}{8(d-2)}\lambda^2 k_d + \frac{1}{d-1}\Lambda\right), && (1.31)\\
k_{d-1} =&\ k_d^2\lambda\frac{d-3}{4(d-2)}, && (1.32)\\
\pi_{\alpha\beta} =&\ \frac{2\tau\tau_{\alpha\beta} - h_{\alpha\beta}\tau^2}{8(d-2)} + \frac{1}{8}\left(h_{\alpha\beta}\tau_{\mu\nu}\tau^{\mu\nu} - 2\tau_{\alpha\gamma}\tau_\beta{}^\gamma\right). && (1.33)
\end{aligned}
$$

This recovers Einstein's equations on brane at low energy regime since $E_{\mu\nu}$ is also negligible in the low energy regime [4]. However, the behaviour could be quite different from GR at high energy scheme. Furthermore, due to the interaction between the brane and the bulk, the equations on the brane do not form a close system. Therefore the attempts to solve the dynamics by evolving only the brane content, is ruled out.

### 1.3.3 Parameter Setting

In order to let $k_{d-1}$ have the correct sign, (1.32) requires that [4]

$$\lambda > 0. \tag{1.34}$$

According to (1.31), $\Lambda_{d-1}$, the equivalent cosmological constant on the brane, can achieve any value by tuning the value of $\lambda$. In the case $\Lambda_{d-1}$ is taken to be zero, (1.31) yields

$$\lambda = \frac{2(d-2)}{\ell\sqrt{k_d}}, \tag{1.35}$$

where $\ell$ is the AdS radius in the bulk, which relates to $\Lambda$ as

$$\ell = \sqrt{-\frac{(d-1)(d-2)}{2\Lambda}}. \tag{1.36}$$

Eq. (1.32) and eq. (1.35) imply

$$G_{d-1} = \frac{\sqrt{(d-3)^2 \cdot 2\pi G_d}}{\ell} G_d, \tag{1.37}$$

where $G_n \equiv k_n/8\pi$ is Newton's gravitational constant in $n$ dimension. This relation clearly shows that $\ell$ is related to the relation between the Newton's constant on the brane with that in the bulk. In the theory of RSII, $\ell$ is a freely adjustable parameter, whose value can be determined by experiments. Taking $d = 5$, (1.37) reduces to $G_4 = G_5 \cdot \sqrt{8\pi G_5}/\ell$. Choosing the unit $k_5 = \ell = 1$, this equation implies $k_4 = 1$.

### 1.3.4 Vacuum Solution

There are a class of solutions when matter is absent [6, 42–44]

$$\mathrm{d}s^2 = e^{-2|y|/\ell} \left( h_{\mu\nu} \mathrm{d}x^\mu \mathrm{d}x^\nu \right) + \mathrm{d}y^2, \tag{1.38}$$

where $y \in (-\infty, \infty)$ is the extra-dimension and the brane is located at $y = 0$. $x^\mu$ is the coordinate in the 4D section. $h_{\mu\nu}$ does not depend on $y$, and can be any 4D vacuum solution of GR [6]. Define $z \equiv \ell e^{y/\ell}$ when $y \geq 0$, and $z \equiv \ell e^{-y/\ell}$ when $y \leq 0$. In coordinate space, this is a two-to-one mapping which maps $\pm y$ to $z = \ell \exp(|y|/\ell) \in [\ell, \infty)$. However, this "two-to-one" feature is only superficial, because the $(x^\mu, y)$ and $(x^\mu, -y)$ are actually the same physical points, according to the $Z_2$ symmetry. Under coordinates $(x^\mu, z)$, (1.38) is changed into

$$\mathrm{d}s^2 = \frac{\ell^2}{z^2} \left[ \left( h_{\mu\nu} \mathrm{d}x^\mu \mathrm{d}x^\nu \right) + \mathrm{d}z^2 \right], \tag{1.39}$$

where $z \geq \ell$ covers the whole physical spacetime, and the brane is located at $z = \ell$.

The simplest case is to let $h_{\mu\nu}$ be 4D Minkowski spacetime. The corresponding 5D space is a part of the Poincaré patch of AdS (Anti-de Sitter) space, and can be regarded as the counter part of Minkowski spacetime in the braneworld (see Sec. 1.4). $h_{\mu\nu}$ can also take black hole solutions, and the corresponding solutions in the braneworld are called black strings.

### 1.3.5 Matter in RSII

Because of the specific form of the RS braneworld II, equation (1.21) and (1.22) now reduce to

$$\mathcal{D}_\mu \mathscr{S}^\mu{}_\nu = 0; \tag{1.40}$$

$$0 = 0, \tag{1.41}$$

where $\mathcal{D}$ is the covariant derivatives associated with $h_{\mu\nu}$. Eq. (1.40) is important since it is the conservation law of the matter on brane. Since the tension part $\lambda h_{\mu\nu}$ satisfies the conservation law, then eq. (1.40) requires that the matter part $\tau_{\alpha\beta}$ must satisfy the conservation law. This is consistent with equation of motion of matter on brane, which takes its form in 4D GR since matter is strictly trapped on the brane and can not directly "feel" the extra dimension.

We will study the gravitational collapse of massless scalar field on the brane. For massless scalar field (denoted by $\Phi$), the conservation law is equivalent to its equation of motion

$$\mathcal{D}^{\mu}\mathcal{D}_{\mu}\Phi = 0. \tag{1.42}$$

The matter's energy momentum tensor is

$$\tau_{\mu\nu} = \mathcal{D}_{\mu}\Phi\mathcal{D}_{\nu}\Phi - \frac{1}{2}h_{\mu\nu}(h^{\alpha\beta}\mathcal{D}_{\alpha}\Phi\mathcal{D}_{\beta}\Phi). \tag{1.43}$$

## 1.4 The Global Structures of AdS Spacetime and the Poincaré Patch

As discussed in Sec. 1.3.4, in the context of the braneworld, the counter part of the Minkowski spacetime is

$$\mathrm{d}s^2 = \frac{\ell^2}{z^2}\Big( - \mathrm{d}t^2 + \mathrm{d}r^2 + r^2\left(\mathrm{d}\theta^2 + \sin^2\theta \mathrm{d}\phi^2\right) + \mathrm{d}z^2\Big), \qquad \text{where } z \geq \ell. \tag{1.44}$$

The spacetime with this metric and $z \geq 0$, is a Poincaré patch [75–77] of the AdS spacetime with AdS radius $\ell$. In this section we will discuss the global structure of AdS spacetime and its Poincaré patch. The discussion is necessary for us to examine whether the RSII braneworld can be defined as an initial value problem, and whether event horizon can be defined. These two aspects are going to be discussed in Sec. 2.3.1.

### 1.4.1 AdS Spacetime and the Poincaré Patch

The $d$ dimensional homogeneous isotropic spacetime satisfying Einstein's equations with a positive (or negative) cosmological constant, is called a *de Sitter* (or *Anti-de Sitter*) spacetime, and is denoted as $\mathrm{dS}_d$ (or $\mathrm{AdS}_d$). In this section, we focus on the Anti-de Sitter spacetime. As discussed in [75–77], $\mathrm{AdS}_d$ is a hyperboloid of radius $\ell$

$$-X_0^2 - X_1^2 + \sum_{i=1}^{d-1} Y_i^2 = -\ell^2, \tag{1.45}$$

embedded in the flat spacetime expanded by $(X_0, X_1, Y_1, ..., Y_{d-1})$ whose metric is

$$\mathrm{d}s^2 = -\mathrm{d}X_0^2 - \mathrm{d}X_1^2 + \sum_{i=1}^{d-1} \mathrm{d}Y_i^2. \tag{1.46}$$

One coordinate system of the AdS spacetime is defined as

$$X_0 = \ell \sec\chi \cos T, \tag{1.47}$$

$$X_1 = \ell \sec\chi \sin T, \tag{1.48}$$

$$Y_i = \ell\ \omega_i \tan\chi, \quad \text{for } i = 1, ..., d-1. \tag{1.49}$$

Here $\omega_i$ satisfy

$$\sum_{i=1}^{d-1} \omega_i^2 = 1.$$

i.e. they form a $(d-2)$ dimensional unit sphere in $(d-1)$ dimensional Euclidean space. Under this coordinate system, the AdS spacetime is

$$\mathrm{d}s^2 = \frac{\ell^2}{\cos^2\chi}\left(-\mathrm{d}T^2 + \mathrm{d}\chi^2 + \sin^2\chi \mathrm{d}\Omega_{d-2}^2\right), \tag{1.50}$$

where $\mathrm{d}\Omega_{d-2}^2$ stands for the line element on a $(d-2)$ dimensional unit sphere in $(d-1)$ dimensional Euclidean space. Eq. (1.50) with $\chi \in [0, \pi/2)$ and $T \in (-\infty, \infty)$ defines the whole AdS spacetime, and is therefore a global cover of the AdS spacetime [75–77]. [2]

Alternatively, another coordinate system is

$$X_0 = \frac{\ell}{2z}\left(\sum_{i=1}^{d-2} x_i^2 - t^2 + z^2 + 1\right), \tag{1.51}$$

$$X_1 = \frac{\ell\ t}{z}, \tag{1.52}$$

$$Y_1 = \frac{\ell}{2z}\left(\sum_{i=1}^{d-2} x_i^2 - t^2 + z^2 - 1\right), \tag{1.53}$$

$$Y_{i+1} = \frac{\ell\ x_i}{z}, \quad \text{for } i = 1, ..., d-2. \tag{1.54}$$

The range of the variables are $z \in (0, \infty)$, $t \in (-\infty, \infty)$. $x_1, x_2, ..., x_{d-2}$ form a $(d-2)$ dimensional

[2] In eq. (1.47) and (1.48), $T$ is periodic with the period $2\pi$, therefore there exist closed timelike curves. The global AdS spacetime is to unfold these closed curves by removing the identification of $T$ with $T + 2\pi$.

Euclidean space. The metric of the AdS spacetime is now

$$\mathrm{d}s^2 = \frac{\ell^2}{z^2}\left(-\mathrm{d}t^2 + \mathrm{d}z^2 + \sum_{i=1}^{d-2}\mathrm{d}x_i^2\right). \tag{1.55}$$

Expressing the space expanded by $x_i$'s under spherical coordinates, this metric is

$$\mathrm{d}s^2 = \frac{\ell^2}{z^2}\left(-\mathrm{d}t^2 + \mathrm{d}z^2 + \mathrm{d}r^2 + r^2\mathrm{d}\Omega_{d-3}^2\right), \tag{1.56}$$

where $\mathrm{d}\Omega_{d-3}^2$ stands for the line element on a $(d-3)$ dimensional unit sphere in $(d-2)$ dimensional Euclidean space. This spacetime is called a Poincaré patch of the AdS spacetime, and the coordinates (1.55) or (1.56) are called Poincaré coordinates. [3]

### 1.4.2 The Penrose Diagram of the AdS$_2$ Spacetime

In this section we mainly focus on the causal structure of the spacetimes, which can be conveniently studied via the tool known as the Penrose diagram. For simplicity, let us start with AdS$_2$—the 2D AdS spacetime. The metric of the global cover is

$$\mathrm{d}s^2 = \frac{\ell^2}{\cos^2\chi}\left(-\mathrm{d}T^2 + \mathrm{d}\chi^2\right), \tag{1.57}$$

and the ranges of the two variables are $T \in (-\infty, \infty)$ and $\chi \in (-\pi/2, \pi/2)$. [4] The Penrose diagram is shown in Fig. 1.3(a), where the $T=$ const and $\chi=$ const curves are simply horizontal and vertical lines. Eq. (1.57) shows that the null curves are straight lines with the slope $\mathrm{d}T/\mathrm{d}\chi = \pm 1$ (e.g. the orange lines in the figure).

An interesting feature of the AdS spacetime is that the lightlike signals can propagate to spatial infinities ($\chi = \pm\pi/2$), and then come back to its departure location within a finite proper time (measured at the departure location). Or precisely speaking, there exist closed causal curves with finite proper length, which connect "local" regions with spatial infinities. As an example, let us consider a lightlike signal departing at point $A$ in Fig. 1.3(a) and propagating along the orange line $\overline{AB}$ to point $B$ (spatial infinity). It then propagates back from $B$ to $C$ along the orange line $\overline{BC}$. For an observer sitting at $\chi = 0$, its proper time lapses $\pi$, which is finite. On the other hand,

[3](1.56) remains unchanged by the scaling: $(t, z, r) \to (ct, cz, cr)$ where $c$ is an arbitrary positive constant, and it can be either dimensionful or dimensionless. In particular, when taking $c = 1/\ell$, the $(t, z, r)$ can be regarded as parameters with length dimension measured by unit $\ell$. In this section (Sec. 1.4) these coordinates will be treated as dimensionless parameters since it is easier to relate to the global cover. In all the other parts of the thesis, we will treat them as parameters with length dimension measured by unit $\ell$.

[4]Or equivalently but more closely to relate to (1.50) and higher dimensional AdS spacetime, we can also take $\chi \in [0, \pi/2)$ with $\psi = 0, \pi$, where $\psi$ is to parametrize $\omega_1$ in (1.49) as $\omega_1 = \cos\psi$.

the proper distance from the observer to $\chi = \pi/2$ in a $T = \text{const}$ hypersurface is

$$\int_0^{\pi/2} \frac{1}{\cos\chi} \mathrm{d}\chi = 2 \text{ arctanh} \left[ \tan\left(\frac{\chi}{2}\right)\right]\Big|_0^{\pi/2} = 2 \text{ arctanh}(1) = \infty. \tag{1.58}$$

Actually the lengths of all the spacelike curves connecting a point at $\chi = \pi/2$ with a point at $\chi \neq \pi/2$ are infinite. i.e. $\chi = \pi/2$ is truly a spatial infinity. For higher dimensional AdS spacetimes with $d > 2$, the above analysis is also valid.

The metric of the Poincaré patch under the Poincaré coordinates is now

$$\mathrm{d}s^2 = \frac{\ell^2}{z^2} \left(-\mathrm{d}t^2 + \mathrm{d}z^2\right), \tag{1.59}$$

where $t \in (-\infty, +\infty)$ and $z \in (0, +\infty)$. To get the Penrose diagram of the Poincaré patch, we relate the coordinates $(T, \chi, ...)$ with the coordinates $(t, z, r, ...)$ by the equality of eq. (1.47∼1.49) and eq. (1.51∼1.54). The coordinates $(T, \chi, ...)$ can then be expressed in terms of $(t, z, r, ...)$. i.e. for every $(t, z, r, ...)$, there is a corresponding $(T, \chi, ...)$, which is a point on the Penrose diagram of the global AdS spacetime. The Penrose diagram of the Poincaré patch, is then a subset of the Penrose diagram of the global AdS spacetime. The Penrose diagram for the Poincaré patch is obtained as Fig. 1.3(b,c). In the figures, the blue lines are $t = \text{const}$, and the black lines are $z = \text{const}$. For the Penrose diagram, the boundary is divided into

(1) Point $i^0$ represents $z = \infty$, *the spatial infinity.*

(2) Point $i^+$ represents $t = +\infty$, *the future timelike infinity.*

(3) Point $i^-$ represents $t = -\infty$, *the past timelike infinity.*

(4) The line $j_z^+$ represents the *future Poincaré horizon* (see, e.g. [39]), which is a Killing horizon associated with the Killing vector $\partial_t$. As explained below in Sec. 1.4.3, this boundary can be reached by timelike curves of finite proper length, therefore these are not infinities. However, if the future oriented null curves are parametrized by the Killing parameter $t$, the null curves will end at this boundary when taking $t \to \infty$. For example, the curve defined by $(t, z) = (t, t + 1)$ is a null curve. When $t \to +\infty$, this curve ends at $j_z^+$. i.e. this boundary appears to be infinity as measured by the Killing parameter $t$. In this sense, this boundary is an analogy of the null infinity of Minkowski spacetime [32]. We will put the word "infinity" into double quotes if it is not a true infinity but appears to be infinity as measured by $t$.

(5) Similarly, $j_z^-$ is the *past Poincaré horizon.*

(6) $z = 0$ (which is also $\chi = -\pi/2$), the *conformal boundary* (see, e.g. [40]) of the Poincaré patch, which is a part of the timelike spatial infinity of the global cover.

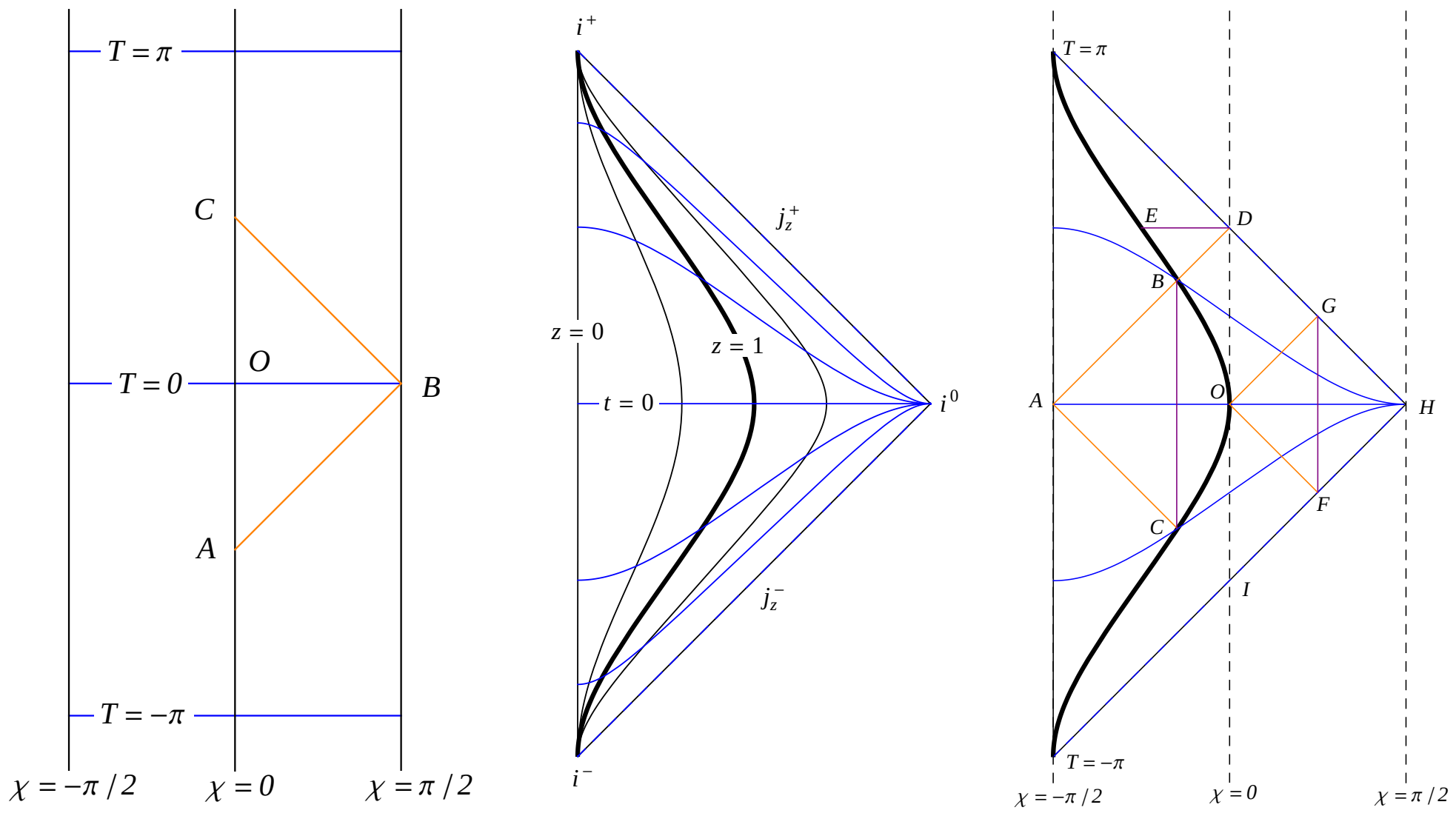

(a) (b) (c)

Figure 1.3: Penrose diagrams of $AdS_2$ and its Poincaré patch. (a) is the Penrose diagram of the whole $AdS_2$. The $T$ = const lines are horizontal, and $\chi$ = const lines are vertical. The range of the two variables are $T \in (-\infty, \infty)$ and $\chi \in (-\pi/2, \pi/2)$. Lightlike curves are locally straight lines with slope $\mathrm{d}T/\mathrm{d}\chi = \pm 1$. (b) is the Penrose diagram of the Poincaré patch, while (c) emphasizes that it is embedded into the global AdS spacetime.

Fig. 1.3(b,c) show that in the Poincaré patch there still exist lightlike signals going to spatial infinity (at the conformal boundary) and coming back whose trajectory can be connected by a timelike curve with finite proper time. e.g. the trajectory $C \to A \to B$ in Fig. 1.3(c) is connected by the purple line $\overline{BC}$ whose proper time is finite.

### 1.4.3 The Infinities of the Poincaré Patch

In the last subsection, we show that at the $i^0, i^\pm$ and $j_z^\pm$, at least one of $t = \infty$ and $z = \infty$ is taken. Now we examine whether these are "true" infinities, by evaluating the proper lengths. The proper length of the $t = 0$ curve is

$$\int_0^\infty \frac{\mathrm{d}z}{z} = \int_0^1 \frac{\mathrm{d}z}{z} + \int_1^\infty \frac{\mathrm{d}z}{z}, \tag{1.60}$$

where both terms on the right hand side are $+\infty$. Actually the lengths of all the spacelike curves connecting a point at $z = \infty$ with a point at $z \neq \infty$ are infinite. Thus $i^0$ is a true spatial infinity. Similarly, $i^{\pm}$ are true timelike infinities.

For the Poincaré horizons which appear to be null "infinities" measured in terms of the Killing parameter $t$, it is easier to perform the calculation under $(T, \chi)$ coordinates. One can see that the proper time of $\overline{FG}$—a curve connecting the past Poincaré horizon with the future Poincaré horizon as shown in Fig. 1.3(c)—is finite, which also means the lightlike signal propagating along $F \to O \to G$ enters from the past Poincaré horizon and escapes via the future Poincaré horizon, within a finite proper time measured by an observer (whose world-line is $\overline{FG}$). This means the information or entities in the space can "vanish" within a finite time. Also, the proper length of spacelike curve $\overline{ED}$ is finite. Therefore $j_z^{\pm}$ are *not* true infinities.

On the other hand, however, the Poincaré patch (1.55) has a timelike Killing vector $\partial_t$, and $j_z^{\pm}$ (the Poincaré horizons [39]) are the Killing horizons associated with this Killing vector. In the literature, $z = \infty$ is often referred as AdS horizon [6]. When approaching $j_z^{\pm}$, the time parameter associated with the Killing vector (which is $t$) approaches infinity. An observer whose world-line is the integral curve of the Killing vector (i.e. his spatial coordinates of the Poincaré coordinates remain constants), is called a Killing observer. For a Killing observer, his proper time takes $(-\infty, \infty)$. $j_z^+$ is the future boundary of the spacetime region that can possibly be observed by any Killing observer, and $j_z^-$ is the past boundary of the spacetime region that can possibly be affected by any Killing observer.

### 1.4.4 The Poincaré Patch of $AdS_3$

$AdS_2$ catches the most important features of the more general $AdS_d$. Also, $AdS_d$ has a translational symmetry in the space expanded by $x_i$'s. Therefore, any finite $x_i$'s can be brought to $x_i = 0$ using this symmetry, where the above discussions regarding $AdS_2$ still apply. On the other hand, however, the behaviour by "turning on" the space expanded by $x_i$'s, might introduce some new features, because the $x_i$ can take infinities. Therefore $AdS_3$ is a better model for the more general $AdS_d$ with $d \geq 3$.

For $AdS_3$, the metric of the global cover is

$$\mathrm{d}s^2 = \frac{\ell^2}{\cos^2\chi}\left(-\mathrm{d}T^2 + \mathrm{d}\chi^2 + \sin^2\chi \mathrm{d}\psi^2\right), \tag{1.61}$$

where $\chi \in [0, \pi/2)$, $\psi \in [0, 2\pi)$ and $T \in (-\infty, \infty)$. The metric of the Poincaré patch is

$$\mathrm{d}s^2 = \frac{\ell^2}{z^2}\left(-\mathrm{d}t^2 + \mathrm{d}z^2 + \mathrm{d}r^2\right), \tag{1.62}$$

where $z \in (0, \infty)$, $t \in (-\infty, \infty)$ and $r \in (-\infty, \infty)$. [5] Eq. (1.47-1.49) and eq. (1.51-1.54) now reads

$$X_0 = \ell \sec\chi \cos T = \frac{\ell}{2z}\left(\sum_{i=1}^{d-2} x_i^2 - t^2 + z^2 + 1\right), \tag{1.63}$$

$$X_1 = \ell \sec\chi \sin T = \frac{\ell\, t}{z}, \tag{1.64}$$

$$Y_1 = \ell\, \omega_1 \tan\chi = \ell\, \cos\psi \tan\chi = \frac{\ell}{2z}\left(\sum_{i=1}^{d-2} x_i^2 - t^2 + z^2 - 1\right), \tag{1.65}$$

$$Y_2 = \ell\, \omega_2 \tan\chi = \ell\, \sin\psi \tan\chi = \frac{\ell\, r}{z}, \tag{1.66}$$

where we have parametrized $\omega_i$'s by $\omega_1 = \cos\psi$ and $\omega_2 = \sin\psi$.

The Penrose diagram of the global cover is a cylinder, where the radius of the cylinder is $\chi$, the angle is $\psi$ and the vertical axis is $T$. Similar to $AdS_2$, the Penrose diagram of the Poincaré patch can be embedded into that of the global cover. Expressing $(T, \chi, \psi)$ by $(t, z, r)$ via eq. (1.63-1.66), and then plotting in terms of $(t, z, r)$, gives the Penrose diagram of the Poincaré patch embedded in the Penrose diagram of the global cover. Please refer to Fig. 1.4. Fig. 1.4(a) is the Penrose diagram of $AdS_3$ and the Penrose diagram of its Poincaré patch with some special lines. The vertical black line is the line of $(\chi, \psi) = (\pi/2, \pi)$ which represents $(t, z, r) = (t, 0, 0)$ with $t \in (-\infty, \infty)$. The other two straight lines (which are half blue half black), correspond to the null "infinities" for $r = \text{const}$ hypersurfaces. These three lines and the plane surrounded by them, form the Penrose digram of the Poincaré patch of the $AdS_2$ discussed in Sec. 1.4.2. The three vertices of this triangle are $i^0$ and $i^\pm$ which represent the spatial infinity and timelike infinities, respectively. The boundary of $AdS_3$ is divided into (see Fig. 1.4)

(1) $i^0$, point $(T, \chi, \psi) = (0, \pi/2, 0)$, the spatial infinity of $z = \infty$ and $r = \pm\infty$.

(2) $i^+$, point $(T, \chi, \psi) = (\pi, \pi/2, \pi)$, the future timelike infinity.

(3) $i^-$, point $(T, \chi, \psi) = (-\pi, \pi/2, \pi)$, the past timelike infinity.

(4) $j_z^+$, straight line described by ($\psi = 0$ or $\pi$, $T = -\chi\cos\psi + \pi/2$), the future null "infinity" with finite $r$.

[5] Again, equivalently but more closely related to higher dimensional spacetime, we can also take $r \in [0, \infty)$ with $\theta = 0, \pi$ where $\omega_3$ is parametrized as $\omega_3 = \cos\theta$.

(5) $j_z^-$, straight line described by ($\psi = 0$ or $\pi$, $T = \chi \cos\psi - \pi/2$), the past null "infinity" with finite $r$.

The subscript $z$ is to label the fact that the null "infinities" are within the $t - z$ plane (since $r =$ const). Similarly, the two crossing lines (half red half black) on the surface of the cylinder, are the null infinities for $z =$ const surfaces, and are denoted as $j_r^\pm$, which are true infinities. i.e.

(6) $j_r^+$, straight line on the side surface of the cylinder described by ($\chi = \pi/2$, $T = \pm\psi$ with $T \geq 0$), the future null infinity with finite $z$.

(7) $j_r^-$, straight line on the side surface of the cylinder described by ($\chi = \pi/2$, $T = \pm\psi$ with $T \leq 0$), the past null infinity with finite $z$.

(8) $z = 0$, the wedge portion of the side surface of the cylinder, the conformal boundary of the Poincaré patch, which is a part of the spatial infinity ($\chi = \pi/2$) of the global cover.

The side surface of the cylinder corresonds to $\chi = \pi/2$. Its gray portion (on Fig. 1.4(a)) is the $z = 0$ hypersurface in the Poincaré patch. Unfolding the cylinder along $\psi = 0$, we get Fig. 1.4(b). The figure shows that the two crossing lines on the surface of the cylinder are actually straight lines with slope $\mathrm{d}T/\mathrm{d}\psi = \pm 1$, which are null curves: taking $\chi = \pi/2$, eq. (1.61) implies straight lines with $\mathrm{d}T/\mathrm{d}\psi = \pm 1$ are null. Fig. 1.4(c) emphasizes the null "infinities" of the Poincaré patch. We take null curves parametrized by the Killing parameter $t$, and these null curves end at the null "infinities" when $t \to \infty$. The orange lines are null curves with finite $r$, which end at $j_z^+$. The green lines are null curves with finite $z$, which end at $j_r^+$ (and start from $j_r^-$). At $j_r^\pm$ (or $j_z^\pm$), $z$ (or $r$) is finite. On the other hand, the purple lines are null curves which end at the surface where all of $t$, $r$ and $z$ are infinite. Therefore these null "infinities" with all of $t$, $r$ and $z$ approaching their infinities are denoted as

(9) $j_{rz}^+$, the future null "infinity" with infinite $r$ and infinite $z$.

(10) $j_{rz}^-$, the past null "infinity" with infinite $r$ and infinite $z$.

Within these null "infinities", $j_z^\pm$ are not true infinities, as discussed above for $AdS_2$. Similarly, $j_{rz}^\pm$ are not true infinities either. On the other hand, $i^0$, $i^\pm$ and $j_r^\pm$ are located at the surface of the cylinder (where $\chi = \pi/2$), and are true infinities. The boundaries that can be reached by null curves are defined as the *future/past Poincaré horizons* [39]

$$j^\pm \equiv j_r^\pm \bigcup j_z^\pm \bigcup j_{rz}^\pm. \tag{1.67}$$

$j^{\pm}$ are the upper/lower boundary surfaces of the Penrose diagram (of the Poincaré patch). The expression for these two surfaces can be obtained as follows [78]. Using eq. (1.63-1.66), $(t, z, r)$ can be expressed in terms of $(T, \chi, \psi)$ as [78]

$$t = \frac{\sin T}{\cos T - \sin\chi\cos\psi}, \tag{1.68}$$

$$z = \frac{\cos\chi}{\cos T - \sin\chi\cos\psi}, \tag{1.69}$$

$$r = \frac{\sin\chi\sin\psi}{\cos T - \sin\chi\cos\psi}. \tag{1.70}$$

Because $z = \infty$ at $j^{\pm}$, the equation for surfaces $j^{\pm}$ is described by requiring the denominator of (1.69) to be zero, which is

$$\cos T - \sin\chi\cos\psi = 0. \tag{1.71}$$

The upper/lower surfaces of Fig. 1.4(a,c) are generated using this equation. The upper surface is confirmed to be the future null "infinity" by Fig. 1.4(c) where the future null curves end at the upper surface. By the time reversal symmetry of the spacetime, the lower surface is also confirmed to be the past null "infinity".

Now we prove that the Poincaré horizons described by (1.71) are indeed the Killing horizons for Killing vector $\partial_t$. We define

$$f \equiv \cos T - \sin\chi\cos\psi, \tag{1.72}$$

then the Poincaré horizons are described by $f = 0$. For these hypersurfaces to be the Killing horizons, we need to prove [38] (1) the Killing vectors $\partial_t$ are orthogonal to these hypersurfaces; (2) the Killing vectors are null on these hypersurfaces.

In general, the hypersurfaces described by $f = \text{const}$ is orthogonal to the vector field $\partial_\alpha f$. $\partial_\alpha f$ is evaluated as

$$\partial_\alpha f = (-\sin T, -\cos\chi\cos\psi, \sin\chi\sin\psi), \tag{1.73}$$

under the coordinates $(T, \chi, \psi)$. A direct evaluation shows that $\partial_\alpha f$ become null *at the hypersurfaces specified by* $f = 0$. On the other hand, a long but otherwise direct calculation shows that *at the hypersurfaces specified by* $f = 0$

$$(\partial_t)^\alpha = \left(\sin T/\cos^2\chi\right)\cdot\partial^\alpha f. \tag{1.74}$$

i.e. at the Poincaré horizons, $\partial_t$ is proportional to $\partial^\alpha f$. Therefore the two conditions for the Poincaré horizons being Killing horizons are met.

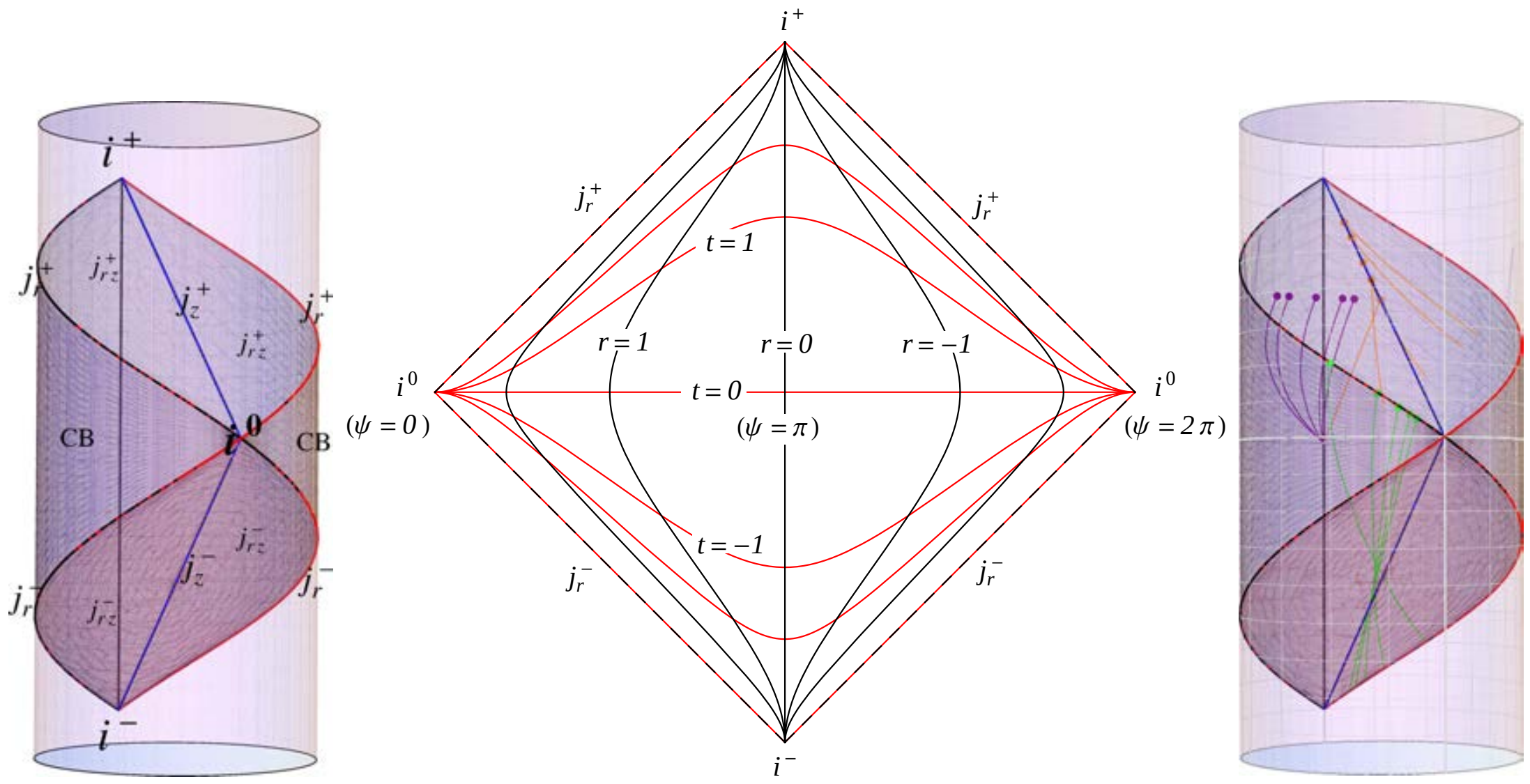

(a) (b) (c)

Figure 1.4: Penrose diagrams of $AdS_3$ and its Poincaré patch. In (a), the cylinder is the Penrose diagram of the whole $AdS_3$, while the portion surrounded by the gray surfaces (with different darkness) is the Penrose diagram for the Poincaré patch. For the notations on the diagram: $i^0$ and $i^{\pm}$ are for points; $j_r^{\pm}$ and $j_z^{\pm}$ are for lines; $j_{rz}^{\pm}$ are for surfaces; and $CB$ stands for the conformal boundary which is the wedge portion of the side surface of the cylinder. Note: $j_{rz}^{\pm}$ appearing on the side of the vertical black line, are actually for the surfaces rather than for the vertical black line. (b) is the conformal boundary (the $z = 0$ hypersurface of the Poincaré patch), which is the part of the Penrose diagram on the side surface of the cylinder. (c) is to emphasize the null "infinities", where the orange lines are the null curves with finite $r$, the green lines are the null curves with finite $z$ and the purple lines are the null curves with all of $t, r, z$ approaching their infinities. When taking $t \to +\infty$, the orange lines end at $j_z^+$, the green lines end at $j_r^+$ and the purple lines end at $j_{rz}^+$.

The upper/lower boundary surfaces have another interesting property. We can put a second patch on top of the existing patch, by requiring that the $i^0$ of the new patch is the $i^+$ of the old patch. The new patch can be regarded as a transformation from the old patch, where the transformation is composed of the following two steps: (1) the patch is rotated by $\pi$ with $\chi = 0$ as the rotational axis. This operation is $\psi \to \psi + \pi$. (2) the patch is moved up vertically by $\pi$, which is $T \to T + \pi$. After these two operations, eq. (1.71) remains unchanged. i.e. it also describes the upper/lower boundaries of the new patch. Therefore, there is no gap between the two patches.

### 1.4.5 The RSII Braneworld and Its Structure

To get a better idea about the structure of the Poincaré patch and the brane, we show the following hypersurfaces with one of the coordinates being constants. Fig. 1.5 shows the $r = \text{const}$ hypersurfaces, where the blue lines are $t = \text{const}$. Or in another word, these blue lines are the integral curves of $\partial_z$, and the black lines are that of $\partial_t$. Fig. 1.6 shows the $t = \text{const}$ hypersurfaces, where the blue lines are the integral curves of $\partial_z$, and the red lines are that of $\partial_r$. Fig. 1.7 shows the $z = \text{const}$ hypersurfaces, where the red lines are the integral curves of $\partial_r$, and the black lines are that of $\partial_t$. In particular, the *brane* in the RSII braneworld is the $z = 1$ hypersurface, which is shown in Fig. 1.7(b).

The following discussion applies to general $d$ case.

As introduced in Sec.1.3.4, the spacetime background of RSII braneworld is to take the $z \geq 1$ portion of the Poincaré patch (which is called the bulk). Therefore the $z = 0$ boundary is eliminated from the Penrose diagram, and the global causal structure is the same as that of Minkowski spacetime.

In fact, any $z = z_0$ (where the constant $z_0 \in (0, \infty)$) can serve as the brane, because the extrinsic curvature of the $z = z_0$ hypersurface can be calculated as $\mathcal{K}_{\mu\nu} = h_{\mu\nu}$, which satisfies the Israel's junction condition (1.29) applied to the vacuum case. Here $h_{\mu\nu}$ is the reduced metric on the hypersurface, and the extrinsic curvature is calculated based on the unit normal vector of the hypersurface, $\mathbb{n}^\mu$, pointing into the bulk (larger $z$ direction). Within the AdS spacetime, the $z = z_0$ hypersurfaces are not geodesic surfaces (in the sense that the extrinsic curvatures of the hypersurfaces do not vanish, see Sec. 2.3.3), but are surfaces with constant acceleration, in the sense that every freely moving massive particle within a $z = z_0$ surface has a constant acceleration as measured by the co-moving inertial observer in the AdS spacetime. Let $d$ be the dimension of the AdS spacetime, and let $v^\mu$ be the $d$-velocity of a massive particle freely moving within a $z = z_0$ hypersurface. The path of the particle is then described by the timelike geodesics generated by the tangent vector $v^\mu$ based on the intrinsic metric of the $z = z_0$ hypersurface, which is $v^\alpha \mathcal{D}_\alpha v^\mu = 0$, where $\mathcal{D}$ is the covariant derivative associated with $h_{\mu\nu}$. Let $u^\mu$ be the $d$-velocity of an observer, then the acceleration of the particle observed by this observer is

$$a^\mu = u^\alpha \nabla_\alpha (v^\mu - u^\mu) . \tag{1.75}$$

If the observer is an inertial observer in the AdS spacetime, his trajectory is then described by a timelike geodesics as $u^\alpha \nabla_\alpha u^\mu = 0$. Furthermore, if the observer's instant velocity is $v^\mu$, he is a co-moving inertial observer. The acceleration of the massive particle moving along a timelike geodesics in the $z = z_0$ hypersurface, as measured by a co-moving observer in the AdS spacetime, is then $v^\alpha \nabla_\alpha v^\mu$, which is described by eq. (2.27) (see Sec. 2.3.4)

$$v^\alpha \nabla_\alpha v^\mu = \mathbb{n}^\mu v^\alpha v^\beta \mathcal{K}_{\alpha\beta} = \mathbb{n}^\mu v^\alpha v^\beta h_{\alpha\beta} = -\mathbb{n}^\mu . \tag{1.76}$$

This acceleration is a constant vector (as measured by the comoving inertial observer in the AdS spacetime) pointing out of the bulk.

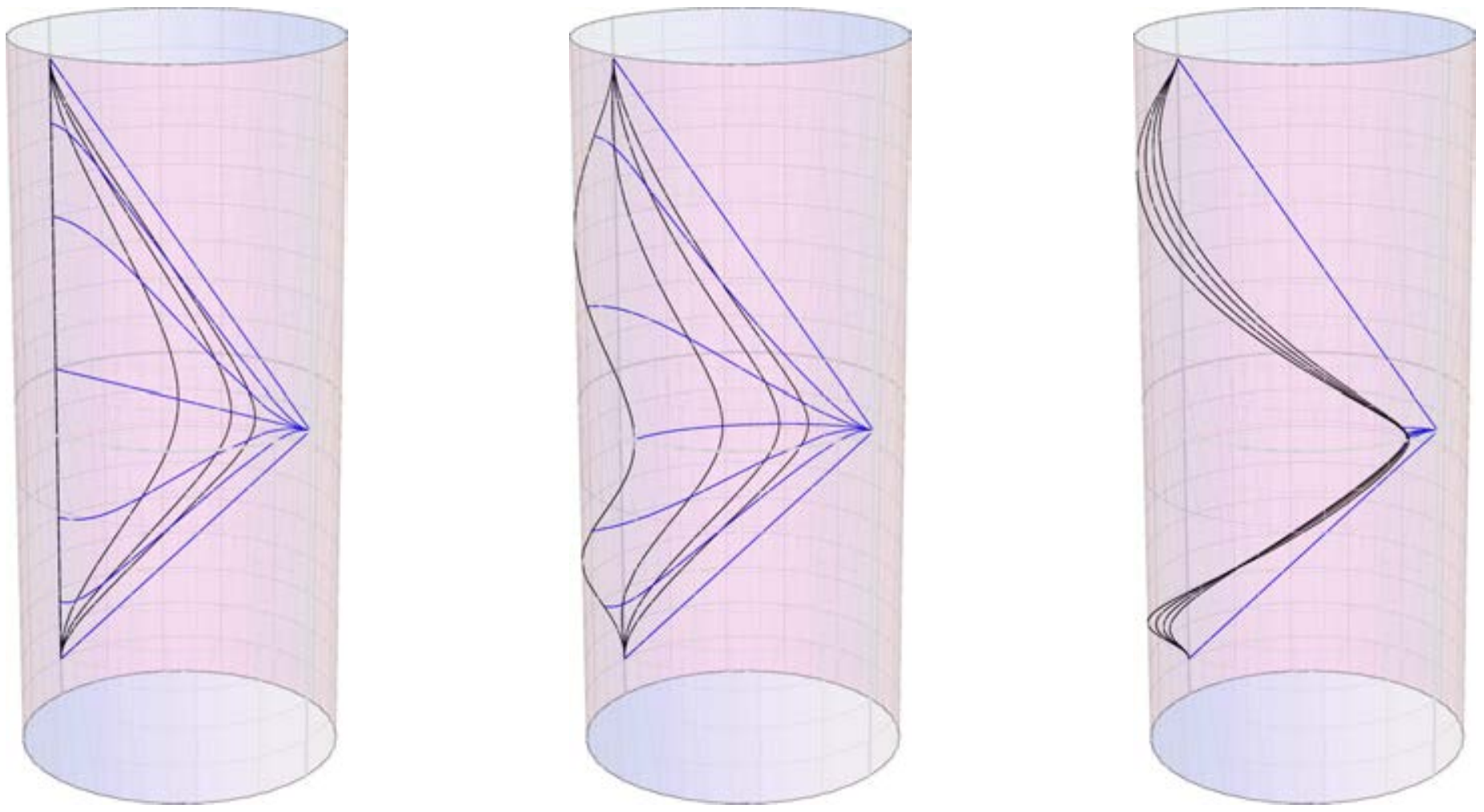

(a) (b) (c)

Figure 1.5: $r =$ const hypersurfaces, where (a) is $r = 0$, which is the AdS$_2$ diagram. (b) is $r = -1$ and (c) is $r = -8$. The blue lines are the integral curves of $\partial_z$, and the black lines are that of $\partial_t$.

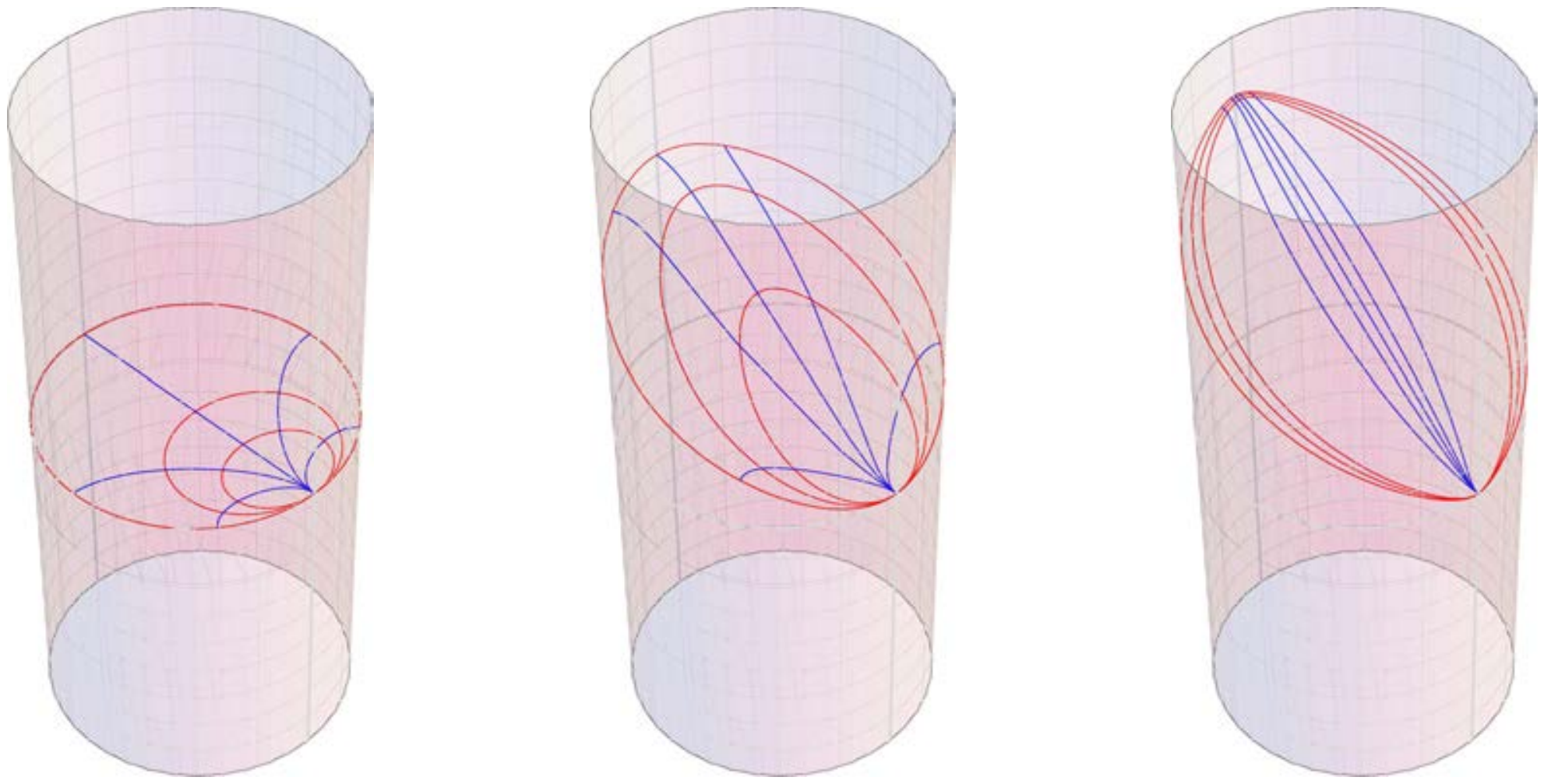

(a) (b) (c)

Figure 1.6: $t = \text{const}$ hypersurfaces, where (a) is $t = 0$. (b) is $t = 2$ and (c) is $t = 8$. The blue lines are the integral curves of $\partial_z$, and the red lines are that of $\partial_r$.

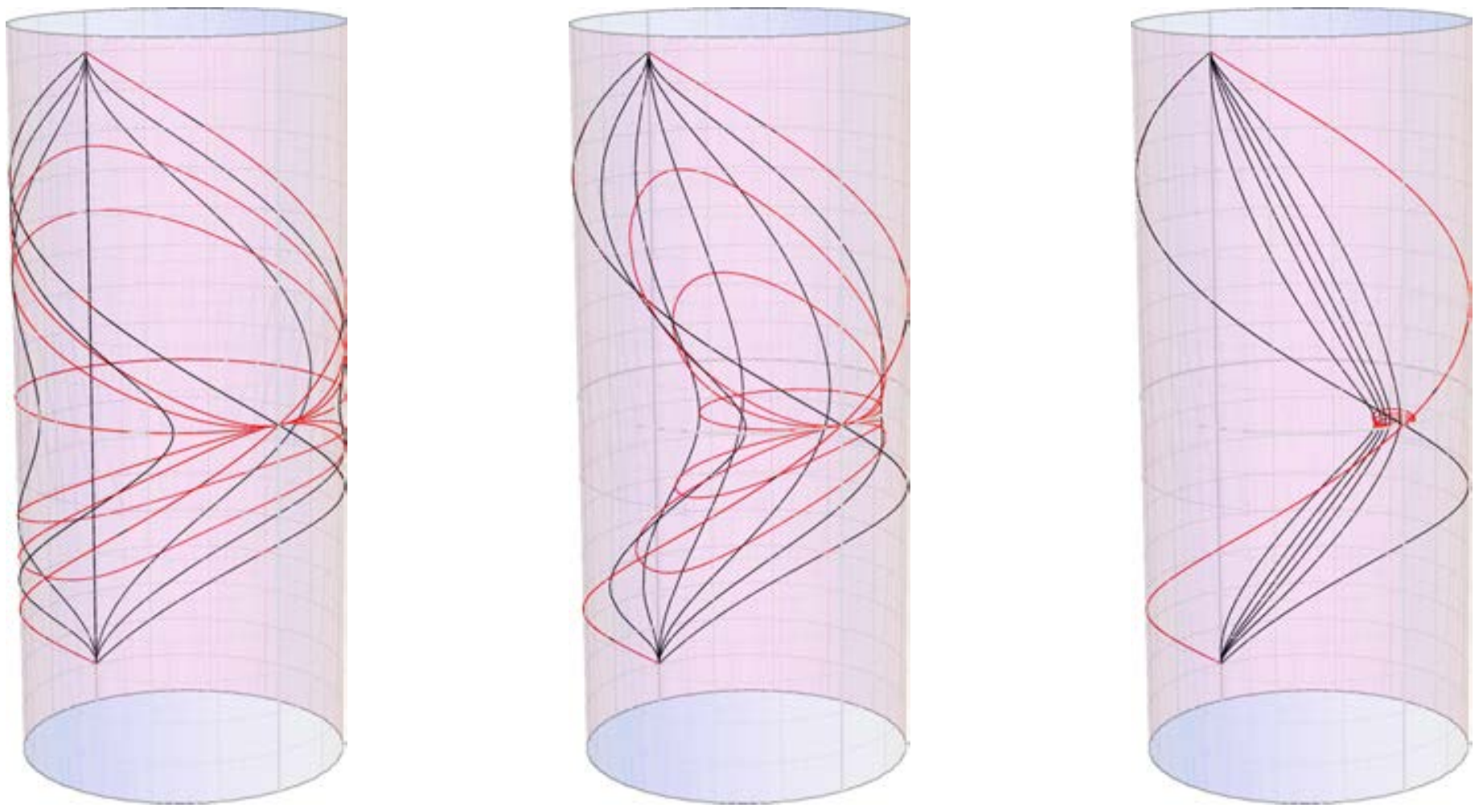

(a) (b) (c)

Figure 1.7: $z = \text{const}$ hypersurfaces, where (a) is $z = 0$, which is opened as Fig. 1.4(b). (b) is $z = 1$, which is the *brane* in the RSII braneworld. (c) is $z = 7$. The red lines are the integral curves of $\partial_r$, and the black lines are that of $\partial_t$.

## 1.5 Evolution Schemes

There exist various formalisms of GR, among which only the ones that are strongly hyperbolic (refer to, for example [34]), can be used as a well-defined formalism of an initial value problem. In this section we only (very) briefly sketch the generalized harmonic formalism since further developments will be present in the next chapter with more details.

The generalized harmonic (GH) formalism [61] uses the gauge source functions

$$H^\alpha \simeq \nabla^\beta \nabla_\beta x^\alpha = -\Gamma^\alpha_{\mu\nu} g^{\mu\nu} \equiv -\Gamma^\alpha, \tag{1.77}$$

as fundamental variables. The notation $\simeq$ means the equation is a constraint relation. Einstein's equations can now be written as

$$-\frac{1}{2} g^{\alpha\beta} g_{\mu\nu,\alpha\beta} - g^{\alpha\beta}{}_{(,\mu} g_{\nu)\beta,\alpha} - H_{(\mu,\nu)} + H_\beta \Gamma^\beta{}_{\mu\nu} - \Gamma^\alpha{}_{\nu\beta} \Gamma^\beta{}_{\mu\alpha} = k_d \left( T_{\mu\nu} - \frac{1}{d-2} g_{\mu\nu} T \right). \tag{1.78}$$

A coordinate gauge choice can now be realized via specifying the $H_\mu$'s. As long as $H_\mu$ does not include derivative of metric functions, the principal part of the above equation $-\frac{1}{2} g^{\alpha\beta} g_{\mu\nu,\alpha\beta}$ is manifestly strongly hyperbolic.

Both the generalized harmonic formalism and the BSSN formalism [34, 45, 46] are widely used in the literature, yet none of them is sufficient to simulate braneworld and we *have to develop* them further. In this thesis, the generalized harmonic formalism will be employed to evolve the branewrold spacetime. Thus we put the introduction and the development of the BSSN formalism to appendix A.

## 1.6 Numerical Methods

The equations of motion of gravitational theory are non-linearly coupled partial differential equations (PDEs). Due to the non-linearity and the complexity, it is not very realistic to study the full dynamics in closed form, especially the behaviour at high energy regime where the fields are so strong that perturbation methods do not apply. We here use a numerical approach. In this section we introduce finite difference approximation (FDA) methods to solve the PDEs. The focus is on the various tests to distinguish numerical solutions from numerical artifacts.

### 1.6.1 Finite Difference Approximation

To demonstrate the concepts in a less abstract way, let us consider the following model problem, which is non-linear wave equation in flat 3+1 dimension spacetime under axisymmetry with source term $f$ (which does not depend on the wave function $\Phi$). This model problem includes a few features that are important for numerical calculation in braneworlds. The equation is assumed to be $(-\partial_{tt} + \partial_{xx} + \partial_{yy} + \partial_{zz})\,\Phi + \Phi^2 = f$ in Cartesian coordinates, or

$$\left(-\partial_{tt} + \partial_{\rho\rho} + \frac{1}{\rho}\partial_{\rho} + \partial_{zz}\right)\Phi + \Phi^2 = f, \tag{1.79}$$

in cylindrical coordinates $(t, \rho, \phi, z)$ that are adapted to the axisymmetry. Therefore the axisymmetry implies $\partial_\phi \Phi = 0$, which has been applied in (1.79). Let us assume the spatial domain is $\rho \in [0, \rho_{\max}], z \in [0, z_{\max}]$.

The whole domain, both spatial and temporal, is divided into discrete grids (or meshes). In principle this division can be arbitrary, as long as the grid/mesh elements are *small*. The meaning of "small" is going to be clear by the discussion in section 1.6.2. To be more specific and to honor simplicity, here let us employ uniform grid. Therefore the spatial domain can be

$$\rho_i = (i-1)\Delta\rho, \quad i = 1, 2, ..., n_\rho \quad \text{where} \quad \Delta\rho = \frac{\rho_{\max}}{n_\rho - 1}; \tag{1.80}$$

$$z_j = (j-1)\Delta z, \quad j = 1, 2, ..., n_z \quad \text{where} \quad \Delta z = \frac{z_{\max}}{n_z - 1}. \tag{1.81}$$

For simplicity, let us choose $\Delta\rho = \Delta z = h$. The time domain is also discretized and the time interval between two subsequent discretized time levels can be expressed as $\Delta t$. $\Delta t/h$ is called the Courant factor.

We use notation

$$\Phi^n_{i,j} \equiv \Phi(t^n, \rho_i, z_j) \equiv \Phi\Big((n-1)\Delta t, (i-1)\Delta\rho, (j-1)\Delta z\Big), \tag{1.82}$$

and similar notation for function $f$. We replace the differential operators by their FDA operators with second order accuracy:

$$\partial_{\rho\rho}\Phi \quad \to \quad \frac{\Phi^n_{i+1,j} - 2\Phi^n_{i,j} + \Phi^n_{i-1,j}}{h^2}, \tag{1.83a}$$

$$\partial_{\rho}\Phi \quad \to \quad \frac{\Phi^n_{i+1,j} - \Phi^n_{i-1,j}}{2h}, \tag{1.83b}$$

$$\partial_{zz}\Phi \quad \to \quad \frac{\Phi^n_{i,j+1} - 2\Phi^n_{i,j} + \Phi^n_{i,j-1}}{h^2}, \tag{1.83c}$$

$$\partial_{tt}\Phi \quad \to \quad \frac{\Phi^{n+1}_{i,j} - 2\Phi^n_{i,j} + \Phi^{n-1}_{i,j}}{(\lambda h)^2}. \tag{1.83d}$$

The FDA operators are obtained by Taylor expansions such as

$$\Phi^n_{i+1,j} = \Phi^n_{i,j} + h\Phi_{,\rho} + \frac{h^2}{2!}\Phi_{,\rho\rho} + \frac{h^3}{3!}\Phi_{,\rho\rho\rho} + \frac{h^4}{4!}\Phi_{,\rho\rho\rho\rho} + O(h^6),$$

which yield

$$\frac{\Phi^n_{i+1,j} - 2\Phi^n_{i,j} + \Phi^n_{i-1,j}}{h^2} = \partial_{\rho\rho}\Phi + \frac{h^2}{12}\Phi_{,\rho\rho\rho\rho} + O(h^4). \tag{1.84}$$

The term $\frac{h^2}{12}\Phi_{,\rho\rho\rho\rho} + O(h^4) = O(h^2)$ is the difference between the exact operator and the FDA operator, which is called truncation error. When $h$ is small (so that the truncation error is not significant), the differential operators can be replaced by their FDA counter parts. Other FDA operators in (1.83) can be obtained similarly. The discretized PDE reads

$$-\frac{\Phi^{n+1}_{i,j} - 2\Phi^n_{i,j} + \Phi^{n-1}_{i,j}}{(\lambda h)^2} + \frac{\Phi^n_{i+1,j} - 2\Phi^n_{i,j} + \Phi^n_{i-1,j}}{h^2} + \frac{1}{\rho_i}\frac{\Phi^n_{i+1,j} - \Phi^n_{i-1,j}}{2h} + \frac{\Phi^n_{i,j+1} - 2\Phi^n_{i,j} + \Phi^n_{i,j-1}}{h^2} + \left(\Phi^n_{i,j}\right)^2 = f^n_{i,j}. \tag{1.85}$$

Now we are ready to introduce the general notations to make the discussion clearer. A set of PDEs, such as equation (1.79), can be collectively denoted as

$$Lu = f, \tag{1.86}$$

where $L$ stands for differential operators and all other operations, $u$ stands for the fundamental variables (the unknown functions) to solve for, and $f$ stands for the terms in the equations that do not include $u$. In equation (1.79), $u = \Phi$, and $Lu = L\Phi = \left(-\partial_{tt} + \partial_{\rho\rho} + \frac{1}{\rho}\partial_\rho + \partial_{zz}\right)\Phi + \Phi^2$.

The discrete FDA operators, such as equation (1.84), can be collectively denoted as

$$\mathcal{A}\Phi = \mathcal{E}\Phi + h^p \cdot E\Phi, \tag{1.87}$$

where $\mathcal{A}$ stands for the FDA version of the exact operator $\mathcal{E}$. $h^p$ means that the approximation level is of $p$-th order in $h$, $E$ stands for the error operator — more specifically, $h^pE\Phi$ is the error. Using (1.87), we can discretize (1.86) as

$$L^h u^h = f^h, \tag{1.88}$$

where $h$ is to label resolution. An example of (1.88) is (1.85).

In (1.85), the approximation is of second order in $h$. Generally the approximation order of $L^h$ is $p$, which can be formally expressed as

$$L^h = L + h^p E. \tag{1.89}$$

From the discussion above, one sees that the validity of FDA needs to be built upon the following two assumptions: (1) the funtion $\Phi$ is smooth; (2) $h$ is small, so that the truncation error is not significant.

However, these two conditions are not sufficient to guarantee that the numerical result $u^h$ is actually a approximation of the exact solution $u$. Therefore systematic test mechanisms need to be developed to distinguish numerical solutions from numerical artifacts.

### 1.6.2 Tests

First, often it is neither practical nor necessary to let equation (1.88) be satisfied exactly. Instead, (1.88) is considered to be satisfied when residual $r^h \equiv L^h u^h - f^h$ is sufficiently small. Again, "small" does not have any measurable meaning yet.

Multiplying equation (1.85) by $\rho_i$, we get the following equation

$$-\frac{\rho_i\left(\Phi^{n+1}_{i,j} - 2\Phi^n_{i,j} + \Phi^{n-1}_{i,j}\right)}{(\lambda h)^2} + \frac{\rho_i\left(\Phi^n_{i+1,j} - 2\Phi^n_{i,j} + \Phi^n_{i-1,j}\right)}{h^2} + \frac{\Phi^n_{i+1,j} - \Phi^n_{i-1,j}}{2h} + \frac{\rho_i\left(\Phi^n_{i,j+1} - 2\Phi^n_{i,j} + \Phi^n_{i,j-1}\right)}{h^2} + \rho_i\left(\Phi^n_{i,j}\right)^2 = \rho_i f^n_{i,j}. \tag{1.90}$$

(1.85) and (1.90) share exactly the same numerical properties, such as convergence, smoothness, regularity, etc. But the two residuals have different numerical values. Therefore, the residual being "small", has *no* absolute meaning.

This feature can be expressed in a more abstract way as: $Lu = f$ and $g \cdot Lu = g \cdot f$ have the same numerical properties. Here $g$ is a non-zero, smooth function over the domain. For example $g$ can be an arbitrary non-zero constant to make the residual take any value. Therefore, the absolute value of residual does not have any meaning. So how to distinguish between a numerical solution and a numerical artifact? And how small is the residual to be considered sufficiently small? These questions will be answered by the following analysis.

Assume the numerical result $u^h$ that satisfies $L^h u^h = f^h + r^h$ is obtained, where $r^h$ is the residual. Generically, $u^h$ is a numerical solution, if the following equation is satisfied when $u^h$ is substituted back into equation (1.86)

$$\lim_{h\to 0} Lu^h - f^h = 0. \tag{1.91}$$

Let us see what it means

$$Lu^h - f^h = L^h u^h - h^p E u^h - f^h = r^h - h^p E u^h = r^h + \mathcal{O}(h^p). \tag{1.92}$$

Therefore (1.91) is satisfied, if $r^h$ is negligible compared to $h^p E u^h$ (in this sense $r^h$ is small).

However, technically it is impossible to apply a continuous operation $L$ to discrete function $u^h$, and then eq. (1.91) can only be understood formally. Instead, $u^h$ is considered a numerical solution, if

$$\begin{aligned} &\lim_{h\to 0} r^h_{\mathrm{I}} = 0, \\ &\text{where } r^h_{\mathrm{I}} \equiv L^h_{\mathrm{I}} u^h - f^h, \text{where } L^h_{\mathrm{I}} \neq L^h \text{ that satisfies } \lim_{h\to 0} L^h_{\mathrm{I}} = L. \end{aligned} \tag{1.93}$$

Since $L^h_{\mathrm{I}}$ is independent of $L^h$ (a different discretization), $r^h_{\mathrm{I}}$ is called *independent residual.*

For the model problem, we can use the following discretization as the independent discretized operators

$$\partial_{rr}\Phi \;\to\; \frac{2\Phi^n_{i,j} - 5\Phi^n_{i+1,j} + 4\Phi^n_{i+2,j} - \Phi^n_{i+3,j}}{h^2}, \tag{1.94a}$$

$$\partial_{r}\Phi \;\to\; -\frac{3\Phi^n_{i,j} - 4\Phi^n_{i+1,j} + \Phi^n_{i+2,j}}{2h}, \tag{1.94b}$$

$$\partial_{zz}\Phi \;\to\; \frac{2\Phi^n_{i,j} - 5\Phi^n_{i,j+1} + 4\Phi^n_{i,j+2} - \Phi^n_{i,j+3}}{h^2}, \tag{1.94c}$$

$$\partial_{tt}\Phi \;\to\; \frac{2\Phi^n_{i,j} - 5\Phi^{n-1}_{i,j} + 4\Phi^{n-2}_{i,j} - \Phi^{n-3}_{i,j}}{(\lambda h)^2}. \tag{1.94d}$$

This discretization is different from (1.83), and is also of the second order accuracy.

In general, the approximation order of $L_{\rm I}^h$ is denoted as $m$, therefore

$$L_{\rm I}^h = L + h^m E_{\rm I} = L^h - h^p E + h^m E_{\rm I}, \tag{1.95}$$

$$r_{\rm I}^h = L_{\rm I}^h u^h - f^h = (L^h - h^p E + h^m E_{\rm I}) u^h - f^h = r^h - h^p E u^h + h^m E_{\rm I} u^h. \tag{1.96}$$

Again, here it is required that $\|r^h\|$ is negligible compared to $\min\left(\|h^p E u^h\|, \|h^m E_{\rm I} u^h\|\right)$, therefore the independent residual $r_{\rm I}^h$ converges to zero at $\min(p, m)$-th order. Here $\|u\|$ is the norm of $u$.

For the model problem, $p = m = 2$, therefore the independent residual behaves as a second order quantity: when $h$ decreases to $h/2$, the independent residual $r_{\rm I}^h$ decreases to $r_{\rm I}^{(h/2)} = \frac{1}{4} r_{\rm I}^h$.

*Note*, the independently discretized operators $L_{\rm I}^h$ can be very different from the discretized operators $L^h$ used to obtain the solution. $L_{\rm I}^h$ and $L^h$ do not need to be of the same method. For example, one can use finite element method or spectrum method to obtain the solution, but use FDA as independent operators to evaluate the independent residual.

### 1.6.3 Tests for General Relativity

For a numerical problem, often there are a certain number of equations to solve, for an equal number of fundamental variables (the unknown functions). If the number of equations is less than the number of unknown functions, in principle there are no unique solutions. On the other hand, in GR, the number of equations is greater than the number of unknown functions. In this case the redundant equations are called constraints.

As an example, in $3 + 1$ formalism of GR, there are six functions $\gamma_{ij}$ to be solved for, by solving the six evolutionary equations. The other four equations are the Hamiltonian constraint and momentum constraints. Analytically, if the constraints are satisfied initially, the consistency (Bianchi identity) guarantees them to be satisfied at all times, as long as the evolutionary equations are satisfied during the evolution. However, numerically there are always small violations to the constraints, and there is no guarantee the violations are controllable. Therefore, for general relativity, the constraints need to be tested as well. i.e. in order to make sure all the components of Einstein's equations are satisfied, *both* the independent residual test *and* the convergence test for constraints are needed.

*Equivalently*, in the case a certain formalism of GR is employed to obtain the numerical results, the results can be substituted into another formalism of GR to produce residuals, and the residuals should converge at the expected order. For example, one can use generalized harmonic formalism to obtain the solution, and then substitute the solution into original Einstein's equations to get residuals, and check whether the residuals converge as expected.

# Chapter 2

# Characteristics in the Braneworld Spacetime

The presence of the brane imposes interesting new physics. This chapter is devoted to develop the formalisms to describe the following topics associated with the brane.

Israel's junction conditions impose cusps in some metric components, which serve as boundary conditions for these metric components. In this chapter we will discuss the boundary conditions of the remaining metric components. We will also discuss other effects of Israel's conditions such as the smoothness of the apparent horizon across the brane.

The main goal of our study is to numerically simulate the process of black hole formation. The definition of a black hole relies on the global causal structure of the spacetime. We will discuss how a black hole can be defined in the braneworld. In the braneworld, the causal structure of the spacetime is determined by the spacetime geometry in 5D, therefore the 4D apparent horizon and event horizon on the brane should play no direct role in braneworlds. However, since the 4D brane is all one can observe, we will study the 4D quantities as well, and compare them with the results in GR to see the observable difference from GR. We will also study the relation between the horizons on the brane and the horizons in the bulk.

Energy in GR is not a locally defined quantity since the energy "density" can be of any value [94]. However, if the system presents asymptotic translational symmetry in time, in certain cases quasi-local energy can be defined, such as ADM energy in asymptotically flat spacetime. A more general formalism of energy obtained from Hamilton-Jacobi analysis [95] [87], will be directly employed in the braneworld to obtain the total energy of the system.

There is also energy exchange between the brane and the bulk. In this chapter we will present our preliminary study on this topic.

## 2.1 Boundary Conditions at the Brane

In this section we discuss the properties and the gauge freedom in the boundary conditions at the brane.

The vacuum solution of the braneworld is eq. (1.38)

$$\mathrm{d}s^2 = e^{-2|y|/\ell}\left(h_{\mu\nu}\mathrm{d}x^\mu \mathrm{d}x^\nu\right) + \mathrm{d}y^2, \tag{2.1}$$

where $y \in (-\infty, \infty)$, and the brane is located at $y = 0$. For the general case (non-vacuum), we setup the coordinate system $(x^a, y)$, where $y$ is the extra dimension. Latin indices $(a, b, \dots)$ are for the coordinates on $y =$ constant surfaces, and their values take $0, 1, 2, 3$. Greek indices $(\mu, \nu, \dots)$ take $0, 1, 2, 3, 4$, and are used for the coordinates of the whole spacetime. Therefore the metric is

$$\mathrm{d}s^2 = g_{ab}\mathrm{d}x^a\mathrm{d}x^b + 2g_{ay}\mathrm{d}x^a\mathrm{d}y + g_{yy}\mathrm{d}y\mathrm{d}y.$$

The coordinate $y$ is set to inherit the following properties: $y = 0$ is where the brane is located, and $y$ is adapted according to the $Z_2$ symmetry, such that the metric components:

$$g_{ab}(x^a, -y) = g_{ab}(x^a, y), \tag{2.2}$$

$$g_{yy}(x^a, -y) = g_{yy}(x^a, y), \tag{2.3}$$

$$\text{and} \qquad g_{ay}(x^a, -y) = -g_{ay}(x^a, y). \tag{2.4}$$

Under this coordinate choice, Israel's first junction condition is simply $g_{ab}|_{y=0^+} = g_{ab}|_{y=0^-} = h_{ab}$, where $h_{ab}$ is the intrinsic metric on the brane (expressed in the coordinates $x^a$), induced from the bulk metric on either sides of the brane. Israel's second junction condition imposes conditions on the extrinsic curvature $\mathcal{K}_{ab}$ (of the brane embedding in the bulk). These conditions can be translated into conditions on $\partial g_{ab}/\partial y$, which will serve as the boundary conditions for $g_{ab}$.

$g_{yy}$ and $g_{ay}$ are not related to Israel's conditions. Rather, since the braneworld spacetimes are "one-sided" (see Sec. 1.3.2 and Sec. 1.3.4), in general there is no generic reasons to require the $y$-coordinate lines (the intersection of the $x^a =$ constant surfaces) to be perpendicular with the $y = 0$ surface (the brane), which means $g_{ay}\big|_{y=0^+} \neq 0$. Taking $y \to 0$ in (2.3) and (2.4), we get

$$g_{yy}\big|_{y=0^-} = g_{yy}\big|_{y=0^+}, \tag{2.5}$$

$$\text{and} \qquad g_{ay}\big|_{y=0^-} = -g_{ay}\big|_{y=0^+}. \tag{2.6}$$

However, since $g_{ay}\big|_{y=0^+} \neq 0$ in general, it means $g_{ay}\big|_{y=0}$ are not defined. This is because only the induced metric on the brane and the extrinsic curvature of the brane are important, while $g_{ay}$ are not needed in defining the induced metric and the extrinsic curvature of the brane. Similarly, $g_{yy}$ is not needed in defining the brane geometry either, which means generically $g_{yy}$ is not defined, although it could be defined as $g_{yy}\big|_{y=0} = g_{yy}\big|_{y=0^-} = g_{yy}\big|_{y=0^+}$. We have generically, $g_{\mu y}$ are not defined on the brane.

Although generally $g_{\mu y}$ are not defined, it is convenient to choose the coordinates at the brane such that the $y$-coordinate lines are perpendicular with the brane. We call this gauge as *perpendicular gauge*. Under this coordinate gauge, we can then define

$$g_{ay}\big|_{y=0} = 0. \tag{2.7}$$

This coordinate gauge has desirable properties such as the smoothness of apparent horizon that is going to be in Sec. 2.2.

Note that $g_{yy}\big|_{y=0}$ is still unconstrained.

## 2.2 Apparent Horizon

An apparent horizon is needed to monitor the evolution of spacetime if a black hole is present during the evolution. This section is devoted to studies on apparent horizons in braneworlds.

### 2.2.1 The Definition

The definition of apparent horizon can be found in standard texts [34, 35, 37]. Let $n_\alpha$ be *future* directed timelike unit vector normal to $t$ = constant hypersurfaces. Let $S$ denote a closed $(d-2)$ dimensional spatial surface within a $t$ = constant hypersurface, and $s^\alpha$ is its unit normal vector pointing towards the outgoing direction, which is within the same $t$ = constant hypersurface. The induced metric on $S$ is then (not to be confused with the $m^\alpha$ defined in (1.3))

$$m^{\alpha\beta} \equiv g^{\alpha\beta} + n^\alpha n^\beta - s^\alpha s^\beta. \tag{2.8}$$

Let $v^\alpha$ be an arbitrary vector field, the *relative* change rate in the area elements of $S$ along $v^\alpha$ is [37]

$$\Theta^{(v)} \equiv \frac{1}{\sqrt{m}}\mathcal{L}_{\boldsymbol{v}}\sqrt{m} = \frac{1}{2}m^{\mu\nu}\mathcal{L}_{\boldsymbol{v}}m_{\mu\nu} = m^{\mu\nu}\nabla_\mu v_\nu, \tag{2.9}$$

where $m$ is the determinant of $m_{\mu\nu}$. The final expression is the expansion of $v^\alpha$ along $S$.

For the $(d-2)$ dimensional spacelike surface $S$, there exist two null curves orthogonal to this surface. Let us denote the two future directed null vectors tangent to these two null curves as ${}^{\pm}l_\alpha$. The relative change rate of the area elements of $S$ along the null geodesics congruences produced by ${}^{\pm}l_\mu$ are then

$$\Theta_\pm \equiv m^{\alpha\beta}\nabla_\alpha\ {}^{\pm}l_\beta. \tag{2.10}$$

A trapped surface is an $S$ whose $\Theta_\pm < 0$, which means the null geodesics congruences produced by both ${}^{+}l_\mu$ and ${}^{-}l_\mu$ drag $S$ to the surfaces with smaller area elements. An $S$ with

$$\Theta_+ = 0 \tag{2.11}$$

is called a marginally outer trapped horizon (MOTH), whose area elements stay the same under the Lie-dragging of ${}^{+}l_\mu$. The MOTH is not unique in a given spacetime since there can be other MOTHs within a MOTH. An apparent horizon is defined as *the outermost MOTH*.

Now let us construct ${}^{\pm}l_\alpha$. Because timelike normal vector $n^\alpha$ is orthogonal to the spacelike normal vector $s^\alpha$, the two future directed null vectors orthogonal to $S$ are

$${}^{\pm}l_\alpha \equiv n_\alpha \pm s_\alpha, \tag{2.12}$$

where ${}^{+}l_\alpha$ is outgoing and ${}^{-}l_\alpha$ is ingoing. Substituting this into (2.10), we obtain

$$\Theta_\pm = m^{\alpha\beta}\nabla_\alpha\left(n_\beta \pm s_\beta\right) = \pm D_\alpha s^\alpha - K + s^\alpha s^\beta K_{\alpha\beta}, \tag{2.13}$$

where $D_\alpha$ denotes the covariant derivative associated with $\gamma_{\alpha\beta}$, and $K_{\alpha\beta}$ is the extrinsic curvature of the $t=$ constant hypersurface. Therefore, the follow equation is satisfied at the apparent horizon

$$\Theta_+ = D_\alpha s^\alpha - K + s^\alpha s^\beta K_{\alpha\beta} = 0. \tag{2.14}$$

The above discussion shows that the definition of apparent horizon relies on the choice of $t=$ constant hypersurfaces. For a given spacetime, different slicing conditions can result in drastically different apparent horizons. Taking Schwarzschild spacetime as an example. In Schwarzschild coordinates or the isotropic coordinates, the apparent horizon coincides with the event horizon, where the physical singularity is inside of the apparent horizon. On the other hand, the Schwarzschild spacetime can be sliced in such a way that there is no apparent horizon [33].

### 2.2.2 Smoothness of the Horizon

Israel's conditions impose cusps to some components of the metric. In this section we will study whether this affects the smoothness of apparent horizon. The smoothness of an apparent horizon can be studied via $s^\alpha$ by asking whether $s^\alpha$ is continuous across the brane. Rewriting $\Theta_+ = 0$ as

$$D_\alpha s^\alpha = K - s^\alpha s^\beta K_{\alpha\beta}, \tag{2.15}$$

we then apply Gauss' theorem (in curved space) on the $t$ = constant hypersurface. By the same procedure to derive the junction condition of electric field across a surface [6], we can find the junction condition for $s^\alpha$ across the brane. If the right hand side of (2.15) is finite (by the $Z_2$ symmetry with respect to the brane, this condition is true if the $t$ = constant hypersurfaces are chosen to be perpendicular with the brane), then the integration over an infinitesimal layer across the brane vanishes. Therefore the component of $s^\alpha$ that is perpendicular to the brane, is continous across the brane. This continuity of the perpendicular component, together with the $Z_2$ symmetry with respect to the brane, imply that the perpendicular component of $s^\alpha$ must be zero. Therefore $s^\alpha$ is continuous. i.e.

> the direction of an apparent horizon is continuous across the brane, if the slicing condition is such that the $t$ = constant hypersurfaces are perpendicular to the brane.

Under the coordinates setting in Sec. 2.1, the slicing condition is expressed as $g_{ty} = 0$. In particular, the perpendicular gauge (2.7) satisfies the slicing condition.

### 2.2.3 Apparent Horizon on the Brane and in the Bulk

Generically the causal structure is determined by 5D geometry. However, only the brane quantities are directly observable, therefore we study the relation between 4D and 5D quantities.

[6]The procedure to derive $\mathbf{E}_+ - \mathbf{E}_- \propto \sigma$ from $\nabla \cdot \mathbf{E} \propto \rho$, where $\rho$ is volume charge density and $\sigma$ is areal charge density of the singular layer. $\mathbf{E}$ is the electric field, and $E_+$ is the electric field on one side of the singular layer, and $E_-$ is the electric field on the other side.

The question we try to address in this subsection is whether the apparent horizon seen on the brane (which is calculated based on the brane geometry only), and the bulk apparent horizon (which is calculated based on the bulk geometry), agree with each other on the brane. This can be studied via the expansions of the outgoing null geodesics congruences on the brane and in the bulk

$$\begin{aligned}\Theta_{\text{brane}} &= \left(h^{\alpha\beta} + {}^{(\text{r})}n^{\alpha}\ {}^{(\text{r})}n^{\beta} - {}^{(\text{r})}s^{\alpha}\ {}^{(\text{r})}s^{\beta}\right)\mathcal{D}_{\alpha}\left({}^{(\text{r})}n_{\beta} + {}^{(\text{r})}s_{\beta}\right)\\ &= {}^{(\text{r})}m^{\alpha\beta}\nabla_{\alpha}\left({}^{(\text{r})}n_{\beta} + {}^{(\text{r})}s_{\beta}\right) && \text{(a)}\\ &= {}^{(\text{r})}m^{\alpha\beta}\nabla_{\alpha}\left(n_{\beta} + s_{\beta}\right) && \text{(b)}\\ &= (m^{\alpha\beta} - \mathbb{n}^{\alpha}\mathbb{n}^{\beta})\nabla_{\alpha}\left(n_{\beta} + s_{\beta}\right)\\ &= \Theta_{\text{bulk}} - \mathbb{n}^{\alpha}\mathbb{n}^{\beta}\nabla_{\alpha}(n_{\beta} + s_{\beta}). && (2.16)\end{aligned}$$

where ${}^{(\text{r})}m^{\alpha\beta} = h^{\alpha\beta} + {}^{(\text{r})}n^{\alpha}\ {}^{(\text{r})}n^{\beta} - {}^{(\text{r})}s^{\alpha}\ {}^{(\text{r})}s^{\beta}$ is the projection operator that projects to the $(d-3)$-surface on the brane, and $\mathcal{D}$ is the covariant derivative associated with the brane metric $h_{\alpha\beta}$. Anything with a superscript (r) is a quantity defined only on the brane. The vector $\mathbb{n}^{\alpha}$ is the unit normal vector that is perpendicular to the brane. Assuming perpendicular gauge (2.7), we have $s_{\beta} = {}^{(\text{r})}s_{\beta}$ and $n_{\beta} = {}^{(\text{r})}n_{\beta}$ on the brane, which are used in deriving eq. (b) from eq. (a). The difference between the two $\Theta$'s, even at the apparent horizon where $\Theta_{\text{bulk}} = 0$, is generally non-zero. i.e. generally these two apparent horizons do not agree. Therefore, we will study the relation between event horizons in the next section.

## 2.3 Event Horizon

### 2.3.1 Event Horizon in the Braneworld

In this subsection we examine whether event horizon is well-defined in the spacetime of the RSII braneworld, and discuss how to define an evolution problem.

The global causal structure of AdS spacetime and its Poincaré patch was introduced in Sec. 1.4. The spacetime background of RSII braneworld is to take the $z \geq 1$ portion of the Poincaré patch. Therefore the $z = 0$ boundary is eliminated from the Penrose diagram, and the global causal structure is the same as that of Minkowski spacetime. In particular, similar to Minkowski spacetime, there is no signal travelling to spatial infinity and then coming back within a finite local proper time, and Cauchy surfaces exist [6]. To discuss the Cauchy surfaces, we define the *future/-past horizons* as $i^0 \bigcup i^{\pm} \bigcup j_r^{\pm} \bigcup j_z^{\pm} \bigcup j_{rz}^{\pm}$ (see Sec. 1.4.4). Considering a $t =$ constant hypersurface (which is spacelike), any future causal curves coming from the past horizon will either go across this hypersurface, or hit the brane and then are reflected to travel back into the bulk (due to the $Z_2$ symmetry with respect to the brane) which eventually go across this hypersurface. Similarly, all the past causal curves of the future horizon go across this hypersurface. Therefore, any $t =$ constant hypersurface is a Cauchy surface, because all the developments of any past event are captured by the surface, and all the future events can be predicted by the data on this surface. Or more precisely, all the inextendible future causal curves of the past horizon and all the inextendible past causal curves of the future horizon, go across the Cauchy surface [32]. Therefore, a Cauchy problem (an initial value problem) is well-defined.

The $z \geq 1$ branch has the horizons which are not true infinities. Accordingly, the definition of event horizon is modified as below. In fact, the definition of event horizon in asymptotically Minkowski spacetime can be directly carried over, while the only change is to replace the notion of *null infinities* in the definitions by the future/past Poincaré horizons defined by eq. (1.67). The event horizon is the boundary of the spacetime region which can not be connected to the external world by future oriented null geodesics. i.e. the event horizon is the collection of "the points of no return". The "external world" needs to be defined. Similar to the case of asympototically Minknowski spacetime, the external world is defined as the past of the future Poincaré horizon, therefore the event horizon is the boundary of the spacetime separating the region that can be connected to the future Poincaré horizon by future null geodesics from the region that can not. The future Poincaré horizon, on the other hand, are defined as the "infinities" (as measured by the Killing parameter $t$) of future null curves departing from the external world. i.e. there is a circular argument in these definitions. The circle can be ended by *physically* identifying certain spacetime region(s) as the external world. Therefore, as long as the external world can be physically identified, the event horizon can be defined, and this argument applies to any spacetime (i.e. it is not limited to the case of the Poincaré patch of the AdS spacetime).

### 2.3.2 Event Horizon on the Brane

The event horizon is defined in Sec. 2.3.1. In the following we will study whether the event horizon based on the brane geometry is well-defined. The causal structure of the braneworld is determined by the 5D geometry, rather than the 4D geometry of the brane. The null geodesics generated by the outgoing, future oriented null vectors on the bulk event horizon, form the boundary (the event horizon) separating the spacetime region that can be connected to the future Poincaré horizon by future oriented null geodesics, from the region that can not. Let us consider the out-going future oriented null vectors within the brane originated from the intersect of the bulk event horizon and the brane. If the future oriented null geodesics generated by these null vectors will stay on the brane forever, then these null geodesics are the boundaries separating the spacetime regions that can be connected (by future oriented null geodesics) with the future Poincaré horizon from the regions that can not. i.e. (1) these null geodesics stay on the brane forever; (2) these null geodesics are the boundaries separating the spacetime regions that can arrive the future Poincaré horizon from the regions that can not. i.e. these null geodesics form the event horizon on the brane. Therefore, the key for the well-definedness of the event horizon on the brane, is to study the relation between the null geodesics produced based on the brane geometry and the null geodesics produced based on the bulk geometry, generated from the same null vector which initially lies within the brane. This relation can be described by the extrinsic curvature of the brane.

### 2.3.3 Extrinsic Curvature as Geodesics Deviation

This subsection (Sec. 2.3.3) is the foundation of the study on the relation of event horizons. The work in this subsection is a "re-discovery" of the *Gaussian curvature* (see, e.g. [80]) and the Gauss-Weingarten equation [81] in differential geometry.

The motivation is as follows. Let us examine how to measure the extrinsic nature of the embedding of a hypersurface $\Sigma$ into a higher dimensional space $M$. $\Sigma$ is considered flatly embedded into $M$, if $\Sigma$ appears to be flat in $M$, in the sense that an arbitrary straight line as seen in $\Sigma$ is also a straight line as seen in $M$. Since straight lines are geodesics, the above statement can be more precisely rephrased as: if the hypersurface is flatly embedded, an arbitrary geodesics in $\Sigma$ (consistent with the hypersurface metric $\gamma_{\alpha\beta}$) will also be a geodesics in $M$ (consistent with the bulk metric $g_{\alpha\beta}$). For non-flat embedding, these two types of geodesics are not the same in general [7]. Therefore, this motivates us to describe the extrinsic curvature as the deviation of the geodesics defined in $\Sigma$ from the geodesics defined in $M$ for a vector initially lying on $\Sigma$. This point of view to describe the extrinsic curvature, is referred as geodesics point of view (GEP for short).

On the other hand, the embedding described via the covariant derivative of the unit normal vector along $\Sigma$, as what has been done in eq. (1.5), is referred as normal vector point of view (NVP for short).

Generally if a $d-C$ dimensional sub-manifold $\Sigma$ is embedded in a $d$ dimensional manifold $M$,

[7] As an example, we can think of a sphere $S^2 : x^2 + (y-1)^2 + z^2 = 1$ embedded in $\mathbb{R}^3$. Take a vector lying on $S^2 : (\partial_x)^\mu$ at $(0,0,0)$. The geodesics produced by it on $S^2$ is the equator, while the geodesics produced by it in $\mathbb{R}^3$ is the $x$-axis.

$C$ is called the co-dimension. In this subsection, we will prove NVP and GEP are equivalent in the $C = 1$ case, in the sense that the embedding studied from GEP (shown below) will also lead to the same definition of the extrinsic curvature as eq. (1.5), which is defined from NVP. In the RSII braneworld, there is only one extra dimension, thus the co-dimension is $C = 1$. For $C > 1$ case, please refer to appendix B.

The basic idea is to study the two geodesics generated by an arbitrary vector $T^\alpha \in \Sigma$, via equations $T^\beta D_\beta T^\alpha = 0$ and $T^\beta \nabla_\beta T^\alpha = 0$, where $D$ is the covariant derivative in $\Sigma$ and $\nabla$ is the covariant derivative in $M$. Then we compare the two geodesics to get the difference, which can describe the embedding nature of $\Sigma$ in $M$. We adopt the following approach: rewrite $T^\beta D_\beta T^\alpha = 0$ as $T^\beta \nabla_\beta T^\alpha =$ leftover, then the *leftover* is the deviation of the two geodesics. Let us denote the unit normal vector of $\Sigma$ as $n_\alpha$ whose length square is $\epsilon = n^\alpha n_\alpha = \pm 1$ where $\epsilon = 1$ if the extra dimension is spacelike and $\epsilon = -1$ if the extra dimension is timelike. The reduced metric of the hypersurface is

$$\gamma_{\alpha\beta} = g_{\alpha\beta} - \epsilon\; n_\alpha n_\beta. \tag{2.17}$$

For a general tensor $T^{\alpha_1 \ldots \alpha_k}{}_{\beta_1 \ldots \beta_l} \in \Sigma$, the covariant derivative associated with the metric $\gamma_{\alpha\beta}$ is [34, 36]

$$D_\gamma T^{\alpha_1 \ldots \alpha_k}{}_{\beta_1 \ldots \beta_l} = \gamma^{\alpha_1}{}_{\delta_1} \ldots \gamma_{\beta_l}{}^{\epsilon_l} \gamma_\gamma{}^\nu \nabla_\nu T^{\delta_1 \ldots \delta_k}{}_{\epsilon_1 \ldots \epsilon_l}. \tag{2.18}$$

Therefore $\forall\; T^\alpha \in \Sigma$, we have

$$D_\alpha T^\mu = \gamma_\alpha{}^\beta \gamma^\mu{}_\gamma \nabla_\beta T^\gamma, \tag{2.19}$$

which tells us that the geodesics generated by $T^\alpha$ in $\Sigma$ is just

$$0 = T^\alpha D_\alpha T^\mu = T^\alpha \gamma_\alpha{}^\beta \gamma^\mu{}_\gamma \nabla_\beta T^\gamma = \gamma^\mu{}_\gamma T^\alpha \nabla_\alpha T^\gamma, \tag{2.20}$$

or

$$\gamma^\mu{}_\nu T^\alpha \nabla_\alpha T^\nu = 0. \tag{2.21}$$

On the other hand, we have the following identity which can be obtained by the fact that $n_\alpha T^\alpha = 0$

$$n^\mu n_\nu T^\alpha \nabla_\alpha T^\nu = -n^\mu T^\alpha T^\nu \nabla_\alpha n_\nu. \tag{2.22}$$

From (2.21) and (2.22), we obtain

$$T^\alpha \nabla_\alpha T^\mu = \left(\gamma^\mu{}_\nu + \epsilon\; n^\mu n_\nu\right) T^\alpha \nabla_\alpha T^\nu = -\epsilon n^\mu T^\alpha T^\beta \nabla_\alpha n_\beta. \tag{2.23}$$

We can now define the deviation of the two types of geodesics equation as the right hand side of equation (2.23). It is then clear that the deviation vanishes, if and only if

$$T^\alpha T^\beta \nabla_\alpha n_\beta = 0. \tag{2.24}$$

This result is for *arbitrary* $T^\alpha$ defined on $\Sigma$, and only the contraction with $T^\alpha$ appears in this expression, which means we can use the following quantity to describe the embedding of $\Sigma$ in $M$

$$K_{\alpha\beta} \equiv -\gamma_\alpha{}^\mu \gamma_\beta{}^\nu \nabla_\mu n_\nu. \tag{2.25}$$

i.e. NVP and GEP are using the same quantity ($K_{\alpha\beta}$) to describe the embedding. The deviation equation (2.23) in terms of the extrinsic curvature is now rewritten as

$$T^\alpha \nabla_\alpha T^\mu = \epsilon\, n^\mu T^\alpha T^\beta K_{\alpha\beta}, \tag{2.26}$$

from which one see that the deviation is in the perpendicular direction ($n^\mu$ direction).

### 2.3.4 The Relation between the Event Horizons

The work in this subsection (Sec. 2.3.4) was first independently carried out in [81].

For the braneworld, the 3 + 1 brane is embedded into the 4 + 1 dimensional bulk. Using the notations in braneworld, the geodesics deviation equation (2.26) is now

$$v^\alpha \nabla_\alpha v^\mu = \mathbb{n}^\mu v^\alpha v^\beta \mathcal{K}_{\alpha\beta}, \tag{2.27}$$

where $v^\alpha$ is the tangent vector of an arbitrary geodesics within the brane. $\mathbb{n}$ is the unit normal vector of the brane, and $\epsilon = \mathbb{n}^\mu \mathbb{n}_\mu = 1$ has been applied.

To study the event horizon relations, we focus on the case where $v^\alpha$ is a null vector. In RSII, $\mathcal{K}_{\alpha\beta}$ is related to brane content by Israel's junction condition (1.29)

$$\mathcal{K}_{\alpha\beta} = \frac{1}{2} k_d \left( \lambda \frac{h_{\alpha\beta}}{d-2} + \tau_{\alpha\beta} - h_{\alpha\beta} \frac{\tau}{d-2} \right), \tag{2.28}$$

which implies the deviation of null geodesics is

$$v^\alpha \nabla_\alpha v^\mu = \frac{1}{2} k_d \mathbb{n}^\mu v^\alpha v^\beta \tau_{\alpha\beta}. \tag{2.29}$$

Therefore the deviation of the two geodesics amounts to whether $v^\alpha v^\beta \tau_{\alpha\beta}$ vanishes. Generally the right hand side of eq. (2.29) does not vanish (otherwise it is a new energy condition for the matter. The discussion of energy condition is beyond the scope of our project). However, in case $\tau_{\mu\nu} = 0$, when the matter on the brane vanishes, the two geodesics coincide. In another word, when the matter is absent, at the intersect of the bulk event horizon and the brane, the future oriented null geodesics that are produced by the null vectors lying in $\Sigma$, will stay on the brane forever. Since these null geodesics are on the event horizon of the bulk, they are the boundaries separating the spacetime regions that can be connected (by future oriented null geodesics) with the future Poincaré horizon from the regions that can not. i.e. (1) these null geodesics stay on the brane forever; (2) these null geodesics are the boundaries separating the spacetime regions that can arrive the future Poincaré horizon from the regions that can not. Therefore, from the brane point of view, they form the event horizon on the brane. i.e.

| |
|---|
| an event horizon on the brane is well-defined when there is no matter around the horizon. |

For gravitational collapse processes, if the systems eventually reach the stationary states that the matters either go into the black holes, or get bounced back and travel towards spatial infinity, then there are no matters at the horizons and the event horizons on the brane are well-defined.

## 2.4 Energy in the Braneworld

In order to quantitatively describe the spacetime evolution, and gravitational interaction between the brane and the bulk, we need to introduce certain quantities, such as energy. However, there is no local definition of energy in general relativity. Instead, there have been many attempts to define quasilocal energy in general relativity, and many of these definitions are only well-defined in a certain background. In this section, the definition developed by Brown and York obtained from a Hamiltonian-Jacobi analysis of the gravitational action [95] (Hawking and Horowitz also gave a similar derivation [87]), is applied to the braneworld.

### 2.4.1 Total Energy

In this subsection, we introduce the energy defined in [87, 95]. In the next subsection, we will apply this definition to the braneworld spacetime.

In general the action of a gravitational system is

$$I(g,\Phi) = \int_M \left[\frac{R}{16\pi G_d} + \mathscr{L}_m(g,\Phi)\right] - \frac{1}{8\pi G_d}\oint_{\partial M} \mathscr{K}, \tag{2.30}$$

where $M$ is the spacetime manifold, $\mathscr{K}$ is the extrinsic curvature of the boundary $\partial M$ embedding into $M$. $\mathscr{L}_m$ is the Lagrangian for all the matter fields and the matter fields are collectively denoted as $\Phi$.

Let us choose the boundary $\partial M = \Sigma_{t_1} \bigcup \Sigma_{t_2} \bigcup \mathscr{B}^Q$, where $\mathscr{B}^Q \equiv \bigcup_{t\in[t_1,t_2]} S_t^Q$ as shown in Fig. 2.1. $S_t^Q$ is the closed $(d-2)$ dimensional spatial surface embedded in each $\Sigma_t$, and $Q$ is the single parameter to characterize the family of the enclosed $(d-2)$ dimensional surfaces in each $\Sigma_t$. When $Q \to \infty$, $S_t^Q$ goes to the spatial infinity boundaries. $\mathscr{B}^Q$ is setup such that its normal unit vector $Q^\mu$ is perpendicular to $n^\mu$ (so that $n^\mu$ lies within $\mathscr{B}^Q$).

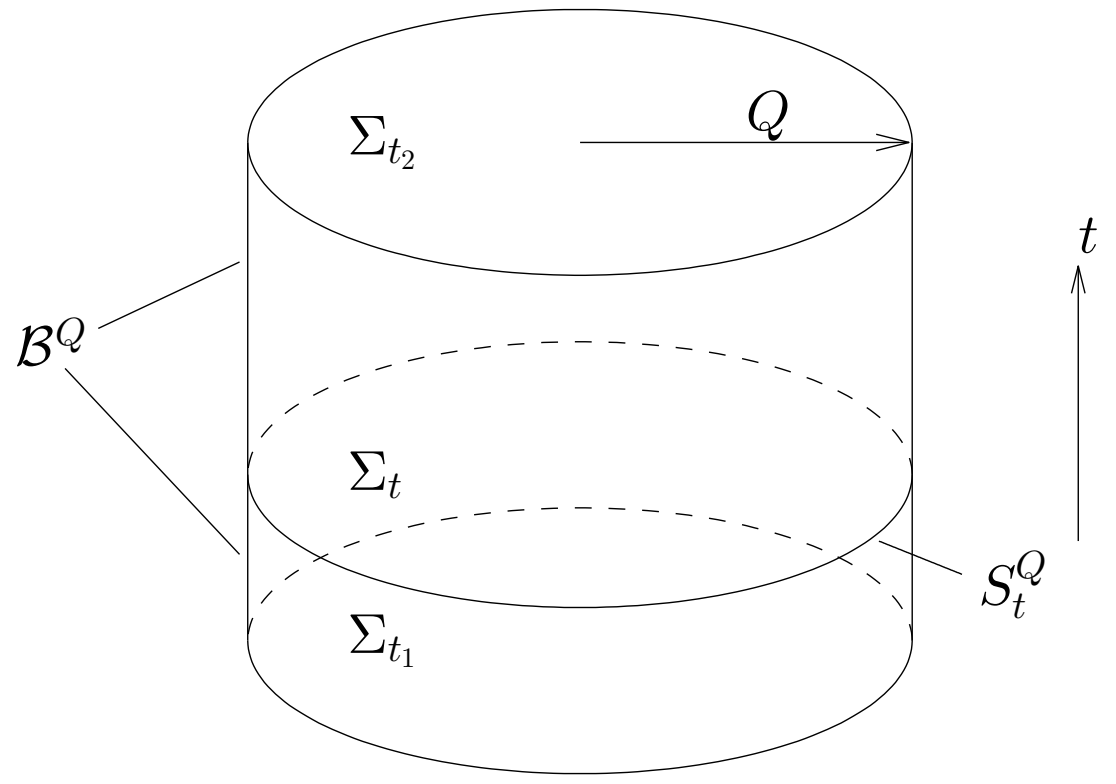

Figure 2.1: Manifold $M$ and its boundaries. The boundary $\partial M$ is composed of $\Sigma_{t_1}$, $\Sigma_{t_2}$, and $\mathscr{B}^Q$. On the diagram, the boundaries are shown to be the upper/lower surface of the cylinder, and the side surface of the cylinder. Here $\mathscr{B}^Q \equiv \bigcup_{t\in[t_1,t_2]} S_t^Q$, where $S_t^Q$ is the enclosed $(d-2)$ dimensional spatial surface embedded in each $\Sigma_t$, and $Q$ is the single parameter to characterize the family of the closed surfaces in each $\Sigma_t$. Choose $\mathscr{B}^Q$ such that its normal unit vector $Q^\mu$ is perpendicular to $n^\mu$ (so that $n^\mu$ lies in $\mathscr{B}^Q$).

The Hamiltonian is then [87, 95]

$$H = \int_{\Sigma_t} (\alpha\mathcal{H} + \beta_\nu \mathcal{M}^\nu) + \frac{1}{8\pi G_d}\oint_{S_t^Q}\left[-\epsilon\; Q_\mu(K\gamma^{\mu\nu} - K^{\mu\nu})\beta_\nu + \alpha\ \cdot\ {}^{(d-2)}k\right], \tag{2.31}$$

where $\epsilon \equiv n_\mu n^\mu = -1$. ${}^{(d-2)}k_{\alpha\beta}$ is the extrinsic curvature of $S_t^Q$ embedded in $\Sigma_t$, and ${}^{(d-2)}k$ is its trace. $\mathcal{H}$ and $\mathcal{M}^\nu$ are Hamiltonian constraint and momentum constraint defined as

$$\mathcal{H} \equiv \frac{-\epsilon}{16\pi G_d}\left(-K^2 + K^{\mu\nu}K_{\mu\nu} + \epsilon \cdot {}^{(d-1)}R + 16\pi G_d\rho\right), \tag{2.32}$$

$$\mathcal{M}^\nu \equiv \frac{\epsilon}{8\pi G_d} D_\mu(K\gamma^{\mu\nu} - K^{\mu\nu}) - S^\nu. \tag{2.33}$$

The Hamiltonian constraint $\mathcal{H}$ and momentum constraints $\mathcal{M}$ are implied by Einstein's equations, therefore vanish for physical configurations, and should be dropped in (2.31).

Hamiltonian (2.31) diverges in general. However, it is the physical Hamiltonian that matters [87]. The physical Hamiltonian is $H - \bar{H}$, where $\bar{H}$ is the Hamiltonian of a background spacetime. We denote background quantities by a *bar* ( $\bar{}$ ). The background is a static solution, then its contribution is only

$$-\frac{1}{8\pi G_d}\oint_{S_t^Q} \alpha \ {}^{(d-2)}\bar{k},$$

where the integration is over a closed surface in the background spacetime that is isometric to the $S_t^Q$ (that is the closed surface chosen in the physical spacetime).

The physical energy is the physical Hamiltonian

$$E = H - \bar{H} = \lim_{Q\to\infty} \frac{1}{8\pi G_d}\oint_{S_t^Q} \left[-\epsilon\, Q_\mu(K\gamma^{\mu\nu} - K^{\mu\nu})\beta_\nu + \alpha\left({}^{(d-2)}k - {}^{(d-2)}\bar{k}\right)\right], \tag{2.34}$$

where the Hamiltonian constraint $\mathcal{H}$ and momentum constraint $\mathcal{M}$ are dropped, since they vanish at physical configurations. Also, at spatial infinities where there is asymptotic time translational symmetry, the lapse function and the shift functions go to the form of the background spacetime.

However, as proved by Shi-Tam [90], the definition does not yield a positive definite energy except for the time symmetric case ($K_{\mu\nu} = 0$). Also, the definition is gauge dependent. Yau and Liu [91–93] defined another formula for energy, which is gauge independent, and can be positive definite under certain conditions. However, our initial data is time symmetric which means the definition by Brown-York and Hawking-Horowitz is sufficient. Also, we can still use the definition by Brown-York and Hawking-Horowitz during evolution since the energy is characterized by the asymptotic behaviour at spatial infinities, which is not affected by finite time evolution (i.e., the local behaviour is not able to propagate to spatial infinities within a finite time evolution).

### 2.4.2 Total Energy in the Braneworld with Axisymmetry

We assume the energy (2.34) is well-defined in the braneworld. In the braneworld, the spacetime background is

$$\mathrm{d}\bar{s}^2 = \frac{\ell^2}{\bar{z}^2}\Big( - \mathrm{d}\bar{t}^2 + \mathrm{d}\bar{r}^2 + \mathrm{d}\bar{z}^2 + \bar{r}^2 \left(\mathrm{d}\bar{\theta}^2 + \sin^2\bar{\theta}\mathrm{d}\bar{\phi}^2\right) \Big), \qquad \text{where } \bar{z} \geq \ell. \tag{2.35}$$

A *bar* is used to denote the quantities associated with the background. For this background, the lapse $\bar{\alpha} = 1/\bar{z}$ and the shift $\bar{\beta}_\nu = 0$. Therefore, (2.34) reduces to

$$E = \lim_{Q\to\infty} \frac{1}{8\pi G_d} \oint_{S_t^Q} \bar{\alpha} \left( {}^{(d-2)}k - {}^{(d-2)}\bar{k} \right), \tag{2.36}$$

which will be the definition of the energy of the braneworld.

To calculate the energy, we need to set a closed surface family $S^Q$ that goes to spatial infinity as $Q \to \infty$. The two requirements on defining the family are: (i) $S^Q$ goes to spatial infinity as $Q \to \infty$, and (ii) the closed surface is smooth to a certain degree so that the extrinsic curvature ${}^{(d-2)}k_{\alpha\beta}$ is well-defined at any point on $S^Q$. In our case where the system has axisymmetry (spherical symmetry on the brane) with coordinates $(t, r, \theta, \phi, z)$ adapted to the symmetry, we may choose the closed surface as, for example, Fig. 2.2(a). Quantities $Q$, $u$ and $v$ are the parameters for defining the closed surfaces. Please refer to the capture of the figure for the details.

Without loss of generality, the spatial metric of any $t =$ constant slice can be

$$\mathrm{d}l^2 = \frac{\ell^2}{z^2}\left[e^{2A+2B}\left(\mathrm{d}r^2 + \mathrm{d}z^2\right) + e^{2A-2B} r^2 \left(\mathrm{d}\theta^2 + \sin^2\theta\mathrm{d}\phi^2\right)\right]. \tag{2.37}$$

The brane is located at $z = \ell$. This is the most general spatial metric for the axisymmetric configuration because, by taking the symmetry into account, the most general form can take

$$\mathrm{d}l^2 = \frac{\ell^2}{z^2}\left[\tilde{\eta}_{rr}\mathrm{d}r^2 + \tilde{\eta}_{zz}\mathrm{d}z^2 + 2\tilde{\eta}_{rz}\mathrm{d}r\mathrm{d}z + r^2\tilde{\eta}_{\theta\theta}\left(\mathrm{d}\theta^2 + \sin^2\theta\mathrm{d}\phi^2\right)\right]. \tag{2.38}$$

For a given $t =$ constant slice, everything only depends on $r$ and $z$, therefore $\tilde{\eta}_{rr}\mathrm{d}r^2 + \tilde{\eta}_{zz}\mathrm{d}z^2 + 2\tilde{\eta}_{rz}\mathrm{d}r\mathrm{d}z$ can transform into a conformally flat form. Lastly, the freedom in the conformal function, can be used to fix the brane at $z = \ell$ [14, 47].

To calculate the energy, we embed $S_t^Q$ into the background spacetime (2.35). In general, it is not guaranteed that such embedding is possible for an arbitrarily chosen closed surface, although it turns out all the closed surfaces considerred in this thesis could be embedded into the background spacetime. The embedding is a mapping from $S_t^Q$ in the physical spacetime to its image in the background spacetime by keeping the intrinsic geometry of the closed surface. i.e. the intrinsic geometry of the image of $S_t^Q$ in the background spacetime is the same as the intrinsic geometry of $S_t^Q$ in the physical spacetime. There is a freedom in this mapping (which is "where we put the image"), and we fix this freedom by naturally mapping the intersection of the closed surface with the brane in the physical spacetime to the brane of the background spacetime

$$\bar{z}|_{z=\ell} = \ell. \tag{2.39}$$

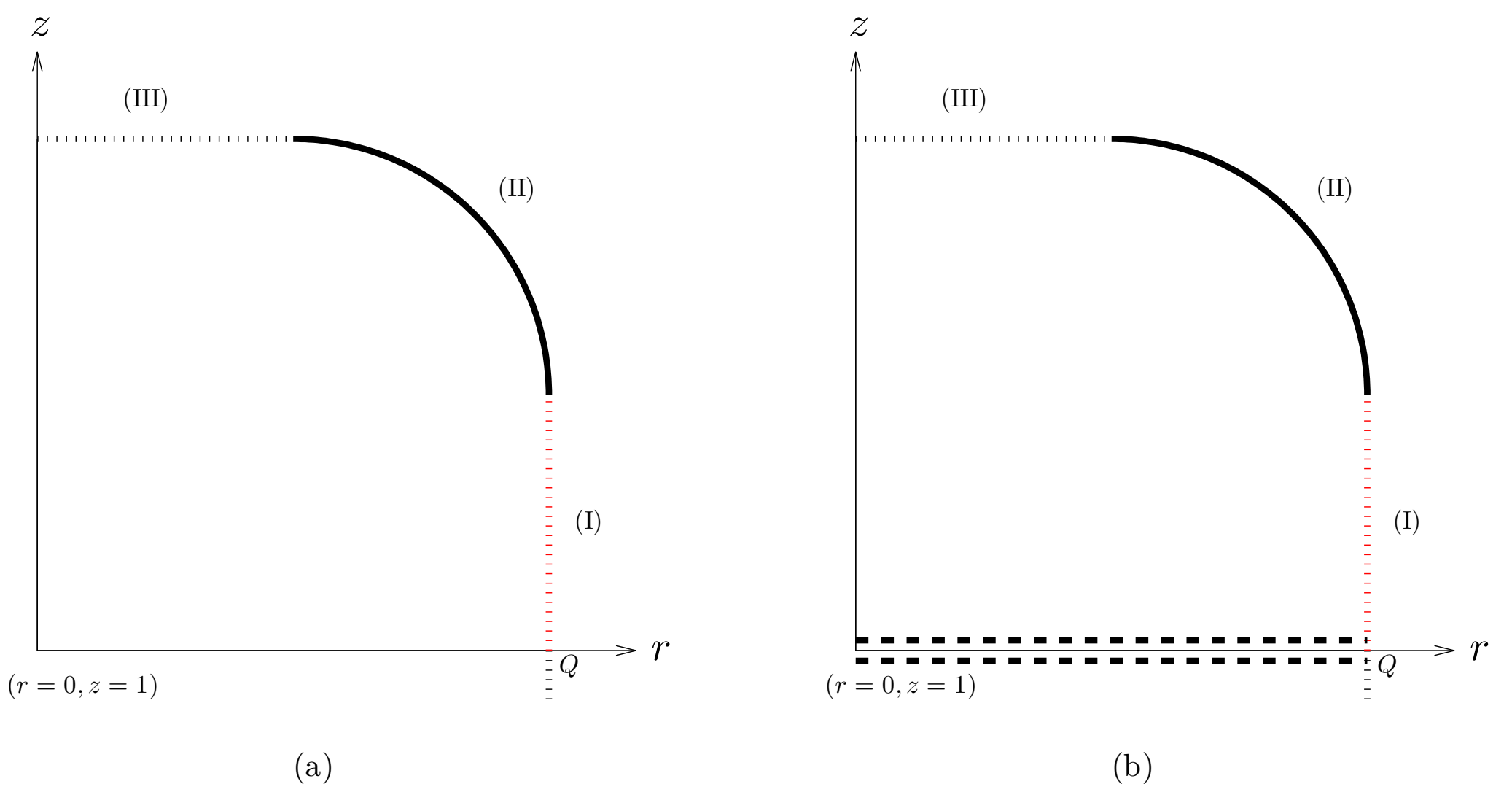

Figure 2.2: The closed surfaces to calculate total energy and energy in the bulk. Diagram (a) is used to calculate the total energy. $z = 1$ is where the brane is located. The system is spherically symmetric on the brane and $r$ is the radius of the spherical coordinate. The closed surface is composed of three segments: segment (I) starts at $r = Q$ and its coordinate length (measured by the coordinate $z$) is $v \cdot Q$. The length of segment (III) is $u \cdot Q$. Here $u \in (0, 1)$ and $v \in (0, 1)$ and their values are *fixed* for a specific closed surface family. Segment (II) is an arc (a quarter of a circle whose radius is $(1 - u) \cdot Q)$) to connect these two segments smoothly. The small segment below the brane is to show that there is another part below the brane which is related to the said part by $Z_2$ symmetry. Note, the closed surface should goes smoothly across the brane (i.e. the closed surface is perpendicular to the brane). In general case this perpendicular surface is not represented by $r =$ constant. When the coordinate gauge at the brane is taken to be perpendicular gauge (2.7) so that $g_{rz}\big|_{z=\ell} = 0$, this surface is represented by $r =$ constant. Diagram (b) is to add two dashed lines (along the brane) onto diagram (a). The use of (b) is going to be explained in Sec. 2.6.

For a closed surface in the physical space with metric (2.38), the embedding into the background

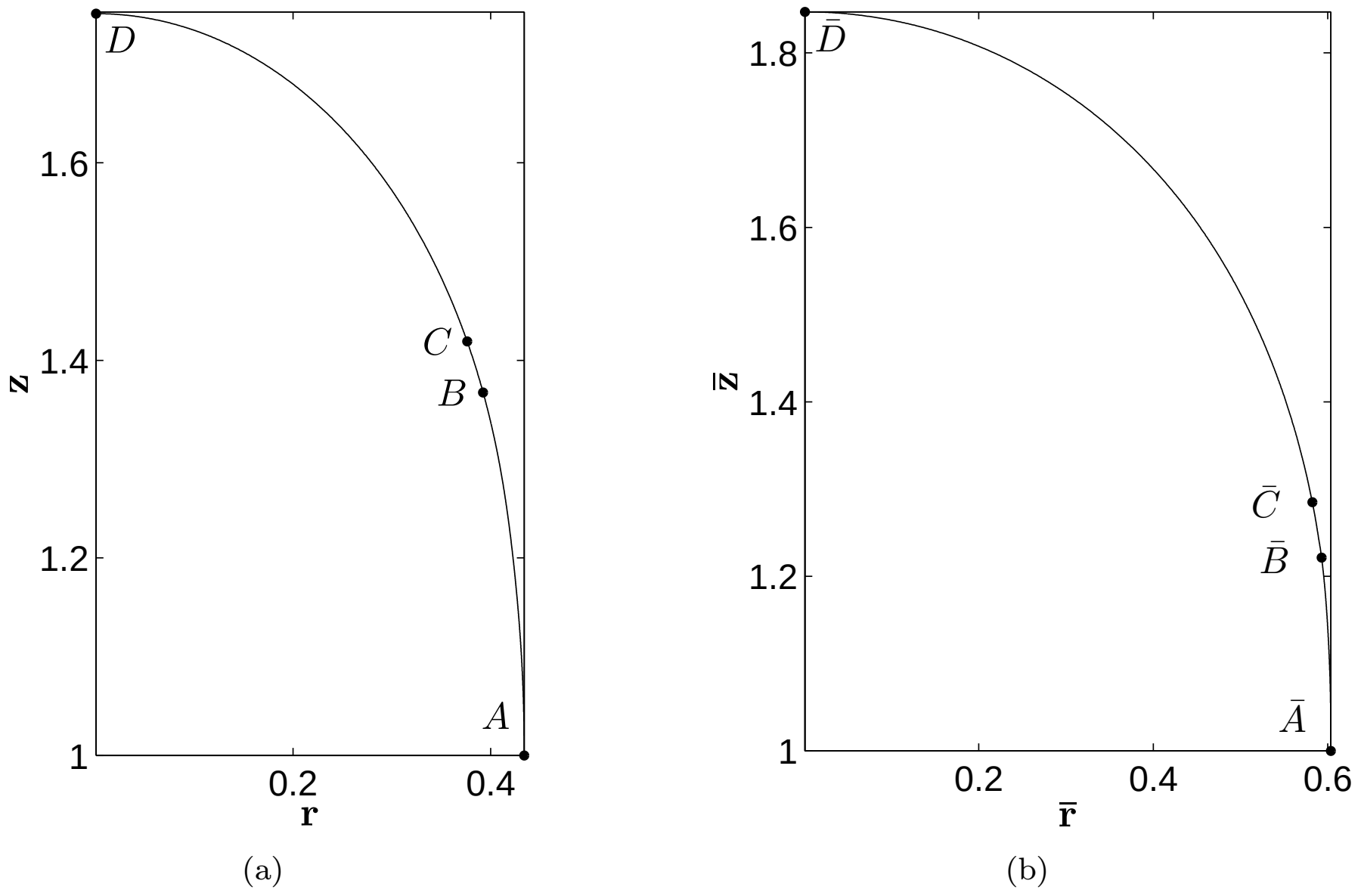

Figure 2.3: The embedding of a closed surface. Fig. (a) is the closed surface in the physical spacetime (2.38), and Fig. (b) shows its embedding into the background spacetime (2.35). The freedom of the embedding is fixed by mapping $A$ to $\bar{A}$.

space (2.35) is demonstrated by Fig. 2.3. The embedding boils down into the following two conditions:

(1) On the $r - z$ plane, an point $B$ (see Fig. 2.3) on the closed surface, represents a 2-sphere expanded by coordinates $(\theta, \phi)$, whose proper area is $4\pi r^2 \tilde{\eta}_{\theta\theta}/z^2$. The area of the 2-sphere at

$\bar{B}$ (the image of $B$ in the background spacetime) is $4\pi\bar{r}^2/\bar{z}^2$, therefore

$$\frac{r^2}{z^2}\tilde{\eta}_{\theta\theta} = \frac{\bar{r}^2}{\bar{z}^2}. \tag{2.40}$$

(2) $C$ is a point on the closed surface that is infinitesimally close to $B$. Let us denote the coordinates of $B$ by $(r, z)$, and denote the coordinates of $C$ by $(r + \mathrm{d}r, z + \mathrm{d}z)$. The square of the length of the infinitesimal line is $\left(\tilde{\eta}_{rr}\mathrm{d}r^2 + \tilde{\eta}_{zz}\mathrm{d}z^2 + 2\tilde{\eta}_{rz}\mathrm{d}r\mathrm{d}z\right)/z^2$. Correspondingly, the coordinates of their images ($\bar{B}$ and $\bar{C}$) are $(\bar{r}, \bar{z})$ and $(\bar{r} + \mathrm{d}\bar{r}, \bar{z} + \mathrm{d}\bar{z})$, and the square of the length of the infinitesimal line is $\left(\mathrm{d}\bar{r}^2 + \mathrm{d}\bar{z}^2\right)/\bar{z}^2$. The equality of these two lengths reads

$$\frac{1}{z^2}\left(\tilde{\eta}_{rr}\mathrm{d}r^2 + \tilde{\eta}_{zz}\mathrm{d}z^2 + 2\tilde{\eta}_{rz}\mathrm{d}r\mathrm{d}z\right) = \frac{1}{\bar{z}^2}\left(\mathrm{d}\bar{r}^2 + \mathrm{d}\bar{z}^2\right). \tag{2.41}$$

Here we emphasize that the infinitesimal line is taken *along the closed surfaces.* The condition "along the closed surface" defines how $\mathrm{d}r$ and $\mathrm{d}z$ are related (also defines how $\mathrm{d}\bar{r}$ and $\mathrm{d}\bar{z}$ are related).

The freedom related to "where to put the images" is fixed by mapping $A$ (the point of the closed surface on the brane) to $\bar{A}$ (the point of the closed surface on the brane).

For metric (2.37), these two conditions reduce to

$$\frac{r^2}{z^2}e^{2A-2B} = \frac{\bar{r}^2}{\bar{z}^2}, \tag{2.42}$$

$$\frac{1}{z^2}e^{2A+2B}\left(\mathrm{d}r^2 + \mathrm{d}z^2\right) = \frac{1}{\bar{z}^2}\left(\mathrm{d}\bar{r}^2 + \mathrm{d}\bar{z}^2\right). \tag{2.43}$$

For background metric (2.35), the lapse funtion to evaluate (2.34) is $\bar{\alpha} = \ell/\bar{z}$ and the shift function is $\bar{\beta}^\mu = 0$. $\oint_{S_t^Q}\left(\alpha \cdot {}^{(d-2)}k\right)$ diverges and the divergence is cancelled by the divergence in $\oint_{S_t^Q}\left(\alpha \cdot {}^{(d-2)}\bar{k}\right)$. Noticing the $\ell^2/z^2$ factor, we examine whether the contribution from segment (II) and segment (III) in Fig. 2.2(a), vanishes as $Q \to \infty$. In fact, one can show by direct calculation that $\oint_{S_t^Q}\left(\alpha \cdot {}^{(d-2)}k\right)$ from segment (II) (the arc) converges to zero as $1/Q$, as long as $A$ and $B$ converge to zero (at any rate) as $Q \to \infty$, so does $\oint_{S_t^Q}\left(\alpha \cdot {}^{(d-2)}\bar{k}\right)$. Therefore $\oint_{S_t^Q}\alpha\left({}^{(d-2)}k - {}^{(d-2)}\bar{k}\right)$, the contribution to the total energy, converge to zero at least as fast as $1/Q$. The same argument also applies to segment (III). Therefore, only the $r = Q$ segment contributes to the total energy. Or

> within the bulk, only the region $r \gg z$, contributes to the total energy.

Now we complete the proof by proving the claim "$\oint_{S_t^Q}\left(\alpha \cdot {}^{(d-2)}k\right)$ from segment (II) and segment (III) converges to zero as $1/Q$". The calculation on segment (III) is similar to the calculation on segment (II). Furthermore, acually segment (III) does not need to exist to define the closed surface, if we take $u = 0$. For conciseness, here we only present the calculation on segment (II) (the arc).

The metric is (2.37), where $r$ and $z$ along the arc can be parametrized as $r = uQ + \rho\cos\chi$ and $z = 1 + vQ + \rho\sin\chi$, where $\rho = (1-u)Q, \chi \in [0, \pi/2]$. A direct calculation gives

$$\begin{aligned}\lim_{Q\to\infty}\oint_{S_t^Q}\left(\alpha \cdot {}^{(d-2)}k\right) &= 8\pi \lim_{Q\to\infty}\int_0^{\pi/2} \mathrm{d}\chi\; e^{2A-2B}\frac{1}{z^3}\left(\xi_1+\xi_2\right) \\ &= 8\pi \lim_{Q\to\infty}\int_0^{\pi/2} \mathrm{d}\chi\; \frac{1}{z^3}\left(\xi_1+\xi_2\right). \end{aligned} \tag{2.44}$$

Here $\xi_1 \equiv -r^2 - 2r\rho\cos\chi + (3r^2\rho\sin\chi)/z$ and it is easy to numerically check

$$\int_0^{\pi/2} \mathrm{d}\chi\; (\xi_1/z^3) \;\propto 1/Q.$$

$\xi_2 \equiv r^2\rho\left(-3A_{,\rho} + B_{,\rho}\right)$, and $A_{,\rho}$ and $B_{,\rho}$ converge to zero at least as fast as $1/\rho$ since this corresponds to $A \sim \log\rho$ and $B \sim \log\rho$, whilst the reality is $A \to 0$ and $B \to 0$ when $\rho \to \infty$. Therefore

$$\int_0^{\pi/2} \mathrm{d}\chi\; (\xi_2/z^3)$$

converges to zero at least as fast as

$$\sim \int_0^{\pi/2} \mathrm{d}\chi\; (r^2/z^3) \propto 1/Q.$$

This completes the proof.

Another potential contribution to the total energy is the "cusp" of the closed surface across the brane, if there is any. Within the physical spacetime, we can properly choose the closed surface such that it crosses the brane smoothly. Under the coordinate gauge $g_{rz}\big|_{z=\ell} = 0$, the smoothness condition is equivalent to $\mathrm{d}r/\mathrm{d}z\big|_{z=\ell} = 0$, where the derivative is taken along the trajectory of the closed surface in the $r - z$ plane. This condition is satisfied by the surface specified by $r = Q$. Within the background spacetime, the smoothness condition is equivalent to $\mathrm{d}\bar{r}/\mathrm{d}\bar{z}\big|_{\bar{z}=\ell} = 0$. If we use $z$ as the parameter of the trajectory in the $\bar{r} - \bar{z}$ plane of the background spacetime, we have

$$\mathrm{d}\bar{r}/\mathrm{d}\bar{z} = \frac{\mathrm{d}\bar{r}/\mathrm{d}z}{\mathrm{d}\bar{z}/\mathrm{d}z},$$

where the derivatives are taken along the trajectory of the closed surface in the $\bar{r} - \bar{z}$ plane of the background spacetime. Therefore the condition $\mathrm{d}\bar{r}/\mathrm{d}\bar{z}\big|_{\bar{z}=\ell} = 0$ is just $\mathrm{d}\bar{r}/\mathrm{d}z\big|_{\bar{z}=\ell} = 0$. If this condition is met, there is no cusp (therefore no contribution to the total energy) across the brane. In fact, since we will take $Q \to \infty$ eventually, it is sufficient to examine whether this condition is satisfied by only keeping the lowest order of $A$ and $B$ in the expressions. However, condition $\mathrm{d}\bar{r}/\mathrm{d}z\big|_{\bar{z}=\ell} = 0$ is generally not gauranteed. A direct calculation from eq. (2.42) and eq. (2.43) yields

$$\mathrm{d}\bar{r}/\mathrm{d}z\big|_{\bar{z}=\ell} \approx r \cdot (A + A_{,z} + B - B_{,z})\big|_{z=\ell}, \tag{2.45}$$

where $\approx$ means this relation is true up to the lowest order in $A$ and $B$. In general, it is hard to say whether this term vanishes. However, in the conformally flat space that we are going to study in the next subsection (Sec. 2.4.3), the smoothness condition is indeed satisfied (see eq. (2.53)). Furthermore, we will obtain a relation between the total energy of the whole braneworld, and the ADM energy calculated based on the brane geometry.

### 2.4.3 Total Energy in Conformally Flat Space of the Braneworld

A general way to realize the embedding is to encode the two conditions (eq. (2.42) and (2.43)) into numerical methods. However, if we take into account the asymptotic behaviours of $A$ and $B$ in the following special case, the embedding and the calculation of total energy can be obtained analytically without using numerical methods.

We consider a special case by assuming $B \ll A$ in $r \gg z$ region [8]. In this region, the spatial metric reduces to

$$\mathrm{d}l^2 = \frac{\ell^2}{z^2} e^{2A} \left(\mathrm{d}r^2 + \mathrm{d}z^2 + \mathrm{d}\theta^2 + \sin^2\theta \mathrm{d}\phi^2\right), \tag{2.46}$$

for which the Hamiltonian constraint is (under the unit $\ell = 8\pi G_5 = 1$, which implies $8\pi G_4 = 1$ via (1.37))

$$(\partial_{rr} + \partial_{zz}) A + \frac{2}{z^2}\left(1 - e^{2A}\right) + \frac{2A_{,r}}{r} - \frac{2A_{,z}}{z} + (A_{,z})^2 + (A_{,r})^2 = 0, \tag{2.47}$$

and Israel's junction condition is

$$A_{,z}\big|_{z=\ell} = 1 - e^{A}. \tag{2.48}$$

In the $r \gg z$ region (which is the only region that contributes to the total energy as proved above),

[8] This condition was *observed* (i.e. a posterior result, rather than a prior assumption) in static star solutions [47]. This condition holds in a solution for small static BHs [14]. This condition is also compatible with Hamiltonian constraint [48]. This condition will also turn out to hold in one version of our initial data (see Chap. 4).

eq. (2.47) reduces to the following by taking only the linear term in $A$

$$(\partial_{rr} + \partial_{zz}) A - \frac{4A}{z^2} + \frac{2A_{,r}}{r} - \frac{2A_{,z}}{z} = 0, \tag{2.49}$$

with linearized Israel's condition as

$$A_{,z}\big|_{z=\ell} = -A. \tag{2.50}$$

The Hamiltonian equation (2.49) subject to boundary condition (2.50) can then have an solution with closed form

$$A \approx \frac{\alpha_1}{rz}, \tag{2.51}$$

here we use $\approx$ to emphasize that this solution is only true at $r \gg z$ region.

*Remark*: this solution satisfies the linearized Hamiltonian constraint subject to the linearized Israel's boundary condition, and the boundary condition at $r \to \infty$. However, there is no proof regarding uniqueness. Therefore, we should justify (2.51) is indeed a solution via the result obtained by numerical calculation, which is going to be carried out in Sec. 4.3.3.

In physical spacetime, the coordinates of the $r = Q$ surface are now $(z, \theta, \phi)$. We will find the metric of the corresponding surface in the background spacetime expressed under coordinates $(z, \theta, \phi)$ as well. The two conditions of the embedding are now

$$\frac{\bar{r}}{\bar{z}} = \frac{r}{z}e^{A}; \tag{2.52a}$$

$$\frac{(\bar{r}')^2 + (\bar{z}')^2}{\bar{z}^2} = \frac{e^{2A}}{z^2}. \tag{2.52b}$$

where $r = Q$ is a fixed value, and $'$ denotes $\mathrm{d}/\mathrm{d}z$. Eq. (2.52) is the ODE set that defines $\bar{r}(z)$ and $\bar{z}(z)$ for fixed $r$, subject to the "initial" value $\bar{z}\big|_{z=1} = 1$. By construction, the $r = Q$ surface in physical space and $r = Q$ surface in background space have the same intrinsic metric, and the coordinates of the surfaces are the same one: $(z, \theta, \phi)$. It is then straightforward to carry out (2.34). Note the shift functions are asymptotically zeros and the lapse function is asymptotically $1/\bar{z}$ for background (2.35).

The solution of the ODE set by keeping the lowest order in $A$ is

$$\bar{z} = z \cdot \exp\left[\frac{\alpha_1(z-1)}{rz}\right]; \ \bar{r} = r \cdot \exp\left(\frac{\alpha_1}{r}\right). \tag{2.53}$$

It is then a routine work to carry out the calculation of total energy using (2.34). Here we only list some important intermediate results

$$
\begin{aligned}
M_{\text{total}} &= \frac{1}{8\pi G_5} \lim_{Q\to\infty} \int_{S_t^Q} \alpha \left( {}^{(d-2)}k - {}^{(d-2)}\bar{k} \right) \\
&= \frac{1}{8\pi G_5} \cdot 2 \cdot 4\pi \cdot \lim_{Q\to\infty} \int_1^{vQ} \mathrm{d}z \; \mathcal{Q} \; \alpha \left( {}^{(d-2)}k - {}^{(d-2)}\bar{k} \right) \\
&= \frac{1}{8\pi G_5} \cdot 24\pi\alpha_1 \int_1^{\infty} \frac{\mathrm{d}z}{z^4} = \alpha_1/G_5. \qquad (2.54)
\end{aligned}
$$

The factor 2 in the second line is due to the $Z_2$ symmetry with respect to the brane in RSII; $4\pi$ is from the integration over $(\theta, \phi)$; $\mathcal{Q}$ is the determinant of the intrinsic metric of the surface: $\mathcal{Q} = e^{3A}r^2/z^3$; and we have used ${}^{(d-2)}k = -2z/r + 5\alpha_1/r^2$, ${}^{(d-2)}\bar{k} = -2z/r + 2\alpha_1/r^2$; $r = Q$; $\alpha = e^A/z$.

Also, from the expressions one see that, the result would have diverged as $\sim Q$, if the background term ${}^{(d-2)}\bar{k}$ is absent.

## 2.5 ADM "Mass" and Hawking "Mass" of the Brane

Due to the equivalence between mass and energy, we use these two words interchangeably.

As will be explained in Sec. 2.6, the masses calculated based on the brane geometry, are not really energies on the brane, from braneworld point of view. Actually they play no direct roles in braneworld, therefore we put "mass" in quotes in the title of this section. The purpose of studying these quantities is to compare the braneworld with GR to examine whether there are observational differences. In previous sections, the dimension of the spacetime is arbitrary. In this section, we only consider 3+1 dimensional spacetime (the brane).

We will first introduce ADM mass and Hawking mass in Sec. 2.5.1 and Sec. 2.5.2, then we will derive the ADM mass in the conformally flat space (2.46).

### 2.5.1 ADM Mass

The ADM mass is introduced in standard texts [29, 32, 36]. One way to obtain it is to use (2.34) by setting the lapse $\alpha = 1$ and the shift $\beta^\mu = 0$ [36]. Note the ADM mass is only defined when the background is asymptotically flat. Applying (2.34) to a asymptotically flat spacetime, the ADM mass reduces to [36]

$$
M_{\text{ADM}} = \frac{1}{16\pi G_4} \lim_{Q\to\infty} \oint_{S_t^Q} \left[ \bar{\mathscr{D}}^i q_{ij} - \bar{\mathscr{D}}_i \left( \bar{q}^{kl} q_{kl} \right) \right] s^i, \qquad (2.55)
$$

where $q_{ij}$ is the spatial metric, and $i, j = 1, 2, 3$ is the spatial coordinate index. $\bar{q}_{ij}$ is the (flat) background metric, and $\bar{\mathscr{D}}$ is the covariant derivative associated with $\bar{q}_{ij}$. $s^i$ is the unit normal vector of $S_t^Q$ pointing outwards, which was denoted as $Q^\alpha$ in (2.34).

Note, only the spatial metric is needed in the definition.

For the spherical symmetric case, the metric can always be rewritten as

$$\mathrm{d}l^2 = g_{rr}\mathrm{d}r^2 + r^2\left(\mathrm{d}\theta^2 + \sin^2\theta\mathrm{d}\phi^2\right), \tag{2.56}$$

in spherical coordinate $(r, \theta, \phi)$, where $r$ is the areal radius. In this situation the ADM mass can be derived to be

$$M_{\mathrm{ADM}} = \lim_{r\to\infty} M_{\mathrm{adm}}, \tag{2.57}$$

where

$$M_{\mathrm{adm}} = \frac{r}{2G_4}\frac{(g_{rr}-1)}{\sqrt{g_{rr}}}, \tag{2.58}$$

which is defined as the integrand in (2.55). In the special case of Schwarzschild metric where $g_{rr} = (1 - 2mG_4/r)^{-1}$, $M_{\mathrm{adm}} = m/\sqrt{1 - 2mG_4/r}$, which has $r$ dependence. Within the horizon, $2mG_4/r > 1$, therefore this quantity is not well-defined within the horizon.

### 2.5.2 Hawking Mass

The Hawking mass is defined on two dimensional surfaces $S$ with spherical topology, and it characterizes the mass in the space enclosed by the surface. The Hawking mass is described in terms of spin-coefficient formalism, for which please refer to [83, 85, 86]. In terms of spin-coefficients, the Hawking mass is defined as

$$M_{\mathrm{H}} = \sqrt{\frac{\mathcal{A}}{16\pi G_4^2}}\left(1 + \frac{1}{2\pi}\oint_S \rho\rho'\mathrm{d}S\right), \tag{2.59}$$

where $\mathcal{A}$ is the area of $S$. The $\rho$ and $\rho'$ are two of the spin-coefficients

$$\rho = \frac{1}{2\sqrt{2}}m^{\alpha\beta}\nabla_\alpha\left(n_\beta + s_\beta\right) = \frac{1}{2\sqrt{2}}\Theta_+, \tag{2.60a}$$

$$\rho' = \frac{1}{2\sqrt{2}}m^{\alpha\beta}\nabla_\alpha\left(n_\beta - s_\beta\right) = \frac{1}{2\sqrt{2}}\Theta_-, \tag{2.60b}$$

where $\Theta_\pm$ are defined in (2.13). Substituting (2.60a) and (2.60b) back into (2.59), we get

$$M_{\mathrm{H}} = \sqrt{\frac{\mathcal{A}}{16\pi G_4^2}} \left(1 + \frac{1}{16\pi} \oint_S \Theta_+ \Theta_- \mathrm{d}S\right). \tag{2.61}$$

At an apparent horizon $\Theta_+ = 0$, therefore the following mass-area relation holds at the apparent horizon

$$M_{\mathrm{H}} = \sqrt{\frac{\mathcal{A}}{16\pi G_4^2}}. \tag{2.62}$$

In the case of spherical symmetric metric (2.56), the Hawking mass is calculated as

$$M_{\mathrm{H}} = \frac{r}{2G_4}\left(1 - (g_{rr})^{-1}\right), \tag{2.63}$$

where we have used

$$\Theta_+ = -\Theta_- = \frac{2}{r}\sqrt{(g_{rr})^{-1}},$$

which are directly calculated by applying the definitions (2.13) to metric (2.56). However, the calculation of $\Theta$'s depends on the time components of the metric too. This result is only valid in case the configuration is time symmetric (where the extrinsic curvature of $t =$ constant hypersurface is zero). This motivates us to give (2.63) a new notation $M_{\mathrm{h}}$, which agrees with $M_{\mathrm{H}}$ in the time symmetric case.

In the special case of Schwarzschild metric where $g_{rr} = (1 - 2mG_4/r)^{-1}$, $M_{\mathrm{H}} = m$ and has no $r$ dependence.

### 2.5.3 The ADM "Mass" of the Conformally Flat Space

Let us now apply the ADM mass or Hawking mass for the conformally flat space (2.46). First let us rewrite it in terms of the areal radius $\tilde{r}$, which is defined as the radius associated with the area of the $r =$ constant surface. Therefore

$$\tilde{r} = re^A. \tag{2.64}$$

Rewrite (2.46) in terms of $\tilde{r}$, we have

$$\mathrm{d}l^2_{\mathrm{brane}} = (1 + r\mathrm{d}A/\mathrm{d}r)^{-2}\,\mathrm{d}\tilde{r}^2 + \tilde{r}^2\left(\mathrm{d}\theta^2 + \sin^2\theta \mathrm{d}\phi^2\right), \tag{2.65}$$

which is the spatial metric on the brane. Substitute in the asymptotic behaviour of $A$ at $z = 1, r \gg 1$: $A \approx \alpha_1/rz = \alpha_1/r$, we have

$$\mathrm{d}l^2_{\text{brane}} = (1 - \alpha_1/r)^{-2}\, \mathrm{d}\tilde{r}^2 + \tilde{r}^2 \left(\mathrm{d}\theta^2 + \sin^2\theta \mathrm{d}\phi^2\right).$$

Substitute this into (2.58) or (2.63), we obtain the brane ADM mass as

$$M_{\text{braneADM}} = \alpha_1/G_4. \tag{2.66}$$

Comparing this with eq. (2.54), we obtain a somewhat surprising result

$$\boxed{M_{\text{total}} = (G_4/G_5) \cdot M_{\text{braneADM}}.} \tag{2.67}$$

## 2.6 A Quest for Brane Energy

There is energy exchange between the bulk and the brane. To quantitatively describe the energy exchange, we need to define the energy in the bulk, and the energy on the brane. Basically only the energy in the bulk, or the energy on the brane, is sufficient since the other can be defined by subtraction from the total energy.

To serve as the energy of the brane, an expression should satisfy:

(1) its value is a part of the total energy. This rules out the ADM mass defined based on the brane geometry since it is equal to the total energy for a class of space configurations. Please refer to eq. (2.67).

(2) it is *not* conserved during the evolutions, because of the energy exchange between the brane and the bulk. This requirement rules out any quasi-local definition on the brane, such as the ADM mass and Hawking mass evaluated from brane geometry. This is because, the quasi-local energies are defined as the limit at spatial infinities utilizing the time translational symmetry at spatial infinities, therefore conserved.

The definition of an energy, especially that for the brane, is subtle and this section is to request the study of energy on the brane, rather than providing a solution. The reader can *skip this section from here* since the following is an attempt (that has conceptual issue) rather than a result, although the attempt turns out to have surprisingly good features exhibiting in the simulations we carried out.

We would like to search the definition from the Hamiltonian-Jacobi analysis described above. Since the energy of a certain region is defined as an integration over the boundary enclosing the region, if this concept could be generalized (which actually can not), we can *tentatively* define the energy of the bulk as the integration of (2.34) over the closed surface of Fig. 2.2(b). i.e. we enclose it along the brane. The difference, which is the integration of (2.34) *along the brane*, can be accordingly defined as *the energy on the brane.*

However, without going into the details, we know this tentative definition is problematic for the following reasons. There is a sharp corner in the surface, where the extrinsic curvature is infinite and the integration of the extrinsic curvature over the corner can be indefinite. In the background spacetime, integrating along the brane makes sense physically. There is a requirement, however, that the surface is embedded into the background so the intrinsic metric of the surface stays the same, which can not be met for any non-trivial brane. Or alternatively, we keep the "embedding" requirement, but then we have to give up the "integrating along the brane" in background spacetime (which also means we give up the freedom fixing condition (2.39).). Also, since this is a definition regarding an integration on the brane, which will fail when there is a physical singularity on the brane. Therefore this definition can not be treated seriously. It is only a (very) rough description. We hope the resulting quantity changes monotonically with the amount of the energy exchanged between the brane and the bulk.

In the simulation (presented in Chap. 5), we will choose the brane to be the surface in the background spacetime. i.e. we give up the requirement that the intrinsic metric of the surfaces in physical spacetime and background spacetime are the same. (2.42) and (2.43) are the two requirements for embedding, and we have to keep one of these to make the definition unambiguous. Here we choose to keep (2.42) because it is simpler and also because the areal radius still makes sense when there is a physical singularity (where condition (2.43) fails).

Considering requirement (2.42), it is natural to express the energy in terms of spherical coordinate $(\tilde{r}, \theta, \phi)$ with $\tilde{r}$ as the areal radius, where the physical metric is $\mathrm{d}l^2 = g_{\tilde{r}\tilde{r}}\mathrm{d}\tilde{r}^2 + \tilde{r}^2\left(\mathrm{d}\theta^2 + \sin^2\theta \mathrm{d}\phi^2\right)$. Since this definition only includes the integration on the brane, we may rewrite it as

$$E_{\text{brane}} = \frac{1}{8\pi G_5}\int_B \alpha\left({}^{(d-2)}k - {}^{(d-2)}\bar{k}\right), \tag{2.68}$$

$$= \lim_{Q\to\infty}\frac{1}{8\pi G_5}\int_0^Q \mathrm{d}\tilde{r}\int_0^\pi \mathrm{d}\theta\int_0^{2\pi}\mathrm{d}\phi\,\sqrt{\det[q]}\,\alpha\left({}^{(d-2)}k - {}^{(d-2)}\bar{k}\right). \tag{2.69}$$

where $\tilde{r} = \bar{r}$ is the areal radius and $\bar{r}$ is the areal radius in the background space. $q$ is the spatial metric of the brane (in physical spacetime) and we have $\sqrt{\det[q]} = \tilde{r}^2\sin\theta\sqrt{g_{\tilde{r}\tilde{r}}}$. $B$ stands for the spatial part of brane manifold (in physical spacetime). Please note it is $G_5$ (rather than $G_4$) in the expression, and $\alpha$ is the lapse in 5D restricted on the brane, rather than the lapse function calculated from brane geometry. ${}^{(d-2)}k$ is the extrinsic curvature of the 3-brane embedded in 4D *spatial* bulk. Note a crucial feature of this energy is that, it is an integration over $B$, and its meaning is the energy over $B$. i.e. it might be possible to think of energy density, which is actually (spatial) coordinate independent on the brane. Noticing the second requirement of the energy on the brane (that the definition should not be semi-local), this definition might have captured a key feature.

# Chapter 3

# Axisymmetric Spacetime with Non-Flat Background

The cost of numerical computation increases dramatically with the number of dimensions. This is usually called *the curse of dimensionality.* To date, it is impractical to directly perform numerical calculations in $4+1$ dimension at a reasonable resolution, or study problems with the demand of high resolution (such as critical phenomena) in more than (effectively) $2+1$ dimension. Therefore, one of the first steps to study a physical system in higher dimensional spacetime, is to consider the system with some kind of symmetry, such as spherical symmetry and axisymmetry, to reduce the effective dimension.

However, in many situations, singularities, instabilities, unavoidable noises or other irregularities, arise in the numerical calculations performed under the coordinates adapted to the symmetry, either at the origin or at the axis of the symmetry ([50] and references therein). On the other hand, this kind of issue does not occur in the simulations of the same system performed under Cartesian coordinates. This kind of behavior is called the regularity problem in non-Cartesian coordinates. To make the discussion and presentation smoother, we will only mention axisymmetry below, which also applies to spherical symmetry.

Where does the irregularity come from? Since the irregularity appears in cylindrical coordinates or spherical coordinates rather than Cartesian coordinates, it is widely believed that the irregularity comes from *coordinate system* choice ([50] and references therein). The terms involving $1/r^n$ (where $r$ symbolically stands for the radius coordinate in spherical coordinates or cylindrical coordinates) often appears in the equations, and it is widely believed that, with or without machine precision playing a role, these terms are responsible for the irregularity [50].

What we do differently in this thesis is to make distinction between the *coordinate system*, and the *fundamental variables*—the unknown functions to be solved for—used in simulation (which are usually the components of tensors and pseudo-tensors), and reveal that the fundamental variables (rather than the coordinate system, or the operators in the coordinate system) are responsible for irregularities. In particular, neither $1/r^n$-terms nor machine precision, is relevant to regularity problem. Therefore the regular results obtained from simulations performed in Cartesian coordinates, are actually due to the fact that the tensor (and pseudo-tensor) components are regular in Cartesian coordinates, rather than the regularity of Cartesian coordinates itself. As one will see in this chapter, actually it does not make sense to talk about the regularity of a coordinate system. The key to avoid irregularity issue in any coordinate system, is to construct tensors (and pseudo-tensors) in terms of regular variables that are compatible with the finite difference approximation (FDA) scheme at the vicinity of the axis (or the origin). There are many ways to express tensors (and pseudo-tensors) in terms of regular variables. To embody the construction, we present a general method to construct regular variables out of Cartesian components (which play the role of regular variables). The method can, for example, enable the generalized harmonic formalism to be used in non-Cartesian coordinates. Then we analyze why certain other formalisms in the literature can avoid regularity issue as well.

Utilizing the knowledge obtained in studying regularity problem, the evolution schemes such as the generalized harmonic formalism and the BSSN formalism of GR can be rewritten under general coordinates which overcome the irregularity issue. On the other hand, there is another problem associated with the braneworld: the asymptotic spacetime background of braneworld is not flat, while the existing knowledge in the literature are only for asymptotically flat spacetimes. It is not clear how to setup the source functions (of the GH formalism) in the braneworld. The second part of this chapter is devoted to solving the non-flat background problem.

## 3.1 Regularity Problem and Our Conjecture

Very often (but not always), the numerical calculations performed in cylindrical coordinates adapted to axisymmetry, generate irregularities in the vicinity of the symmetry axis, but the same calculations performed in Cartesian coordinates do not generate irregularities. This phenomena is called the regularity problem. The irregularity, can appear as: (i) singularity in certain fundamental variables used in numerical calculations in which situation the code would crash and the fundamental variables would not converge, or (ii) non-smoothness of certain fundamental variables, or (iii) smooth fundamental variables which do not converge at the expected order — i.e. they can not pass the independent residual tests.

In this section, we will first introduce the existing analysis of regularity problem, then identify some of the existing methods that can overcome regularity problem, then we will declare a conjec-ture regarding the true source of the problem. We back up the conjecture by a detailed analysis of the deviation between the numerical results and the analytical results, which reveals the second key element to yield regular results.

### 3.1.1 The Existing Analysis and The Existing Solutions

Let us take a specific example as the carrier of the existing analysis: the wave equation of a scalar field $\Phi$ in 3+1 (flat) spacetime with axisymmetry. $(x, y, z)$ are the Cartesian coordinates, and $(\rho, \phi, z)$ are the cylindrical coordinates. The equation of motion is

$$\left(\nabla^2 - \partial_{tt}\right)\Phi = \left(\partial_{xx} + \partial_{yy} + \partial_{zz} - \partial_{tt}\right)\Phi = \left(\partial_{\rho\rho} + \frac{1}{\rho}\partial_\rho + \partial_{zz} - \partial_{tt}\right)\Phi, \tag{3.1}$$

where we have applied the fact that the space is axisymmetric so that the $\phi$-derivatives are zeros. It is "obvious" that the term $(1/\rho)\partial_\rho\Phi$ *was* singular and it was widely believed ([50] and references therein) that terms like this were responsible for regularity problems.

Based on the analysis, the crucial step to cure irregularity was to modify the differential operators so that terms like $1/r^n$ did not appears. There are quite a few widely-used techniques to cure the regularity problem. Among these techniques, a class of methods called Cartoon methods [50, 61], utilize the observation that "there is no regularity issue in Cartesian coordinates" to modify the operators in a way so that the operators are effectively Cartesian.

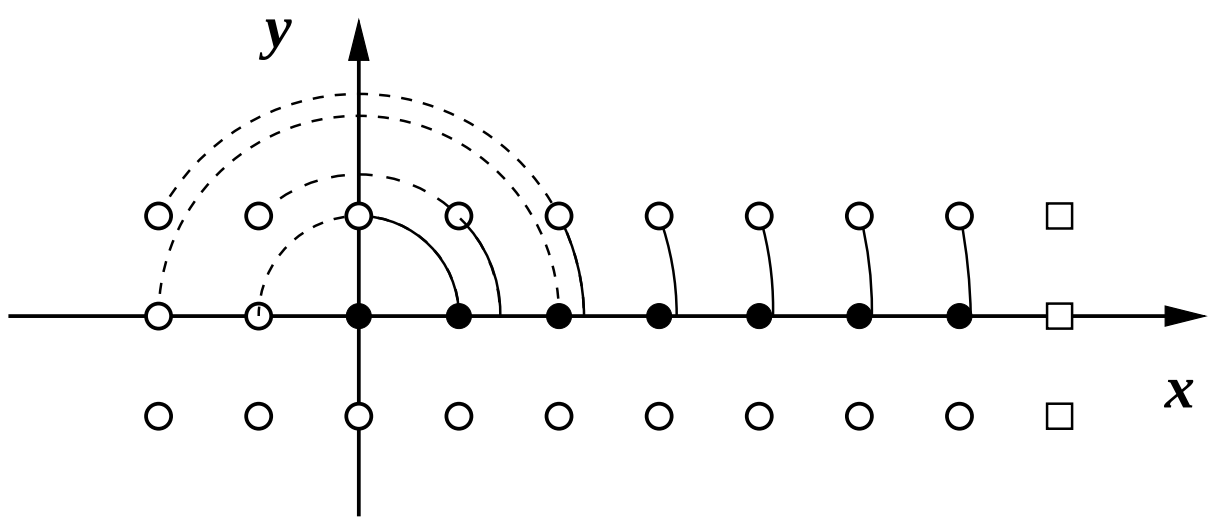

Figure 3.1: The demonstration of the cartoon method. The system is axisymmetric with respect to $z$ axis. For stencils which involve three grid points, only three slices are needed: the $y = 0$ slice, the one above it and the one below it. Note, this diagram is from [50].

#### The Original Cartoon Method

The original cartoon method [50] was proposed by Alcubirre et al in 1999. Here I will again use a scalar field $\Phi$ to demonstrate the basic ideas. The physical space is three dimensional space, with $(x, y, z)$ being the Cartesian coordinates, and the system is axisymmetric with respect to $z$ axis. Analytically, only one slice, say the $y = 0$ slice, is needed to include all the information because of the axisymmetry. Numerically, the cartoon method is trying to use only $y = 0$ slice as well. The method can be implemented via the following four steps.

(1) Write the code using equations expressed in Cartesian coordinates.

(2) Use only the $y = 0$ slice. Of course, this is impossible in Cartesian coordinates, since the discretization of derivatives with respect to $y$ needs more than one slices in the $y$ direction. To be specific, let us assume the equations consist of second order derivatives at most, and the discretization is of the second order accuracy. In this situation, the stencil only needs three grid points to do the finite difference approximation for the differential operators (Fig. 3.1). However, the function values on the upper and the lower slices are not known.

(3) Obtain the function values on the upper and the lower slices by numerical interpolation constructed from the values within $y = 0$ slice, utilizing the axisymmetry. The process is shown in Fig. 3.1.

(4) Replace the function values on the upper and the lower slices, by the interpolated values obtained in step (3), and then substitute into the discretization.

From these four steps, operators such as $1/r^n$ are avoided, and only Cartesian operators are directly used.

**The Lie Derivative Cartoon Method**

In the third step of the original cartoon method, in order to get derivatives with respect to $y$, numerical interpolation is used. The Lie derivative cartoon method [61] improves this, by analytically replacing these derivatives by those within the $y = 0$ slice. For the particular example, using the fact $x\partial_y - y\partial_x$ being the Killing vector, we have

$$\partial_y \Phi = \frac{y}{x}\partial_x \Phi.$$

Taking derivatives with respect to $y$ on both sides, we have

$$\partial_{yy}\Phi = \frac{1}{x}\partial_x \Phi + \frac{y}{x}\partial_{xy}\Phi.$$

The laplacian $\nabla^2\Phi$ can then be evaluated at the $y=0$ plane as

$$\nabla^2\Phi = \left(\partial_{xx} + \partial_{yy} + \partial_{zz}\right)\Phi = \left(\partial_{xx} + \frac{1}{x}\partial_x + \frac{y}{x}\partial_{xy} + \partial_{zz}\right)\Phi\bigg|_{y=0} = \left(\partial_{xx} + \frac{1}{x}\partial_x + \partial_{zz}\right)\Phi. \quad (3.2)$$

The derivatives in $y$ direction are replaced by the derivatives within the $y=0$ slice, on the analytical level (rather than numerical level in the original cartoon method).

**Background Removal Method**

The cartoon method is a general approach to solve regularity problem, which can be used in *any* formalism of GR. On the other hand, Brown [97] and Gourgoulhon [36] developed a method to rewrite the BSSN and generalized harmonic formalisms, by a background removal method, which will be described in detail in the following sections. The same method was also applied in [21] to solve static problems using Ricci-DeTurck flow methods. The simulation using this rewritten BSSN formalism in spherical coordinates "turned out to be" regular [56]. However, the authors did not analyze why this method is regular. Sorkin-Choptuik [54, 55] used a different method to remove the background effect, which again "turned out to be" regular, without analyzing the reason.

### 3.1.2 The Conjecture: Variables Rather Than Coordinates

Various other attempts on the solution to regularity problem had been proposed (see [50] and references therein) and they were not very successful until the cartoon method appeared. The cartoon method is a general method that solves the regularity problem, and there exist other methods which overcome the regularity problem as well[36, 71, 97]. Here we will take a closer look and provide a deeper understanding of the topic. The understanding can serve as a criteria and a guideline to develop regularized formalisms, and can potentially completely solve the regularity problem associated with coordinates.

We start from the facts rather than speculations:

(1) The Lie derivative cartoon method works well [61–63].

(2) $1/x$ still appears in eq. (3.2) which was produced based on Lie derivative cartoon method.

(3) Eq. (3.2) is the same as (3.1) (equation in cylindrical coordinates), if $x$ is identified with $\rho$.

(4) Certain background removal methods (performed in *non-Cartesian coordinates*) [36, 54, 55, 97] are free from regularity issues.

(5) There exist many successful simulations with $1/r^n$ terms (many references, e.g: [59, 61–63, 71]).

(6) All in all, a multiplication by $1/r^n$ terms does not change the numerical feature. $(\rho \cdot \partial_{\rho\rho} + \partial_\rho)\,\Phi$ is not problematic, therefore $(\partial_{\rho\rho} + \partial_\rho/\rho)\,\Phi$ should be as safe.

These facts lead me to ask whether the spherical/cylindrical coordinates were the sources of the problem, and ask the question: does the regularity issue come from the terms $(1/r^n)$ in the operators associated with the coordinate system, or somewhere else such as the fundamental variables being used in numerical simulation?

In fact, if we take a closer look at the difference between the past simulations in cylindrical coordinates and those in Cartesian coordinates, the situation is either to use cylindrical components (of tensors and pseudo-tensors) under cylindrical coordinate, or to use Cartesian components under Cartesian coordinates. i.e. whenever one switches from cylindrical coordinates to Cartesian coordinates, he *also* "naturally" switches the fundamental variables from cylindrical components to Cartesian components. i.e. he has performs two changes: change of coordinates, and change of fundamental variables. It is not clear whether the regular results obtained from simulations under Cartesian coordinates, are due to the change of coordinates, or due to the change of fundamental variables. All the previous studies, including those yielding regular results, did not make the distinction between the effects from coordinates and the effects from fundamental varibles. Here we will make the distinction and make the conjecture: The regularity issue come from the *fundamental variables* used in simulation, rather than the *operators* associated with coordinate systems.

According to the conjecture, the fact that simulations in Cartesian coordinates with Cartesian components are regular, is because the Cartesian components (of tensors and pseudo-tensors) are regular, rather than "Cartesian coordinates are regular", or "operators in Cartesian coordinates are regular" [9]. According to the conjecture, the key to avoid regularity issue, is to express various tensors and pseudo-tensors in the equations in terms of regular functions[10]. In the following, we will first construct regular functions out of Cartesian components, then construct an example as a direct test to our conjecture.

[9] According to the conjecture, the fact that the cartoon methods produce regular simulations, is because the Cartesian components (of tensors and pseudo-tensors) are used as fundamental variables, rather than that the operators are made effectively Cartesian.

[10] According to the conjecture, eq. (3.1) should be free from regularity problem. The simulation confirms this claim.

### 3.1.3 Cartesian Components Method

It is fairly easy to express a tensor in terms of regular components and there are multiple ways to achieve this. As a specific example, here we adopt a "canonical" approach: since the Cartesian components are regular, one way to express tensors in terms of regular components, is to rewrite tensors in cylindrical coordinates in terms of the (regular) Cartesian components, using the basic tensor transformation relation. Since there is a geometrical relation between Cartesian coordinates and cylindrical coordinates, we will do the coordinate transformation at certain region of the space, such that the radius coordinate $\rho$ (or $r$) coincides with one of the Cartesian coordinates (refer to Fact 3 in Lie derivative cartoon method above), and the *functional forms* in the cylindrical coordinates, are the same as that in the Cartesian coordinates. Let us call this procedure as Cartesian component method.

To clarify the concept and demonstrate the usage, let us take metric function in $3+1$ dimension as an example. We use indices "cylin" and "Cart" to indicate cylindrical and Cartesian, respectively. The procedure to express tensors in terms of their Cartesian components in cylindrical coordinates, comprises the following four steps.

(1) write down the most general form of the metric (according to axisymmetry) in cylindrical coordinates $(t, \rho, \phi, z)$

$$g^{(\mathrm{cylin})} = \begin{pmatrix} g_{tt}^{(\mathrm{cylin})} & g_{t\rho}^{(\mathrm{cylin})} & 0 & g_{tz}^{(\mathrm{cylin})} \\ g_{t\rho}^{(\mathrm{cylin})} & g_{\rho\rho}^{(\mathrm{cylin})} & 0 & g_{\rho z}^{(\mathrm{cylin})} \\ 0 & 0 & g_{\phi\phi}^{(\mathrm{cylin})} & 0 \\ g_{tz}^{(\mathrm{cylin})} & g_{\rho z}^{(\mathrm{cylin})} & 0 & g_{zz}^{(\mathrm{cylin})} \end{pmatrix}, \tag{3.3}$$

(2) apply coordinate transformation

$$g_{\mu\nu}^{(y)} = \frac{\partial x^\alpha}{\partial y^\mu}\frac{\partial x^\beta}{\partial y^\nu} g_{\alpha\beta}^{(x)}, \tag{3.4}$$

to transform this metric into its Cartesian coordinates $(t, x, y, z)$ at the location $(y = 0, x \geq 0)$, which is $\phi = 0$ half plane in cylindrical coordinates, where the positive half of the $x$ axis *coincides* with $\rho$. Now the metric in Cartesian coordinates at $(y = 0, x \geq 0)$ is

$$g^{(\mathrm{Cart})} \equiv \begin{pmatrix} \eta_{tt} & \eta_{tx} & 0 & \eta_{tz} \\ \eta_{tx} & \eta_{xx} & 0 & \eta_{xz} \\ 0 & 0 & \eta_{yy} & 0 \\ \eta_{tz} & \eta_{xz} & 0 & \eta_{zz} \end{pmatrix} \tag{3.5}$$

$$= \begin{pmatrix} g_{tt}^{(\mathrm{cylin})} & g_{t\rho}^{(\mathrm{cylin})} & 0 & g_{tz}^{(\mathrm{cylin})} \\ g_{t\rho}^{(\mathrm{cylin})} & g_{\rho\rho}^{(\mathrm{cylin})} & 0 & g_{\rho z}^{(\mathrm{cylin})} \\ 0 & 0 & g_{\phi\phi}^{(\mathrm{cylin})}/x^2 & 0 \\ g_{tz}^{(\mathrm{cylin})} & g_{\rho z}^{(\mathrm{cylin})} & 0 & g_{zz}^{(\mathrm{cylin})} \end{pmatrix}, \tag{3.6}$$

where the first matrix is defined by

$$\eta_{\mu\nu}(t, x, z) \equiv g_{\mu\nu}^{(\mathrm{Cart})}(t, x, y, z)\Big|_{y=0, x\geq 0},$$

which is merely the restriction of Cartesian components onto $(y = 0, x \geq 0)$ half plane. The second matrix is the calculation result of (3.3) via coordinate transformation relation (3.4).

(3) compare the components in (3.5) and (3.6), and rewrite the tensor in terms of Cartesian components. Take the $tt$ component as an example: the above relation tells us that $\eta_{tt}(t, x, z) = g_{tt}^{(\mathrm{cylin})}(t, \rho, z)$ for *all* $\rho = x \geq 0$, therefore $\eta_{tt}$ and $g_{tt}^{(\mathrm{cylin})}$ have the same value at every $(t, \rho, z)$, which means they have the same *function form* in terms of $x$ (and $\rho$). Since $g_{\mu\nu}^{(\mathrm{Cart})}(t, x, y, z)$ is regular, its restriction to $(y = 0, x \geq 0)$, $\eta_{tt}(t, x, z)$, must be regular, which means $g_{tt}^{(\mathrm{cylin})}(t, \rho, z)$ is regular in cylindrical coordinates. Actually all functions are of the same form as the Cartesian components, except for $\phi\phi$ component, for which we have $g_{\phi\phi}^{(\mathrm{cylin})}/x^2 = \eta_{yy}$ (and remember, $x = \rho$), and this relation suggests to rewrite the $\phi\phi$ component as $\rho^2\eta_{yy}$.

(4) Finally we assemble components together to express the metric in cylindrical coordinates in terms of Cartesian components. The metric in cylindrical coordinates can be written as

$$g^{(\mathrm{cylin})} = \begin{pmatrix} \eta_{tt} & \eta_{t\rho} & 0 & \eta_{tz} \\ \eta_{t\rho} & \eta_{\rho\rho} & 0 & \eta_{\rho z} \\ 0 & 0 & \rho^2\eta_{\phi\phi} & 0 \\ \eta_{tz} & \eta_{\rho z} & 0 & \eta_{zz} \end{pmatrix}, \tag{3.7}$$

which is the same as

$$\begin{pmatrix} \eta_{tt} & \eta_{tx} & 0 & \eta_{tz} \\ \eta_{tx} & \eta_{xx} & 0 & \eta_{xz} \\ 0 & 0 & \rho^2\eta_{yy} & 0 \\ \eta_{tz} & \eta_{xz} & 0 & \eta_{zz} \end{pmatrix}, \tag{3.8}$$

with the subindices renamed. In the future we will denote this process roughly as

$$\eta_{\mu\nu}(t,\rho,z) \equiv g^{(\mathrm{Cart})}_{\mu\nu}\Big|_{y=0=\phi, x=\rho}(t,\rho,0,z).$$

To this point the procedure of "expressing (metric) tensors in terms of their Cartesian components" is complete. Afterwards, Einstein's equations (and other tensorial equations) are expressed in *cylindrical* coordinates to perform numerical calculations, which are regular.

Let us study how this method is related to Lie derivative cartoon method. The residuals of Einstein's equations are $\Re_{\mu\nu} \equiv G_{\mu\nu} - k_d T_{\mu\nu}$. The following two equations are obtained by direct calculation, via Lie derivative cartoon method and our Cartesian components method, respectively.

$$\Re^{\mathrm{LDC}} = \begin{pmatrix} \Re^{\mathrm{LDC}}_{tt} & \Re^{\mathrm{LDC}}_{tx} & 0 & \Re^{\mathrm{LDC}}_{tz} \\ \Re^{\mathrm{LDC}}_{tx} & \Re^{\mathrm{LDC}}_{xx} & 0 & \Re^{\mathrm{LDC}}_{xz} \\ 0 & 0 & \Re^{\mathrm{LDC}}_{yy} & 0 \\ \Re^{\mathrm{LDC}}_{tz} & \Re^{\mathrm{LDC}}_{xz} & 0 & \Re^{\mathrm{LDC}}_{zz} \end{pmatrix}, \tag{3.9}$$

$$\Re^{\mathrm{CC}} = \begin{pmatrix} \Re^{\mathrm{CC}}_{tt} & \Re^{\mathrm{CC}}_{t\rho} & 0 & \Re^{\mathrm{CC}}_{tz} \\ \Re^{\mathrm{CC}}_{t\rho} & \Re^{\mathrm{CC}}_{\rho\rho} & 0 & \Re^{\mathrm{CC}}_{\rho z} \\ 0 & 0 & \Re^{\mathrm{CC}}_{\phi\phi} & 0 \\ \Re^{\mathrm{CC}}_{tz} & \Re^{\mathrm{CC}}_{\rho z} & 0 & \Re^{\mathrm{CC}}_{zz} \end{pmatrix} = \begin{pmatrix} \Re^{\mathrm{LDC}}_{tt} & \Re^{\mathrm{LDC}}_{tx} & 0 & \Re^{\mathrm{LDC}}_{tz} \\ \Re^{\mathrm{LDC}}_{tx} & \Re^{\mathrm{LDC}}_{xx} & 0 & \Re^{\mathrm{LDC}}_{xz} \\ 0 & 0 & \rho^2\Re^{\mathrm{LDC}}_{yy} & 0 \\ \Re^{\mathrm{LDC}}_{tz} & \Re^{\mathrm{LDC}}_{xz} & 0 & \Re^{\mathrm{LDC}}_{zz} \end{pmatrix}, \tag{3.10}$$

where LDC stands for Lie derivative cartoon method, and CC stands for our Cartesian components method. LDC is expressed under Cartesian coordinates, while CC is expressed under cylindrical coordinates. The result means, with identifying $x = \rho$, the two residuals are the same, except for a multiplication of $\rho^2$ in the $\phi\phi$ components in cylindrical coordinates. However, as explained in fact (6), or Sec. 1.6, a multiplication of a smooth, non-zero function onto a *residual* equation, does not change the numerical properties of the numerical calculation. Therefore, our Cartesian component method is the same as Lie derivative cartoon method via a different approach under different philosophy.

This agreement is not surprising. We can actually "prove" it as the following: both the Lie derivative cartoon method and our Cartesian components method, are expressing the same tensor (the Einstein tensor) under the same coordinates (with identifying $x$ with $\rho$), at the same physical location ($y = 0 = \phi$, $x = \rho \geq 0$), using the same variables (Cartesian components), therefore the results from the two methods must be the same.

In the end, we reiterate that there are multiple ways to express tensors in terms of their regular components (such as the example with a superficially singular metric in Sec. 3.4, or background removal method in generalized harmonic formalism and BSSN formalism to be introduced below). Here expressing the metric (and other tensors or pseudo-tensors) in terms of their Cartesian components using the transformation relation of tensors and pseudo-tensors, is just a specific example to obtain regular components.

Also, the above example can actually be improved: using local flatness [52] at $\rho = 0$, the above $\eta_{\phi\phi}$ can be replaced by $\eta_{\rho\rho} + \rho W$ with $W|_{\rho=0} = 0$ (or replaced by $\eta_{\rho\rho} + \rho^2 W$ with $W_{,\rho}|_{\rho=0} = 0$). The local flatness condition can also be obtained by the procedure in Sec. 3.1.4.

### 3.1.4 Results for General Symmetric Tensors

In GR, the governing equations $G_{\mu\nu} = k_d T_{\mu\nu}$ are symmetric tensors of $(0, 2)$ type (i.e. tensors with two "downstairs" indices). As long as we know how to deal with symmetric tensors of $(0, 2)$ type, we know how to deal with Einstein's equations — we know how to deal with GR.

Let us consider a general symmetric tensor of $(0, 2)$ type which in general has the following expression in Cartesian coordinates $(t, x, y, z)$

$$T^{(\mathrm{Cart})} = \begin{pmatrix} T_{tt} & T_{tx} & T_{ty} & T_{tz} \\ T_{tx} & T_{xx} & T_{xy} & T_{xz} \\ T_{ty} & T_{xy} & T_{yy} & T_{yz} \\ T_{tz} & T_{xz} & T_{yz} & T_{zz} \end{pmatrix}. \tag{3.11}$$

Now we will express $T$ in cylindrical coordinates. Since the final expression should be independent of $\phi$, one can express $T$ at any value of $\phi$. e.g. let us choose $\phi = 0$. At this "location" we have $x = \rho$, $y = 0$. [11]

The axisymmetry is expressed as

$$\mathcal{L}_{(-y\partial_x + x\partial_y)} T = 0. \tag{3.12}$$

[11] Again, we emphasize that the expression in cylindrical coordinates does not depend on the value of $\phi$, and here we just take advantage of this fact and do the calculation at $\phi = y = 0$.

Opening up this expression, we obtain the following relations at $y = 0$

$$T_{ty} = T_{xy} = T_{yz} = 0, \tag{3.13}$$

$$T_{tt,y} = T_{tx,y} = T_{tz,y} = T_{xz,y} = T_{xx,y} = T_{zz,y} = T_{yy,y} = 0, \tag{3.14}$$

$$T_{tx} = x \cdot U,\ T_{xz} = x \cdot V,\ T_{yy} - T_{xx} = x \cdot W,\ \text{where}\ W|_{x=0} = 0, \tag{3.15}$$

where $(U, V, W)$ are regular expressions in terms of the $T_{\mu\nu}$, which specific forms are irrelevant at this point.

Performing a coordinate transformation from Cartesian coordinates $(t, x, y, z)$ to cylindrical coordinates $(t, \rho, \phi, z)$ and using (3.13) (and take $\phi = 0$), we obtain

$$T^{(\text{cylin})} = \begin{pmatrix} \tau_{tt} & \tau_{t\rho} & 0 & \tau_{tz} \\ \tau_{t\rho} & \tau_{\rho\rho} & 0 & \tau_{\rho z} \\ 0 & 0 & \rho^2 \cdot \tau_{\phi\phi} & 0 \\ \tau_{tz} & \tau_{\rho z} & 0 & \tau_{zz} \end{pmatrix}, \tag{3.16}$$

where $\tau_{\mu\nu}(t, \rho, z) \equiv T_{\mu\nu}|_{y=\phi=0, x=\rho}(t, \rho, 0, z)$, which are guaranteed regular if the Cartesian components are regular.

The condition (3.15) now reads

$$\tau_{t\rho} = \rho \cdot U, \tag{3.17}$$

$$\tau_{\rho z} = \rho \cdot V, \tag{3.18}$$

$$\tau_{\phi\phi} = \tau_{\rho\rho} + \rho \cdot W \text{ such that } W|_{\rho=0} = 0. \tag{3.19}$$

The first two can be alternatively expressed as $\tau_{t\rho}|_{\rho=0} = \tau_{\rho z}|_{\rho=0} = 0$ [12], which are parity conditions, and the last condition is called local flatness. To complete the parity conditions at $\rho = 0$, similar to the derivation of (3.14) at $y = 0$, now opening up (3.12) at $x = 0$, and then renaming the indices and notations, we get

$$\tau_{tt,\rho}\big|_{\rho=0} = \tau_{tz,\rho}\big|_{\rho=0} = \tau_{\rho\rho,\rho}\big|_{\rho=0} = \tau_{\phi\phi,\rho}\big|_{\rho=0} = \tau_{zz,\rho}\big|_{\rho=0} = 0. \tag{3.20}$$

i.e. the local flatness condition and the parity conditions about the metric tensor obtained in Sec. 3.1.3, actually apply to any symmetric tensors of (0,2) type.

[12] For functions that can be expressed as Taylor expansion $f = \sum_{i=0}^{\infty} f_i r^i$, the condition $f|_{r=0} = 0$ implies $f_0 = 0$. Therefore $f = \sum_{i=1}^{\infty} f_i r^i = r \cdot \sum_{i=0} f_{i+1} r^i \equiv r \cdot V$.

### 3.1.5 Test of the Conjecture

Here we study a testing problem: massless scalar field evolution in 5D. By the same procedure as what is used in 4D, the metric in cylindrical coordinates $(t, r, \theta, \phi, z)$ is

$$g^{(4+1)}_{\text{cylin}} = \begin{pmatrix} \eta_{tt} & \eta_{tr} & 0 & 0 & \eta_{tz} \\ \eta_{tr} & \eta_{rr} & 0 & 0 & \eta_{rz} \\ 0 & 0 & \eta_{\theta\theta} r^2 & 0 & 0 \\ 0 & 0 & 0 & \eta_{\theta\theta} r^2 \sin^2\theta & 0 \\ \eta_{tz} & \eta_{rz} & 0 & 0 & \eta_{zz} \end{pmatrix}. \tag{3.21}$$

Einstein's equations in terms of (3.21) are equivalent to that from Lie derivative cartoon method, and the simulation is, without surprise, regular.

Now, to directly test our conjecture, we also perform the simulation using the following metric representation

$$g^{(4+1)}_{\text{singular}} = \begin{pmatrix} \eta_{tt} & \eta_{tr} & 0 & 0 & \eta_{tz} \\ \eta_{tr} & \xi \cdot \frac{r+1}{r} & 0 & 0 & \eta_{rz} \\ 0 & 0 & \eta_{\theta\theta} r^2 & 0 & 0 \\ 0 & 0 & 0 & \eta_{\theta\theta} r^2 \sin^2\theta & 0 \\ \eta_{tz} & \eta_{rz} & 0 & 0 & \eta_{zz} \end{pmatrix}. \tag{3.22}$$

i.e. we purposely introduce a "singular" term $\xi \cdot (r+1)/r$ (or, we have defined $\xi = \frac{r}{r+1} \cdot \eta_{rr}$), which would be troublesome from the conventional point of view. However, if our conjecture is correct, then the simulation in terms of (3.22) would be just as good as the simulation in terms of (3.21), since $\xi = \frac{r}{r+1} \cdot \eta_{rr}$ is regular.

To test the claim, we performed the simulation in terms of both metric representations. It turns out that the results of the two simulations are the same (which means the superficially singular term $\xi \cdot \frac{r+1}{r}$ does not cause trouble). Therefore, this simulation supports our conjecture.

For the details of the simulation, please refer to Sec. 3.4.

### 3.1.6 Validity and the Extension of the Conjecture

The conjecture is simple, and its usage and validity have been demonstrated though examples. However, the demonstration is not a proof. In this section we try to give a proof (in a physicist's sense) and we will address the following three questions:

(1) Why terms like $1/r^n$ do not cause singularity problems?

(2) When taking only regular functions, are there still any other kind of regularity issues (other than the singularity issue)? And if yes, how to cure it?

(3) What role does machine precision play in the regularity problem?

To answer these questions, we study at what degree the finite difference approximation (FDA) can represent a derivative operator. It is clear to use an example. For a function $f(x)$ with $h$ as the spacing of the grids, let us consider the following FDA

$$\frac{f_{i+1}-f_{i-1}}{2h} = f'(x) + h^2 C_2 f''' + h^4 C_4 f''''' + \cdots = \sum_{n=0}^{\infty} h^{2n} C_{2n} f^{(2n+1)},$$

where $C_n$ are constants which specific values are irrelevant here, and $f^{(n)}$ is the $n$-th derivative of $f$. The expression shows the FDA of a first order derivative, discretized at a finite difference scheme that is of the second order accuracy. Generally, the FDA ($\mathcal{A}$) of a $d$-th derivative operator, discretized at a finite difference scheme that is of $a$-th approximation order, is

$$\mathcal{A}f^{(d)} = f^{(d)} + h^a C_a f^{(a+d)} + h^{2a} C_{2a} f^{(2a+d)} + \cdots = \sum_{n=0}^{\infty} h^{na} C_{na} f^{(na+d)}, \tag{3.23}$$

where $h$ is the grid spacing.

In GR, when $r$ is small, the leading orders of the Cartesian components of symmetric tensors of (0,2) type (denoted as $f$) are asymptotically either linear or quadratic in $r$ (refer to Sec. 3.1.4). Therefore, as long as $a + d \geq 3$, we have $f^{(na+d)} = 0$ $(n = 1, 2, ...)$ in (3.23), therefore

$$\mathcal{A}f^{(d)} = f^{(d)}. \tag{3.24}$$

i.e. the FDA approximation with $a + d \geq 3$, is *exact* when $f$ is linear or quadratic in $r$. In general, when the FDA is exact in the vicinity of the symmetry axis, we say that the FDA is *compatible* with the asymptotic behaviour of fundamental variables. Therefore, as long as the equations are well-behaved at the continuous limit, they are also well-behaved at the discrete level. In particular, terms with multiplications of $1/r^n$ are well-behaved since the discretization results of the functions are exact, and multiplication operation is (almost) exact as well.

The above equation explained why our solution to regularity issue works, and it can also predict the failure of FDA. From equation (3.23) one can see that if $f^{(a+d)} \neq 0$, then $f^{(d)}$ and $h^a C_a f^{(a+d)}$ are comparable at small $r$ (because $(a+d) - a = d$), then the FDA is a bad approximation. i.e. the FDA *has to be exact*, otherwise the error ($h^a C_a f^{(a+d)}$) is of the same order as the value ($f^{(d)}$), which invalidates the FDA. For example, when $f = r^3$ and $a = 2$, then the FDA for $f'$ has trouble. Given the asymptotic behaviour of $f$ being $f \sim r^3$, one can find ways to cure, such as

(1) Use higher order FDA.

(2) $f' \to \text{diff}(f, r^3) \times 3r^2$, then only discretize the "diff". Here the notation "diff" is the partial derivative: $\text{diff}(A, B) \equiv \partial A/\partial B$.

(3) Rewrite $f$ as $f = r \cdot F$ and let $F$ be the fundamental variable instead. The criterion to rewrite $f$ in terms of $F$ is that the asymptotic behaviour of $F$ is compatible with the FDA in the vicinity of the symmetric axis.

These method can be easily generalized to other asymptotic behaviours (for example, using the third method, a function with asymptotic behaviour as $f \sim r^{5/2}$ can be rewritten as $f = r^{1/2} \cdot F$). The first two methods have been used in the literature and proved to be successful. Overall, our opinion is that, as long as the asymptotic behaviours are known, there is always a way to easily fix these issues. [13]

We end this section by analyzing machine precision and show that it is not related to the irregularity issues. In the above analysis, we only analyzed truncation error (caused by the finiteness of $h$) with machine precision $\varepsilon$ ($\sim 10^{-16}$ in double precision) being ignored and we concluded that both $1/r^n$ terms and non-$(1/r^n)$ terms are exact, as long as the FDA scheme is compatible with the asymptotic behaviours of the fundamental variables in the vicinity of $r = 0$. Now we restore $\varepsilon$ and analyze its effects. The concern regarding $\varepsilon$ is that the $1/r^n$ terms would amplify the effect of machine precision so that the errors in $1/r^n$ terms might be significantly larger than the errors in non-$(1/r^n)$ terms. In the following, we will show that the errors in $1/r^n$ terms are *not* larger than the errors in non-$(1/r^n)$ terms.

For specificity, let us consider the FDA of

$$f_{,rr} + f/r^2 - 1/r^2,$$

[13] However, in the literature there is yet another type of "cure" to regularity problem. Let us take $f = r^3$ using second order FDA as the example. The basic idea of the cure is to still discretize $f'$ as $f' \to (f_{i+1} - f_{i-1})/2h$, which is actually wrong according to our analysis above since the truncation error is comparable to the value itself. But it is easy to prove that, for any fixed $r$, the truncation error convergences to zero as the resolution increases. However, for any resolution, there is always a region in the vicinity of $r = 0$, where the error is still comparable to the value itself. In particular, the values at the first few grids in the vicinity of $r = 0$ will always be wrong. Since the mistake only happens in a finite region (which size shrinks as the resolution increases), the result in other regions is still correct, as long as the simulation is stable. Therefore the focus of this kind of "cure" is to make the simulation stable. Our opinion is that, these methods start from a mistake and try to do something to cover the mistake so that it is controllable within a shrinking region (as the resolution increases), and even if the mistake is controllable, the behaviour at $r = 0$ is always problematic (for example, the independent residual tests would not be passed at the point). Instead of searching for any "cure" of this kind, our suggestion is to develop methods based on our two criteria (see Sec. 3.1.7) on regularity, such as the three ways presented above.

where the asymptotic behaviour of $f(r)$ in the vicinity of $r = 0$ is $f \sim 1 + r^2$. The concern is that $f_{,rr} \sim f \sim 1$, but $f/r^2$ would be of order $1/\varepsilon$ when $r$ is as small as $\sim \sqrt{\varepsilon}$, therefore the addition operation in $f_{,rr} + f/r^2$ is not accurate. Symbolically, we denote this concern of non-accuracy as

$$f_{,rr} + f/r^2 \sim 1 + 1/\varepsilon = \varepsilon^{-1} \cdot (1 + \varepsilon), \tag{3.25}$$

whose accuracy is the same as $1 + \varepsilon$ since multiplication operation does not lose accuracy. We will analyze what it takes to let this non-accuracy in $f_{,rr} + f/r^2$ show up. For the expression $f_{,rr} + f/r^2 - 1/r^2$, the discretized equation at the second order FDA is $(f_{i+1} - 2f_i + f_{i-1})/h^2 + (f - 1)/r^2$. Representing $f_{,rr}$ as $(f_{i+1} - 2f_i + f_{i-1})/h^2$, relies on the validity of Taylor expansions such as $f_{i+1} = f_i + hf_{,r} + (h^2/2)f_{,rr} + (h^3/6)f_{,rrr} + ....$ It is then crucial to examine whether the higher orders in these expansions can be expressed accurately by floating point numbers. To let $(f_{i+1} - 2f_i + f_{i-1})/h^2$ be of second order accuracy, it is required that $(h^3/6)f_{,rrr}$ in the expansion must be representable by floating point number. Even if we give up the desire for second order accuracy, it is required *at least* $(h^2/2)f_{,rr}$ is representable in the expansion by floating point number. i.e. it is required the value of $(h^2/2)f_{,rr}$ is not lost in $f_i + hf_{,r} + (h^2/2)f_{,rr}$, which can be symbolically expressed as $\sim 1 + \sqrt{\varepsilon} + \varepsilon$. Therefore in $(f_{i+1} - 2f_i + f_{i-1})/h^2$, the operation is symbolically $(1 + \sqrt{\varepsilon} + \varepsilon - 2 + 1 - \sqrt{\varepsilon} + \varepsilon)/\varepsilon$. The accuracy level is determined by the largest and the smallest values involved in the addition (subtraction) operations, which are $1/\varepsilon$ and $\varepsilon/\varepsilon$. i.e. the accuracy level of this FDA is

$$\varepsilon^{-1} \cdot \big(1 + \varepsilon\big). \tag{3.26}$$

i.e. in order to let the FDA $(f_{i+1} - 2f_i + f_{i-1})/h^2$ represent $f_{,rr}$ at the *lowest order* accuracy, the $\varepsilon$ is required to be representable in the operation $1 + \varepsilon$. Comparing (3.26) with (3.25) one sees that, the accuracy level (due to machine precision) of non-$(1/r^n)$ terms is *the same* as that of the $1/r^n$ terms. If the operations involved $1/r^n$ terms (such as $f_{,rr} + f/r^2$) lose their accuracy due to machine precision, then the FDAs of non-$(1/r^n)$ terms can not represent the derivative operators any more (such as $(f_{i+1} - 2f_i + f_{i-1})/h^2$ fails to represent $f_{,rr}$ at any grid point, rather than merely a few grids in the vicinity of $r = 0$). i.e. what it takes to fail $f_{,rr} + f/r^2$, is the failure of FDA method.

Or equivalently, provided the FDA is still valid (so that finite difference method can be used), then $f_{,rr} + f/r^2$ does not lose its accuracy (therefore machine precision can still be ignored). i.e. for the aspect of machine precision, the $1/r^n$ terms are *not worse* than the non-$(1/r^n)$ terms. Therefore machine precision plays no role in regularity. The discussion above reveals that the truncation error (due to the finiteness of $h$) is fundamentally different from the machine error. The former is a generic feature associated with the FDA method which has nothing to do with machines, while the latter is due to the limited ability of machines. One can improve the latter by using more accurate numbers such as quadruple-precision floating point numbers with the machine precision $\varepsilon \sim 10^{-34}$, while the truncation error can not be improved in this way.

### 3.1.7 Summary of Regularity

This subsection is the summary of the regularity discussion, which should be treated as a rephrase of our conjecture: the irregularity has nothing to do with coordinate system (and operators in the coordinate system) or machine precision. The simulation in cylindrical coordinates (or any other coordinates) is regular, provided the following two conditions are met:

(1) The fundamental variables being used in numerical calculations, are regular functions;

(2) The FDA scheme is compatible with the asymptotic behaviours of the fundamental variables at the symmetry axis (or the origin or any other special locations in the coordinate system).

## 3.2 The Generalized Harmonic Formalism in Non-Cartesian Coordinates

In the previous sections, we described how to apply the Cartesian components method to express the metric functions in terms of regular functions. In this section, we apply the same method to the generalized harmonic formalism (and to BSSN in appendix A).

### 3.2.1 An Introduction to the Generalized Harmonic Formalism

Now we briefly introduce the generalized harmonic formalism [61] of GR. The source functions $H^\mu$ are defined as

$$H^\alpha \equiv \nabla^\beta \nabla_\beta x^\alpha = -\Gamma^\alpha_{\mu\nu} g^{\mu\nu}, \tag{3.27}$$

and $H_\mu \equiv g_{\mu\nu} H^\nu$. In terms of $H_\mu$, Einstein's equations

$$R_{\mu\nu} = k_d \left( T_{\mu\nu} - \frac{1}{d-2} g_{\mu\nu} T \right),$$

reduce to [61]

$$-\frac{1}{2}g^{\alpha\beta}g_{\mu\nu,\alpha\beta} - g^{\alpha\beta}{}_{(,\mu}g_{\nu)\beta,\alpha} - H_{(\mu,\nu)} + H_\beta\Gamma^\beta{}_{\mu\nu} - \Gamma^\alpha{}_{\nu\beta}\Gamma^\beta{}_{\mu\alpha} = k_d\left(T_{\mu\nu} - \frac{1}{d-2}g_{\mu\nu}T\right). \quad (3.28)$$

The generalized harmonic formalism [61] is to lift $H_\mu$ as *fundamental variables*, then the defining equations (3.27) are not defining equations any more. Instead, they are now constraints

$$\mathcal{C}^\alpha \equiv H^\alpha - \Gamma^\alpha_{\mu\nu}g^{\mu\nu} \simeq 0, \quad (3.29)$$

where $\simeq$ means the equations are constraint relations. i.e. in the generalized harmonic formalism, the fundamental variables are the metric functions $g_{\mu\nu}$ and the source functions $H_\mu$, where $H_\mu$ are subject to the constraints (3.29). Since $H_\mu$ are now fundamental variables, the principal parts of (3.28) are now $-\frac{1}{2}g^{\alpha\beta}g_{\mu\nu,\alpha\beta}$, which are manifestly *strongly hyperbolic.*

Using the Bianchi identities, it can be proved [61] that the constraints (3.29) will always be satisfied during an evolution, as long as they are satisfied at any instant of the evolution and the Hamiltonian constraint together with the momentum constraints are also satisfied at the instant. In numerical simulations, however, there are always modes violating the constraints, and the violation modes might grow with time (even exponentially), which will produce unphysical results. Gundlach et al. then suggested [58] to add constraint damping terms to the left hand side of eq. (3.28), which can be

$$Z_{\mu\nu} \equiv \kappa\left(n_{(\mu}\mathcal{C}_{\nu)} - \frac{1+p}{2}g_{\mu\nu}n^\beta\mathcal{C}_\beta\right), \quad (3.30)$$

where $\kappa > 0$, $-1 \le p \le 0$. If the constraints are satisfied, the damping terms vanish. If small violation modes develop, the damping terms can damp out the modes [58] and drive the numerical results back to physical configurations.

Since $H_\mu$ are now fundamental variables, their equations of motion need to be *imposed.* The constraints $H^\alpha \simeq -\Gamma^\alpha_{\mu\nu}g^{\mu\nu} = \nabla^\beta\nabla_\beta x^\alpha$ show that the $H_\mu$ are related to coordinate gauge choices. There are a lot of freedom to impose these equations of motion. For example, $H_\mu$ can even be functions of coordinates and metric functions such as $H_\mu = g_{\mu\nu}x^\nu$. But generally it is required that $H_\mu$ do not include the derivatives of metric functions (which means the equations of motion such as $H_\mu = g_{\mu\nu,\alpha}g^{\nu\alpha}$ are generally not recommended), so that the principal parts of (3.28) are not affected. We do not intend to extensively discuss the gauge choices in this section. Please refer to Sec. 5.3 for the details of some popular gauges being used in the literature.

To simplify future discussions, we define

$$\Gamma^{\alpha} \equiv \Gamma^{\alpha}_{\mu\nu} g^{\mu\nu}.$$

### 3.2.2 The Generalized Harmonic Formalism in Cylindrical Coordinates

Numerical evolutions using GH in cylindrical coordinates in terms of $H_\mu$, are unsuccessful due to the fact that the $H_\mu$'s are singular. Instead, our approach is again to use Cartesian components method: employ the coordinate transformation relation for Christoffel symbols

$$^{(y)}\Gamma^{\kappa}_{\alpha\beta} = \frac{\partial x^\mu}{\partial y^\alpha}\frac{\partial x^\nu}{\partial y^\beta}{}^{(x)}\Gamma^{\gamma}_{\mu\nu}\frac{\partial y^\kappa}{\partial y^\gamma} + \frac{\partial y^\kappa}{\partial x^\mu}\frac{\partial^2 x^\mu}{\partial y^\alpha \partial y^\beta}, \tag{3.31}$$

to express $H_\mu$'s in terms of their Cartesian components. By the same procedure, we obtain

$$\begin{pmatrix} H_t \\ H_\rho \\ H_\phi \\ H_z \end{pmatrix} = \begin{pmatrix} h_t + (1/\rho)\,(\eta_{t\rho}/\eta_{\phi\phi}) \\ h_\rho + (1/\rho)\,(\eta_{\rho\rho}/\eta_{\phi\phi}) \\ 0 \\ h_z + (1/\rho)\,(\eta_{\rho z}/\eta_{\phi\phi}) \end{pmatrix}, \tag{3.32}$$

where $h_\alpha(t,\rho,z) \equiv {}^{(\mathrm{Cart})}H_\alpha(t, x=\rho, y=0, z)$ with sub-index renamed, therefore are regular. Specifying the gauge amounts to choosing appropriate $h_\alpha$ (rather than $H_\alpha$). e.g. the harmonic gauge now reads $h_\mu = 0$, which is indeed the harmonic gauge in Cartesian coordinates. From (3.32), one can see why the simulations with $H_\mu$ as fundamental variable fail—because $H_r$ is singular, which violates our first criterion for regularity.

By the same procedure, the metric in spherical coordinates $(t, r, \theta, \phi)$, with spherical symmetry, is

$$g^{(3+1)}_{\mathrm{sphere}} = \begin{pmatrix} \eta_{tt} & \eta_{tr} & 0 & 0 \\ \eta_{tr} & \eta_{rr} & 0 & 0 \\ 0 & 0 & \eta_{\theta\theta} r^2 & 0 \\ 0 & 0 & 0 & \eta_{\theta\theta} r^2 \sin^2\theta \end{pmatrix}. \tag{3.33}$$

Again, $\eta_{\theta\theta}$ is replaced by $\eta_{rr} + rW$, considering local flatness. The source functions are

$$\begin{pmatrix} H_t \\ H_r \\ H_\theta \\ H_\phi \end{pmatrix} = \begin{pmatrix} h_t + (2/r)\,(\eta_{tr}/\eta_{\theta\theta}) \\ h_r + (2/r)\,(\eta_{rr}/\eta_{\theta\theta}) \\ \cot\theta \\ 0 \end{pmatrix}. \tag{3.34}$$

For 5D, the metric in cylindrical coordinates $(t, r, \theta, \phi, z)$ is

$$g^{(4+1)}_{\text{cylin}} = \begin{pmatrix} \eta_{tt} & \eta_{tr} & 0 & 0 & \eta_{tz} \\ \eta_{tr} & \eta_{rr} & 0 & 0 & \eta_{rz} \\ 0 & 0 & \eta_{\theta\theta} r^2 & 0 & 0 \\ 0 & 0 & 0 & \eta_{\theta\theta} r^2 \sin^2\theta & 0 \\ \eta_{tz} & \eta_{rz} & 0 & 0 & \eta_{zz} \end{pmatrix}, \tag{3.35}$$

where $\eta_{\theta\theta}$ is replaced by $\eta_{rr} + rW$, considering local flatness. And the source functions are

$$\begin{pmatrix} H_t \\ H_r \\ H_\theta \\ H_\phi \\ H_z \end{pmatrix} = \begin{pmatrix} h_t + (2/r)\,(\eta_{tr}/\eta_{\theta\theta}) \\ h_r + (2/r)\,(\eta_{rr}/\eta_{\theta\theta}) \\ \cot\theta \\ 0 \\ h_z + (2/r)\,(\eta_{rz}/\eta_{\theta\theta}) \end{pmatrix}. \tag{3.36}$$

### 3.2.3 Background Removal in the Literature

In the literature, there exist a class of methods for solving the regularity issue of the $H_\mu$'s through the use of the source functions with a background term subtracted, which can all be called background removal methods.

In [54, 55], the authors used (for example in the axisymmetric case in 5D)

$$H_r = \hat{H}_r + \frac{2}{r}, H_t = \hat{H}_t, H_z = \hat{H}_z,$$

where $2/r$ is the value of $H_r$ in flat spacetime background, and the variables with a hat ( $\hat{}$ ) are the variables the authors used as fundamental variables. From the discussion above, we obtain

$$\hat{H}_r = h_r + \frac{2}{r}\left(\frac{\eta_{rr}}{\eta_{\theta\theta}} - 1\right) = h_r + \frac{2}{r}\left(\frac{\eta_{\theta\theta} - rW}{\eta_{\theta\theta}} - 1\right) = h_r - \frac{2W}{\eta_{\theta\theta}},$$

which is related to $h_r$ by a regular term, therefore is regular. Similarly, their

$$\hat{H}_t = h_t + \frac{2}{r}\frac{\eta_{tr}}{\eta_{\theta\theta}}$$
$$\text{and}\ \ \hat{H}_z = h_z + \frac{2}{r}\frac{\eta_{rz}}{\eta_{\theta\theta}},$$

are regular too (because $\eta_{tr}$ and $\eta_{rz}$ are asymptotically linear in $r$ as discussed in section 3.1.4).

On the other hand, there is another way to remove the background [36, 97]. Instead of using $H_\mu$'s, the authors developed the formalisms to use

$$\hat{H}_\mu \simeq -g_{\mu\nu}\left(\Gamma^\nu_{\alpha\beta} - \bar{\Gamma}^\nu_{\alpha\beta}\right) g^{\alpha\beta}$$

as fundamental variables, where the "bar" means that the associated quantities are for the background. $\hat{H}_\mu$ is actually a vector (while $H_\mu$ is not). Although no simulations using this method have been reported[14], this method should be able to generate regularized simulations if the background is chosen correctly. The reason is as follows: if the background is chosen to be flat, in Cartesian coordinates (where $\bar{\Gamma}^\nu_{\alpha\beta} = 0$) we have $\hat{H}_\mu = -g_{\mu\nu}\Gamma^\nu_{\alpha\beta}g^{\alpha\beta} = h_\mu$, where $h_\mu$ is the Cartesian component of the source function, which is what is used in our Cartesian components method. Since $\hat{H}_\mu$ is a vector, its transformation from Cartesian coordinates to cylindrical coordinates at $(y = 0, x \geq 0)$ is simply $\hat{H}_t^{(\text{cylin})} = \hat{H}_t^{(\text{Cart})} = h_t$, $\hat{H}_r^{(\text{cylin})} = \hat{H}_x^{(\text{Cart})} = h_r$, and $\hat{H}_z^{(\text{cylin})} = \hat{H}_z^{(\text{Cart})} = h_z$, while other components are zeros. Therefore $\hat{H}_\mu = h_\mu$ holds in cylindrical coordinates as well. i.e. when the background is flat, the fundamental variables used by the background removal method in [36, 97], are the same as the ones used in Cartesian components method. Therefore, by our conjecture, this method has no regularity issues, as long as the Cartesian components are regular and the FDA scheme is compatible with the asymptotic behaviour of the fundamental variables.

This formalism can solve the regularity issue (by choosing the background properly), and the source function is now a vector. Furthermore, it can remove any background (especially non-flat backgrounds) [15]. We will develop it further in Sec. 3.3.

[14] We will use a generalization of this method in our simulation of the braneworld.

[15] However, when the background is not flat, the regularity of the source functions is not guaranteed, in which case the background needs to be analyzed in a case-by-case basis.

## 3.3 The Generalized Harmonic Formalism in Non-Flat Backgrounds

The spacetime background for RSII is

$$\mathrm{d}s^2 = \frac{\ell^2}{z^2}\Big( -\mathrm{d}t^2 + \mathrm{d}r^2 + r^2\left(\mathrm{d}\theta^2 + \sin^2\theta \mathrm{d}\phi^2\right) + \mathrm{d}z^2\Big). \tag{3.37}$$

For this background, the $h$'s from eq. (3.36) are

$$(h_t, h_r, h_z) = \left(0, 0, -\frac{3}{z}\right). \tag{3.38}$$

Normally setting up the gauge choices is to let $h_\mu$ take specific values or satisfy specific conditions. For example, in the case for the harmonic gauge, the $h_\mu$ are zeros. However, the spacetime background is now non-trivial, in the sense that $h_z$ is not zero for the background, and it is not clear how to let gauges, such as harmonic gauge, make sense. In other words, it is not clear how to setup $h_\mu$.

Generally, in the current literature, the gauge choices for GH are all in Cartesian coordinates with a flat spacetime as the background. In the last section, we have developed the GH in non-Cartesian coordinates, where we have used the assumption that Cartesian components are perfect. For the braneworld, however, Cartesian components might not be sufficient in the sense that it is not clear how to easily employ the existing gauge choices used in the literature. i.e. actually Cartesian components are not perfect and we need to find a way to "get rid of" the background.

### 3.3.1 Tensorial Source Functions

The first way is to use the "background removal" methods mentioned above. For example, following [97], we employ the following functions to serve as the fundamental variables

$$h_\mu \simeq -\, g_{\mu\nu}\left(\Gamma^{\nu}{}_{\alpha\beta} - \bar{\Gamma}^{\nu}{}_{\alpha\beta}\right) g^{\alpha\beta}, \tag{3.39}$$

where the *bar* ( $\bar{\ }$ ) stands for quantities and operations associated with the background. As discussed in section 3.2.3, when the background is taken to be flat, $h_\mu$ in (3.39) is the same as the $h_\mu$ used in "Cartesian components" methods. Yet, (3.39) can be applied to non-flat background. Another good feature is that $h_\mu$ is now a tensor (vector), which geometrically makes more sense. For the braneworld, if (3.37) is taken as the background in (3.39), we have $h_\alpha = 0$ for the background.

However, still this can not be easily linked with the existing work in the literature where the background is flat. i.e. it is not clear what gauge should be given to the $h_\alpha$.

### 3.3.2 Conformal Transformation

Noticing the background metric (3.37) is conformally flat, we can do a conformal transformation $\tilde{g}_{\mu\nu} \equiv (z^2/\ell^2) g_{\mu\nu}$ to obtain $\tilde{g}_{\mu\nu}$ whose background is flat. Actually we can use a general conformal factor $\Psi$ in the conformal transformation as $\tilde{g}_{\mu\nu} \equiv \Psi^{-2} g_{\mu\nu}$. The only requirements on this $\Psi$ are: (a) $\Psi = \ell/z$ when the metric is the background metric; and (b) $\Psi$ goes to $\ell/z$ asymptotically at spatial infinities. Or even more generally, we define conformal transformation

$$\tilde{g}_{\mu\nu} \equiv \Psi^{-q} g_{\mu\nu}, \tag{3.40}$$

where $q$ is a numerical factor that can be set as any value for convenience. A tilde ($\tilde{}$) is used for the quantities and operations associated with $\tilde{g}_{\mu\nu}$. For example, the Christoffel symbols are

$$\tilde{\Gamma}^{\alpha}{}_{\mu\nu} = \frac{1}{2} \tilde{g}^{\alpha\beta} \left( \tilde{g}_{\beta\nu,\mu} + \tilde{g}_{\mu\beta,\nu} - \tilde{g}_{\mu\nu,\beta} \right). \tag{3.41}$$

Repeating the derivations in [32] or [36] while keeping $q$ and $d$ general, we obtain

$$R_{\mu\nu} = \tilde{R}_{\mu\nu} + R^{(\Psi)}_{\mu\nu}, \tag{3.42}$$

where

$$\begin{aligned} R^{(\Psi)}_{\mu\nu} \equiv & - \frac{q(d-2)}{2} \tilde{\nabla}_\mu \tilde{\nabla}_\nu \ln \Psi - \frac{q}{2} \tilde{g}_{\mu\nu} \tilde{\nabla}_\alpha \tilde{\nabla}^\alpha \ln \Psi \\ & + \frac{q^2(d-2)}{4} \left( \tilde{\nabla}_\mu \ln \Psi \tilde{\nabla}_\nu \ln \Psi - \tilde{g}_{\mu\nu} \tilde{\nabla}_\alpha \ln \Psi \tilde{\nabla}^\alpha \ln \Psi \right). \end{aligned} \tag{3.43}$$

Substituting this into Einstein's equations

$$R_{\mu\nu} - \frac{1}{2} g_{\mu\nu} R = k_d T_{\mu\nu} \tag{3.44}$$

$$\Leftrightarrow \quad R_{\mu\nu} = k_d \left( T_{\mu\nu} - \frac{1}{d-2} g_{\mu\nu} T \right). \tag{3.45}$$

we have

$$\tilde{R}_{\mu\nu} = k_d \left( T_{\mu\nu} - \frac{1}{d-2} \tilde{g}_{\mu\nu} \tilde{T} \right) - R^{(\Psi)}_{\mu\nu}, \tag{3.46}$$

where $\tilde{T} \equiv \tilde{g}^{\mu\nu} T_{\mu\nu}$. The equations are Einstein's equations with a modified right hand side.

Depending on how the left hand side is rewritten, one can proceed with either one of the following two approaches. The first approach is to define the source function as

$$\tilde{H}^{\beta} \equiv -\tilde{\Gamma}^{\beta}{}_{\alpha\delta} \tilde{g}^{\alpha\delta}, \tag{3.47}$$

which results in the following GH formalism with a conformal function

$$-\frac{1}{2}\tilde{g}^{\alpha\beta}\tilde{g}_{\mu\nu,\alpha\beta} - \tilde{g}^{\alpha\beta}{}_{(,\mu}\tilde{g}_{\nu)\beta,\alpha} - \tilde{H}_{(\mu,\nu)} + \tilde{H}_{\beta}\tilde{\Gamma}^{\beta}{}_{\mu\nu} - \tilde{\Gamma}^{\alpha}{}_{\nu\beta}\tilde{\Gamma}^{\beta}{}_{\mu\alpha} = k_d \left( T_{\mu\nu} - \frac{\tilde{g}_{\mu\nu}\tilde{T}}{d-2} \right) - R^{(\Psi)}_{\mu\nu}. \tag{3.48}$$

For braneworld simulation, this formalism is sufficient since the conformal transformation let the background of $\tilde{g}_{\mu\nu}$ be flat which enables us to borrow the existing results regarding gauge specification in the literature. Eq. (3.48) can be solved using (3.35) as the metric and (3.36) as the source function (every quantity needs to have a tilde though).

Alternatively, $\tilde{R}_{\mu\nu}$ can be rewritten using tensorial source functions. This approach does better in the following aspects: it is not limited to the case where the background of $\tilde{g}_{\mu\nu}$ is flat, and the source function forms a tensor. The background spacetime of $\tilde{g}_{\mu\nu}$ is denoted as $\bar{g}_{\mu\nu}$. We further define

$$\tilde{C}^{\alpha}{}_{\mu\nu} \equiv \tilde{\Gamma}^{\alpha}{}_{\mu\nu} - \bar{\Gamma}^{\alpha}{}_{\mu\nu}, \tag{3.49}$$

$$\Delta\tilde{\Gamma}^{\beta} \equiv \tilde{C}^{\beta}{}_{\alpha\delta}\tilde{g}^{\alpha\delta}, \tag{3.50}$$

$$\tilde{h}^{\beta} \equiv -\Delta\tilde{\Gamma}^{\beta} = -\left(\tilde{\Gamma}^{\beta}{}_{\alpha\delta} - \bar{\Gamma}^{\beta}{}_{\alpha\delta}\right)\tilde{g}^{\alpha\delta}. \tag{3.51}$$

Repeating the derivation in [36] (for the BSSN derivation though), while keeping $d$ general and *without* requiring $\tilde{g} = \bar{g}$, we obtain the final equation as

$$\begin{aligned}
&-\frac{1}{2}\tilde{g}^{\alpha\beta}\bar{\nabla}_{\alpha}\bar{\nabla}_{\beta}\tilde{g}_{\mu\nu} - \bar{\nabla}_{\alpha}\tilde{g}_{\beta(\mu}\bar{\nabla}_{\nu)}\tilde{g}^{\alpha\beta} - \bar{\nabla}_{(\mu}\tilde{h}_{\nu)} + \tilde{h}_{\alpha}\tilde{C}^{\alpha}{}_{\mu\nu} - \tilde{C}^{\alpha}{}_{\mu\beta}\tilde{C}^{\beta}{}_{\alpha\nu} \\
&= k_d\left(T_{\mu\nu} - \frac{\tilde{g}_{\mu\nu}\tilde{T}}{d-2}\right) - \bar{R}_{\mu\nu} - R^{(\Psi)}_{\mu\nu}.
\end{aligned} \tag{3.52}$$

### 3.3.3 The Implementation of Generalized Harmonic Formalism

The discussions in this subsection regarding the implementation apply to all the GH formalisms (3.28), (3.48) and (3.52). Here we use (3.52) as a specific example.

The $(t, \mu)$ components of equation (3.52) are the Hamiltonian constraint and momentum con-straints. Eq. (3.51) are extra constraints due to the introduction of $\tilde{h}$. In this sense, performing the evolution using all the components in (3.52), is a full-evolution scheme with Hamiltonian and momentum constraints being satisfied. Constraints (3.51) are driven to be satisfied by adding the constraint violation damping terms [58] into (3.52). Therefore we give it the name full-evolution with source driving gauge. This is what has been adopted in the literature. However, it is not very clear how the coordinates condition is imposed via $\tilde{h}_\mu$ (or even $H_\mu$).

Alternatively, we can adopt the same strategy as that in BSSN (therefore we give it the name BSSN-like method): the lapse and shift functions are specified in the "ordinary way" (such as maximal slicing, $1 + \log$ slicing, Gamma freezing, etc), while the Hamiltonian/momentum constraints are left un-evolved. Also, an evolution equation for $\tilde{h}_\mu$ will be derived, rather than using $\tilde{h}_\mu$ to drive lapse and shift. Using $\bar{\Gamma}^\beta{}_{\beta\alpha} = \frac{1}{2}\partial_\alpha \ln \bar{g}$ (and $\tilde{\Gamma}^\beta{}_{\beta\alpha} = \frac{1}{2}\partial_\alpha \ln \tilde{g}$), we obtain $\bar{\nabla}_\alpha \tilde{g}^{\alpha\beta} = \tilde{h}^\beta - \tilde{g}^{\alpha\beta}\bar{\nabla}_\alpha F$, where $F \equiv \frac{1}{2}\ln(\tilde{g}/\bar{g})$, which is a scalar. Taking derivative with respect to $t$, we have

$$\bar{\nabla}_t \tilde{h}_\mu = -\bar{\nabla}_t \tilde{g}^{\alpha\beta}\bar{\nabla}_\alpha \tilde{g}_{\beta\mu} - \tilde{g}^{\alpha\beta}\bar{\nabla}_t\bar{\nabla}_\alpha \tilde{g}_{\beta\mu} + \bar{\nabla}_t\bar{\nabla}_\mu F, \tag{3.53}$$

within which there are terms $\bar{\nabla}_t\bar{\nabla}_t\tilde{g}_{\alpha\beta}$, which will be replaced by quantities with first-order and zeroth-order in $\bar{\nabla}_t$, via eq. (3.52). As for the Hamiltonian/momentum constraints, the damping terms for their violation modes could be constructed. Beyond these, $\tilde{h}_\mu + \Delta\tilde{\Gamma}_\mu \simeq 0$ are still constraints (which are similar to the constraints $\tilde{\Gamma}^i \simeq \tilde{\gamma}^{jk}\left({}^{(d-1)}\tilde{\Gamma}^i{}_{jk} - {}^{(d-1)}\bar{\Gamma}^i{}_{jk}\right)$ in BSSN formalism).

Note, we have not implemented a code using BSSN-like methods yet. Performing simulations using this method is part of our future plans.

The BSSN method is widely used in numerical relativity, and we have simply borrowed its spirit, in deriving the generalized harmonic formalism in non-Cartesian coordinates in non-flat background. For completeness, we also generalized the BSSN formalism to non-flat background with non-Cartesian coordinates. However, since we will not use the BSSN in simulating the braneworld, the development of the BSSN method is put in appendix A.

## 3.4 Evolution of Massless Scalar Field under Cylindrical Coordinates

In this section, we study a model problem: the massless scalar field collapse in 5D under axisymmetry. The reason for choosing the collapse in 5D under axisymmetry is that this system is close to our main project — the massless scalar field collapse in the braneworld, and is therefore an instructive model problem. We will use this section to demonstrate the use of the various methods developed in this chapter. Namely: we will test the Cartesian components methods of generalized harmonic formalism. Then we will test our conjecture by introducing a seemingly singular term into the metric, while the fundamental variables are known to be regular.

### 3.4.1 Initial Data

At the initial instant, we assume the metric takes the following form

$$\mathrm{d}s^2 = e^{2A}\Big(-\mathrm{d}t^2 + \mathrm{d}r^2 + r^2\left(\mathrm{d}\theta^2 + \mathrm{d}\phi^2 \sin^2\theta\right) + \mathrm{d}z^2\Big), \tag{3.54}$$

under cylindrical coordinates $(t, r, \theta, \phi, z)$, with $r = 0$ being the symmetry axis. The range of the coordinates are of course $r \geq 0, 0 \leq \theta \leq \pi, 0 \leq \phi < 2\pi, -\infty < z < \infty$. In this simulation we will keep the spacetime symmetric about $z = 0$ plane, therefore the range of $z$ (in the code) is $[0, \infty)$.

The initial data is obtained by solving the momentum constraints and the Hamiltonian constraint. We take time symmetric initial data, therefore the momentum constraints are satisfied automatically, which leaves only the Hamiltonian constraint to be solved. The Hamiltonian constraint reads

$$A_{,rr} + A_{,zz} + \frac{2A_{,r}}{r} + (A_{,r})^2 + (A_{,z})^2 + \frac{1}{6}\Big((\Phi_{,r})^2 + \left(\Phi_{,z}{}^2\right)\Big) = 0, \tag{3.55}$$

where $\Phi$ is the massless scalar field. We choose its initial configuration to be

$$\Phi(0, r, z) = a \cdot \exp\left(-\frac{\left(\sqrt{z^2 + r^2} - r_0\right)^2}{\sigma^2}\right), \tag{3.56}$$

which is symmetric about $z = 0$. Here we purposely choose the spherically symmetric initial configuration, to test the quality of our axisymmetric code by examining whether a spherically symmetric initial data will remain spherically symmetric. For the testing, we take $a = 0.1, \sigma = 0.25, r_0 = 1$.

The time derivatives of spatial components of the metric are taken as zeros since the initial data configuration is time symmetric. The other metric components are taken trivially $g_{tr} = g_{tz} = 0$, but the time derivatives of $g_{tt}, g_{tr}, g_{tz}$ are chosen such that $h_t = h_r = h_z = 0$ (ref. eq. (3.58)) at the initial instant [62, 63].

### 3.4.2 Evolution

The evolution is performed using our Cartesian components method applied to the generalized harmonic formalism (eq. (3.28)), therefore the metric representation is (3.35)

$$g = \begin{pmatrix} \eta_{tt} & \eta_{tr} & 0 & 0 & \eta_{tz} \\ \eta_{tr} & \eta_{rr} & 0 & 0 & \eta_{rz} \\ 0 & 0 & \eta_{\theta\theta} r^2 & 0 & 0 \\ 0 & 0 & 0 & \eta_{\theta\theta} r^2 \sin^2\theta & 0 \\ \eta_{tz} & \eta_{rz} & 0 & 0 & \eta_{zz} \end{pmatrix}, \tag{3.57}$$

where $\eta_{\theta\theta}$ is replaced by $\eta_{rr} + rW$, considering local flatness condition at $r = 0$, as discussed in Sec. 3.1.3. The gauge sources are (3.36)

$$\begin{pmatrix} H_t \\ H_r \\ H_\theta \\ H_\phi \\ H_z \end{pmatrix} = \begin{pmatrix} h_t + (2/r)\,(\eta_{tr}/\eta_{\theta\theta}) \\ h_r + (2/r)\,(\eta_{rr}/\eta_{\theta\theta}) \\ \cot\theta \\ 0 \\ h_z + (2/r)\,(\eta_{rz}/\eta_{\theta\theta}) \end{pmatrix}. \tag{3.58}$$

During the evolution, the gauge is chosen to be $h_t = h_r = h_z = 0$, which is equivalent to harmonic gauge in Cartesian coordinates.

A constraint damping term [58] is added to the left hand side of eq. (3.28)

$$Z_{\mu\nu} \equiv \kappa \left( n_{(\mu}\mathcal{C}_{\nu)} - \frac{1+p}{2} g_{\mu\nu} n^\beta \mathcal{C}_\beta \right), \tag{3.59}$$

where $\kappa > 0$, $-1 \le p \le 0$. $\kappa = 6$ and $p = 0$ were chosen during the simulation.

The simulation is performed in compactified coordinates $\hat{R} \equiv r/(r+1)$ and $\hat{Z} \equiv z/(z+1)$. In this way the spatial infinities, $r = \infty$ and $z = \infty$ are mapped to $\hat{R} = 1$ and $\hat{Z} = 1$. Since we focus on the behaviour in the region that is close to $r = z = 0$, where the resolution in terms of $(\hat{R}, \hat{Z})$ and $(r, z)$ are comparable, the compactification does not cause loss in resolution. The boundary conditions at $r = \infty$ or $z = \infty$ are trivial Dirichlet condition such that the metric is asymptotically flat. At $z = 0$, the spacetime remains symmetric about $z = 0$. The boundary conditions at $r = 0$ are parity conditions and local flatness condition which are consequences of axisymmetry: $\partial_r \eta_{tt}|_{r=0} = \partial_r \eta_{zz}|_{r=0} = \partial_r \eta_{tz}|_{r=0} = \partial_r \eta_{rr}|_{r=0} = \eta_{tr}|_{r=0} = \eta_{rz}|_{r=0} = W|_{r=0} = \partial_r \Phi|_{r=0} = 0$.

The simulations are performed by "standard" FDA of second order accuracy

$$\partial_t f \to \frac{f^{n+1}_{i,j} - f^{n-1}_{i,j}}{2 \cdot \Delta t}, \tag{3.60}$$

$$\partial_{\hat{R}} f \to \frac{f^{n}_{i+1,j} - f^{n}_{i-1,j}}{2 \cdot \Delta \hat{R}}, \tag{3.61}$$

$$\partial_{\hat{Z}} f \to \frac{f^{n}_{i,j+1} - f^{n}_{i,j-1}}{2 \cdot \Delta \hat{Z}}, \tag{3.62}$$

$$\partial_{tt} f \to \frac{f^{n+1}_{i,j} - 2 f^{n}_{i,j} + f^{n-1}_{i,j}}{(\Delta t)^2}, \tag{3.63}$$

$$\partial_{\hat{R}\hat{R}} f \to \frac{f^{n}_{i+1,j} - 2 f^{n}_{i,j} + f^{n}_{i-1,j}}{\left(\Delta \hat{R}\right)^2}, \tag{3.64}$$

$$\partial_{\hat{Z}\hat{Z}} f \to \frac{f^{n}_{i,j+1} - 2 f^{n}_{i,j} + f^{n}_{i,j-1}}{\left(\Delta \hat{Z}\right)^2}, \tag{3.65}$$

$$\partial_{t\hat{R}} f \to \frac{1}{4 \cdot \Delta t \cdot \Delta \hat{R}} \left( f^{n+1}_{i+1,j} - f^{n+1}_{i-1,j} + f^{n-1}_{i-1,j} - f^{n-1}_{i+1,j} \right), \tag{3.66}$$

$$\partial_{t\hat{Z}} f \to \frac{1}{4 \cdot \Delta t \cdot \Delta \hat{Z}} \left( f^{n+1}_{i,j+1} - f^{n+1}_{i,j-1} + f^{n-1}_{i,j-1} - f^{n-1}_{i,j+1} \right), \tag{3.67}$$

$$\partial_{\hat{R}\hat{Z}} f \to \frac{1}{4 \cdot \Delta \hat{R} \cdot \Delta \hat{Z}} \left( f^{n}_{i+1,j+1} - f^{n}_{i-1,j+1} + f^{n}_{i-1,j-1} - f^{n}_{i+1,j-1} \right). \tag{3.68}$$

where the index $i$ (or $j$) is to characterize the grid position in $\hat{R}$ (or $\hat{Z}$) direction, and $\Delta\hat{R}$ (or $\Delta\hat{Z}$) is the (uniform) spacing of the grids in $\hat{R}$ (or $\hat{Z}$) direction. In the simulation, we set $\Delta\hat{R} = \Delta\hat{Z}$. The superscript $n, n+1, n-1$ is to characterize the discretized time levels, while $\Delta t$ is the spacing. Since compactified coordinates are used, the Courant factor is defined as $\Delta t / \min(\Delta r) = \Delta t / \Delta\hat{R}$, where min is taken over all $\Delta r$. The Courant factor is set to be 0.5 in the simulation.

The residual equations are the discretized equation (3.28) (the GH formalism of GR). The update scheme is to obtain quantities at time level $n+1$ from given quantities at level $n$ and $n-1$, by solving the residual equations. We solve the residual equations by pointwise Newton-Gauss-Seidel iteration, until residuals are smaller than a "small" threshold.

A Kreiss-Oliger [73] style numerical dissipation is adopted to control high frequency numerical noises. Since we are using second order FDA, a fourth order KO dissipation is sufficient. Before obtaining the advanced time level $n+1$, the dissipation is applied to both $n$ and $n-1$ time levels [61].

### Tests and Validation of the Numerical Scheme

The first test we perform is to examine whether spherical initial data evolved under axisymmetric code, can remain spherically symmetric. Fig. 3.2 shows instants during the process when the majority of $\Phi$ field is being bounced back at the centre. The figure shows that the spherical symmetry is indeed preserved during the evolution.

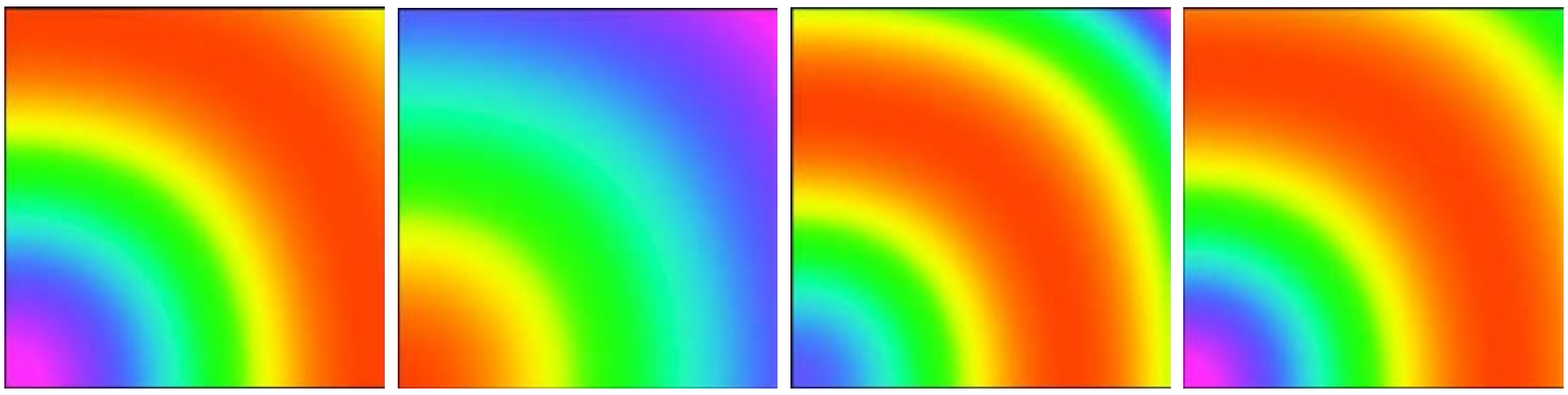

(a) (b) (c) (d)

Figure 3.2: The preservation of spherical symmetry monitored by $\Phi$. The horizontal axis is $r$, and the vertical axis is $z$. The simulation was carried out by 16 cores, and these graphs show the results at the $r = z = 0$ corner. Since compactified coordinates are used, the spherical configuration will appear to be non-spherical. However, when $r$ and $z$ are small, the distortion due to the compactification is small. This is why we choose to show the $r = z = 0$ corner. The four graphs show four instants when the pulse travels towards the center, and then gets bounced back and travels outwards. The graphs show that the spherical symmetry is indeed preserved during the evolution.

By the general theory of numerical solutions, the numerical result is a numerical solution only if the independent residual tests and the convergence tests of the constraints are passed. The independent residuals are obtained by evaluating the residuals resulting from a different discretization which is to discretize the $\hat{R}$ and $\hat{Z}$ derivatives by forward finite difference approximation of second order accuracy, while the time derivative is discretized by backwards finite difference approximation with second order accuracy. Fig. 3.3 shows the convergence of independent residual of the $tt$ component of (3.28), and the convergence of the $t$ component of the constraint equation. The results of other independent residual tests and convergence tests are similar, therefore omitted. These tests show that the simulation produces valid physical results, which confirms that our Cartesian components method does not have the regularity issue.

Another way to justify that the solution is indeed a physical solution, is to directly evaluate the residuals of original Einstein's equations $R_{\mu\nu} = k_d \left(T_{\mu\nu} - g_{\mu\nu}\frac{T}{d-2}\right)$, without referring to the source functions at all. i.e. we use the generalized harmonic formalism to obtain the solutions, then substitute them into original Einstein's equations to get the residuals. Fig. 3.4 shows the $tt$ and

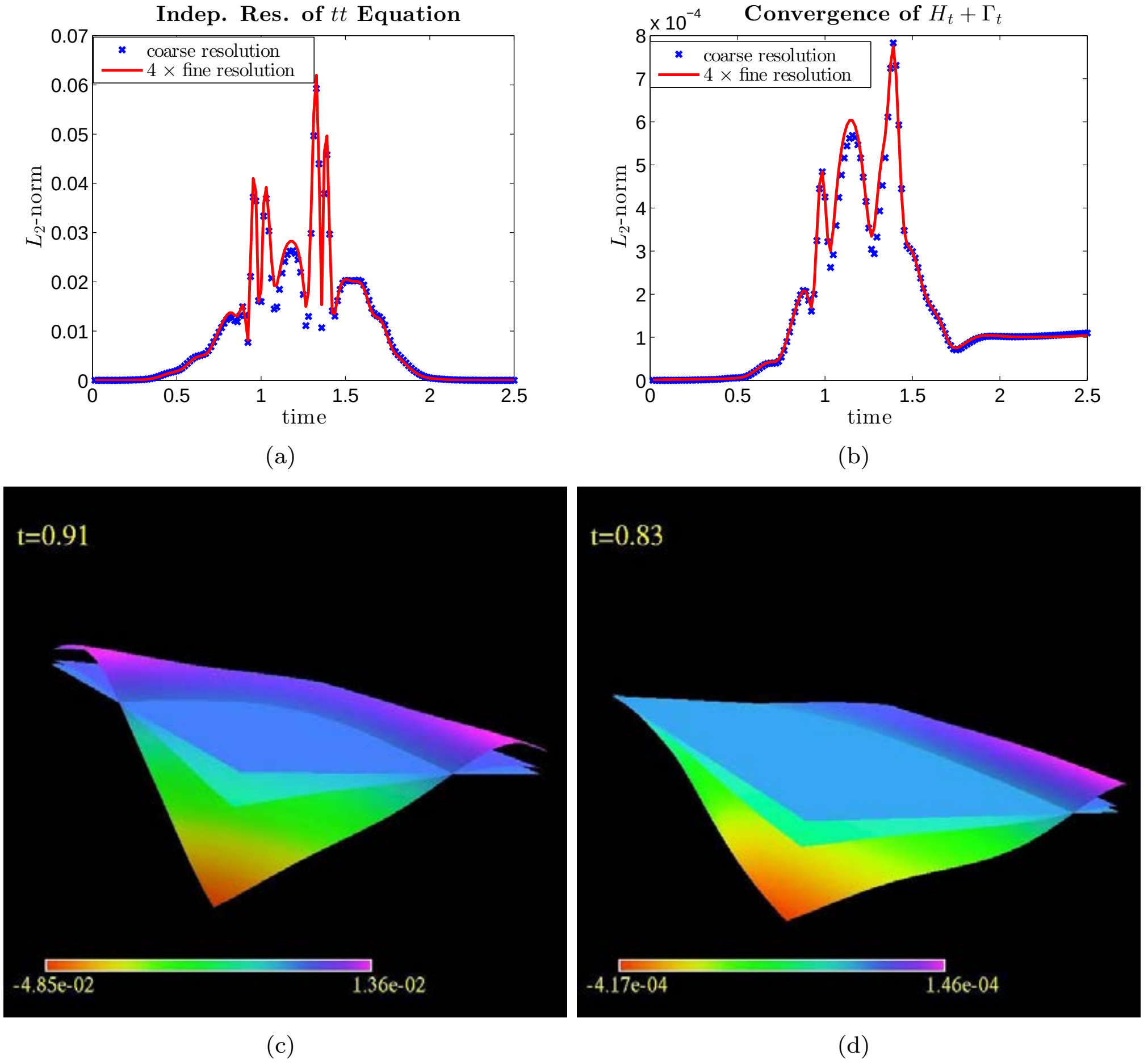

Figure 3.3: The independent residual tests and the constraint tests. Results from two resolutions are needed to perform the tests. Here the finer resolution's grid spacing is one half of that in the coarser resolution. If the numerical results are actually numerical solutions, the independent residuals obtained at finer level are one-fourth of the independent residuals obtained at the coarser resolution, since these results are obtained from the FDA with second order accuracy, and the independent residuals are also discretized at second order accuracy. Fig. (a) is the independent residual test of the $(tt)$ component of Einstein's equations (3.28). Fig. (b) is the test of the $t$-component of the constraint equations $H_\mu + \Gamma_\mu \simeq 0$. Fig. (c) is the snapshot of the $(tt)$ component of the independent residual at the $(r = 0, z = 0)$ corner, at some instant. The function with zero value is shown as a reference, which appears as a plane with uniform colour (light blue). The one taking larger value is obtained from the coarser grid, which value *should* be 4 times of the value obtained from the finer grid at all grid points at all instants. We purposely show the $(r = 0, z = 0)$ corner which is the location suffering the most severe irregularity (if there is). This figure shows that there is no irregularity in our simulation, and the independent residual indeed convergences as expected at every grid point. Fig. (d) is the snapshot of the residual of constraint $H_t + \Gamma_t$ at the $(r = 0, z = 0)$ corner, at some instant.

$rz$ components respectively, as two examples. Indeed they converge as second order quantities.

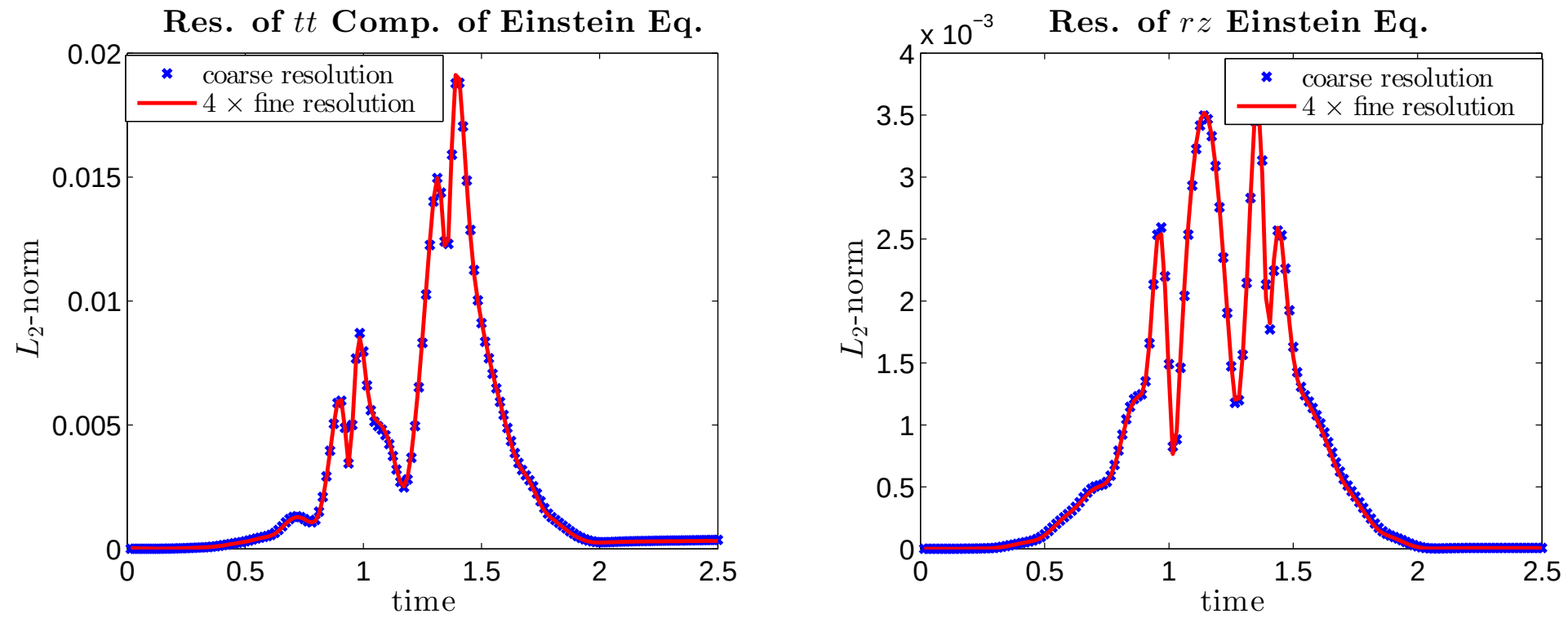

(a) (b)

Figure 3.4: The convergence tests using the residuals obtained from the original Einstein's equations. Fig. (a) is the convergence of the residual obtained from evaluating the ($tt$) component of the original Einstein's equations $R_{\mu\nu} = k_d \left( T_{\mu\nu} - g_{\mu\nu} \frac{T}{d-2} \right)$. Fig. (b) is the test from the ($rz$) component.

### 3.4.3 Superficially Singular Metric Representation

To test our conjecture, instead of using (3.57), here we perform the evolution in terms of the following representation of metric which includes a superficially singular term

$$g_{\text{singular}} = \begin{pmatrix} \eta_{tt} & \eta_{tr} & 0 & 0 & \eta_{tz} \\ \eta_{tr} & \xi \cdot \frac{r+1}{r} & 0 & 0 & \eta_{rz} \\ 0 & 0 & \eta_{\theta\theta} r^2 & 0 & 0 \\ 0 & 0 & 0 & \eta_{\theta\theta} r^2 \sin^2\theta & 0 \\ \eta_{tz} & \eta_{rz} & 0 & 0 & \eta_{zz} \end{pmatrix}, \tag{3.69}$$

where $\eta_{\theta\theta}$ is replaced by $\xi \cdot \frac{r+1}{r} + rW$. The boundary conditions of $\xi$ are $\xi|_{r=0} = 0$, $\xi|_{r=\infty} = 1$, $\partial_z \xi|_{z=0} = 0$, $\xi|_{z=\infty} = r/(r+1)$.

The source functions are of course

$$\begin{pmatrix} H_t \\ H_r \\ H_\theta \\ H_\phi \\ H_z \end{pmatrix} = \begin{pmatrix} h_t + (2/r)\,(\eta_{tr}/\eta_{\theta\theta}) \\ h_r + \left(2/r^2\right)\cdot\xi\cdot(r+1)/\eta_{\theta\theta} \\ \cot\theta \\ 0 \\ h_z + (2/r)\,(\eta_{rz}/\eta_{\theta\theta}) \end{pmatrix}. \qquad (3.70)$$

The metric representation $g_{\text{singular}}$ should be problematic due to $\xi/r$ terms, according to the conventional point of view in [50]. However, according to our conjecture, as long as $\xi$ is regular, and its asymptotic behaviour is compatible with the relevant FDA (in the sense that the FDA is able to accurately represent the derivatives of $\xi$ at small $r$), then the result would be regular (i.e. not problematic). Since $\eta_{rr} \sim \text{const} + r^2$ at small $r$, one should expect $\xi \sim r \cdot \eta_{rr} \sim r$ at small $r$, and the second order finite difference approximation to its derivatives should be exact. Therefore the simulation in terms of $\xi$ should be regular, if our conjecture holds.

The independent residual test and constraint convergence test of the simulation in terms of (3.69) and (3.70) are shown in Fig. 3.5. It shows the simulation produces valid physical results as well. The superficially singular term $\xi/r$ does not cause problem.

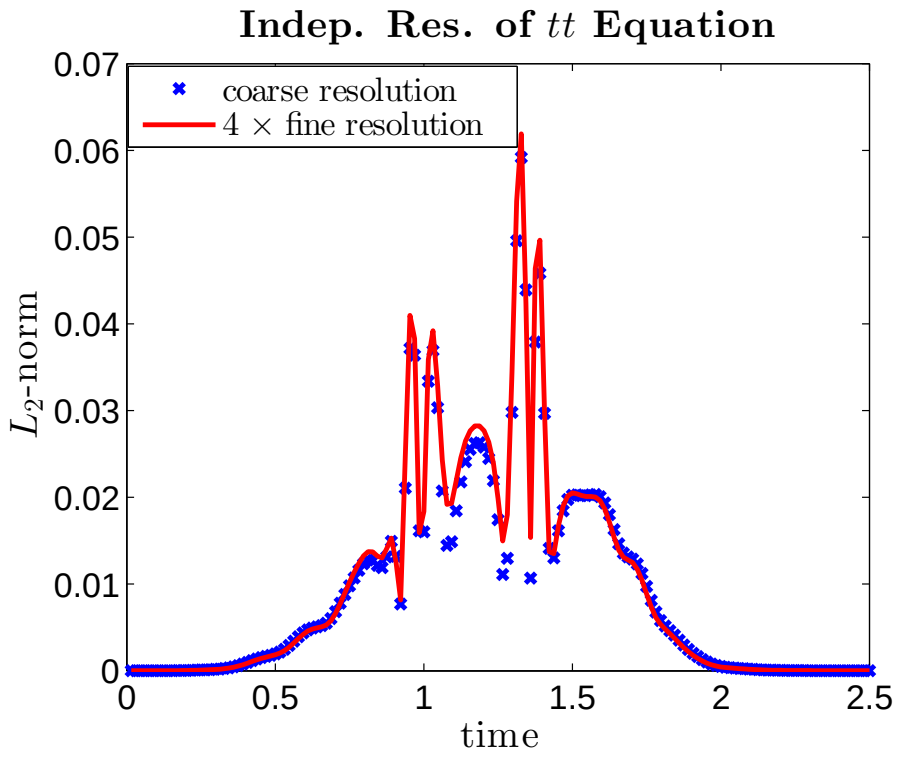

(a)

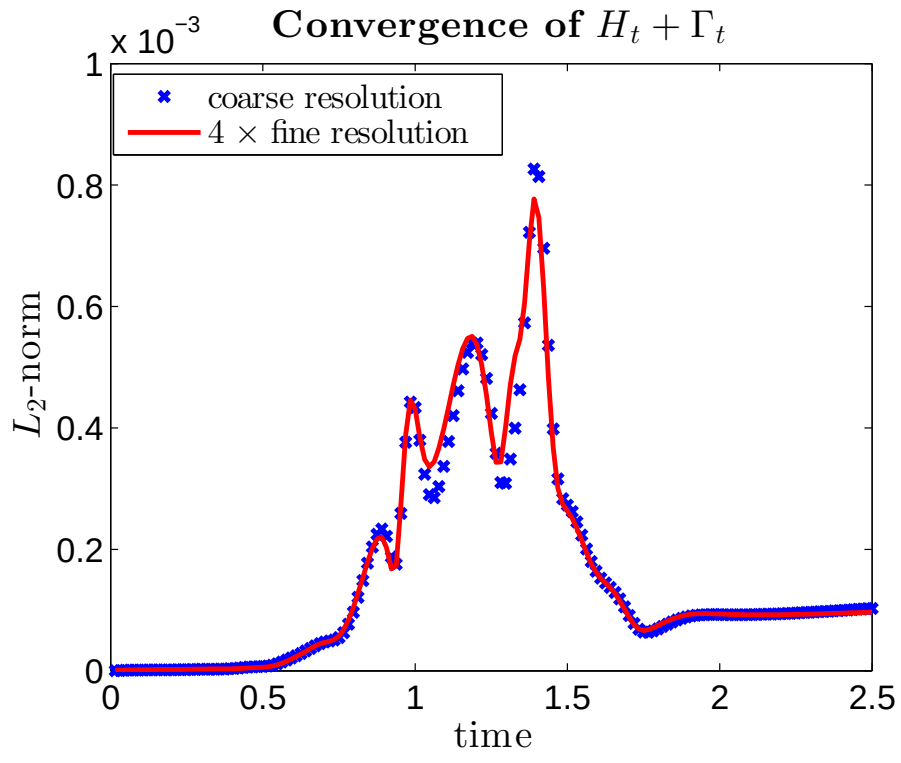

(b)

Figure 3.5: The independent residual tests and constraint tests for the simulation using the supervifially singular metric (3.69). Fig. (a) is the independent residual test of the $(tt)$ component of Einstein's equations (3.28). Fig. (b) is the test of the $t$-component of the constraint equations $H_\mu + \Gamma_\mu \simeq 0$. These two tests show that the simulation using (3.69) and (3.70) indeed satisfies Einstein's equations.

Now we examine whether these two simulations produce the same result. We perform this test by comparing the same quantities resulting from two simulations. Here we show the $L_2$-norms of $(\alpha - 1)$ (where $\alpha$ is the lapse function) and $\Phi$ in Fig. 3.6, which shows that the two simulations indeed produce the same results.

Therefore, our conjecture passed this challenging test.

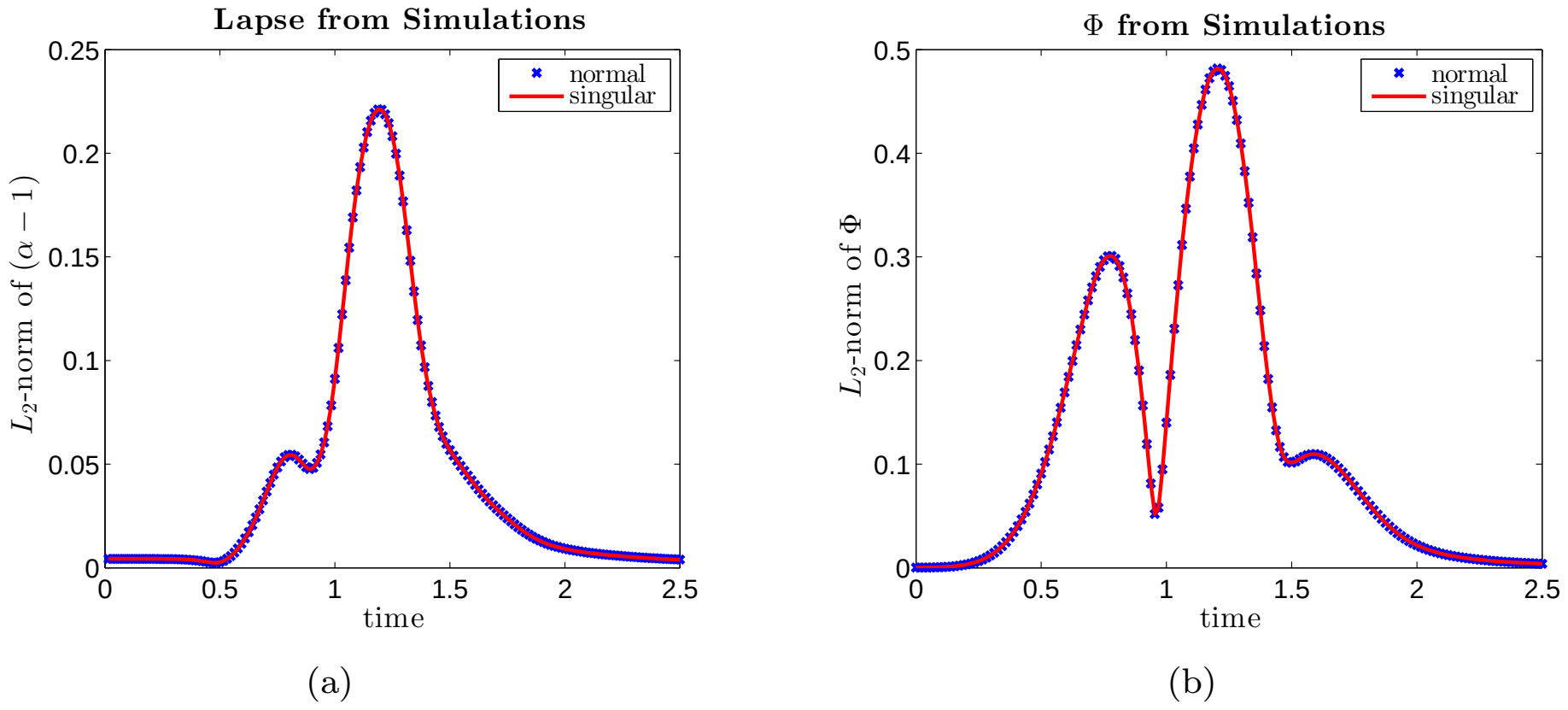

Figure 3.6: The comparison of the results from two the simulations. In the legend, "normal" means the result obtained from the simulation using metric representation (3.57) and "singular" means that obtained from (3.69) where there is a superficially singular term in the metric representation. (a) shows the $L_2$-norms of $(\alpha - 1)$ from the two simulations, and (b) shows the $L_2$-norms of $\Phi$ from the two simulations. These two graphs show that the two simulations indeed produces the same results.

## 3.5 Remark about Regularity

We have a conjecture regarding the regularity issue: provided the fundamental variables used in a numerical simulation are regular functions, and the FDA scheme is compatible with the asymptotic behaviours of the fundamental variables at the symmetric axis (or the origin), there is no irregularity issue.

According to the conjecture, the key to overcome irregularity issues is to express tensors (and pseudo tensors) in terms of regular functions that are compatible with the FDA. One way to do it is to carry the regular Cartesian components into cylindrical coordinates (by coordinate transformation relations) and let them serve as the fundamental variables.

There exist formalisms in the literature that are able to generate simulations without regularity issue, according to our conjecture. Among these specific methods, our Cartesian components method and Lie derivative cartoon method are equivalent (although the philosophies underlying them are different). Both methods use Cartesian components, and are general ways to rewrite equations in non-Cartesian coordinates using Cartesian components, which apply to any formalism of GR. The background removal methods with tensorial source functions [36, 97], however, only apply to the formalisms where Christoffel symbols (or other connection coefficients) play fundamental roles, such as generalized harmonic formalism and BSSN formalism. In these formalisms, when the background is chosen to be flat, this background removal method generates the same results as those of our Cartesian components method (and Lie derivative cartoon method). One advantage of the background removal method is that it can be applied in non-flat background, where our Cartesian components method (and Lie derivative cartoon method) need to be modified. The gauge choice in terms of the tensorial source functions needs further study. The background removal method used in [54, 55], however, is different from any of the above. It is also regular, although it is not clear how to setup coordinate gauges in this formalism.

Above all, I emphasize the following: for the aspect of regularity, our conjecture is *more fundamental* than any specific formalisms in this chapter. For example, Cartesian components are not guaranteed to be singularity free—an example is the behaviour in the vicinity of the physical singularities. In this case the Lie derivative cartoon or Cartesian component method will not help. In background removal methods, the modified source function serving as fundamental variable, is the subtraction of two terms. In principle it is possible that both terms are singular but the function resulting from the subtraction is regular, and in this case the formalism can still generate regular results. All in all, before using any formalisms, one need to check the two aspects of the conjecture: regular functions being used as fundamental variables, and FDA being compatible with the asymptotic behaviours of the fundamental variables in the vicinity of the axis/origin/other special points.

# Chapter 4

# Initial Data

The evolution of the braneworld is specified by the evolution equation(s) with the initial data and the boundary conditions. The governing equations of the braneworld are Einstein's equations in the 5D bulk: $R_{\mu\nu} - \frac{1}{2} g_{\mu\nu} R = -\Lambda g_{\mu\nu}$. We will use the formalism of Einstein's equations with the conformal factor (3.46). For the conformal factor $\Psi$, we simply choose $\Psi = \ell/z$ and $q = 2$. Therefore the conformal metric is

$$\tilde{g}_{\alpha\beta} = \Psi^{-q} g_{\alpha\beta} = \frac{z^2}{\ell^2} g_{\alpha\beta}. \tag{4.1}$$

According to the knowledge we obtained from the regularity discussion in Chap. 3, the most general yet regular metric can be

$$\tilde{g} = \begin{pmatrix} \tilde{\eta}_{tt} & \tilde{\eta}_{tr} & 0 & 0 & \tilde{\eta}_{tz} \\ \tilde{\eta}_{tr} & \tilde{\eta}_{rr} & 0 & 0 & \tilde{\eta}_{rz} \\ 0 & 0 & \tilde{\eta}_{\theta\theta} r^2 & 0 & 0 \\ 0 & 0 & 0 & \tilde{\eta}_{\theta\theta} r^2 \sin^2\theta & 0 \\ \tilde{\eta}_{tz} & \tilde{\eta}_{rz} & 0 & 0 & \tilde{\eta}_{zz} \end{pmatrix}, \tag{4.2}$$

where $\tilde{\eta}_{\theta\theta}$ is replaced by $\tilde{\eta}_{rr} + r\tilde{W}$.

In this chapter and the next, we choose the unit $\ell = k_5 = 1$ (which implies $8\pi G_4 = 8\pi G_5 = 1$). In this chapter, the initial data is obtained by solving the Hamiltonian constraint, subject to the boundary conditions imposed by Israel's junction condition.

The initial data is specified on a spacelike hypersurface. Because the evolution equations are partial differential equations with second order time derivatives, the initial data is specified by the values of the evolving quantities (to be specified below) and their first order time derivatives. The lapse and the shift functions are related to the gauge freedom in choosing the coordinates which can be chosen arbitrarily, therefore the initial data is the combination of spatial $(d-1)$-metric $g_{ij}$, matter distribution, and their first order time derivatives that satisfy the Hamiltonian constraint and momentum constraints. For time symmetric initial data, the momentum constraints are satisfied automatically, which leaves the Hamiltonian constraint to be solved. In most situations of GR, we can then choose a spatial metric ansatz with only one unknown function such as $\mathrm{d}l^2 = A\left(\mathrm{d}r^2 + r^2(\mathrm{d}\theta^2 + \sin^2\theta \mathrm{d}\phi^2) + \mathrm{d}z^2\right)$. The Hamiltonian constraint is then solved for $A$. We will examine whether this situation will still be true in the braneworld.

## 4.1 Formulation

In the braneworld, Israel's boundary condition can not be met if there is only one unknown variable in the spatial metric. To "absorb" this "inconsistency", there must be at least two unknown variables in the spatial metric. The spatial metric can be taken to have a conformal form in $(r, z)$ coordinates

$$\mathrm{d}l^2 = \frac{\ell^2}{z^2}\left[e^{2A+2B}\left(\mathrm{d}r^2 + \mathrm{d}z^2\right) + e^{2A-2B}r^2\left(\mathrm{d}\theta^2 + \sin^2\theta \mathrm{d}\phi^2\right)\right]. \tag{4.3}$$

The brane is located at $z = \ell$, while $z = \infty$ is the spatial infinity. Eq. (4.3) is actually the most general spatial metric for the axisymmetric (in the bulk) case, as discussed in Sec. 2.4.2. The background spacetime (vacuum) corresponds to $A = B = 0$.

The Hamiltonian constraint is

$$\begin{aligned} &\frac{3}{2}\left(\nabla^2 A\right) - \frac{1}{2}\left(\nabla^2 B\right) + \frac{1-e^{4B}}{2r^2} + \frac{3}{z^2}\left(1 - e^{2(A+B)}\right) \\ &+ \frac{3U_{,r}}{r} - \frac{3U_{,z}}{z} + \frac{3}{2}\left(U_{,z}\right)^2 + \frac{3}{2}\left(U_{,r}\right)^2 = 0, \end{aligned} \tag{4.4}$$

where $\nabla^2 \equiv \partial_{rr} + \partial_{zz}$, and $U \equiv A - B$. The domain within which to solve the equations is $(r, z) \in [0, \infty) \times [1, \infty)$. Let $\Phi$ be the massless scalar field that lives on brane, Israel's junction condition (1.29) is then translated into

$$A_{,z} = 1 - e^{A+B} - \frac{1}{6}e^{-A-B}\left(\Phi_{,r}\right)^2, \tag{4.5a}$$

$$B_{,z} = -\frac{1}{4}e^{-A-B}\left(\Phi_{,r}\right)^2, \tag{4.5b}$$

from which one see that a vanishing $B$ is only possible when there is no matter ($\Phi_{,r} = 0$) on the brane.

The boundary conditions at the symmetric axis $r = 0$ are simply the parity condition and the local flatness condition which translate into $A_{,r}\big|_{r=0} = B_{,r}\big|_{r=0} = B\big|_{r=0} = \Phi_{,r}\big|_{r=0} = 0$. When $z \to \infty$, the spacetime should approach the background: $A|_{z\to\infty} = B|_{z\to\infty} = 0$. If the

matter is localized to finite $r$, then the spacetime should approach the background as $r \to \infty$: $A|_{r\to\infty} = B|_{r\to\infty} = 0 = \Phi|_{r\to\infty}$.

There is only one equation (the Hamiltonian constraint) for two variables $A$ and $B$, therefore there remains freedom in the initial data. We fix the freedom by *imposing* the form of $B$. Based on the specific forms of $B$, we used two specifications: Laplacian specification and direct specification.

### 4.1.1 Laplacian Specification

Since there is no other requirement on $B$ except for the boundary conditions, we will *impose* an equation for $B$, for example, $\nabla^2 B = 0$. However, it turns out that there are numerical instability issues at spatial infinities, which suggests us to carry the asymptotic behaviour at these boundaries by some factors. I.e. we define $Y$ as $B \equiv f_B Y$ where the asymptotic behaviour at spatial infinities is carried by $f_B$. Since we do not know the asymptotic behaviour yet, we will try to choose $f_B$ in an experimental manner. There is actually a guidance we can start with. A similar problem is the static star configuration studied in [47], where the boundary condition at $z \to \infty$ is asymptotically

$$B \sim \frac{\text{const}}{z^3}. \tag{4.6}$$

If the matter is localized at finite $r$, the boundary condition at $r \to \infty$ is asymptotically [47]

$$B \sim \frac{\text{const}}{r}. \tag{4.7}$$

As discussed in Chap. 2, the asymptotic bahaviour of $A$ at $r \gg z$ is supposed to be $A \sim 1/rz$ for conformally flat space. We can then introduce factors like

$$A = \frac{X}{(r+1)z}, \tag{4.8a}$$

$$B = \frac{Yr^2}{(r+z)^3}, \tag{4.8b}$$

where the $r^2$ in equation (4.8b) is to let the boundary conditions $B\big|_{r=0} = B_{,r}\big|_{r=0} = 0$ be satisfied automatically. In the numerical method, we will directly solve for $X$ and $Y$ instead of $A$ and $B$. The equation of $B$ can be imposed via $Y$ as

$$\nabla^2 Y - aY^b = 0. \tag{4.9}$$

The principle part $\nabla^2$ plays the role as a smoother, and the second part is to serve as an amplitude damper (to be explained below) when both $a$ and $b$ are set to be positive values, such as $a = b = 3$. The functionality of the damping term is to prevent the amplitude of $Y$ from getting too large. The damping term is needed because instabilities arise at the brane when the amplitude of $Y$ becomes too large.

### 4.1.2 Direct Specification

Since the only requirements on $B$ are the boundary conditions among which only the boundary condition at the brane is non-trivial, we may directly set $B$ to take the following simple form so that (4.5b) holds

$$B(r,z) = \left.\frac{\partial B}{\partial z}\right|_{z=1} \cdot f(z), \tag{4.10}$$

where $f(z)$ satisfies the condition: $f_{,z}\big|_{z=1} = 1$. In case we want the magnitude of $B$ to be small, we can also let $f(z)$ satisfy a second condition: $f|_{z=1} = 0$. There is still a lot of freedom to choose $f$. e.g. we can choose it to be $f(z) = (z-1)/z^{p_z}$, where $p_z = 4$ if we want $B$ to satisfy (4.6) (not necessary though). Or we can let $B$ die off faster by choosing

$$f(z) = (z-1) \cdot \exp\left[-(z-1)^2/{\sigma_z}^2\right], \tag{4.11}$$

where $\sigma_z$ can be chosen arbitrarily. In all the numerical results presented in this chapter, $\sigma_z = 2\sigma_r$ is chosen, where $\sigma_r$ is going to be defined in (4.14). Because (4.10) and (4.11) yield $B|_{z=1} = 0$, we have

$$B(r,z) = -\frac{1}{4}e^{-A_0}\left(\Phi_{,r}\right)^2 \cdot f, \tag{4.12}$$

where $A_0(r,z) = A(r,z)\big|_{z=1}$. It is easy to check all the boundary conditions of $B$ are satisfied.

It turns out the numerical behaviour of the direct specification is better than that of the Laplacian specification, so we will mainly focus on the direct specification method.

## 4.2 Numerical Methods

In order to let the spatial infinities be a part of our computational domain, compactified coordinates $(\hat{R}, \hat{Z})$ are used

$$\hat{R} \equiv \frac{r}{r + r_0}; \quad \hat{Z} \equiv \frac{z - \ell}{z - \ell + z_0}, \tag{4.13}$$

where $r_0$ and $z_0$ are parameters to control the compactification. $r_0 = z_0 = 1$ is used for the numerical results that are going to be shown in this chapter. The discretized grids are uniform in $\hat{R}$ and $\hat{Z}$, and the equations are discretized by finite difference approximation with second order central stencil, which are shown by eq. (3.60) to eq. (3.68). Then Newton-Gauss method is employed iteratively to solve for the numerical solution. After the numerical solutions are obtained, we validate the solutions by the mean of independent residual test.

## 4.3 The Numerical Solution

### 4.3.1 The Solution and Apparent Horizon

We used the following profile for the matter field

$$\Phi = \mathscr{A} \cdot \exp\left[-\frac{(r - x_0)^2}{\sigma_r^2}\right]. \tag{4.14}$$

The condition $\Phi_{,r}\big|_{r=0} = 0$ needs to be satisfied in the initial data, therefore we either choose $x_0 = 0$, or choose $x_0$ to be at least several $\sigma_r$.

The independent residual, calculated as the residual of equation (4.4) under a different discretization scheme (other than eq. (3.60) to eq. (3.68)) that is of second order accuracy, should behave like a second order quantity everywhere in the domain if a numerical solution is obtained. To show the independent residuals compactly, we use their $L_2$ norms. As an example, we perform the independent residual test to the numerical results obtained from the calculation with $(\mathscr{A}, \sigma_r, x_0) = (2.8, 0.15, 0)$ using the direct specification. The $L_2$ norms of the independent residual at resolutions $64 \times 64$ (the domain in $\hat{R}$ direction and $\hat{Z}$ direction are uniformly divided into 64 intervals), $128 \times 128$ and $256 \times 256$, are $0.0799, 0.0194$ and $0.00498$, respectively. The $L_2$ norm shrinks by a factor of 3.9-4.1 when the grid spacing decreases by a factor of 2. The independent residual indeed converges to zero at second order, therefore the numerical scheme is justified.

The results of the numerical calculation with $(\mathscr{A}, \sigma_r, x_0) = (3.2, 0.3, 0)$ using the direct specification, are shown in Fig. 4.1. The scalar field is strong enough to produce an apparent horizon which is shown in Fig. 4.2.

### 4.3.2 Brane Geometry as Seen by a Brane Observer

In this subsection we focus on the geometry of the brane, since the bulk is not directly observable. In the next subsection, we will study the geometry of the bulk. Again, here we reiterate that the masses of the spacetime obtained from the brane geometry, play no direct roles in the braneworld. Especially, they do not really represent the masses of the brane. The purpose of studying these masses, is to compare with GR and obtain some observable difference from GR.

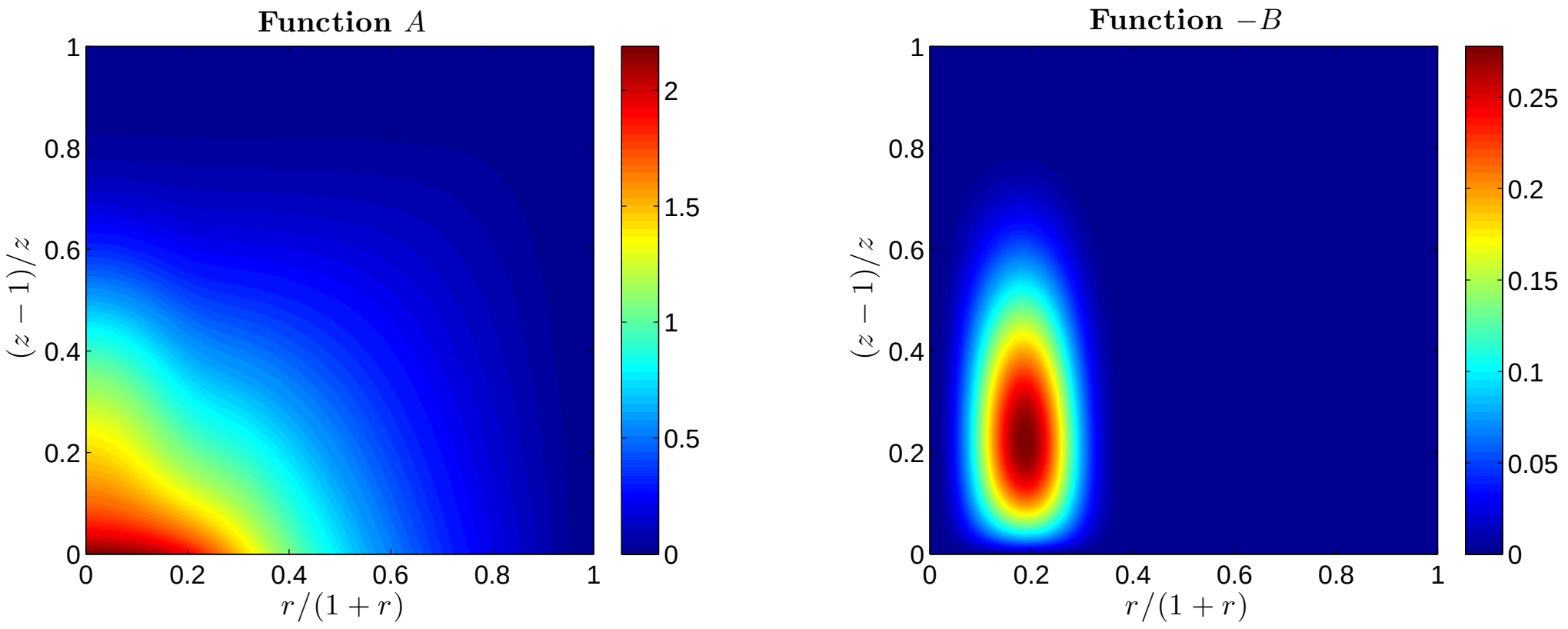

(a) function $A$ (b) function $-B$

Figure 4.1: Function $A$ and $-B$ in the initial data metric (4.3) obtained from the parameters $(\mathscr{A}, \sigma_r, x_0) = (3.2, 0.3, 0)$

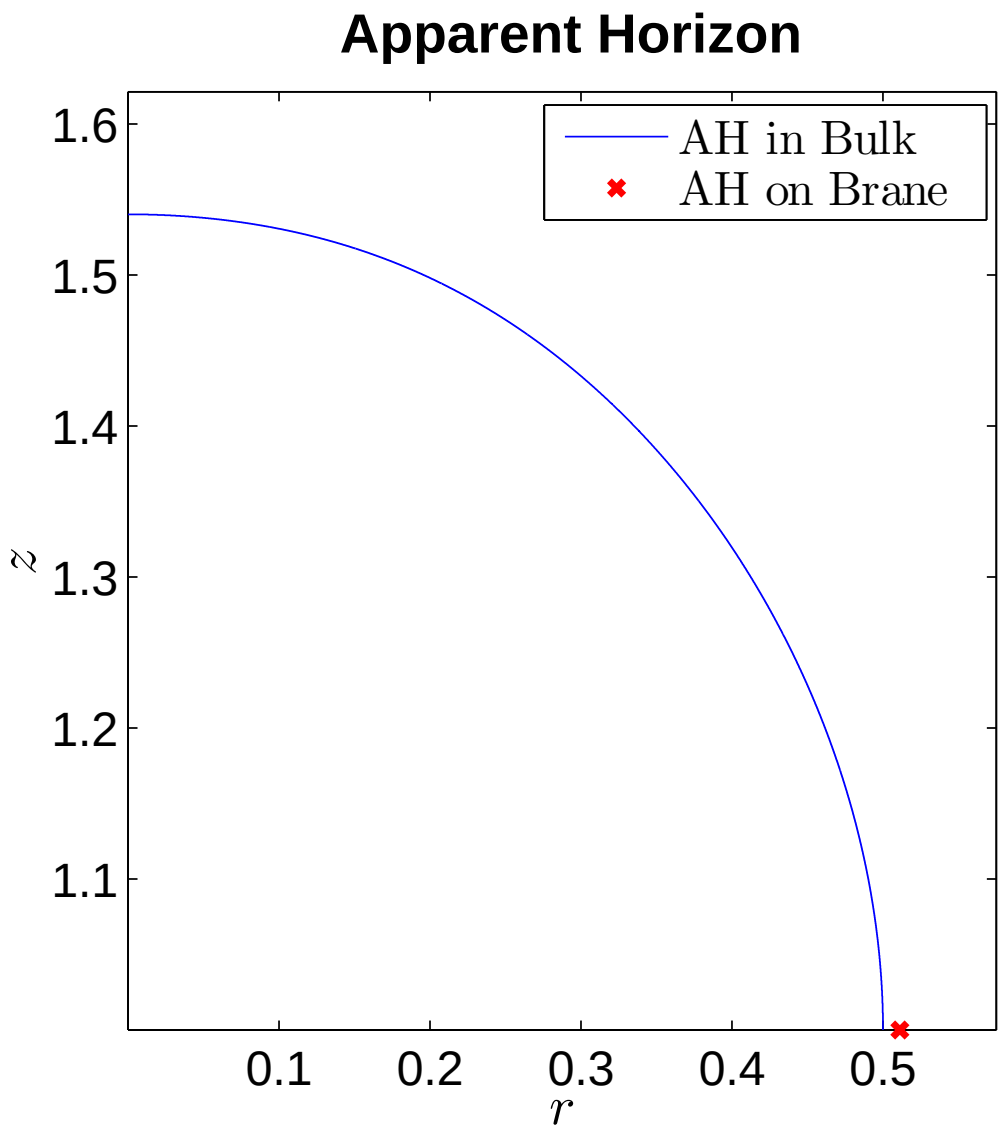

Figure 4.2: The apparent horizons for the configuration with parameters $(\mathscr{A}, \sigma_r, x_0) = (3.2, 0.3, 0)$. The red **x** in the figure is the apparent horizon calculated using the brane geometry only. It does not coincide with the intersect of the brane and the bulk apparent horizon.

We will study two different masses defined on the brane—brane ADM mass and the Hawking mass (see Sec. 2.5.1 and 2.5.2). The proper area radius $\tilde{r}$ on the brane is

$$\tilde{r} = e^{A-B} r, \tag{4.15}$$

so that the metric on the brane is now

$$\mathrm{d}s^2_{\mathrm{brane}} = g_{\tilde{r}\tilde{r}} \mathrm{d}\tilde{r}^2 + \tilde{r}^2 \mathrm{d}\Omega^2, \tag{4.16}$$

where $g_{\tilde{r}\tilde{r}} = e^{4B} \left[1 + r\left(\mathrm{d}A/\mathrm{d}r - \mathrm{d}B/\mathrm{d}r\right)\right]^{-2}$.

In terms of $g_{\tilde{r}\tilde{r}}$ and $\tilde{r}$, the Hawking mass is

$$M_{\mathrm{H}} = \frac{\tilde{r}}{2G_4} \left(1 - (g_{\tilde{r}\tilde{r}})^{-1}\right), \tag{4.17}$$

which has $\tilde{r}$ dependence. Then the $\tilde{r}$ dependence of the Hawking mass is an observable difference from 4D GR. One example is shown in Fig. 4.3.

The Hawking mass goes to ADM mass as $\tilde{r} \to \infty$

$$M_{\mathrm{ADM}} = \lim_{\tilde{r}\to\infty} \frac{\tilde{r}}{2G_4} \left(1 - (g_{\tilde{r}\tilde{r}})^{-1}\right) = \frac{\alpha_1 + \beta_1}{G_4}. \tag{4.18}$$

$\alpha_1$ and $\beta_1$ are defined via the asymptotic behaviour of $A$ and $B$ at large $r$, which are assumed to be $A\big|_{z=1} \approx \alpha_1/r$ and $B\big|_{z=1} \approx \beta_1/r$. As it is shown in Sec. 4.3.3, the asymptotic behaviour of $B$ actually implies $\beta_1 = 0$.

The masses for the results of the configuration with $(\mathscr{A}, \sigma_r, x_0) = (3.2, 0.3, 0)$, are shown in Fig. 4.3. In the next section we will study the total mass in the bulk. To distinguish different masses, the ADM mass on brane will be referred as $M_{\mathrm{braneADM}}$.

### 4.3.3 Asymptotic Behaviour

The asymptotic behaviour at $r \gg z$ is crucial for the calculation of total energy (see Sec. 2.4.2). The construction of $B$ in the direct specification according to eq. (4.12), directly implies $|B| \ll |A|$ at the $r \gg z$ region. When $B$ is negligible, in Sec. 2.4.2 we showed that a solution at $r \gg z$ region

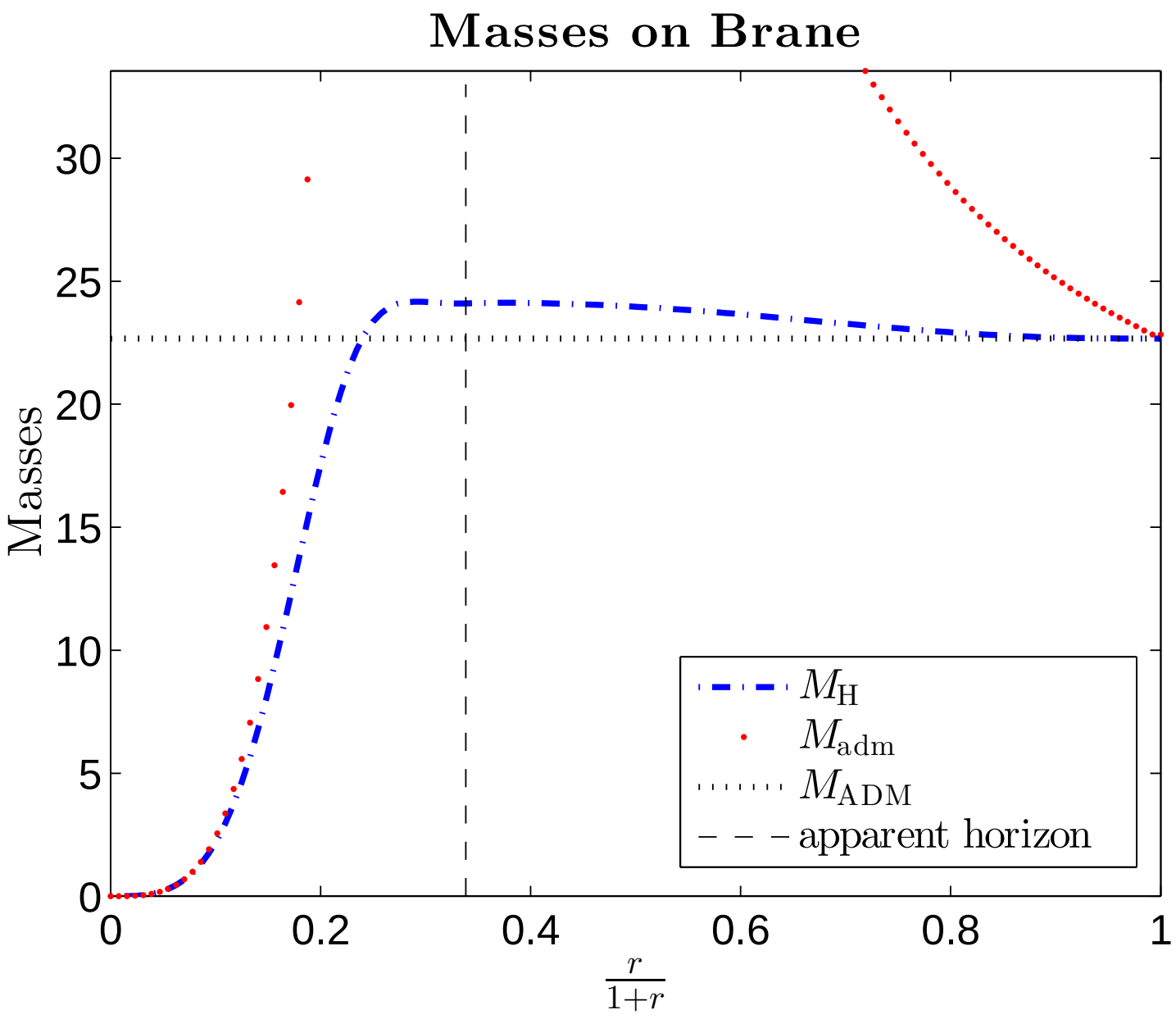

Figure 4.3: Brane Masses from the initial data configuration with parameters $(\mathscr{A}, \sigma_r, x_0) = (3.2, 0.3, 0)$. The result is $M_{\mathrm{ADM}} = 22.67$. The graph shows that the Hawking mass is not a constant in the braneworld, in contrast to 4D GR. Hawking mass agrees with ADM mass only at $r \to \infty$. $M_{\mathrm{adm}}$ (defined in equation (2.58)) still blows up around the apparent horizon.

was

$$A \approx \frac{\alpha_1}{rz} \quad \text{when } r \gg z. \tag{4.19}$$

However, as discussed in that section, there is no proof of the uniqueness of this solution. Therefore in this section we need to test whether the solution is indeed given by eq. (4.19). Please refer to Fig. 4.4 for the results of a series of simulations from the family $(\mathscr{A}, \sigma_r, x_0) = (\mathscr{A}, 0.3, 0)$ using the direct specification. The graphs show that the asymptotic behaviour is indeed described by eq. (4.19).

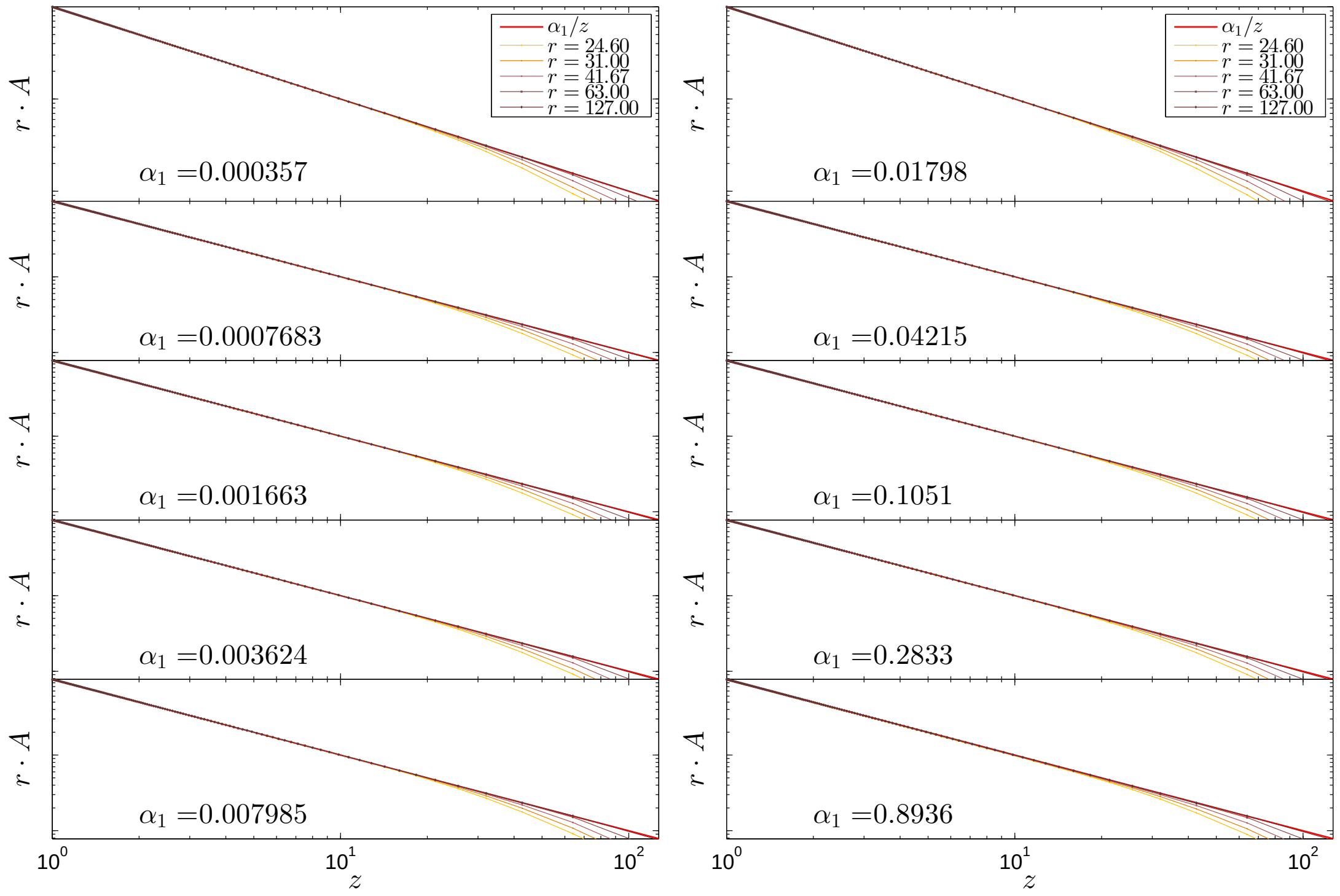

Figure 4.4: The asymptotic behaviour of $A$ at $r \gg z$ region, shown by the simulations from the family $(\mathscr{A}, \sigma_r, x_0) = (\mathscr{A}, 0.3, 0)$. In the first diagram, we plot $r \times A$ versus $z$ in log-log scale. The red line, is plotted assuming $r \times A = \alpha_1/z$. If $A \approx \alpha_1/rz$, one would expect the plots for *different* $r$ should follow the red line. Other diagrams are plotted in the same way, but with different $\alpha_1$ resulted from different $\mathscr{A}$. The diagrams look almost identical, despite that $\alpha_1$ has changed over three orders of magnitude. It means that the asymptotic behaviour of $A$ at $r \gg z$ region is indeed $A \approx \alpha_1/(rz)$.

## 4.4 The Total Energy

The asymptotic behaviour of eq. (4.19) is confirmed by Fig. 4.4, therefore the calculation in Sec. 2.4.2 applies. The calculation shows that the total energy ($M_{\text{total}}$) in the whole spacetime is

$$M_{\text{total}} = \alpha_1/G_5 = 8\pi\alpha_1. \tag{4.20}$$

Consequently, as shown by eq. (2.67), the relation between the brane "energy" and the total energy is then simply

$$M_{\text{braneADM}} = M_{\text{total}}. \tag{4.21}$$

### 4.4.1 The Relation with the Area of Apparent Horizon

In 4D GR case, there is a simple relation between the Hawking energy of the spherically symmetric vacuum spacetime and the area of the horizon of the black hole sitting at the symmetric center, which is eq. (2.62). In the braneworld, we would like to examine whether the total energy can also be characterized by the area of apparent horizon in 5D braneworld. To study this relation, we choose the configurations where the matter outside of the apparent horizon are negligible.

Fig. 4.5 shows the relation between the total energy $M_{\text{total}}$, and $\sqrt{4\pi\mathcal{A}_{\text{bulk}}}$, where $\mathcal{A}_{\text{bulk}}$ is the area of the apparent horizon in the *bulk*. The figure shows these two quantities are equal. Therefore we have

$$M_{\text{total}} = \sqrt{4\pi\mathcal{A}_{\text{bulk}}}. \tag{4.22}$$

How to understand this relation? Eq. (4.21) tells us that the mass can also be understood as the ADM mass calculated on the brane. The size of BH—the areal radius of the intersect of the horizon with the brane—is $r_{\text{a}} \sim (0.7, 2.0)$ according to the area-versus-radius relation shown by Fig. 1.1, where $r_{\text{a}}$ is the areal radius of the horizon on the brane. These BHs are then considered to be medium to large. For large BHs, the area of the horizon in the bulk reduces to the area of black string, which is equal to the area of the horizon on the brane, and the brane geometry (without matter) reduces to 4D GR. Eq. (4.22) might be just

$$M_{\text{braneADM}} = \sqrt{4\pi\mathcal{A}_{\text{brane}}}. \tag{4.23}$$

On the other hand, given there are many BHs of medium size in this data set, the knowledge of large BHs (that the areas reduces to those for black strings) might not apply. Therefore the relation might be generically about total energy and the area of apparent horizon in the bulk.

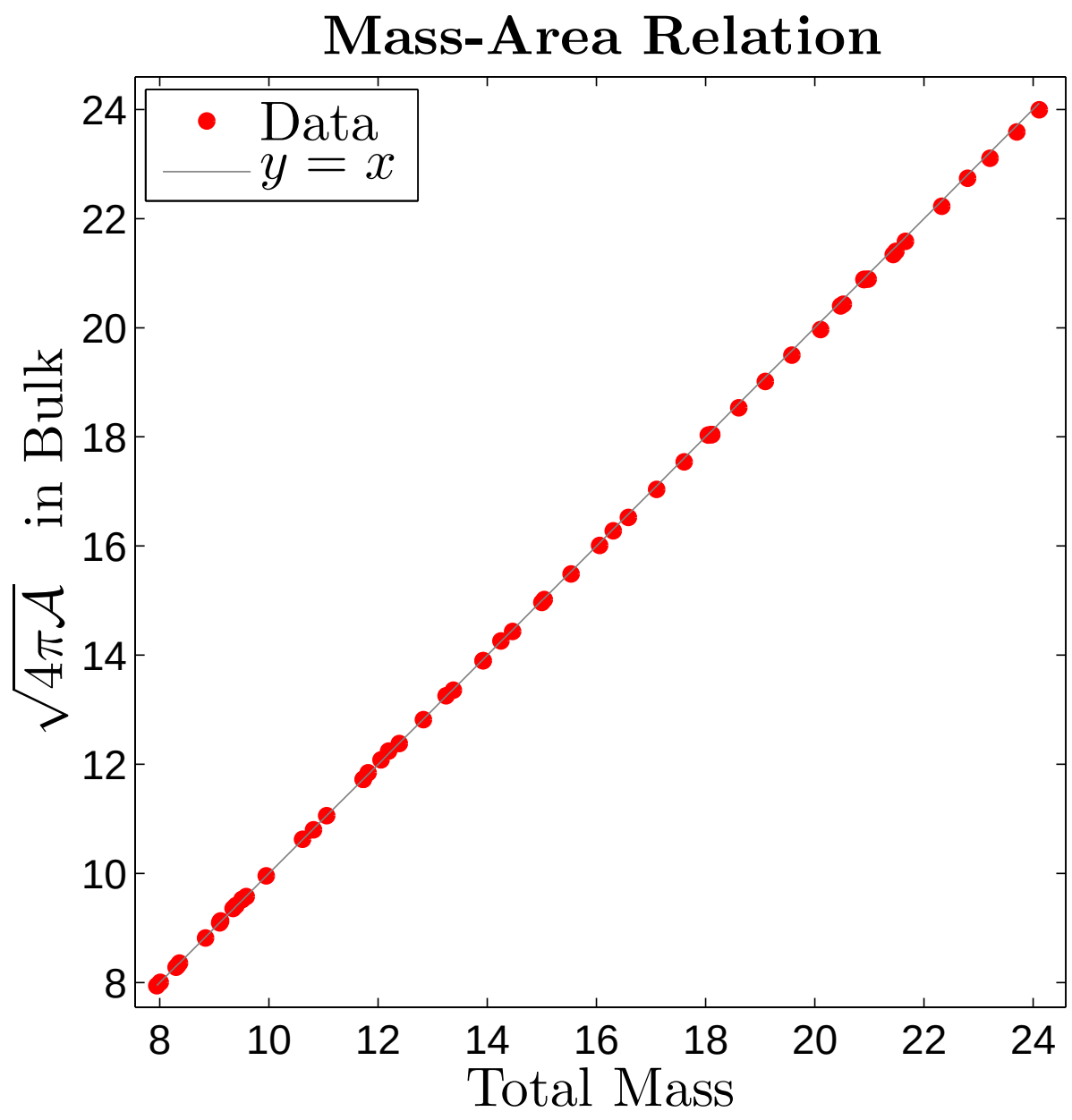

Figure 4.5: The mass-area relation of apparent horizon. This plot is produced from simulations with various parameters $(\mathscr{A}, \sigma_r, x_0)$, using both direct method and Laplacian method. The only criteria is that the configurations defined by the parameters are able to produce apparent horizons in the bulk.

At the end, we emphasize that the range of the size of the BHs is merely $r_{\mathrm{a}} \sim (0.7, 2.0)$, and Fig. 4.5 exhibits a small but systematic deviation between the data and the line of a linear fit. Therefore further study is needed.

## 4.5 Discussion: the Results of Laplacian Specification

The total energy is obtained in eq. (4.20), which was derived based on the condition that $|B| \ll |A|$ in $r \gg z$ region. By the construction of the "direct specification", $|B| \ll |A|$ is met automatically. On the other hand, without *pre-setting* this condition, this condition *emerged* in the star solution [47] and the small black hole solution [14]. In our initial value problem, there is no such a *pre-set* condition in Laplacian specification method. It is then interesting to see whether this condition can still *emerge*. Fig. 4.6 shows the results of a series of simulations from the family $(\mathscr{A}, \sigma_r, x_0) = (\mathscr{A}, 0.15, 0.5)$ using Laplacian specification, where we see that indeed $A \approx \alpha_1/rz$ emerges. The validity of the discussion leading to $M_{\text{total}} = \alpha_1/G_5$ is based on the assumption $A \approx \alpha_1/rz$ and $|B| \ll |A|$. Since $B \approx c/r^{2.3}$ and $A \approx \alpha_1/rz$, $\forall$ (small) $\varepsilon > 0$, $\exists\, Q = (cv/\alpha_1\varepsilon)^{10/3}$ such that $|B/A| \leq \varepsilon$ for $r = Q, z \in [1, v \cdot Q]$ (which is the region to calculate the total energy, as shown in Fig. 2.2(a) and discussed in Sec. 2.4.2). Therefore the assumptions hold, and we still have $M_{\text{total}} = \alpha_1/G_5$.

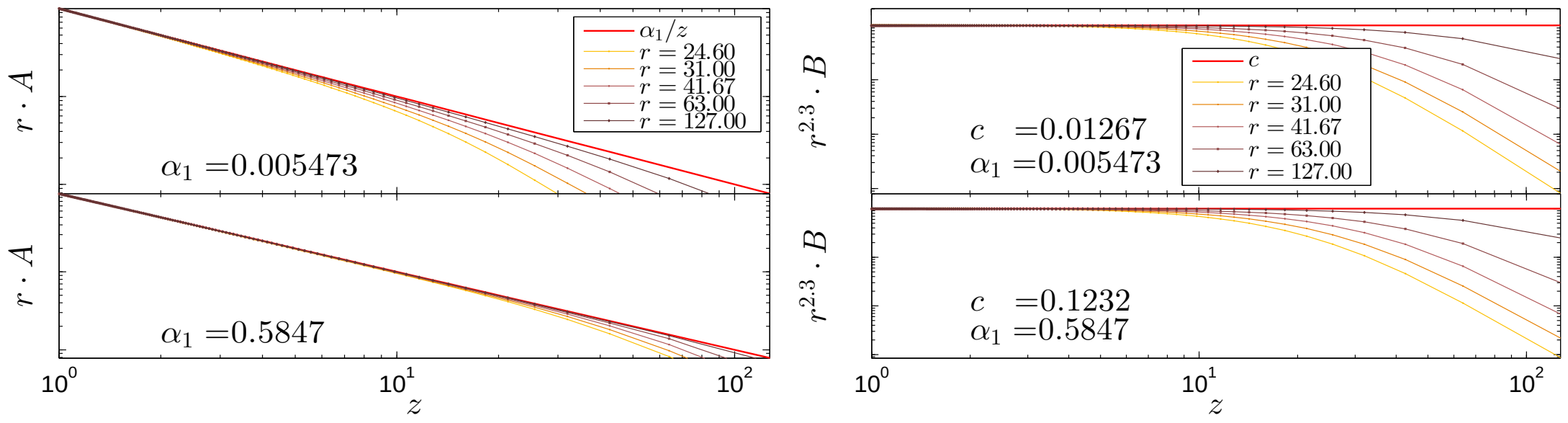

Figure 4.6: The asymptotic behaviour of $A$ at $r \gg z$ region obtained via Laplacian Specification, shown by simulations from the family $(\mathscr{A}, \sigma_r, x_0) = (\mathscr{A}, 0.15, 0.5)$. This diagram shows that the asymptotic behaviour of $A$ is indeed $A \approx \alpha_1/(rz)$. Instead of almost identical configurations as in Fig. 4.4, this diagram shows interesting pattern: 1) the asymptotic behaviour of stronger data is more like $A \approx \alpha_1/rz$; 2) disregard the trend with the strength of the data, as long as $r$ is big enough, there is always $A \approx \alpha_1/rz$. The asymptotic behaviour of $B$ is observed to be $B \approx c/r^{2.3}$. The fact that $c/\alpha_1$ decreases with $\alpha_1$ explains that $A$ become more like $\alpha_1/(rz)$ with the increase of $\alpha_1$.

## 4.6 Prepare the Initial Data for Evolution

Often the lapse $\tilde{\alpha}$ and the shifts $\tilde{\eta}_{ti}$ are freely specifiable in initial data. However, for the braneworld, the Israel's junction condition imposes non-trivial boundary conditions on the lapse function and the shift functions. For our specific initial data ansatz with time symmetric initial configuration, the boundary conditions for the shift functions are automatically satisfied by setting $\tilde{\eta}_{ti} = 0$. The lapse function, on the other hand, has a non-trivial boundary condition. We denote the $tt$ component of the metric as $g_{tt} = -\left(L^2/z^2\right) e^{2a}$, then Israel's boundary condition for $a$ is

$$a_{,z} = 1 - e^{A+B} + \frac{1}{12} e^{-A-B} \left(\Phi_{,r}\right)^2 . \tag{4.24}$$

Using the direct specification, $a$ can take

$$a = f \cdot \left(1 - e^{A} + \frac{1}{12} e^{-A} \left(\Phi_{,r}\right)^2\right), \tag{4.25}$$

where $f$ is the same $f$ used in eq. (4.10) to specify $B$.

### 4.6.1 Prepare the Initial Data for Specific Initial Source Functions

Given the values of the metric functions and their time derivatives, the constraints for source functions $\tilde{h}_\mu + \Delta\tilde{\Gamma}_\mu \simeq 0$ (which is satisfied at initial instant), can give us the initial values of the source functions. Alternatively, as used by Pretorius [63], if one would like to set a certain gauge at the initial instant by setting the value of $\tilde{h}_\mu$, then one can use the constraint equations to work backwards to get the values of $\partial_t \tilde{\eta}_{t\mu}$. In this way the initial values of $\partial_t \tilde{\eta}_{t\mu}$ is expressed in terms of the value of the metric components and their derivatives. Let us denote the obtained *expression* as $\tilde{\eta}_{t\mu,t}(\tilde{\eta}, \partial\tilde{\eta})$.

This procedure works well in ordinary spacetime (such as the 5D massless scalar field studied in section 3.4). However, in the braneworld, generally $\tilde{\eta}_{tz,t}(\tilde{\eta}, \partial\tilde{\eta})$ is not continuous at the brane under the *perpendicular gauge (2.7)*. This is because, $\partial_t \tilde{\eta}_{tz}$ should be zero at the brane, due to the boundary condition $\tilde{\eta}_{tz}\big|_{z=1} = 0$ at all time. On the other hand, $\tilde{\eta}_{tz,t}(\tilde{\eta}, \partial\tilde{\eta})$ will include $z$ derivatives of other metric functions (i.e. $\tilde{\eta}_{tt,z}, \tilde{\eta}_{rr,z}, \tilde{\eta}_{zz,z}, \tilde{\eta}_{\theta\theta,z}$), which generally results in a non-zero $\tilde{\eta}_{tz,t}(\tilde{\eta}, \partial\tilde{\eta})$ at the brane, due to Israel's junction condition. i.e. $\tilde{\eta}_{tz,t}$ is not continuous at the brane.

This discontinuity problem can be solved by properly adjusting $\tilde{\eta}_{zz}$. If we do not choose $\tilde{\eta}_{zz} = \tilde{\eta}_{rr}$ at the initial instant, there is the freedom to setup the value of $\tilde{\eta}_{zz}$ since neither Israel's junction condition nor parity condition impose any requirement on $\tilde{\eta}_{zz}$. We can use this freedom to adjust $\tilde{\eta}_{zz}$ such that $\tilde{\eta}_{tz,t}(\tilde{\eta}, \partial\tilde{\eta}) = 0$ at the brane. In this case the simplest ansatz for the *initial metric* is

$$\mathrm{d}s^2 = \frac{\ell^2}{z^2}\Big( - e^{2a}\mathrm{d}t^2 + e^{2A+2B}\mathrm{d}r^2 + e^{2A-2B}\left(\mathrm{d}\theta^2 + \sin^2\theta \mathrm{d}\phi^2\right) + e^{2b}\mathrm{d}z^2\Big), \tag{4.26}$$

for which the Hamiltonian constraint reads

$$\begin{aligned}
&-\frac{-1+e^{4B}}{2r^2} - \frac{3e^{2(A+B)}}{z^2} + \frac{1}{2z^2}e^{2(A-b+B)}\bigg(6 + 6z^2\left(A_{,z}\right)^2 + 3zB_{,z} + 2z^2\left(B_{,z}\right)^2 \\
&+ zb_{,z}\left(3 + zB_{,z}\right) - zA_{,z}\left(9 + 3zb_{,z} + 4zB_{,z}\right) + 3z^2A_{,zz} - z^2B_{,zz}\bigg) \\
&+ \frac{1}{2}\left(A_{,r}\right)^2 + \frac{1}{2}A_{,r}b_{,r} + \frac{1}{2}\left(b_{,r}\right)^2 + \frac{2A_{,r} + b_{,r} - 4B_{,r}}{r} \\
&- 3A_{,r}B_{,r} - \frac{3}{2}b_{,r}B_{,r} + \frac{5}{2}\left(B_{,r}\right)^2 + A_{,rr} + \frac{1}{2}b_{,rr} - B_{,rr} = 0.
\end{aligned} \tag{4.27}$$

The boundary conditions of $a, A, B$ are now

$$a_{,z} = \frac{1}{12}e^{-2A-2B}\left(\Phi_{,r}\right)^2,\ A_{,z} = -\frac{1}{6}e^{-2A-2B}\left(\Phi_{,r}\right)^2,\ B_{,z} = -\frac{1}{4}e^{-2A-2B}\left(\Phi_{,r}\right)^2. \tag{4.28}$$

The only requirement on $b$ (from solving $\tilde{\eta}_{tz,t}(\tilde{\eta}, \partial\tilde{\eta}) = 0$) is its derivative at the brane

$$b_{,z} = -\left(\tilde{h}_z + \frac{1}{6}e^{-2A-2B}\left(\Phi_{,z}\right)^2\right), \tag{4.29}$$

so that $\partial_t\tilde{\eta}_{tz}$ is continuous on the brane. The $\tilde{h}_z$ in (4.29) is the target initial value of $\tilde{h}_z$ that we try to achieve at the initial instant. For example, $\tilde{h}_z = 0$ if the harmonic gauge is chosen to be imposed at the initial instant.

Similar to the direct specification method applied to $B$ above, we can adopt direct specification method to specify $(a, B, b)$

$$a = f\cdot\left(\frac{1}{12}e^{-2A}\left(\Phi_{,r}\right)^2\right),\ B = f\cdot\left(-\frac{1}{4}e^{-2A}\left(\Phi_{,r}\right)^2\right),\ b = f\cdot\left(-\left(\tilde{h}_z + \frac{1}{6}e^{-2A}\left(\Phi_{,z}\right)^2\right)\right). \tag{4.30}$$

Or

$$a = f\cdot\left(\frac{1}{12}e^{-2A_0}\left(\Phi_{,r}\right)^2\right),\ B = f\cdot\left(-\frac{1}{4}e^{-2A_0}\left(\Phi_{,r}\right)^2\right),\ b = f\cdot\left(-\left(\tilde{h}_z + \frac{1}{6}e^{-2A_0}\left(\Phi_{,z}\right)^2\right)\right), \tag{4.31}$$

where $A_0 \equiv A|_{z=1}$, is the value of $A$ on the brane. While both of (4.30) and (4.31) work, the numerical experiments show that (4.31) enables the Hamiltonian constraint equation (which is an elliptic equation in term of $A$) converges faster when performing numerical relaxation. Therefore we adopt (4.31).

For an example of the initial data, please refer to Fig. 4.7.

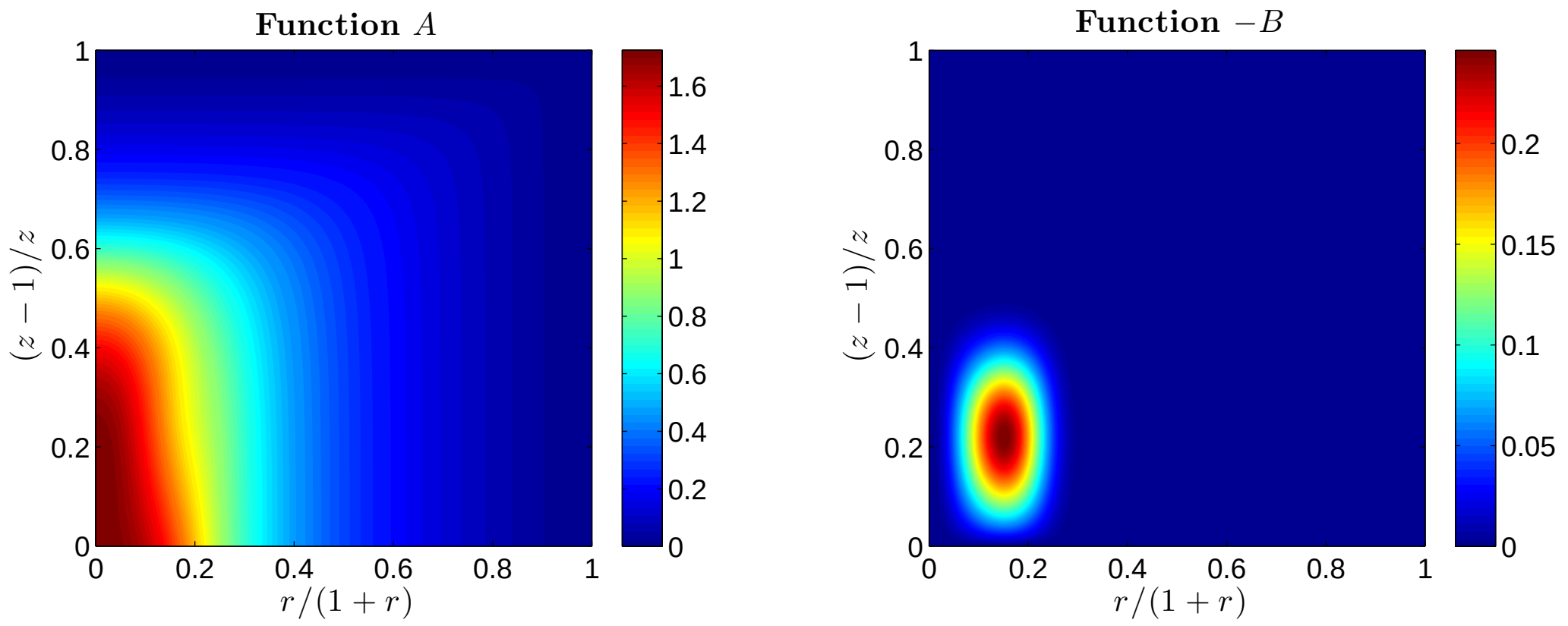

(a) Initial Data: function $A$

(b) Initial Data: function $-B$

Figure 4.7: Function $A$ and $-B$ in initial data metric (4.26) from the initial data configuration with parameters $(\mathscr{A}, \sigma_r, x_0) = (2.5, 0.2, 0)$

### 4.6.2 Total Energy

The total energy from the metric (4.26) can be obtained by the same method presented in Sec. 2.4.2. For this metric, according to equation (4.30) or (4.31), we have $b \sim 0$ and $B \sim 0$ at large $r$ (or large $z$) region. Therefore the embedding conditions (2.40) and (2.41) reduce to

$$\frac{\bar{r}}{\bar{z}} = \frac{r}{z} e^{A}; \tag{4.32a}$$

$$\frac{(\bar{r}')^2 + (\bar{z}')^2}{\bar{z}^2} = \frac{1}{z^2}, \tag{4.32b}$$

where $' \equiv \mathrm{d}/\mathrm{d}z$.

Only the asymptotic behaviour of $A$ at $r \gg z$ region contributes to the calculation of total energy. The asymptotic behaviour is (shown by Fig. 4.8)

$$A \approx \frac{\alpha_1}{r} \quad \text{at} \quad r \gg z. \tag{4.33}$$

This asymptotic behaviour gives the solution to (4.32) at the lowest order of $A$ as

$$\bar{z} \approx z, \quad \bar{r} \approx r \cdot \exp\left(\frac{\alpha_1}{r}\right), \tag{4.34}$$

which yield the total energy

$$M_{\text{total}} = \alpha_1 / G_5 = 8\pi\alpha_1. \tag{4.35}$$

Therefore, the equality of the total energy with the ADM "mass" calculated on the brane, also holds for metric (4.26).

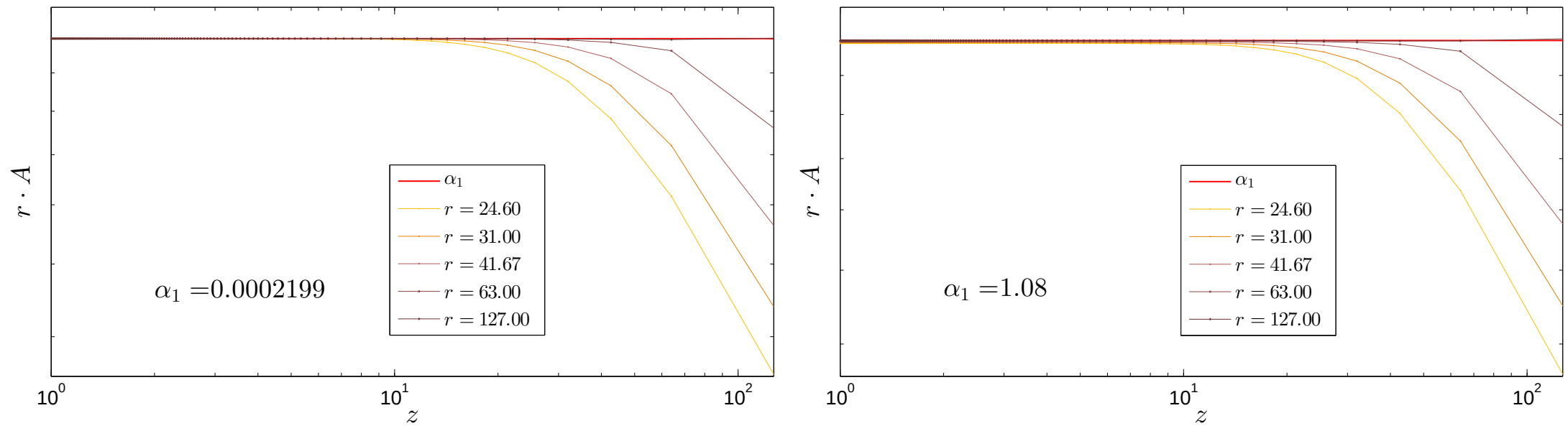

Figure 4.8: The asymptotic behaviour of $A$ at $r \gg z$ region, shown by numerical solutions for the family $(\mathscr{A}, \sigma_r, x_0) = (\mathscr{A}, 0.15, 0.5)$. The diagrams are almost identical, and here we only show the one with the smallest $\alpha_1$ and the one with the largest $\alpha_1$. These diagrams show that the asymptotic behaviour of $A$ is $A \approx \alpha_1/r$ for $\alpha_1$ over a wide range of magnitude. Therefore, the conclusion of $A \approx \frac{\alpha_1}{r}$ at $r \gg z$, is robust.

# Chapter 5

# Evolution

## 5.1 The Evolution as an Initial Value Problem

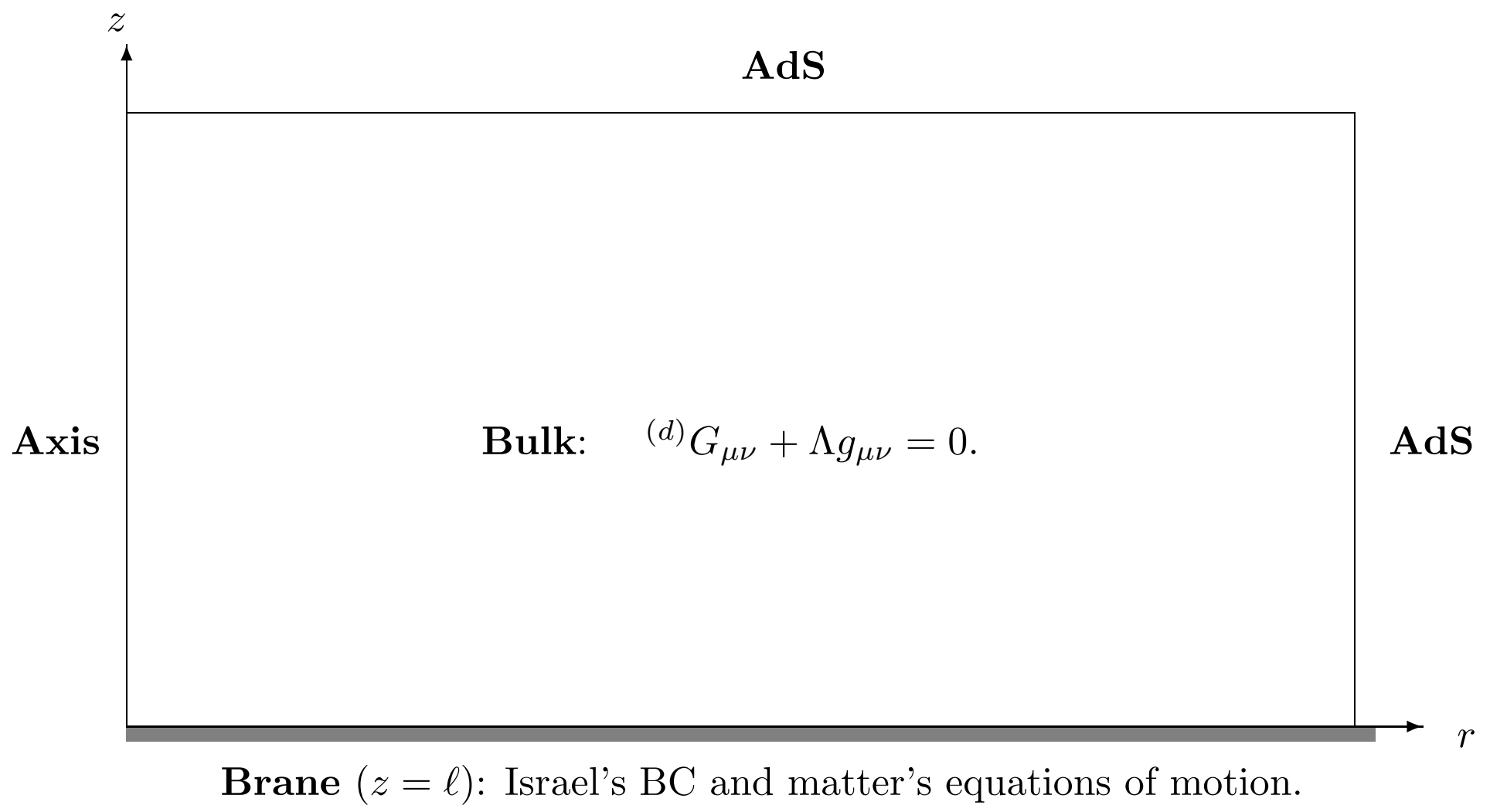

Figure 5.1: The evolution as an initial value problem with boundary conditions. $z \in [\ell, \infty)$ is the extra dimension, while the brane is located at $z = \ell$ (expressed by the shadowed line in the diagram). $r \in [0, \infty)$ is the radius coordinate, and $r = 0$ ($z$ axis) is the symmetry axis of the SO(3) symmetry.

The evolution of the braneworld is specified by the evolution equation(s) with initial data and the boundary conditions. The initial data was discussed in Chap. 4. Here we present the evolution equation and the boundary conditions.

The governing equations are composed of the 5D Einstein's equations in the bulk and the matter's equation of motion on the brane. The 5D Einstein's equations are

$$G_{\alpha\beta} = -\Lambda g_{\alpha\beta} =: k_d T_{\alpha\beta} \quad \Leftrightarrow \quad R_{\alpha\beta} = k_d \left( T_{\alpha\beta} - g_{\alpha\beta} \frac{T}{d-2} \right). \tag{5.1}$$

Its form in terms of the conformal metric $\tilde{g}_{\alpha\beta}$ is

$$\tilde{R}_{\alpha\beta} = k_d \left( T_{\alpha\beta} - \tilde{g}_{\alpha\beta} \frac{\tilde{T}}{d-2} \right) - R^{(\Psi)}_{\alpha\beta}, \tag{5.2}$$

where $\tilde{g}$, $\Psi$, $R^{(\Psi)}_{\alpha\beta}$ and $\bar{R}_{\alpha\beta}$ were introduced in section 3.3. For the conformal function and the conformal parameter, we take $q = 2$ and $\Psi = 1/z$ in our calculations.

The most general conformal metric in terms of regular functions $\tilde{\eta}$, is (see eq. (3.35))

$$\tilde{g}_{\alpha\beta} = \begin{pmatrix} \tilde{\eta}_{tt} & \tilde{\eta}_{tr} & 0 & 0 & \tilde{\eta}_{tz} \\ \tilde{\eta}_{tr} & \tilde{\eta}_{rr} & 0 & 0 & \tilde{\eta}_{rz} \\ 0 & 0 & \tilde{\eta}_{\theta\theta} r^2 & 0 & 0 \\ 0 & 0 & 0 & \tilde{\eta}_{\theta\theta} r^2 \sin^2\theta & 0 \\ \tilde{\eta}_{tz} & \tilde{\eta}_{rz} & 0 & 0 & \tilde{\eta}_{zz} \end{pmatrix}; \tag{5.3}$$

where $\tilde{\eta}_{\theta\theta}$ is taken as equal to $\tilde{\eta}_{rr} + r\tilde{W}$, which makes the local flatness condition $\tilde{\eta}_{\theta\theta}\big|_{r=0} = \tilde{\eta}_{rr}\big|_{r=0}$ be satisfied automatically.

The evolution equation (5.2) can be solved in two ways. The first way, is to use (3.48), in terms of the following source functions (see eq. (3.36))

$$\begin{pmatrix} \tilde{H}_t \\ \tilde{H}_r \\ \tilde{H}_\theta \\ \tilde{H}_\phi \\ \tilde{H}_z \end{pmatrix} = \begin{pmatrix} \tilde{h}_t + (2/r)\,(\tilde{\eta}_{tr}/\tilde{\eta}_{\theta\theta}) \\ \tilde{h}_x + (2/r)\,(\tilde{\eta}_{rr}/\tilde{\eta}_{\theta\theta}) \\ \cot\theta \\ 0 \\ \tilde{h}_z + (2/r)\,(\tilde{\eta}_{rz}/\tilde{\eta}_{\theta\theta}) \end{pmatrix}. \tag{5.4}$$

Alternatively, a second way, is to use the generalized harmonic formalism with conformal function and tensorial source functions, which is equation (3.52)

$$\begin{aligned} &-\frac{1}{2}\tilde{g}^{\alpha\beta}\bar{\nabla}_\alpha\bar{\nabla}_\beta\tilde{g}_{\mu\nu} - \bar{\nabla}_\alpha\tilde{g}_{\beta(\mu}\bar{\nabla}_{\nu)}\tilde{g}^{\alpha\beta} - \bar{\nabla}_{(\mu}\tilde{h}_{\nu)} + \tilde{h}_\alpha\tilde{C}^\alpha{}_{\mu\nu} - \tilde{C}^\alpha{}_{\mu\beta}\tilde{C}^\beta{}_{\alpha\nu} \\ &= k_d\left(T_{\mu\nu} - \frac{1}{d-2}\tilde{g}_{\mu\nu}\tilde{T}\right) - \bar{R}_{\mu\nu} - R^{(\Psi)}_{\mu\nu}. \end{aligned} \tag{5.5}$$

When the background in (5.5) is taken to be flat spacetime (which is what we adopted in the numerical calculations presented in this chapter), the two ways are *equivalent*. For discussion and presentation purpose, we will adopt the second, eq. (5.5).

A second part of the governing equations is the equation of motion of the matter on the brane, which only "feels" the 4D metric. For massless scalar field, the equation of motion is

$$\mathcal{D}^{\alpha}\mathcal{D}_{\alpha}\Phi = 0, \tag{5.6}$$

where $\mathcal{D}$ is the covariant derivative associated with the brane metric $h_{\alpha\beta}$.

Next we look at boundary conditions. The boundary conditions at the spatial infinities ($r \to \infty$ or $z \to \infty$) are Dirichlet conditions such that the metric takes the form of the background

$$\mathrm{d}s^2 = \frac{\ell^2}{z^2}\Big( -\mathrm{d}t^2 + \mathrm{d}r^2 + r^2\left(\mathrm{d}\theta^2 + \sin^2\theta\mathrm{d}\phi^2\right) + \mathrm{d}z^2\Big), \tag{5.7}$$

which is a Poincaré patch of the AdS spacetime (see Sec. 1.4). The background of RSII braneworld is the $z \geq \ell$ portion. In terms of the conformal metric components, the boundary conditions at $r \to \infty$ and $z \to \infty$ are $\tilde{\eta}_{rr} = 1, \tilde{\eta}_{tt} = -1, \tilde{\eta}_{zz} = 1, \tilde{\eta}_{tr} = 0, \tilde{\eta}_{tz} = 0, \tilde{\eta}_{rz} = 0, \tilde{W} = 0$, and $\Phi = 0$ for the matter at $r \to \infty$.

The boundary conditions on the symmetry axis are parity conditions, and the local flatness condition, which are $\tilde{\eta}_{tt,r}|_{r=0} = 0$, $\tilde{\eta}_{rr,r}|_{r=0} = 0$, $\tilde{\eta}_{tz,r}|_{r=0} = 0$, $\tilde{\eta}_{zz,r}|_{r=0} = 0$, $\tilde{\eta}_{tr}|_{r=0} = 0$, $\tilde{\eta}_{rz}|_{r=0} = 0, \tilde{W}|_{r=0} = 0$, $\Phi_{,r}|_{r=0} = 0$.

The boundary conditions at the brane are Israel's junction condition (1.29)

$$\mathcal{K}^{+}_{\alpha\beta} = -\mathcal{K}^{-}_{\alpha\beta} = \frac{1}{2}k_d\left(\lambda\frac{h_{\alpha\beta}}{d-2} + \tau_{\alpha\beta} - h_{\alpha\beta}\frac{\tau}{d-2}\right), \tag{5.8}$$

which translates into conditions on $\tilde{\eta}_{tt,z}, \tilde{\eta}_{tr,z}, \tilde{\eta}_{rr,z}, \tilde{W}_{,z}$. The expressions are long and the specific forms do not matter at this stage, and are thus not written here. The expressions are in terms of metric functions, their first order derivatives with respect to $r$ and $t$, and $\Phi_{,r}$, $\Phi_{,t}$. As discussed in Sec. 2.1, generically there is no boundary condition for $\tilde{\eta}_{\mu z}$, yet we need their boundary conditions in the numerical calculation. These conditions are more subtle and will be discussed below.

The massless scalar field $\Phi$ only exists at $z = \ell$. Therefore it is not defined in the bulk.

## 5.2 Details of the Numerical Implementation

For spatial coordinates, we use "compactified coordinates" to include spatial infinities into the calculation domain, and numerical grid is uniform in $\hat{R}$ and $\hat{Z}$:

$$\hat{R} \equiv \frac{r}{r + r_0}; \qquad \hat{Z} \equiv \frac{z - \ell}{z - \ell + z_0}, \tag{5.9}$$

where $r_0$ and $z_0$ are parameters to control the scale of the compactification. The derivatives in the equations need to be changed accordingly. For example, $f_{,r} = f_{,\hat{R}} \cdot \frac{\partial \hat{R}}{\partial r}$, etc. The numerical calculations are performed using the central stencils of finite difference approximation operators of second order accuracy, whose explicit forms are

$$\partial_t f \to \frac{f_{i,j}^{n+1} - f_{i,j}^{n-1}}{2 \cdot \Delta t}, \tag{5.10}$$

$$\partial_{\hat{R}} f \to \frac{f_{i+1,j}^{n} - f_{i-1,j}^{n}}{2 \cdot \Delta \hat{R}}, \tag{5.11}$$

$$\partial_{\hat{Z}} f \to \frac{f_{i,j+1}^{n} - f_{i,j-1}^{n}}{2 \cdot \Delta \hat{Z}}, \tag{5.12}$$

$$\partial_{tt} f \to \frac{f_{i,j}^{n+1} - 2f_{i,j}^{n} + f_{i,j}^{n-1}}{(\Delta t)^2}, \tag{5.13}$$

$$\partial_{\hat{R}\hat{R}} f \to \frac{f_{i+1,j}^{n} - 2f_{i,j}^{n} + f_{i-1,j}^{n}}{\left(\Delta \hat{R}\right)^2}, \tag{5.14}$$

$$\partial_{\hat{Z}\hat{Z}} f \to \frac{f_{i,j+1}^{n} - 2f_{i,j}^{n} + f_{i,j-1}^{n}}{\left(\Delta \hat{Z}\right)^2}, \tag{5.15}$$

$$\partial_{t\hat{R}} f \to \frac{1}{4 \cdot \Delta t \cdot \Delta \hat{R}} \left( f_{i+1,j}^{n+1} - f_{i-1,j}^{n+1} + f_{i-1,j}^{n-1} - f_{i+1,j}^{n-1} \right), \tag{5.16}$$

$$\partial_{t\hat{Z}} f \to \frac{1}{4 \cdot \Delta t \cdot \Delta \hat{Z}} \left( f_{i,j+1}^{n+1} - f_{i,j-1}^{n+1} + f_{i,j-1}^{n-1} - f_{i,j+1}^{n-1} \right), \tag{5.17}$$

$$\partial_{\hat{R}\hat{Z}} f \to \frac{1}{4 \cdot \Delta \hat{R} \cdot \Delta \hat{Z}} \left( f_{i+1,j+1}^{n} - f_{i-1,j+1}^{n} + f_{i-1,j-1}^{n} - f_{i+1,j-1}^{n} \right). \tag{5.18}$$

where the index $i$ and $j$ are grid indices in $\hat{R}$ and $\hat{Z}$ directions, and $\Delta \hat{R}$ and $\Delta \hat{Z}$ are the (uniform) spacing of the grids in $\hat{R}$ and $\hat{Z}$ directions. The superscripts $n, n+1, n-1$ are the discretized time levels, where $\Delta t$ is the spacing. In the simulations performed in this chapter, we set $r_0 = z_0$ and the resolution in $\hat{R}$ is the same as that in $\hat{Z}$. Courant factor $\Delta t / \min(\Delta r) = \Delta t / (r_0 \cdot \Delta \hat{R})$ is set to be 0.5.

The residual equations are the discretized eq. (5.5) and eq. (5.6). Let $n$ denote the current time level. The update scheme is to obtain time level $n+1$ from given quantities at level $n$ and $n-1$, by solving the residual equations. We solve the residual equations by pointwise Newton-Gauss-Seidel iteration in a black-red manner (see, e.g. [73]) until residuals are smaller than a "small" threshold.

A Kreiss-Oliger (KO) [73] style numerical dissipation is added to control high frequency numerical noises. Since we use second order FDA to discretize eq. (5.5) and eq. (5.6), a fourth order KO dissipation (see [61] for the specific form) is employed. Following [61], the dissipation is applied to both $n$ and $n-1$ time levels before solving for the advanced time level $n+1$.

Adaptive Mesh Refinement (AMR) is used to reach high resolution (when needed). We used PAMR/AMRD [65] as the tool to realize the parallelization of the code. Both of the KO style dissipation and AMR are built into PAMR/AMRD. The simulations producing BHs in this chapter, however, were obtained using only one CPU with uniform grid structure, because a shooting method is employed to locate apparent horizons, which is not parallelized, nor adapted to AMR. Our plan for the future is to upgrade the code to use flow method to locate apparent horizons, which can be parellelized and adapted to AMR [62].

## 5.3 Gauge Freedom

As discussed in section 3.3, there are two ways to perform evolutions using GH formalism: the BSSN-like method and the source function driven method. Here we adopt the source function driven method. Therefore we need to consider how to impose gauges via the source functions, and how to make sure the $\tilde{h}_\mu + \Delta\tilde{\Gamma}_\mu \simeq 0$ constraints are preserved during the evolutions. In this section we focus on imposing gauges. Please refer to section 5.4 for the study of the constraint preservation.

### 5.3.1 Fixing Gauges via Source Functions

There are gauge freedoms in gravitational theories, which are about the coordinate choices. It is important to properly specify coordinates in numerical relativity to avoid coordinate pathology, to evolve spacetime with strong fields, and to deal with physical singularities. For the generalized harmonic formalism, the following gauges are generally adopted in the simulations in the literature.

(1) The simplest gauge is the harmonic gauge

$$\tilde{h}_\mu = 0. \tag{5.19}$$

(2) The lapse driving gauge used by Pretorius [63]:

$$\tilde{\Box}\tilde{h}_t + c_1 \frac{\tilde{\alpha} - \tilde{\alpha}_0}{\tilde{\alpha}^s} - c_2 \tilde{h}_{t,\mu}\tilde{n}^\mu = 0, \tag{5.20}$$

where $s \geq 0, c_1 > 0, c_2 > 0$. It is generalized from the damped harmonic equation $x_{,tt} + c_1 \cdot x + c_2 \cdot x_{,t} = 0$. With $\tilde{\Box}$, the equation has the functionality as a smoother. The effect of this gauge is to drive the lapse function towards its desired value $\tilde{\alpha}_0$. For example, Pretorius found that the instability tends to happen at the apparent horizon excision boundary when the value of the lapse function is too small. He then chose $\tilde{\alpha}_0 = 1$, the value of the lapse function in flat spacetime.

One can also apply this method to the spatial components $\tilde{h}_i$, to achieve the desired gauges.

(3) Damped wave gauge [60, 98]. In our case, the gauge reads

$$\tilde{h}_\mu = c_1 \log\left(\frac{\tilde{\eta}^P}{\tilde{\alpha}}\right) \tilde{n}_\mu - c_2\ \tilde{\eta}_{\mu i}\tilde{\beta}^i/\tilde{\alpha}, \tag{5.21}$$

where $\tilde{\eta} = \left[\tilde{\eta}_{rr}\tilde{\eta}_{zz} - (\tilde{\eta}_{rz})^2\right](\tilde{\eta}_{\theta\theta})^2$ is the determinant of the spatial (conformal) metric in Cartesian coordinate. $P, c_1, c_2$ are positive parameters. The effect is to damp out the dynamics in spatial coordinates on the time scale $1/c_2$, and to suppress the growth in $\sqrt{\tilde{\eta}}/\tilde{\alpha}$ (when $P = 1/2$) on time scale $1/c_1$.

In ours simulations, (3) was adopted for the simulations that produce small black holes. For small black holes, it is crucial to set both $r_0$ and $z_0$ much less than 1 to let the black hole boundaries include many grid points, so that the small black holes are well-resolved. (2) was adopted for the simulations that produce large black holes. Pretorius' group successfully performed many simulations [62, 66, 67] using (5.20) as the slicing condition, and $\tilde{h}_i = 0$ for spatial coordinates, which are also the conditions we use. $\tilde{\alpha}_0 = 1$ is chosen. For large BHs, (1) also works well. In any case, after BHs are obtained and the evolution is stablized, we *gradually* change the gauge into (3) to drive the value of $\tilde{\alpha}$ towards $\tilde{\alpha}_0 = 1$ so that the lapse rate of the coordinate time is comparable to that of the proper time, which enables us to define and to monitor the apparently stationary state that is going to be introduced in Sec. 5.7.1.

## 5.4 Constraint Violation and the Cure

### 5.4.1 Constraint Damping

The generalized harmonic formalism, has the constraints $\mathcal{C}_\mu \equiv H_\mu + \Gamma_\mu \simeq 0$ (or $\tilde{\mathcal{C}} \equiv \tilde{h}_\mu + \Delta\tilde{\Gamma}_\mu \simeq 0$). During the evolutions, there are always numerically errors that violate the constraints, and often the modes of the deviation from the constraint equations grow with time, and drive the evolution away from Einstein's equations. The phenomena, numerical violation from constraint equations growing with time, is a very common challenge for numerical relativity (see, e.g. [58]).

One way to improve the performance, is to add the following constraint damping terms [58] to the left hand side of the evolution equations (3.28)

$$Z_{\mu\nu} \equiv \kappa \left( n_{(\mu}\mathcal{C}_{\nu)} - \frac{1+p}{2} g_{\mu\nu} n^\beta \mathcal{C}_\beta \right), \tag{5.22}$$

where $\kappa > 0$ and $p > -1$ are constant parameters. For evolution equations (5.5), the damping terms are changed to

$$\tilde{Z}_{\mu\nu} \equiv \kappa \left( \tilde{n}_{(\mu}\tilde{\mathcal{C}}_{\nu)} - \frac{1+p}{2} \tilde{g}_{\mu\nu} \tilde{n}^\beta \tilde{\mathcal{C}}_\beta \right), \tag{5.23}$$

where $\tilde{\mathcal{C}}_\mu \equiv \tilde{h}_\mu + \Delta\tilde{\Gamma}_\mu$. These damping terms have been proven to be useful in keeping the constraints satisfied for simulations in "ordinary" spacetimes (e.g. the simulation in 5D carried out in Sec. 3.4). However, for the braneworld, it turns out that these damping terms are not sufficient — the constraints do not converge to zero as the resolution increases. In fact, the residuals of the constraints almost stay the same when the resolution increases.

We experimented with many changes to the damping term—some worked better than others. One version was to make $\kappa$ spacetime dependent. We tried a few choices and it turns out the following choice worked to a certain degree

$$\kappa \to \kappa \frac{z^n}{(z - c\,\ell)^n}, \tag{5.24}$$

where $c$ was chosen to be, for example, 0.9 or 0.95, which effectively puts more damping near the brane. This change enhanced (a little) the convergence of the constraints. However, the convergence criterion is still not perfectly satisfied.

It turns out the following choice is more successful, although we can not explain why at this moment. More work is needed.

$$\tilde{Z}_{\mu\nu} = \kappa \left( \tilde{n}_{(\mu}\tilde{\mathcal{C}}_{\nu)} - F \cdot \frac{1+p}{2} \tilde{g}_{\mu\nu} \tilde{n}^{\beta} \tilde{\mathcal{C}}_{\beta} \right), \tag{5.25}$$

where $F$ may depend on spacetime. i.e. only the second term of $\tilde{Z}_{\mu\nu}$ is multiplied by $F$, which is a spacetime dependent function. We experimented with different forms of $F$, and the following family worked the best so far (among all our trials)

$$F = \frac{z^n}{z^n - c\,\ell^n}, \tag{5.26}$$

where $c$ is close to 1, which again adds more damping on the brane. A choice was $n = 4, c = 0.99$ (refer to Fig. 5.2). This damping was successful, since the constraints converged at the expected order.

The success of this method means that the second term in constraint (5.23) and the first term may have very different effects.

The constraint damping is very important. We do not have specific guidance to give the forms of such damping parameter settings. A survey over the parameters showed that the effect of damping is very *sensitive* to parameters. Also, even for the most successful result (Fig. 5.2), the independent residuals (and the residual of constraint equations) increase rapidly after certain time. i.e. *very long term* simulations seems to be problematic.

### 5.4.2 Imposing Constraints on the Brane

From the result of Fig. 5.2, we learn that 1) the gauge choice $\tilde{h}_\mu = 0$ works; 2) the damping should be larger at the brane, which motivates us to exactly impose constraints $\tilde{h}_\mu + \Delta\tilde{\Gamma}_\mu = 0$ on the brane.

How to impose constraints $\tilde{h}_\mu + \Delta\tilde{\Gamma}_\mu = 0$? What one usually does in successful simulations in ordinary spacetimes (non-braneworld, such as the 5D simulation in Sec. 3.4), is to set $\tilde{h}$'s to certain pre-set values (e.g. in the case of the harmonic gauges, we have $\tilde{h} = 0$). In this situation, the effect of constraint damping is to drive the *metric components* to the values that satisfy the constraint equations, rather than doing something to the *source functions* $\tilde{h}_\mu$. i.e. to impose constraints, it is the metric (rather than $\tilde{h}$'s) that should be guided.

How does this guidance happen? Let us define ${}^{\text{should}}\tilde{h}_\mu \equiv -\tilde{g}_{\mu\nu}\left(\tilde{\Gamma}^{\nu}{}_{\alpha\beta} - \bar{\Gamma}^{\nu}{}_{\alpha\beta}\right)\tilde{g}^{\alpha\beta}$. i.e. they are what the $\tilde{h}$'s should be, if the constraints are exactly satisfied. The ${}^{\text{should}}\tilde{h}$'s are functions of the metric components and their derivatives. One can set the value of metric components or their derivatives at the brane to let ${}^{\text{should}}\tilde{h}_\mu = \tilde{h}_\mu$ (therefore imposing constraints), which can be done in multiple ways. Here we adopt the following. ${}^{\text{should}}\tilde{h}_\mu = \tilde{h}_\mu$ can be equivalently expressed by setting the values of $\tilde{\eta}_{tz,z}, \tilde{\eta}_{rz,z}$ and $\tilde{\eta}_{zz,z}$, which in principle can serve as the boundary conditions for $\tilde{\eta}_{\mu z}$—as shown in Sec. 2.1, generically there is no boundary condition for $\tilde{\eta}_{\mu z}$.

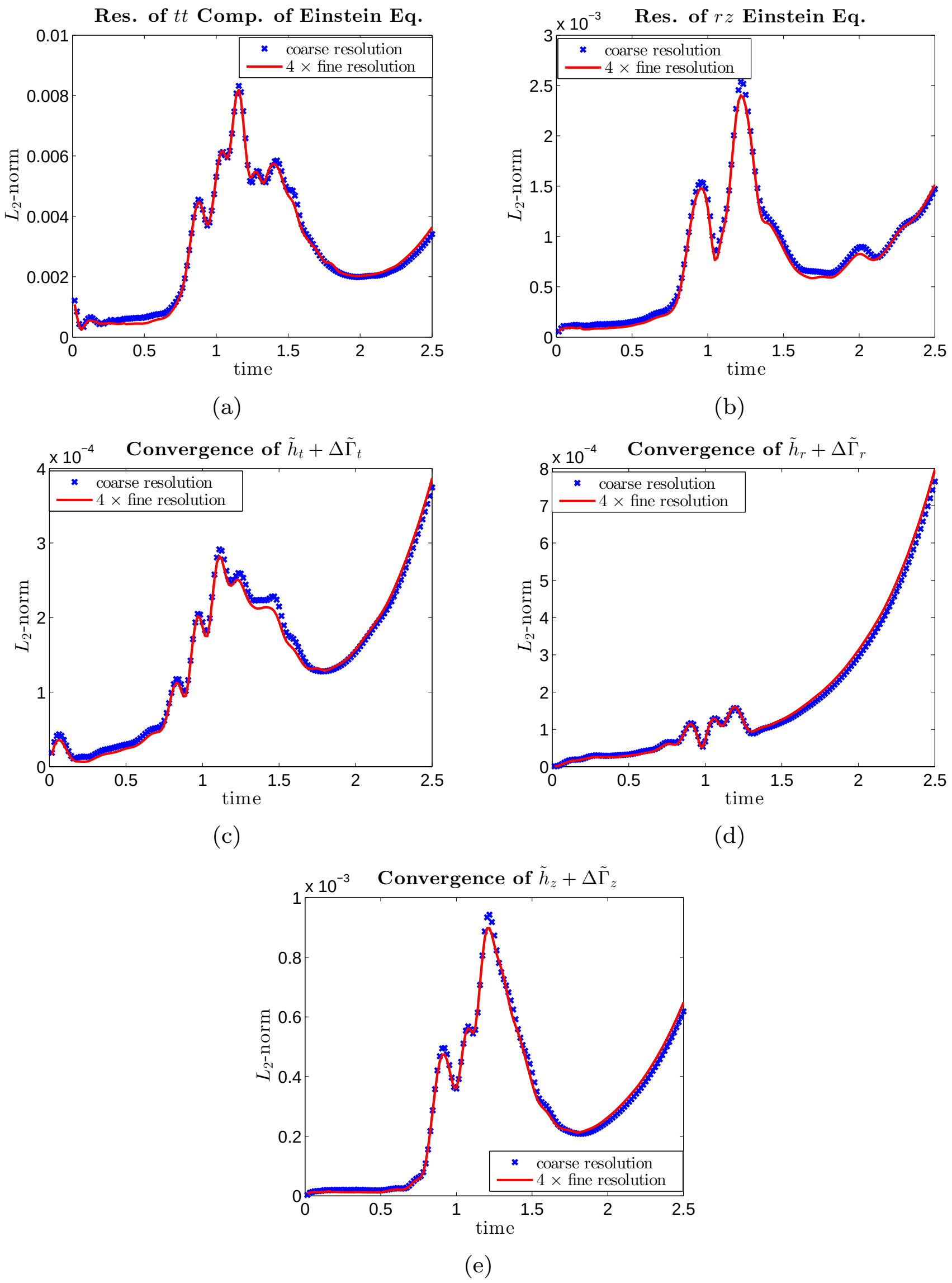

Figure 5.2: This convergence test is for the results obtained from the simulation using harmonic gauge $\tilde{h}_\mu = 0$, and constraint damping $\tilde{Z}_{\mu\nu} = \kappa\Big(\tilde{n}_{(\mu}\tilde{C}_{\nu)} - \frac{1+p}{2}\tilde{g}_{\mu\nu}\tilde{n}^\alpha\tilde{C}_\alpha \cdot \frac{z^4}{z^4-0.99}\Big)$. The initial data is $\Phi = \mathscr{A}\exp\left[-(r-x_0)^2/\sigma_r^2\right]$ with $(\mathscr{A}, r_0, \sigma_r) = (0.05, 1, 0.25)$. After the results are obtained by the generalized harmonic formalism with a conformal function, the results are substituted into the original Einstein's equations in terms of the original metric functions without the conformal function, to get the residuals. The residuals should converge to zero as second order quantities, if the results are numerical solutions. Fig. (a) is the convergence of the residual obtained from the $(tt)$ component of Einstein's equations $R_{\mu\nu} = k_d\left(T_{\mu\nu} - g_{\mu\nu}\frac{T}{d-2}\right)$. Fig. (b) is the test from the $(rz)$ component. Fig. (c,d,e) show the convergence of constraints: $\tilde{h}_\mu + \Delta\tilde{\Gamma}_\mu$ with $\mu = t, r, z$, respectively. The residuals converge at the expected order. However, this is the result obtained from very extreme damping, and all the residuals have up-climbing tails.

In addition, we impose the perpendicular gauge at the brane ($\tilde{\eta}_{tz}\big|_{z=\ell} = \tilde{\eta}_{rz}\big|_{z=\ell} = 0$) since this gauge gives the smoothness of the apparent horizon across the brane, as discussed in Sec. 2.2.2. Now we have two conditions for $\tilde{\eta}_{tz}$ (and two conditions for $\tilde{\eta}_{rz}$): one is constraint imposing condition which is a condition on the value of $\tilde{\eta}_{tz,z}$ (and $\tilde{\eta}_{rz,z}$), the other is the perpendicular gauge condition which is a condition on the value of $\tilde{\eta}_{tz}$ (and $\tilde{\eta}_{rz}$). Imposing two boundary conditions on one function is achieved by the following trick: let one of the conditions be satisfied automatically by choosing the forms of the functions:

$$\tilde{\eta}_{tz} \equiv (z-\ell)\cdot\tilde{\xi}_t, \tag{5.27}$$

$$\tilde{\eta}_{rz} \equiv (z-\ell)\cdot\tilde{\xi}_r. \tag{5.28}$$

In this way, the $\tilde{\eta}_{tz}|_{z=\ell} = \tilde{\eta}_{rz}|_{z=\ell} = 0$ are automatically satisfied. The conditions on $\tilde{\eta}_{tz,z}$ and $\tilde{\eta}_{rz,z}$ are now converted into the conditions on the values of $\tilde{\xi}_t$ and $\tilde{\xi}_r$. We then use $\tilde{\xi}_t$ and $\tilde{\xi}_r$ as fundamental variables instead of $\tilde{\eta}_{tz}$ and $\tilde{\eta}_{rz}$.

It turns out that this method works well, and there is *no need* to use special damping. Beyond the constraints imposed on the brane, the "normal and plain" damping term $\tilde{Z}_{\mu\nu} = \kappa\left(\tilde{n}_{(\mu}\tilde{\mathcal{C}}_{\nu)} - \frac{1+p}{2}\tilde{g}_{\mu\nu}\tilde{n}^\alpha\tilde{\mathcal{C}}_\alpha\right)$ is still employed to control "ordinary" violation modes in the bulk. The tests of the method are shown in Fig. 5.3.

## 5.5 The Evolution with an Apparent Horizon

During the evolution, sometimes singularities are formed and the code crashes. There are at least two types of singularities. The first is the coordinate singularity due to pathological coordinate gauges, which can be avoided by properly choosing coordinate gauges (which is non-trivial). The second is the physical singularity. However, Penrose's cosmic censorship hypothesis [84] states that, the physical singularity is always hidden behind an event horizon. While there is no proof of this hypothesis, it does seem to be satisfied in many cases.

Event horizon is the boundary in spacetimes that separates the events which can be causally connected with future Poincaré horizon by future oriented null geodesics, from the events which can not. The interior surrounded by event horizons, by definition, can not affect the regions outside of the event horizons. Therefore, one way to avoid the physical singularity in the calculation domain is to perform the evolution without referring to the interior of the event horizons (so that the interior of the event horizons is excised from the calculation domain).

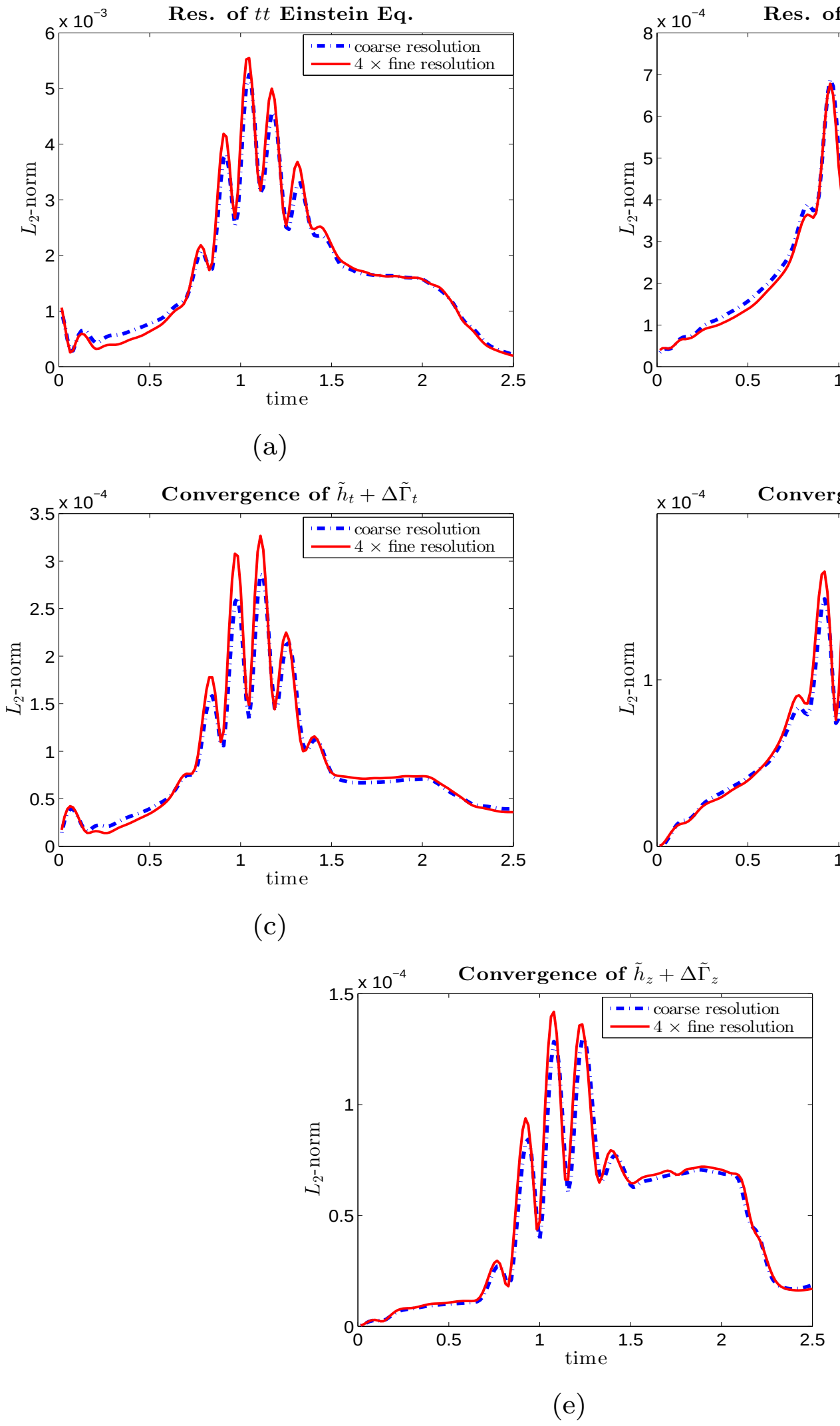

Figure 5.3: The convergence Tests for the simulation using gauge $\tilde{h}_\mu = 0$. The initial data is $\Phi = \mathscr{A}\exp\left[-(r-x_0)^2/\sigma_r^2\right]$ with $(\mathscr{A}, r_0, \sigma_r) = (0.05, 1, 0.25)$. After the results are obtained by the generalized harmonic formalism with a conformal function, the results are substituted into the original Einstein's equations in terms of the original metric functions without the conformal function, to get the residuals. The residuals should converge to zero as second order quantities, if the results are numerical solutions. Fig. (a) is the convergence of the residual from the $(tt)$ component of Einstein's equations $R_{\mu\nu} = k_d\left(T_{\mu\nu} - g_{\mu\nu}\frac{T}{d-2}\right)$. Fig. (b) is the test for the $(rz)$ component. Fig. (c,d,e) show the convergence of constraints: $\tilde{h}_\mu + \Delta\tilde{\Gamma}_\mu$ with $\mu = t, r, z$, respectively. In all the tests shown in the figures, the spacing of the coarser grid is $\Delta\hat{R} = \Delta\hat{Z} = 1/256$ and the spacing of the finer grid is $\Delta\hat{R} = \Delta\hat{Z} = 1/512$. The simulations were performed using 16 CPUs and the test is performed on the result obtained by the CPU at the $\hat{R} = \hat{Z} = 0$ corner, which is the region that would suffer from the most severe problems (if there was). The convergences are shown to be good. After $t \sim 2$, the residuals suddenly decreased, because the interesting dynamics propagated out of the region where we evaluate the residuals.

Event horizon, however, is a concept based on the global structure of the whole spacetime, which is therefore not quite useful during the evolution since one can not tell the event horizon until the full evolution is completed. But the full evolution is not able to be obtained without knowing the event horizon to excise the physical singularities. Fortunately there is the concept of apparent horizon which is locally (in time) defined. Apparent horizons are not guaranteed to exist in a certain evolution. But, if they do exist, they are inside of the event horizons. Therefore the interior of an apparent horizon can not affect the exterior of the event horizon. Since the apparent horizon lies inside the event horizon, and sometimes by a long way, the Penrose hypothesis does not guarantee that singularities can not exist outside the apparent horizon. But again, often the physical singularities (to be formed in a future instant) are inside of the apparent horizon. Therefore, if an apparent horizon appears, we may excise the interior of the apparent horizon to get rid of the physical singularities from the calculation domain. This idea (black hole excision) was proposed by W. G. Unruh [70], and had become a common practice to deal with physical singularities in numerical relativity.

To perform black hole excision, one needs to locate the apparent horizon.

### 5.5.1 Smoothness of Apparent Horizons in the Braneworld

In the braneworld, the Israel's junction condition at the brane essentially imposes cusp conditions to certain metric functions. This raises the question of whether apparent horizons will be non-smooth across the brane. By the discussion of the smoothness of apparent horizons in Sec. 2.2.2, the apparent horizons in the braneworld is smooth under the perpendicular gauge (2.7). Under gauge $g_{rz}\big|_{z=1} = 0$ (perpendicular gauge), the smoothness can be simply expressed as $(\mathrm{d}r/\mathrm{d}z)\big|_{z=1} = 0$, where the derivative is evaluated along the apparent horizons.

### 5.5.2 Apparent Horizon Finder

In this subsection we introduce a method to locate apparent horizons [64].

We introduce polar coordinates $(\rho, \chi)$ via $r = \rho \sin\chi, z = 1 + \rho\cos\chi$ (length dimensions are in the unit of $\ell$). In this coordinate system, the symmetric axis $r = 0$ is $\chi = 0$, and the brane $z = 1$ is $\chi = \pi/2$. On a $t =$ constant hypersurface, we define function

$$f \equiv \rho - \varrho(\chi), \tag{5.29}$$

and let the apparent horizon be the one with $f = 0$. The form of $\varrho$ is going to be determined by the apparent horizon equation (2.14). The spacelike outpointing unit vector normal to $f =$ constant surfaces is now

$$s^\alpha = \frac{p^\alpha}{\sqrt{p^\beta p_\beta}}, \quad \text{where } p_\alpha \equiv \gamma_\alpha^{\ \beta} \partial_\beta f. \tag{5.30}$$

Substituting $s^\alpha$ into (2.14), we get a second order ordinary differential equation (ODE) of $\varrho$ with respect to $\chi$. The boundary condition at $\chi = 0$ (the symmetric axis $r = 0$) is $\mathrm{d}\varrho/\mathrm{d}\chi = 0$; the boundary condition at $\chi = \pi/2$ (the brane) is $\mathrm{d}\varrho/\mathrm{d}\chi = 0$ under the perpendicular gauge at the brane ($\tilde{\eta}_{rz} = \tilde{\eta}_{tz} = 0$), because of the $Z_2$ symmetry of the braneworld and the smoothness of the apparent horizon. Apparent horizon is obtained via solving this second order ODE subject to the boundary conditions. Numerically, we use a version of the shooting method to obtain the apparent horizon. There are multiple ways to implement shooting methods. The basic idea is to start the trajectory (implied by the ODE) at certain initial point ($\chi = 0$), and then solving the ODE subject to the initial conditions yields a value of $\mathrm{d}\varrho/\mathrm{d}\chi$ at the final point ($\chi = \pi/2$) of the trajectory. We then adjust the inital point accordingly until $\mathrm{d}\varrho/\mathrm{d}\chi = 0$ at final point. One version is implemented as the following. At $\chi = 0$, we pick up certain value of $z(0)_p$ (which is $\varrho(\chi = 0)$) as the initial value of $z$ and solve the ODE to obtain the trajectory to $\chi = \pi/2$. Not losing generality, let us assume the value of $\mathrm{d}\varrho/\mathrm{d}\chi$ at $\chi = \pi/2$ is positive. Then we try different initial value of $z(0)$ until we find an initial value $z(0)_n$ such that $\mathrm{d}\varrho/\mathrm{d}\chi$ is negative at $\chi = \pi/2$. Now we have found a bracket of the initial guesses: ($z(0)_p$, $z(0)_n$). Then we can use the following binary search to find the apparent horizon. We use `z_p` for $z(0)_p$, `z_n` for $z(0)_n$ and `Rp` for $\mathrm{d}\varrho/\mathrm{d}\chi$ in the pseudo code.

```
eps = pre-set small value
Rp = 10 * eps
do while ( abs(Rp) > eps )
z_m = (z_p + z_n) / 2
solve ODE to get Rp at chi = pi/2, by initial value z = z_m
if (Rp > 0) then
     z_p = z_m
else if (Rp < 0) then
     z_n = z_m
```

```
end if
end do
z(0) = z_m
```

In performing shooting method above, we need to solve the ODE, for which we used the Runge-Kutta method of second order accuracy.

### 5.5.3 Dissipation at the Excision Boundary

To perform the excision, which is to ignore the interior of the apparent horizon during numerical evolutions, is conceptually simple and neat. However, technically it is tricky and probably messy to deal with the excision boundary. To ignore the interior of the apparent horizon, we only evolve the exterior, and the boundary condition at the excision boundary is "no boundary condition". Instabilities and noises tend to happen at the excision boundary, which should be removed by certain numerical dissipation. One way to do it, is to reconstruct the interior from the exterior, such that the reconstructed functions are smooth across the excision boundary. Then we apply Kreiss-Oliger dissipation to remove the noises. This process, while sounding trivial, is very difficult to realize. There is no good way to reconstruct smooth functions across the excision boundary. This method is able to evolve the spacetime with excision for "a while", so its success is limited. Another way, which is the best among the methods we tried, is to use the one-side dissipation at the excision boundary, which was developed by Calabrese et al. [72], and adopted in PAMR/AMRD [65]. This method significantly improves the performance.

## 5.6 Tests and the Validation of the Numerical Scheme

As pointed out in Sec. 1.6.2, the tests after the numerical results are obtained, are essential to make sure the results are actually *numerical solutions* rather than *numerical artifacts*. To this end, the independent residual tests need to be performed. For systems with constraints, convergence tests for constraint residuals need to be performed as well.

For GR, one can either perform independent residual tests together with constraint convergence tests, or perform convergence tests for residuals obtained from a different formalism of GR. Given generalized harmonic formalism (5.5) was used to obtain the solutions, where the conformally transformed metric and source functions were fundamental variables, now we perform the convergence tests of the residuals obtained from the original Einstein's equations (5.1). i.e. the numerical solutions are obtained via the generalized harmonic formalism with the conformally transformed metric as the fundamental variables. Then the solutions are transformed back to obtain the physical metric, and are then substituted into the original Einstein's equations to get residuals. These residuals should converge to zero at the expected order (which is of order 2 in our simulations), if the numerical solutions are obtained. The tests are shown in Fig. 5.3.

Beyond this, we also performed independent residuals, which are perfect thus omitted here. Beyond these, we also performed convergence tests for the residuals of the constraints, which are shown in Fig. 5.3.

## 5.7 The Numerical Solution

The scheme can be used to study a wide range of dynamical processes, such as critical phenomena, the evolution problem in cosmology, gravitational wave from collapse, etc. But we will limit our attention to the end states of gravitational collapse at this time.

We performed a series of simulations from the initial data metric (4.26) with the initial matter field as

$$\Phi = \mathscr{A} \cdot \exp\left[-(r - x_0)^2/\sigma_r^2\right] , \tag{5.31}$$

from different initial data families. Within each family, only amplitude $\mathscr{A}$ changes, while Gaussian parameters ($\sigma_r$ and $x_0$), $\sigma_z$ (the parameter in the "direct specification" function $f$ in (4.11), which is used in eq. (4.31)) and the compactification parameters ($r_0$ and $z_0$), are fixed.

### 5.7.1 The Evolution Process and Apparently Stationary State

The initial data represents a localized Gaussian pulse. Since the initial data is time symmetric, the pulse evolves into two pulses: the ingoing pulse and the outgoing pulse. For weak data, the ingoing pulse is bounced back from $r = 0$ to travel outwards, which is the same phenomena as that in GR. The unique phenomena in RSII, is the interaction between the brane and the bulk, which mainly appears as the energy leaking into the bulk from the brane. Please refer to Fig. 5.4 and 5.5.

Sufficiently strong data will lead to BHs. The spacetime with BH can continue to evolve using BH excision techniques. The properties of the BHs are studied via apparent horizons in the bulk. Apparent horizon is generally different from event horizon. However, at the end of an evolution, the system reaches its stationary state and its apparent horizon coincides with its event horizon [89]. In the braneworld, when matter is absent around the horizon, the intersection of the bulk event horizon with the brane, is actually a well-defined event horizon on the brane as proved in Sec. 2.3, therefore observable, in principle.

We monitor $r_{\rm a}$, $\mathcal{A}_{\rm bulk}$ and $C_5$ during an evolution, where $r_{\rm a}$ is the areal radius of the intersection of the apparent horizon with the brane, $\mathcal{A}_{\rm bulk}$ is the area of the apparent horizon in the bulk, and $C_5$ is the length of the circumstance of the horizon (restricted on the $r-z$ plane) in the bulk. For the simulations that can reach their end states or their apparently stationary state (to be defined below), the quantities reach the values that are almost constants. Please refer to Fig. 5.6 and Fig. 5.7 for an evolution which produces a BH of medium size, Fig. 5.8 for an evolution which produces a BH at a smaller size, Fig. 5.9 for an evolution which produces a small BH and Fig. 5.10 for an evolution which produces a large BH. Fig. 5.4 and 5.5 show the evolutions from two initial data families, monitored by the Kretschmann scalar $R_{\mu\nu\alpha\beta}R^{\mu\nu\alpha\beta}$.

To obtain the end state of a system, it is necessary to let the evolution continue sufficiently long so that the system settles down. Even so, it is non-trivial to recognize the end state in a given simulation, due to coordinate effects. When a system reaches its end state, the system has a Killing vector that is asymptotically timelike, which corresponds to the time translational symmetry. If the slicing condition is such that the $t=$ constant slices are Lie-dragged by the Killing vector, it is easy to overcome the spatial coordinate effects, by embedding the apparent horizons into a fixed space, such as the background space. During a particular evolution, the apparent horizon evolves with time, but its embedding in the fixed space will evolve into a time-independent state, if the slicing condition is adapted according to the Killing vector.

However, it is non-trivial to impose such slicing conditions. Such a slicing condition is currently not imposed in our numerical schemes [16]. As a result, we do not have a time-independent state at the end of the evolutions. In the simulations, instead, we can obtain *apparently stationary states*— the states with BHs whose horizons (embedded into a fixed space) appear to be stationary for a finite time. By this definition, the apparently stationary state that stays stationary for infinite time is actually the end state.

To recognize the apparently stationary state of a system, we monitor the evolution of the apparent horizon by its embedding into the "vacuum" background

$$ds^2 = \frac{\ell^2}{\bar{z}^2}\Big( - d\bar{t}^2 + d\bar{r}^2 + \bar{r}^2 \left(d\bar{\theta}^2 + \sin^2\bar{\theta} d\bar{\phi}^2\right) + d\bar{z}^2\Big). \tag{5.32}$$

[16] In the literature, there are the so-called symmetry seeking coordinate conditions [118–120] towards the conditions which can evolve into the coordinate configurations such that the time coordinate $t$ is adapted to the Killing vector associated with the time translational symmetry of a stationary system. It will be our future work to implement such slicing conditions in our code.

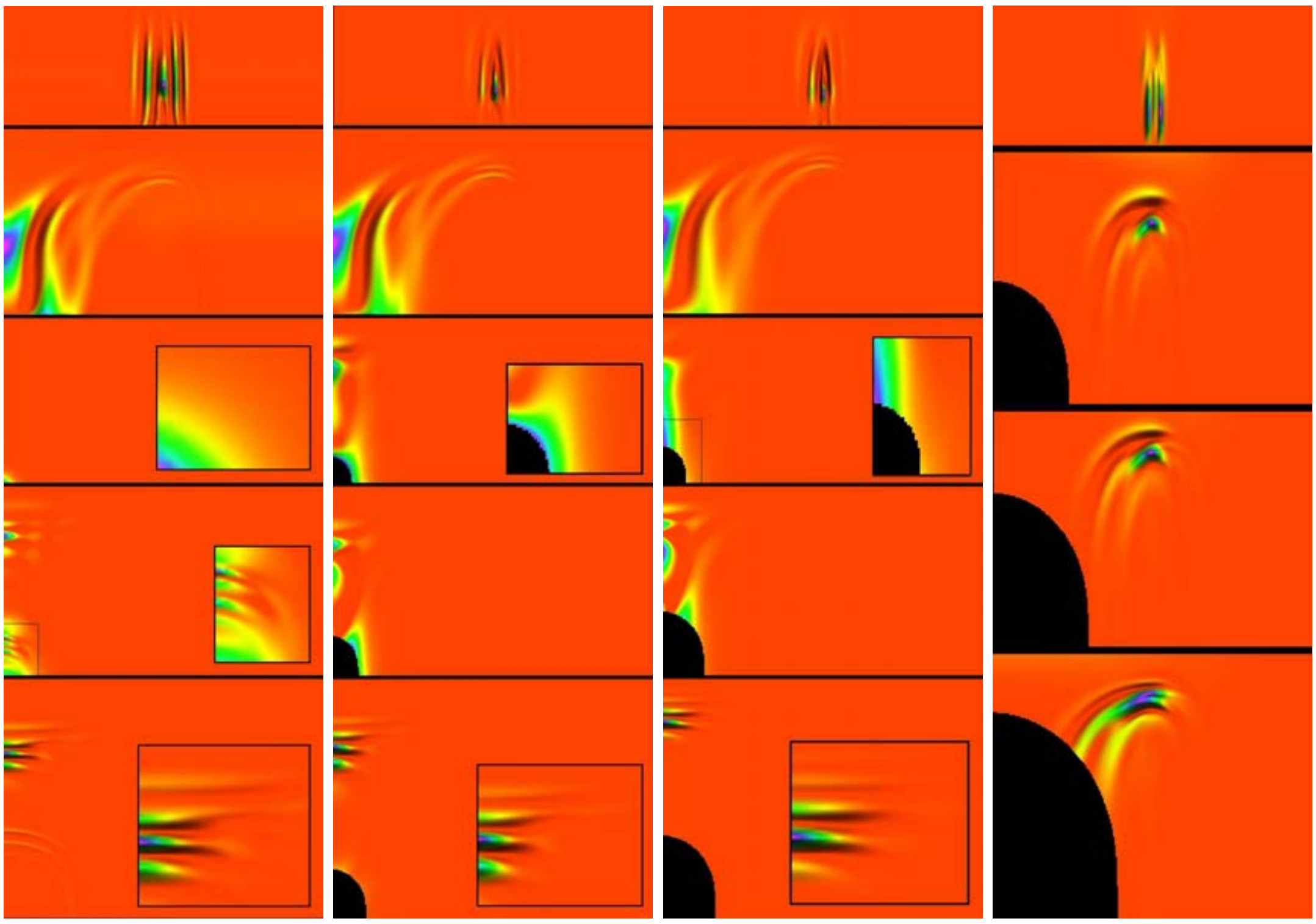

Figure 5.4: The snapshots of four evolutions resulting in BHs with different sizes, monitored via Kretschmann scalar $R_{\mu\nu\alpha\beta}R^{\mu\nu\alpha\beta}$. Each small panel represents an instant of the space, where the horizontal and vertical axes are $r$ and $z$ axes expressed in compactified coordinates $r/(r+r_0)$ and $(z-\ell)/(z-\ell+z_0)$, therefore the bottom of each panel is $z=\ell$, the brane. The complete space in $r$ direction is shown. In $z$ direction, only the part with interesting dynamics is shown. The evolutions are from the family with $x_0=2, \sigma_r=0.2, \sigma_z=0.4$, and $r_0=z_0=2$ are chosen. They produce no-BH, BH with $r_{\rm a}=0.29\ell, 0.61\ell, 3.78\ell$ from $\mathscr{A}=0.04, 0.15, 0.24, 0.49$ respectively. The two smaller BHs are apparently stationary states, and the largest BH is not since the configuration has not settled down to stationary state after evolving for a *long* time, and eventually the code crashes. In general it is harder for an evolution to settle down if the excision surface is going across the "wiggling" regions. This appears to be a technical issue related to the black hole excision at relatively coarse resolutions. The black ellipses in the figures are the excision surfaces inside the apparent horizons. The evolutions clearly show how the energy flows from the brane into the bulk, and flow from the exterior towards the symmetric axis. Part of the energy is captured by black holes, the remaining energy continue to flow towards the "far end" of the bulk.

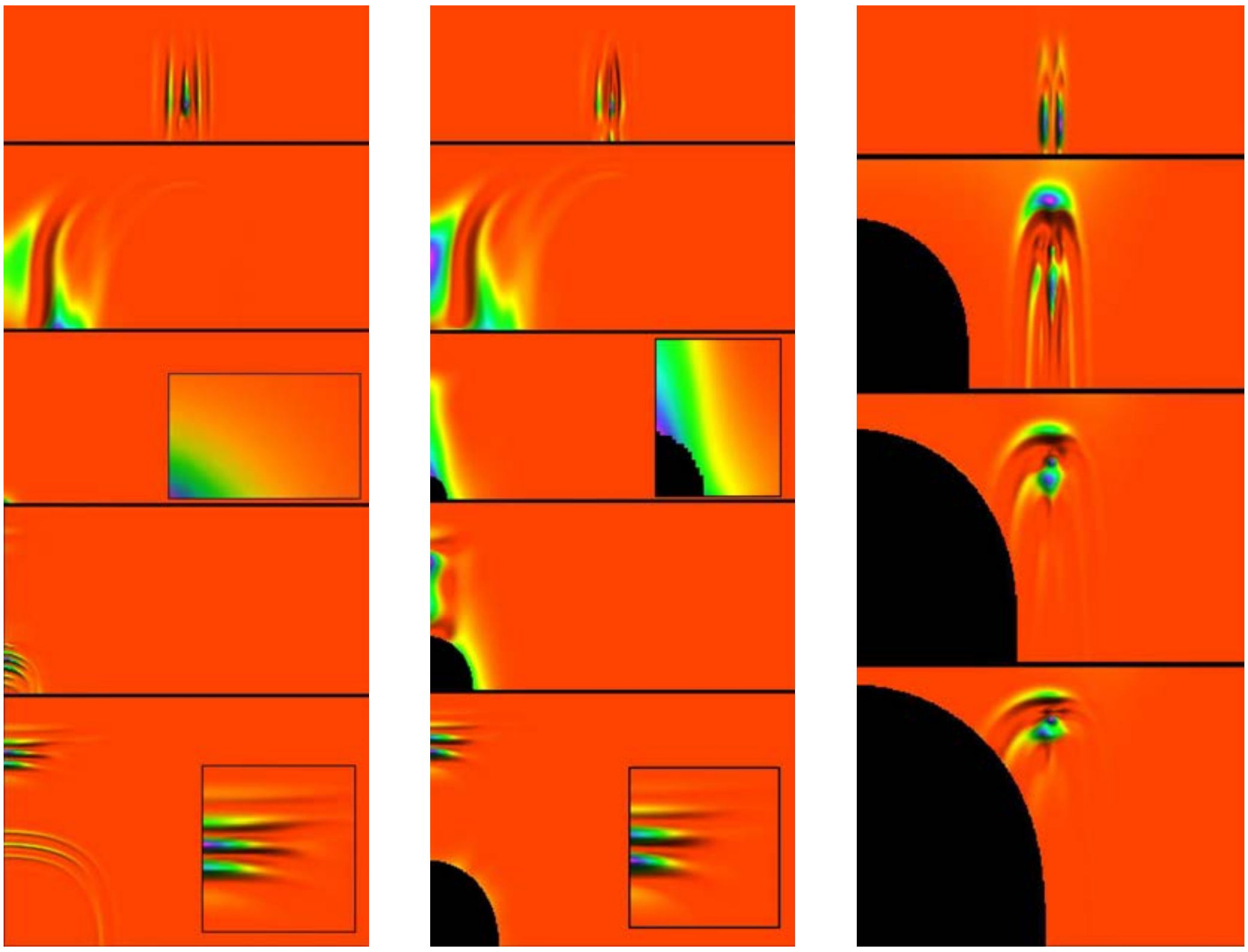

Figure 5.5: The snapshots of three evolutions resulting in BHs with different sizes, monitored via Kretschmann scalar $R_{\mu\nu\alpha\beta}R^{\mu\nu\alpha\beta}$. The evolutions are from the family with $x_0 = 1, \sigma_r = 0.1, \sigma_z = 0.2$, and $r_0 = z_0 = 1$ are chosen. They produce no-BH, BH with $r_{\rm a} = 0.22\ell, 2.9\ell$ from $\mathscr{A} = 0.03, 0.16, 0.54$ respectively. The smaller BH is a apparently stationary state, and the larger BH is not since the configuration is not settled down to stationary state after evolving for a *long* time. Again, presumbly this is a technical issue related to the black hole excision.

Here a bar ($\bar{\ }$) is used to emphasize that this spacetime is fixed, and the $\bar{z} \geq 1$ portion is actually the background of the braneworld spacetime. The embedding is demonstrated by Fig. 2.3 (also eq. (2.40) and eq. (2.41)). The apparently stationary state appears as an "accumulating" curve in the embedding plot. Please refer to Fig. 5.6, Fig. 5.7, Fig. 5.8 and Fig. 5.9 as examples where the apparently stationary states are obtained. The processes of the settlement into time-independent states, show that the portion of the apparent horizons that is close to the brane, gets settled first, when the remaining portion might be still dynamical.

Similar to apparent horizon, the existance of apparently stationary state is not guaranteed, and its relation with end state is not clear. In our simulations, as it turns out, apparently stationary states can be easily obtained by long term evolutions, as long as the apparent horizons do not cross the regions with interesting dynamics (the "wiggling" regions), which can be realized by properly choosing the initial data such that the wiggles either finally travel away from the apparent horizons, or are captured by the black holes. Furthermore, the plots of the quantities of apparently stationary states (such as the plots of $\mathcal{A}_{\mathrm{bulk}}$-versus-$r_{\mathrm{a}}$ and $C_5$-versus-$r_{\mathrm{a}}$ shown by Fig. 5.11), generated from the evolutions of the initial data profiles from different *families*, exhibite certain trends, whilst the same plots with horizons that are not apparently stationary state do not have trends. Also, the $\mathcal{A}_{\mathrm{bulk}}$-versus-$r_{\mathrm{a}}$ plot (the upper panel of Fig. 5.11) agrees perfectly with that obtained in the static system studied by Figueras-Wiseman in [20]. Therefore, we conjecture that the apparently stationary state is close to the end state, and we use apparently stationary state to approximate the end state.

### 5.7.2 Black Holes as the Result of Gravitational Collapse

For the BHs as apparently stationary states of gravitational collapse, we focus on the following aspects: the topology, the size and the shape.

For the topology, one can see that the BHs appear to be localized on the brane with finite extension into the bulk.

For the sizes, please refer to the results of all the simulations we performed, which are shown by Fig. 5.11, and table 5.1 for the results from some selected simulations. We did not try extremely hard to find the largest/smallest BHs possible. But within the simulations we performed, we obtained BHs with

$$r_a \in (0.04\ell, 19.6\ell). \tag{5.33}$$

At the end of an evolution, the matter has either fallen into the BH, or escaped to infinity, which makes the brane tension be the only content associated with the brane. The strength of the

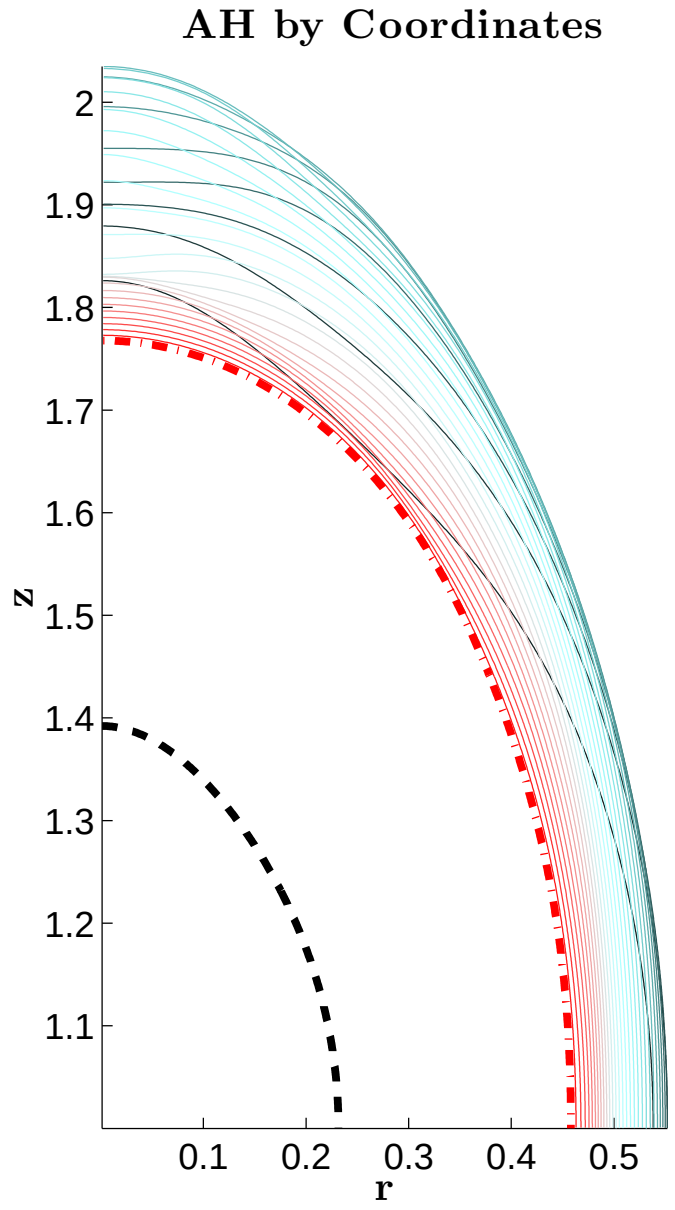

(a) AH evolution by coordinates

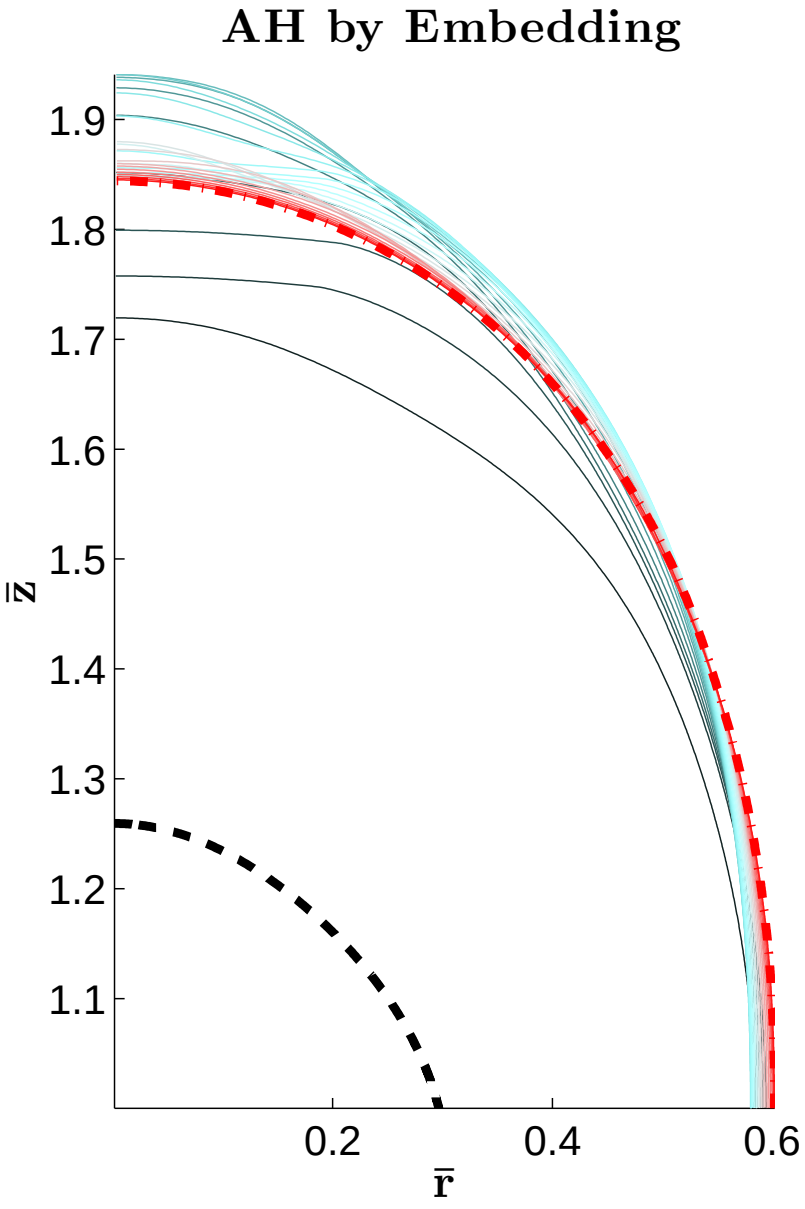

(b) AH evolution by embedding

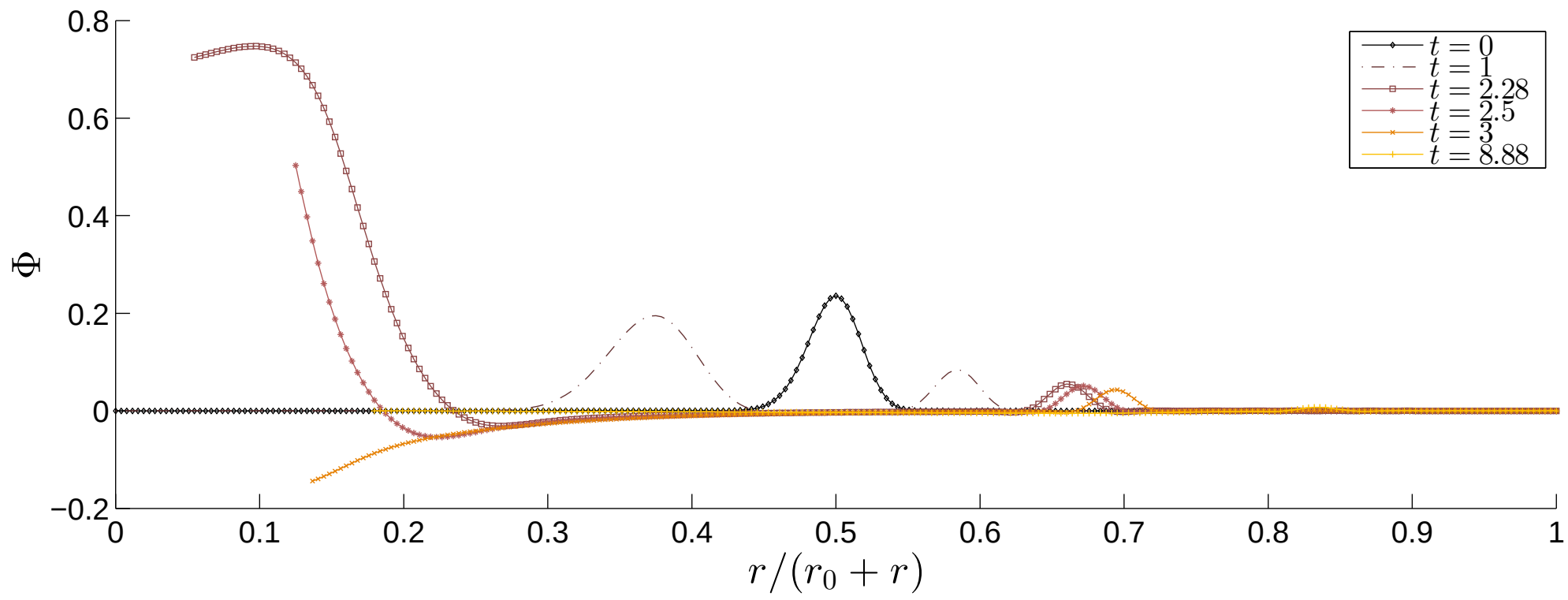

(c) The evolution of massless scalar field $\Phi$

Figure 5.6: This figure and Fig. 5.7 show an evolution that produces apparently stationary state as medium BH with $(r_{\rm a}, \mathcal{A}_{\rm bulk}, C_5) = (0.601, 2.785, 3.164)$, using parameters $(\mathscr{A}, x_0, \sigma_r, \sigma_z, r_0, z_0) = (0.24, 2, 0.2, 0.4, 2, 2)$ which defines the initial data and the compactification. (a) is the evolution of the apparent horizon, where $r$ and $z$ are coordinates. The black dashed line is the first apparent horizon that appears during the evolution, and the red "-." line is the apparently stationary state of this specific evolution of gravitational collapse. Other lines are apparent horizons at intermediate instants, which color changes continuously from black to red. To better study the evolution without coordinate distortion effects, we embed each apparent horizon into the background spacetime: $\mathrm{d}s^2 = \frac{\ell^2}{\bar{z}^2}\Big(-\mathrm{d}\bar{t}^2 + \mathrm{d}\bar{r}^2 + \bar{r}^2\left(\mathrm{d}\bar{\theta}^2 + \sin^2\bar{\theta}\mathrm{d}\bar{\phi}^2\right) + \mathrm{d}\bar{z}^2\Big)$. The embedding is demonstrated by Fig. 2.3. The embedding plot is shown in (b). This graph shows more clearly that the apparently stationary state is obtained: the shape of the apparent horizon has settled down, since the coordinate time $t \sim 5.5$. (c) shows the evolution of the massless scalar field that only lives on the brane. The initial profile is a Gaussian pulse, which splits into ingoing and outgoing branches. The ingoing branch is gradually captured by the BH.

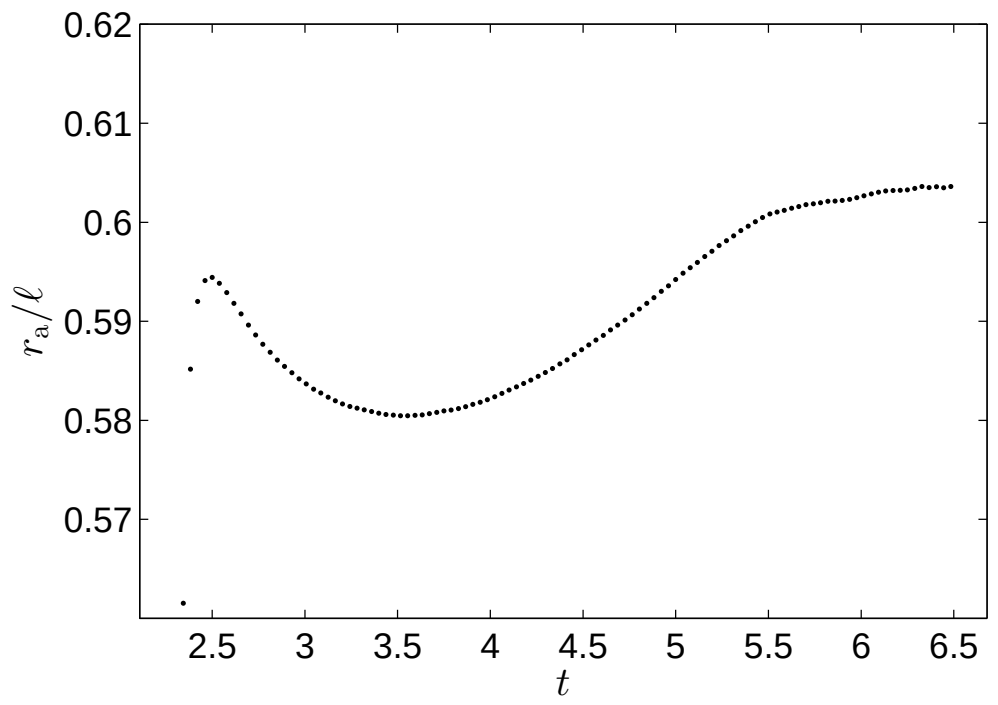

(a) The evolution of $r_a$

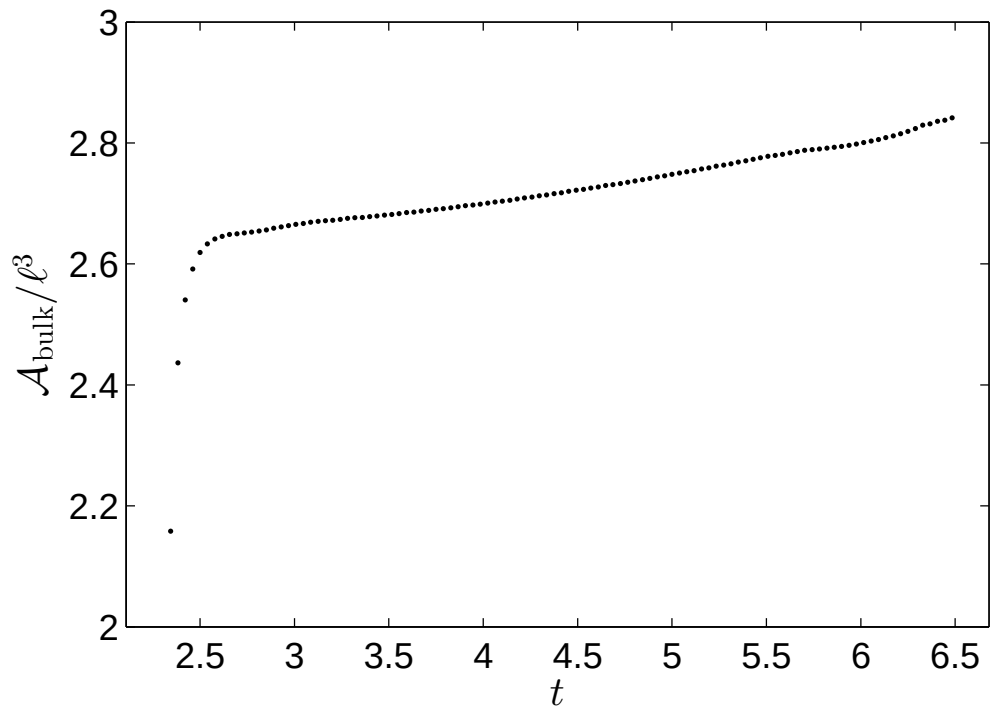

(b) The evolution of $\mathcal{A}_{\rm bulk}$

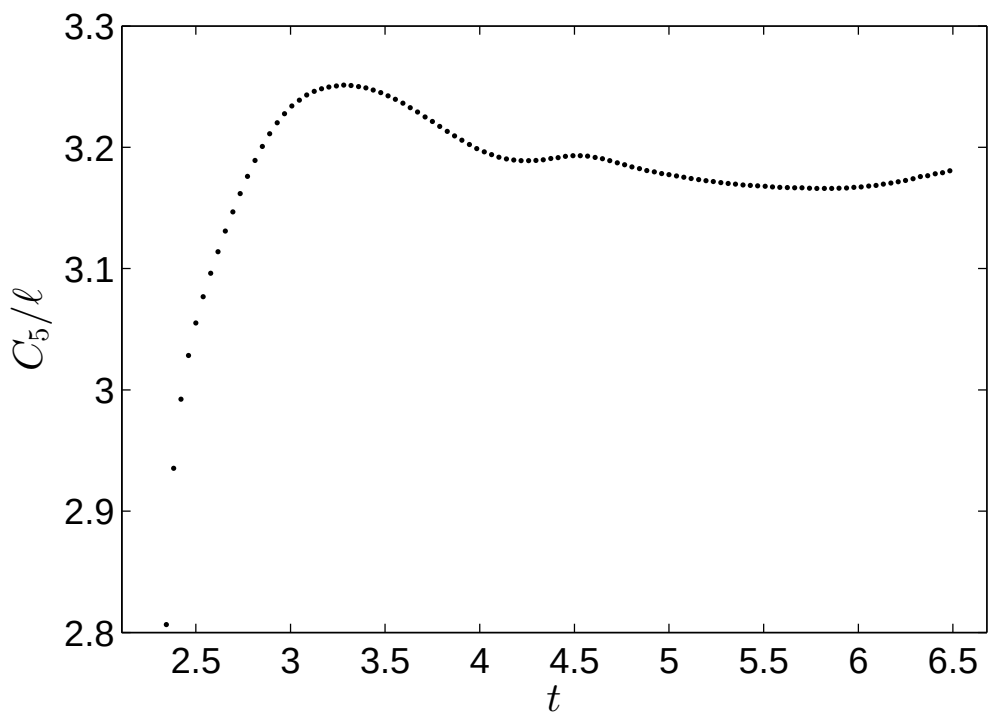

(c) The evolution of $C_5$

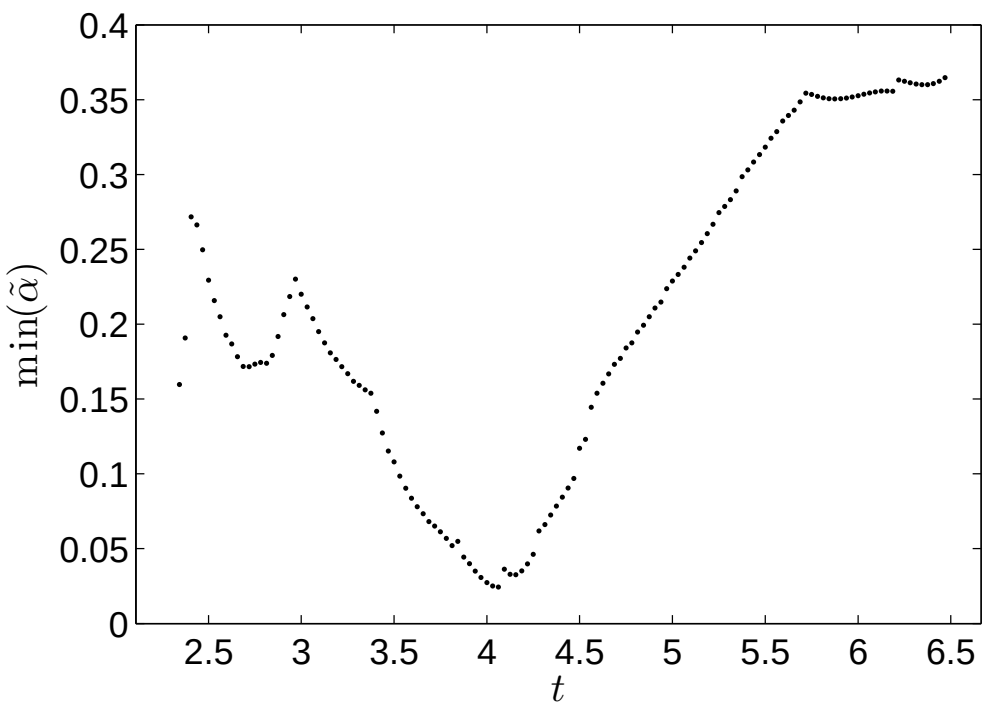

(d) The minimum of $\tilde{\alpha}$

Figure 5.7: This is the continuation of Fig. 5.6. Here we show the evolution of $r_{\rm a}$, $\mathcal{A}_{\rm bulk}$, $C_5$ and the *minimum* of lapse function $\tilde{\alpha}$ over the whole calculation domain. One can see the apparently stationary state is obtained by sub-figure (b) in Fig. 5.6, since $t \sim 5.5$. The quantities approaches quasi-constant since then. However, the quantities do not stay strictly at constants. Combining the plot of $\tilde{\alpha}$, finally the values at $t = 5.8$ are recorded as the data for the apparently stationary state. $\min(\tilde{\alpha}) \sim 0.35$ means the lapsing rate of coordinate time is comparable with that of proper time.

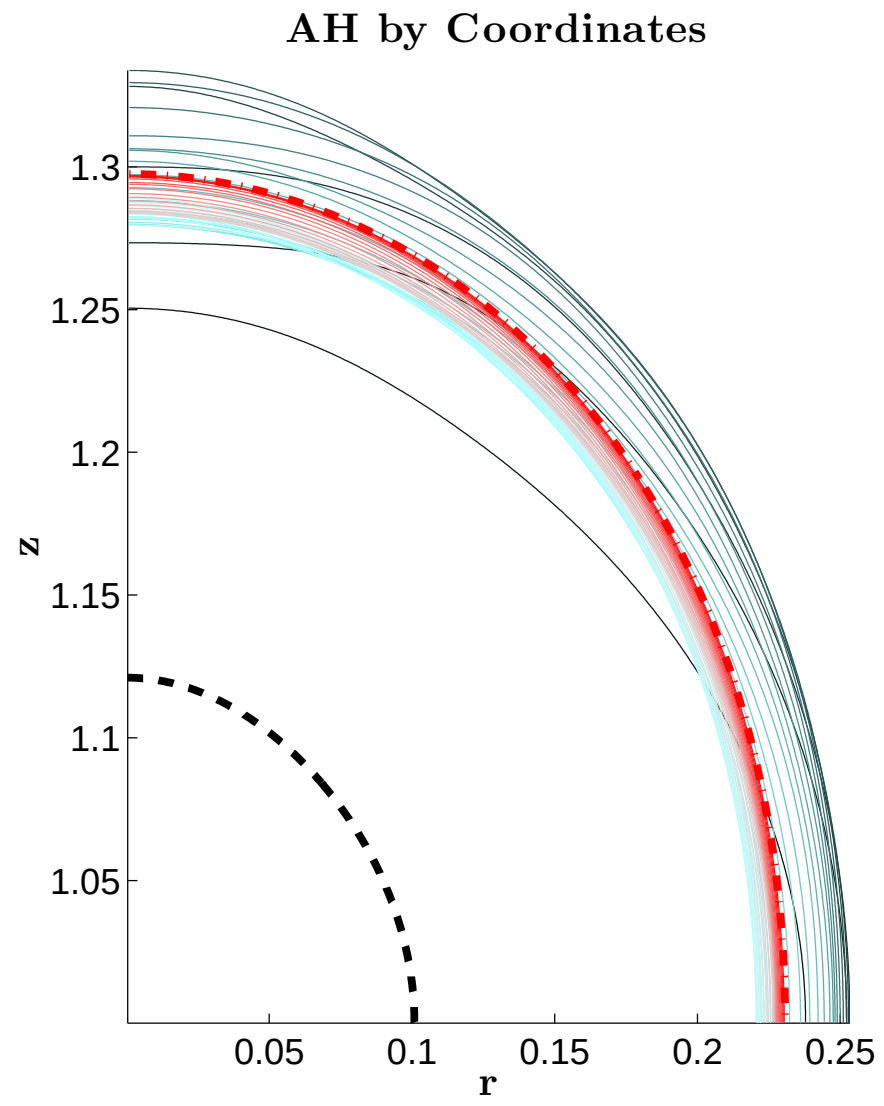

(a) AH evolution by coordinates

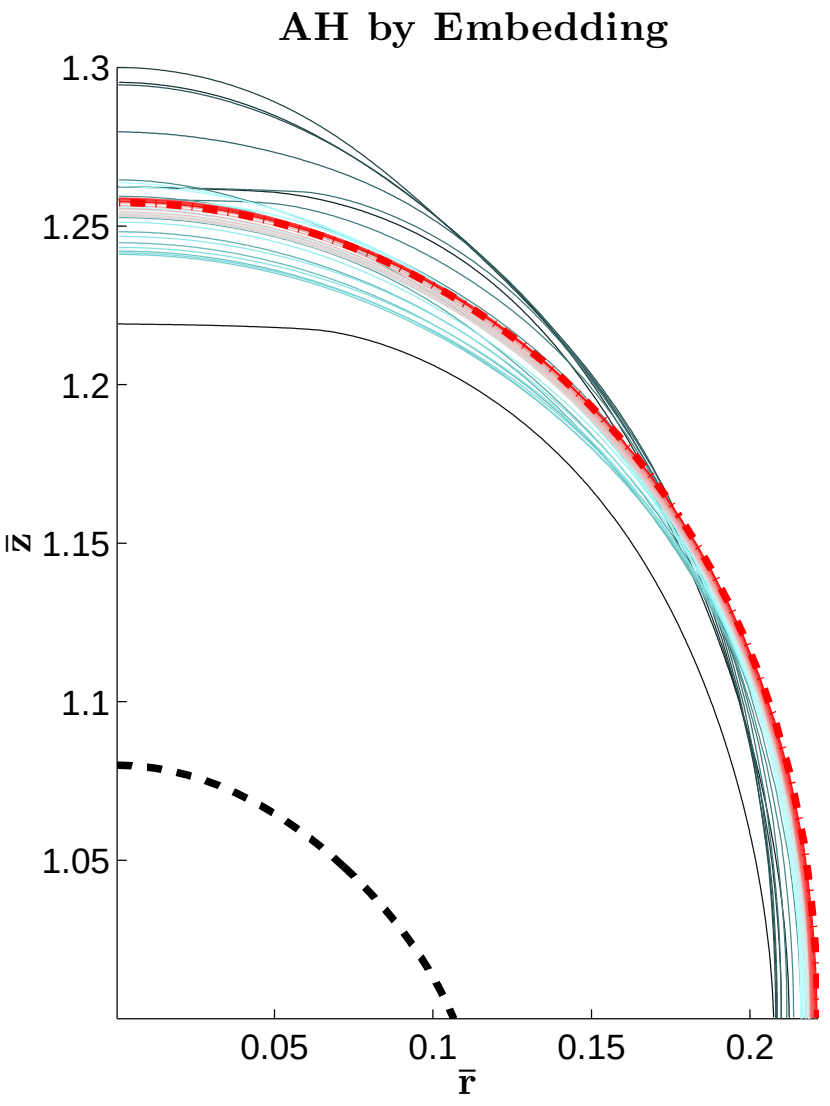

(b) AH evolution by embedding

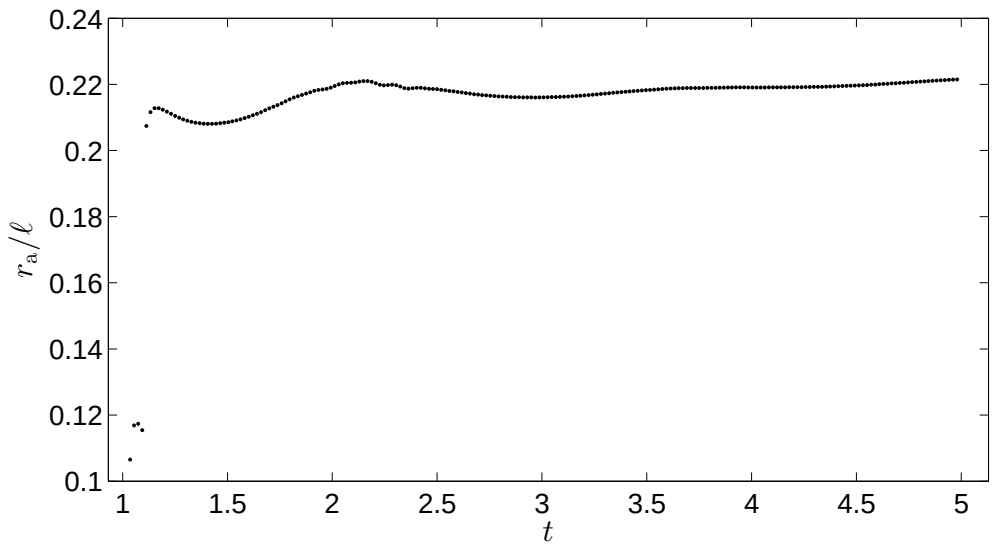

(c) The evolution of $r_{\rm a}$

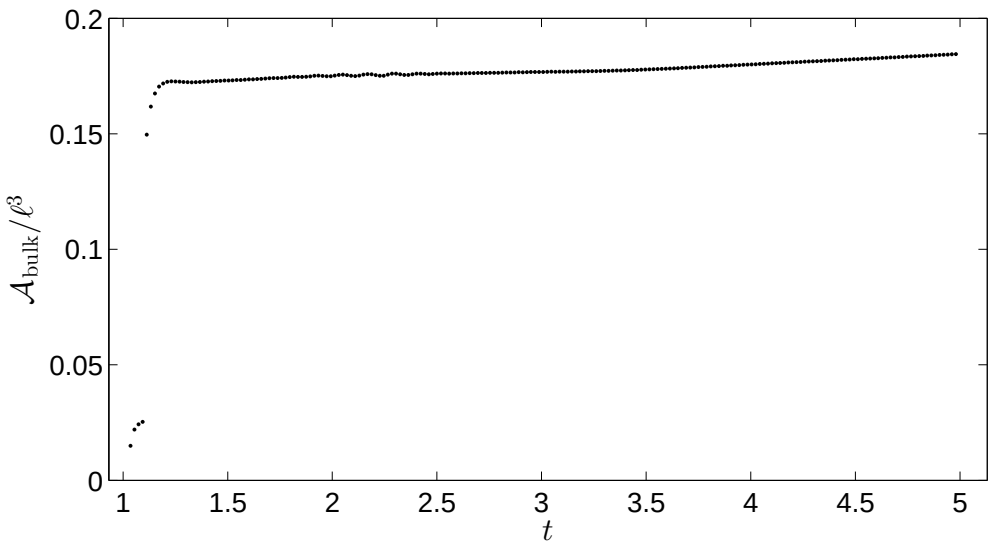

(d) The evolution of $\mathcal{A}_{\rm bulk}$

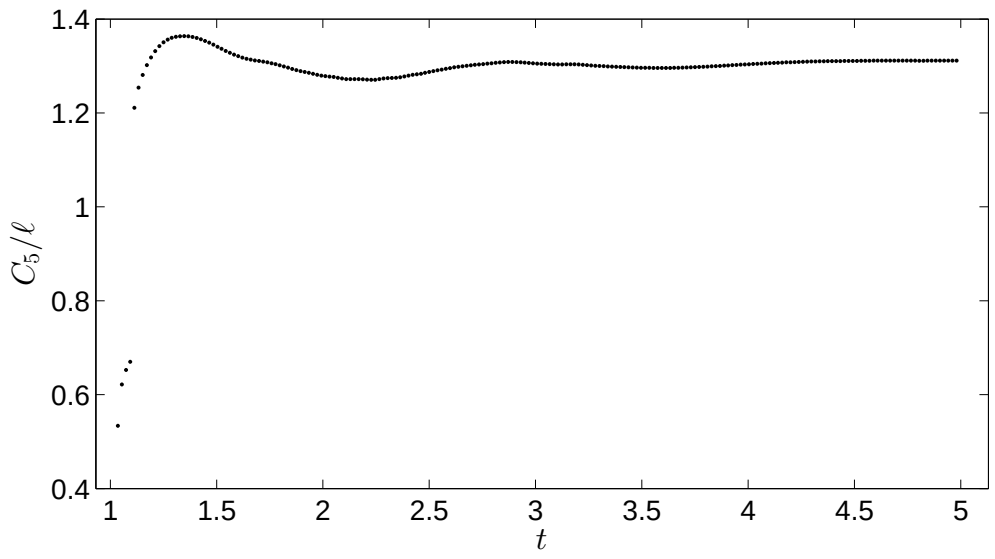

(e) The evolution of $C_5$

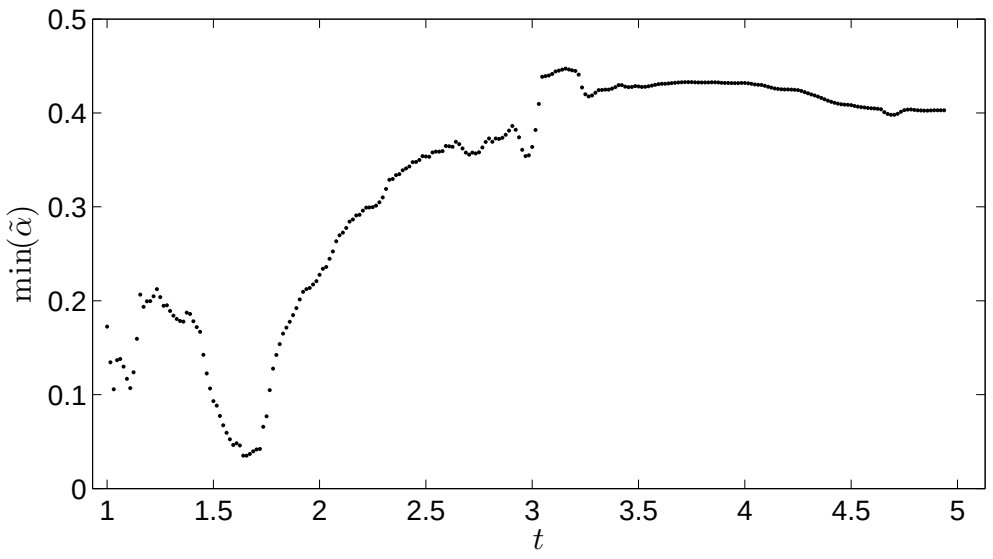

(f) The minimum of $\tilde{\alpha}$

Figure 5.8: This figure shows an evolution that produces apparently stationary state as small BH with $(r_{\rm a}, \mathcal{A}_{\rm bulk}, C_5) = (0.221, 0.184, 1.311)$, using parameters $(\mathscr{A}, x_0, \sigma_r, \sigma_z, r_0, z_0) = (0.16, 1, 0.1, 0.2, 1, 1)$.

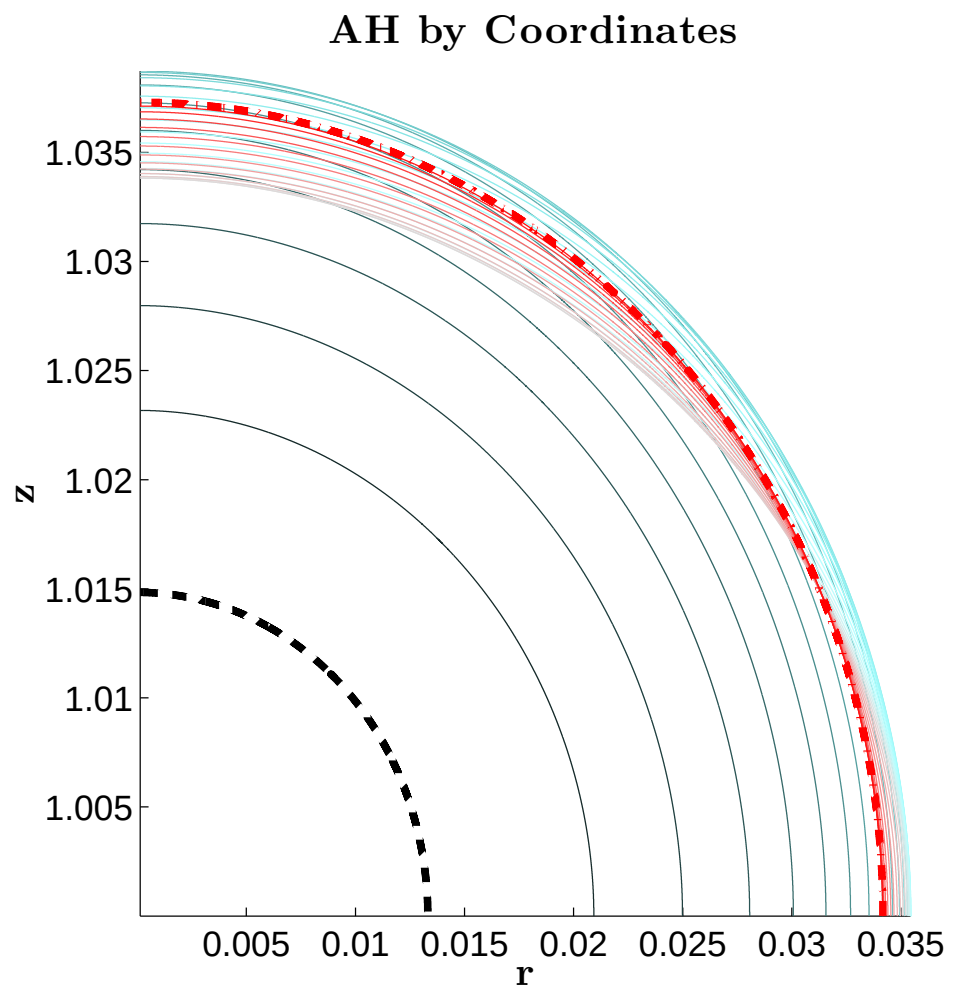

(a) AH evolution by coordinates

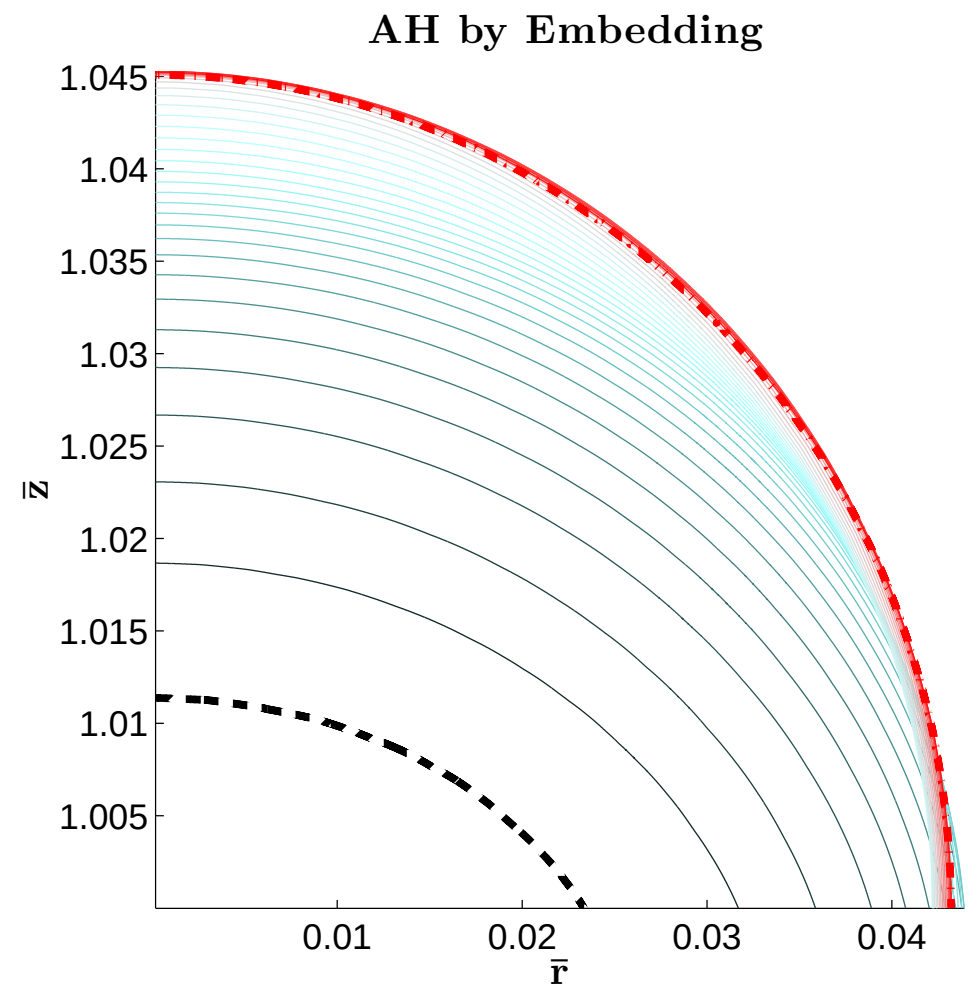

(b) AH evolution by embedding

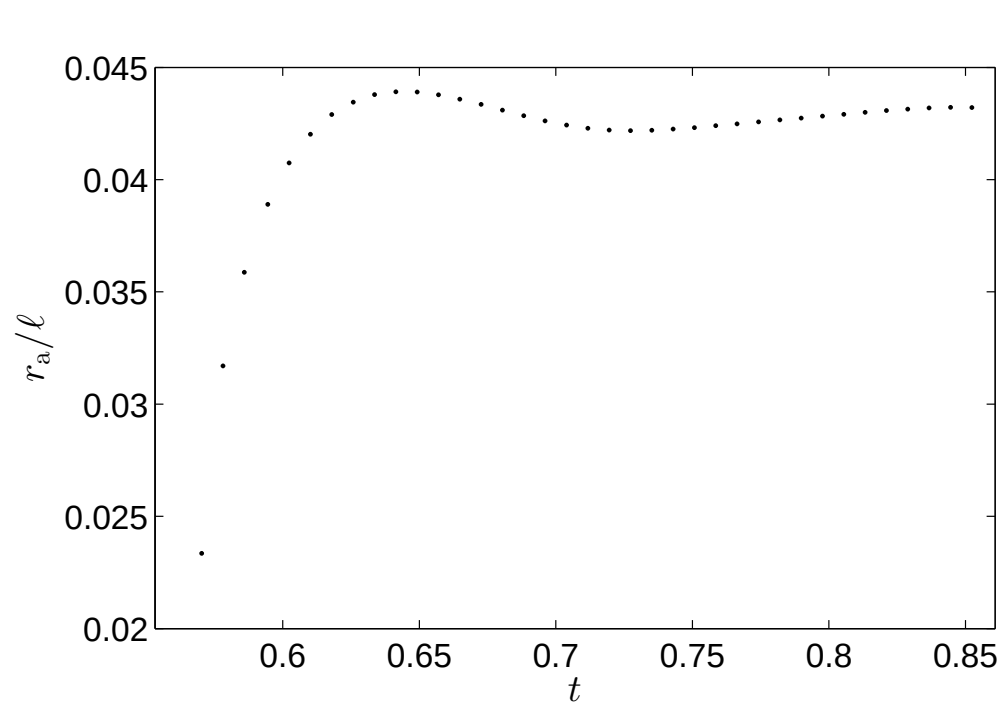

(c) The evolution of $r_{\mathrm{a}}$

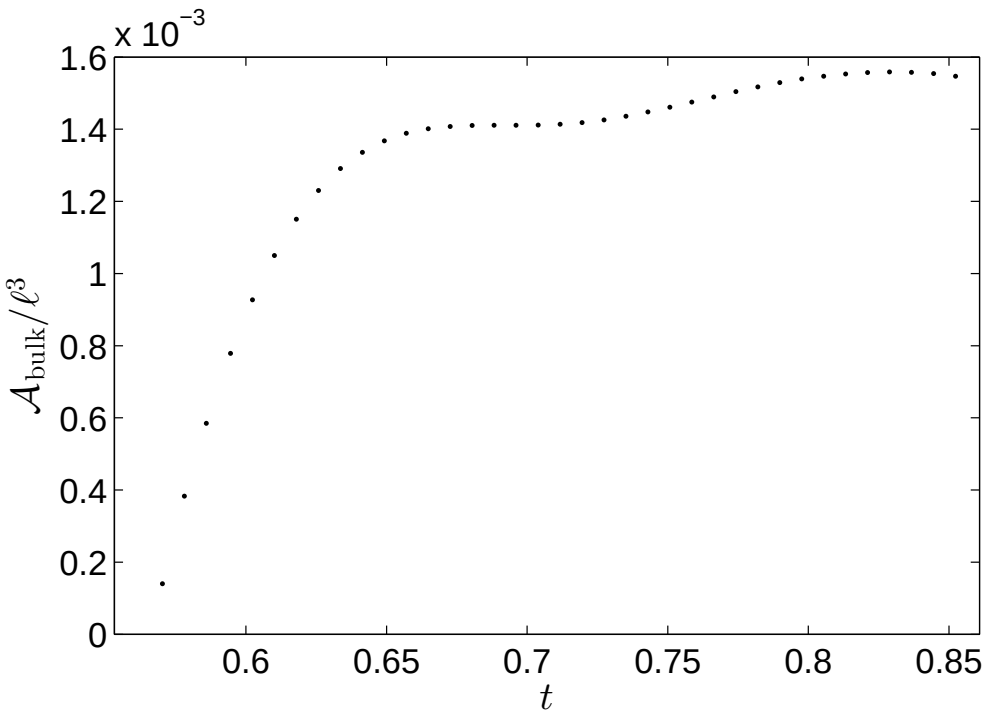

(d) The evolution of $\mathcal{A}_{\mathrm{bulk}}$

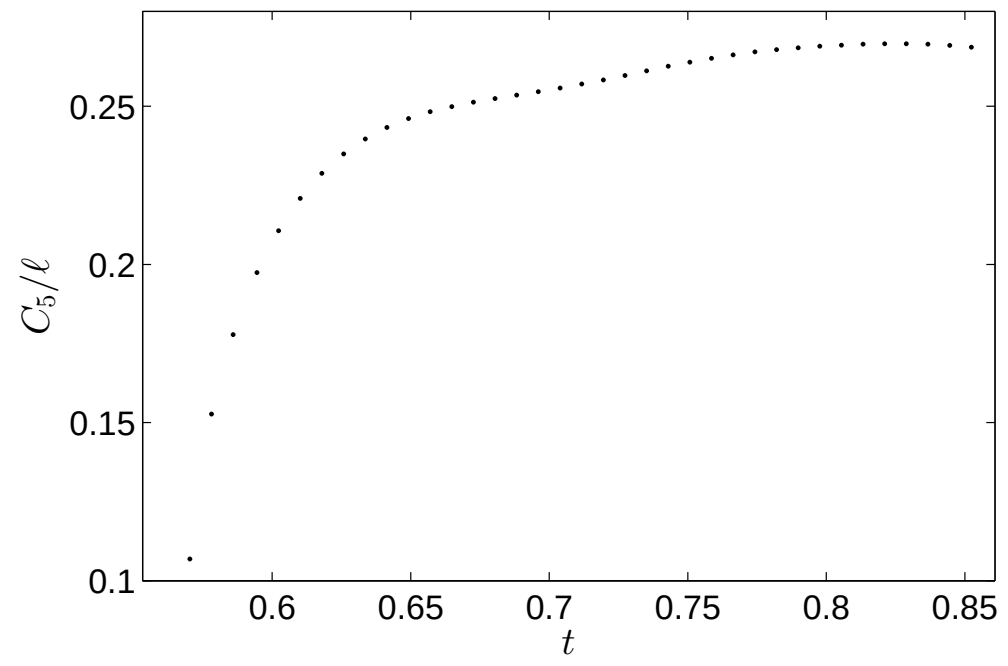

(e) The evolution of $C_5$

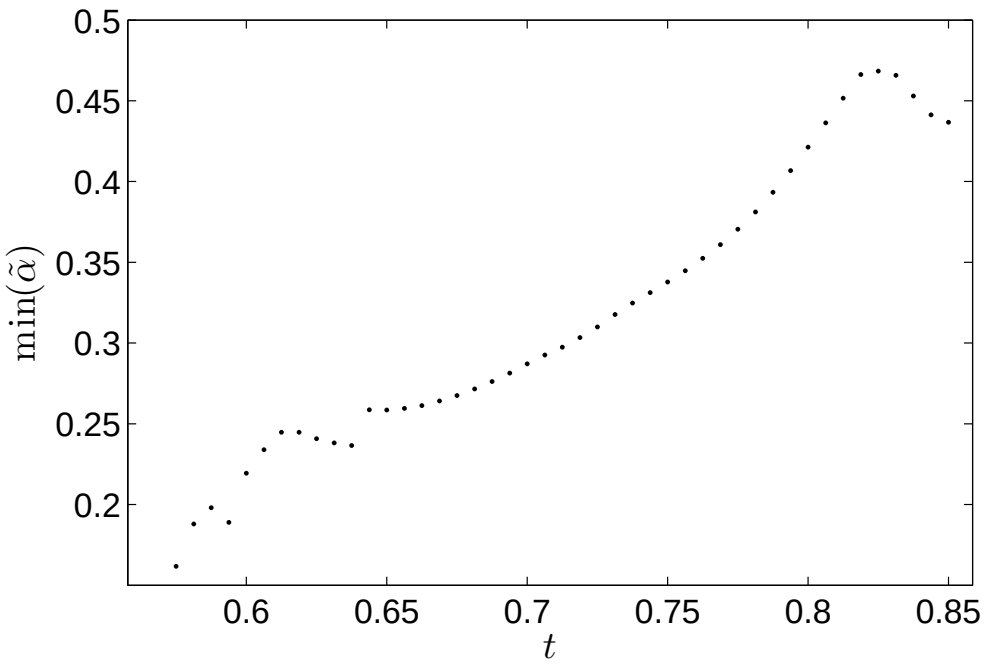

(f) The minimum of $\tilde{\alpha}$

Figure 5.9: This figure shows an evolution that produces apparently stationary state as small BH with $(r_{\mathrm{a}}, \mathcal{A}_{\mathrm{bulk}}, C_5) = (0.0432, 0.00155, 0.269)$, using parameters $(\mathscr{A}, x_0, \sigma_r, \sigma_z, r_0, z_0) = (0.08, 0.5, 0.1, 0.2, 0.5, 0.5)$.

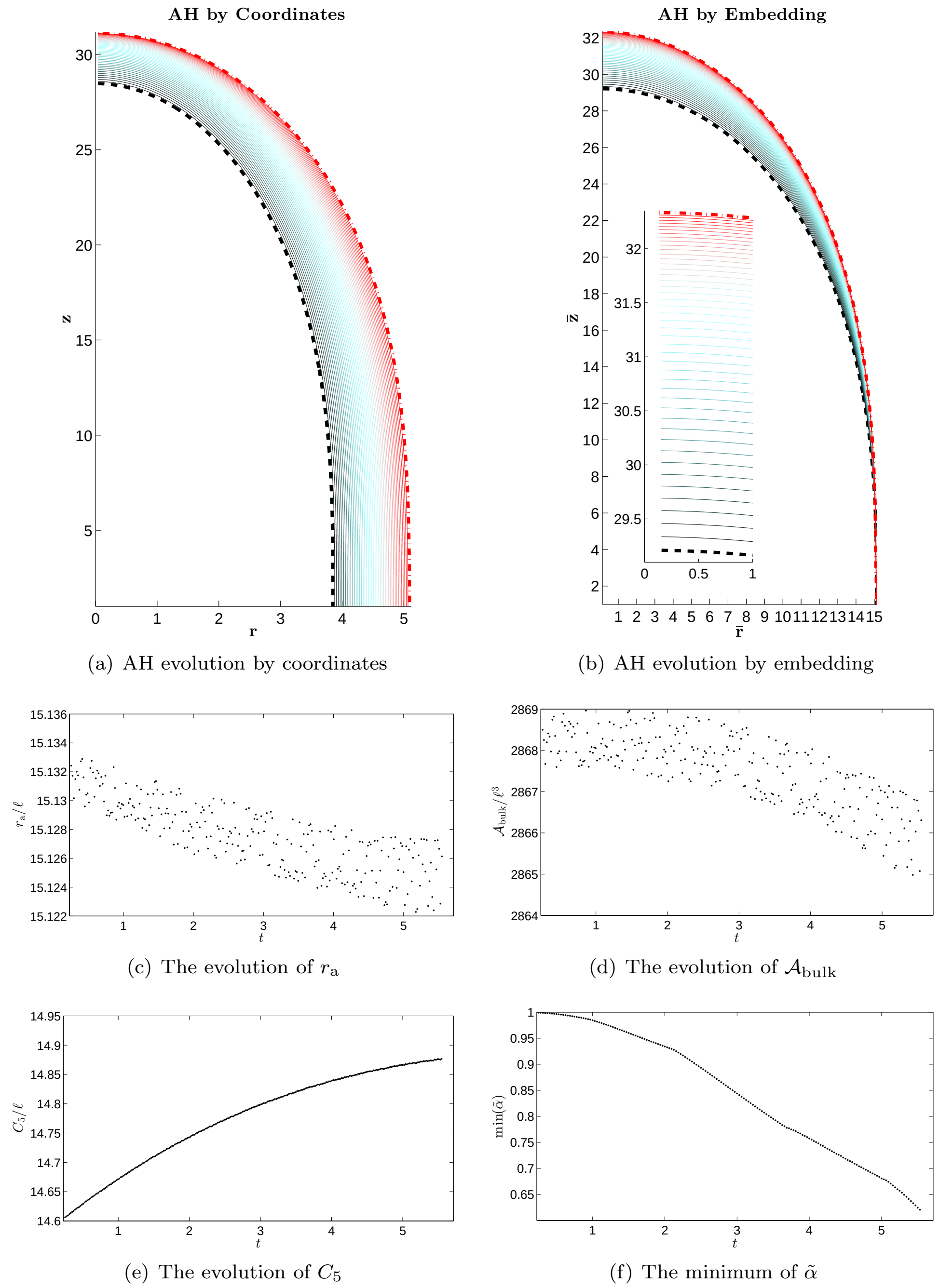

(a) AH evolution by coordinates

(b) AH evolution by embedding

(c) The evolution of $r_{\rm a}$

(d) The evolution of $\mathcal{A}_{\rm bulk}$

(e) The evolution of $C_5$

(f) The minimum of $\tilde{\alpha}$

Figure 5.10: This figure shows an evolution that produces large BH as apparently stationary state with $(r_{\rm a}, \mathcal{A}_{\rm bulk}, C_5) = (15.1, 2866, 14.9)$, using parameters $(\mathscr{A}, x_0, \sigma_r, \sigma_z, r_0, z_0) = (1.05, 2, 0.5, 1, 2, 2)$. This is the case that data is so strong that all the interesting dynamics is captured by the BH. Note the plots of $r_a$ and $\mathcal{A}_{\rm bulk}$ are noisy, which we can not explain. This might be caused by the KO dissipation—the dissipation appears to have stronger effects at larger spatial coordinates, and its perturbing effect on larger horizon is also stronger. As a result, the energy of the BH gradually decreases, which causes the decrease in $\mathcal{A}_{\rm bulk}$.

| $r_0$ | $z_0$ | $\sigma_z$ | $x_0$ | $\sigma_r$ | $\mathscr{A}$ | $r_{\rm a}/\ell$ | $\mathcal{A}_{\rm bulk}/\ell^3$ | $C_5/\ell$ |
|---|---|---|---|---|---|---|---|---|
| 0.5 | 0.5 | 0.2 | 0.5 | 0.1 | 0.08 | 0.0432 | 0.00155 | 0.269 |
| 1 | 1 | 0.3 | 0 | 0.3 | 2.05 | 0.600 | 2.80 | 3.18 |
| 1 | 1 | 0.2 | 1 | 0.1 | 0.16 | 0.221 | 0.18 | 1.31 |
| 2 | 2 | 0.4 | 2 | 0.2 | 0.24 | 0.601 | 2.79 | 3.16 |
| 2 | 2 | 1 | 2 | 0.5 | 1.05 | 15.1 | 2866 | 14.9 |

Table 5.1: BHs produced from different initial data from different families. The initial data profile for the massless scalar field on the brane is $\Phi = \mathscr{A} \cdot \exp\left[-(r-x_0)^2/\sigma_r^2\right]$. $\sigma_z$ is the parameter to set up metric functions via "direct specification" eq. (4.11), which is then substituted into (4.31). The spatial metric for initial data is eq. (4.26). $r_0$ and $z_0$ are simply compactification parameters defined in eq. (5.9).

brane tension is proportional to $1/\ell$, therefore invisible to small BHs whose size $r_{\rm a} \ll \ell$. These BHs will be asymptotically 5D Schwarzschild, therefore $\mathcal{A}_{\rm bulk} = 2\pi^2 r_{\rm a}^3$ and $C_5 = 2\pi r_{\rm a}$. Please refer to Fig. 5.11.

For the shape of large BHs, please refer to Sec. 5.7.3.

### 5.7.3 The Relation with Black Strings

The black string solution is $\mathrm{d}s^2 = \frac{\ell^2}{z^2}\left(h_{ab}\mathrm{d}x^a\mathrm{d}x^b + \mathrm{d}z^2\right)$, where $a = 0,1,2,3$ and $x^a$ stands for a coordinate other than $z$, and $h_{ab}$ is a BH solution of vacuum Einstein's equations in 4D, which does not depend on coordinate $z$. This could be called black cone instead of black string, if we had considered the intrinsic geometry of the horizon of black strings. It is named *string*, in the sense that the $\left(h_{ab}\mathrm{d}x^a\mathrm{d}x^b + \mathrm{d}z^2\right)$ part is a string. The "string" shape can be revealed by embedding the horizon into the background spacetime (5.32). Therefore, the embedding of the BHs into this background, can give a direct comparison with the black strings. The embedding of the BHs (represented by the apparently stationary states) of different sizes is shown in Fig. 5.12. Please refer to the caption of the figure for more details.

Fig. 5.12 shows that small BHs are almost spherical. The BHs gradually change into a cigar-shape as the size increases, which suggests to call those BHs, *black cigars*, as first suggested in [6]. As the size of BH increases, the portion of the horizon that is close to the brane gradually changes into a black string. Also, the $\ell^2/\bar{z}^2$ factor in the background metric (5.32), contributes to $\mathcal{A}_{\rm bulk}$ as $\ell^3/\bar{z}^3$, which means the contribution from large $\bar{z}$ region become relatively negligible. As a result, for large BHs, only the portion that is close to the brane actually contributes to $\mathcal{A}_{\rm bulk}$, regardless the behaviours in the large $\bar{z}$ region. This explains why the relation $\mathcal{A}_{\rm bulk}$-versus-$r_{\rm a}$ for large BHs (shown in the upper panel of Fig. 5.11), behaves as that of black strings. On the other hand, this also means that the relation of $\mathcal{A}_{\rm bulk}$-versus-$r_{\rm a}$ can not reveal the difference between the BHs and black strings.

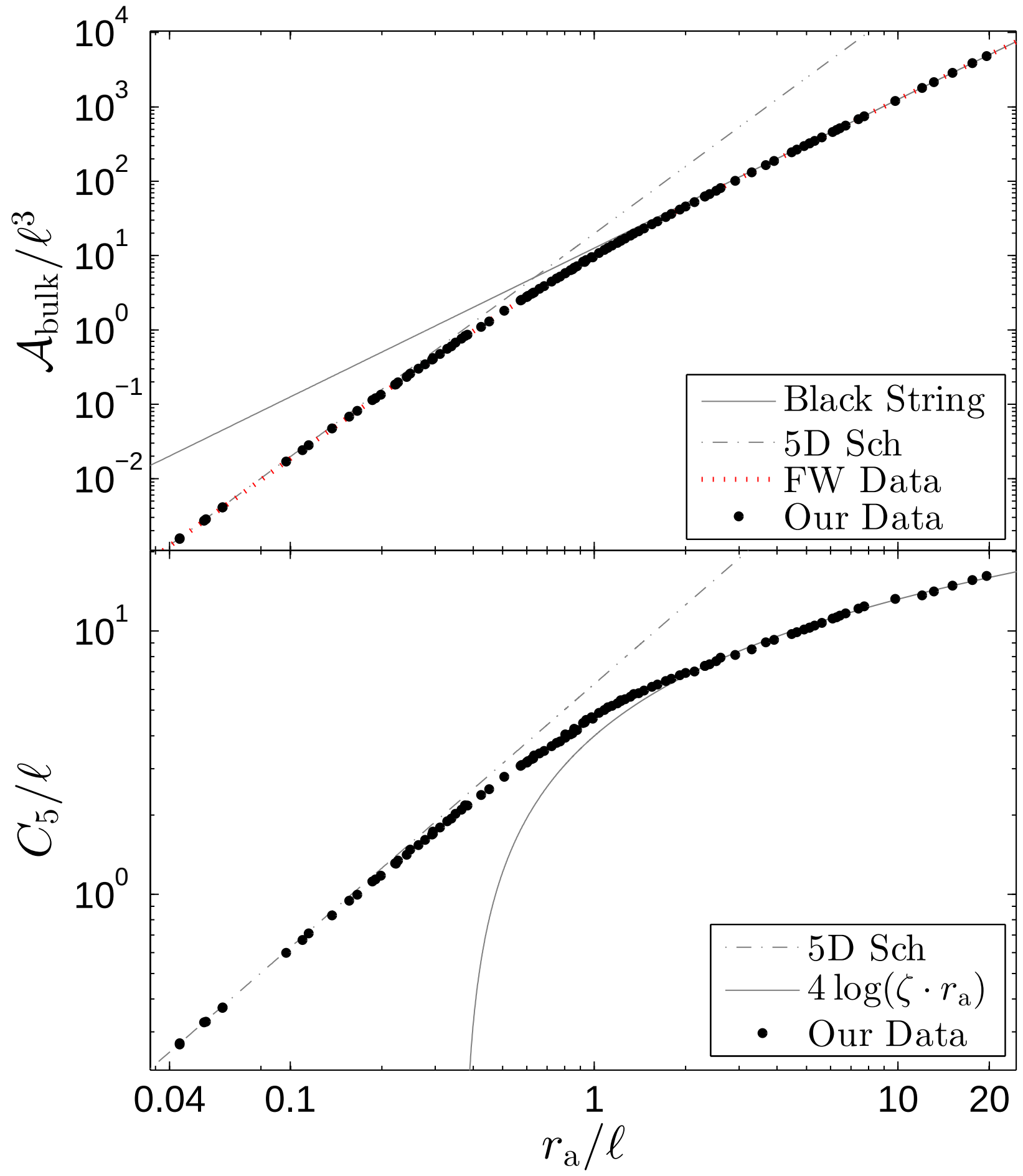

Figure 5.11: The shape of BHs with all sizes. Small BHs are asymptotically 5D Schwarzschild. The area-radius relation for large BHs is that of black string, which is also 4D Schwarzschild according to $AdS_5/CFT_4$ [20, 22]. The area-radius relation is consistent with the one obtained from a static problem by Figueras-Wiseman. $C_5$ for large BHs follows $C_5 = 4\log(\zeta\cdot r_{\rm a})$ [23], where $\zeta = 2.71$ is the best fit of our data.

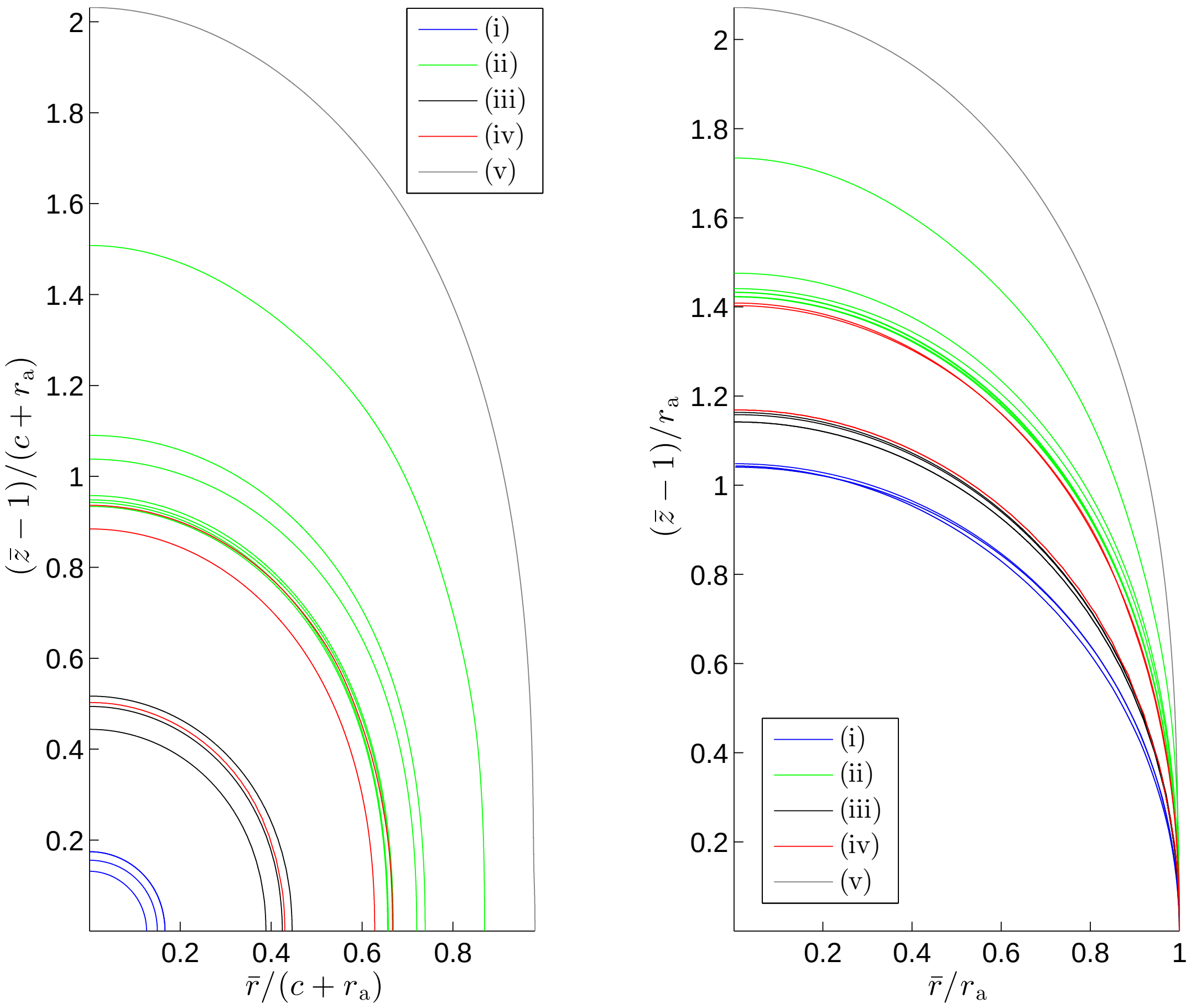

Figure 5.12: The shape of BHs with all sizes shown by embedding into the background spacetime (5.32). The horizons are scaled by a factor of$(c + r_{\rm a})$. Note both coordinates $\bar{r}$ and $(\bar{z} - 1)$ are scaled by the same factor, therefore the shape is not affected by the scaling. The scaling is to bring BHs with very different sizes (here $r_{\rm a} \sim (0.04\ell, 15\ell)$) onto comparable plotting scale. Here the scaling parameter on the left panel is $c = 0.3$, and the right panel corresponds to $c = 0$. The left panel still carries the information of sizes, which emphasizes on the uniqueness feature. The right panel emphasizes on how the shape changes with size, which can also be read-off from the left panel.
Small BHs are asymptotically 5D Schwarzschild, which are almost spherical. They then change into cigars (seen from the embedding point of view) as the size increases, which suggest to call them black cigars as firstly suggested in [6].
More importantly, the BHs are apparently stationary states from different families. For family (i), the parameters specifying the family are $(x_0, \sigma_r, \sigma_z, r_0, z_0) = (0.5, 0.1, 0.2, 0.5, 0.5)$. These parameters for other families are: (ii) $(0, 0.3, 0.3, 1, 1)$; (iii) $(1, 0.1, 0.2, 1, 1)$; (iv) $(2, 0.2, 0.4, 2, 2)$; (v) $(2, 0.5, 1, 2, 2)$. From the left panel we see that BHs obtained from different families with the same size, almost agree with one another. This indicates the detail of initial data is lost, and the solution is unique. The right panel shows the relative extension into the bulk increases with the size, and the portion (of the horizons) that is close to the brane looks more and more like black string as the size increases.

Therefore, we study $C_5$ (the length of the circumstance of the horizon restricted on the $r-z$ plane) versus $r_{\rm a}$, since $C_5$ is infinite for black strings, but finite for BHs. Furthermore, Fig. 5.12 shows that the relative extension into the bulk (as seen from the point of view of the background metric (5.32)), increases with the size on the brane. It is not very clear whether the relative extension has an upper limit. On the other hand, by the uniqueness of the BH solutions that is going to be studied in Sec. 5.7.4, and the comparison with Figueras-Wiseman solution that is going to be shown in Sec. 5.7.5, there is strong indication that our solutions generated from evolution systems, are the same as those were obtained by Figueras-Wiseman in their static systems. For the Figueras-Wiseman solution, as shown in the upper figure of Fig. 5.14, the large BHs has a limiting shape which is the $AdS_5/CFT_4$ solution [20, 21]. Generally, for configurations with *fixed shape* in the space (5.32), there is the following property for large $r_{\rm a}$

$$C_5 = 4\log(\zeta \cdot r_{\rm a}), \tag{5.34}$$

regardless what the shape is, as long as the shape is fixed. This statement can be justified by direct numerical experiments for a few shapes. $\zeta$ is determined by the specific shape. $C_5$ versus $r_{\rm a}$ is plotted as the lower panel of Fig. 5.11, which supports the assumption that large BHs have *the same shape*, and the best fit of $\zeta$ is 2.71. $\zeta \sim 2.8$ was first independently found by Figueras-Wiseman [23], and eq. (5.34) was proposed by Toby Wiseman in [23].

Large BHs have the *same shape* (black cigar) in the background space (5.32), and black strings also appear as strings (rather than cones) in this space, whilst $C_5/r_{\rm a}$ merely appears as $\log(r_{\rm a})/r_{\rm a}$ (black pancake) which is not as great as a fixed shape (personal tastes), therefore it makes more sense to name BHs as *black cigars* as first suggested in [6], rather than black pancakes.

### 5.7.4 The No-hair Feature

We purposely performed the simulations from distinct initial data families. Because the result of a well-posed numerical simulation depends smoothly on its initial data, if only one family is considered (only one parameter $\mathscr{A}$ in the initial data profile is changed), the quantities (such as $\mathcal{A}_{\rm bulk}$, $C_5$ and $r_{\rm a}$) will depend smoothly on $\mathscr{A}$. Therefore, for a given family, relations between quantities such as $\mathcal{A}_{\rm bulk}$-versus-$r_{\rm a}$, will emerge. For a different family, in principle the relation of $\mathcal{A}_{\rm bulk}$-versus-$r_{\rm a}$ might be different from the relation obtained from the previous family. Should this happens, the BH solutions are not unique. In reality, however, the relations of $\mathcal{A}_{\mathrm{bulk}}$-versus-$r_{\mathrm{a}}$ obtained from different families are the same: the $\mathcal{A}_{\mathrm{bulk}}$-versus-$r_{\mathrm{a}}$ relation plotted using the data obtained from the evolutions from *different* initial data families is shown in the upper panel of Fig. 5.11, which appears to be *one single* curve. Similarly, $C_5$-versus-$r_{\mathrm{a}}$ is plotted as the lower panel of Fig. 5.11, which also appears to be one single curve. i.e. Fig. 5.11 shows that the BHs with the same sizes have the same areas and the same circumferences, regardless which families the results are generated from. This indicates that the shape of the horizon is solely determined by the size $r_{\mathrm{a}}$, regardless which initial data family the BHs are generated from. Therefore a no-hair theorem of the BH solution in RSII is suggested. In general, however, the BH solutions in AdS spacetimes may not be unique (see, e.g. [117]). Therefore, the uniqueness of the BHs is limited to the RSII spacetimes studied in this thesis—these spacetimes are axisymmetric without angular momentum and non-gravitational charges. In this situation, the shapes of the horizons are directly studied in Fig. 5.12, where one can see that the BHs with the same size produced from different families actually have almost the same shape, which is an indication that the detail of the initial data is lost in the final state, and the geometry of a BH is solely determined by its size. Therefore a no-hair theorem (the uniqueness of BH solution) is suggested to hold for BHs in RSII.

### 5.7.5 The Comparison with Figueras-Wiseman Solution

If the BH solutions in RSII are unique in axisymmetric spacetimes without angular momentum and non-gravitational charges, then the BH solutions should be the same, regardless how the BH solution is obtained. In particular, the BHs produced as the end states of *evolutionary* systems should agree with the ones obtained from the *static* problem studied in [20, 22]. Here we still use apparently stationary states to approximate end states. Following [20, 22], we plot our data of $\mathcal{A}_{\mathrm{bulk}}$ versus $r_{\mathrm{a}}$ on top of the Figueras-Wiseman data in the upper panel of Fig. 5.11. The figure shows the agreement with Figueras-Wiseman solution as illustrated by $\mathcal{A}_{\mathrm{bulk}}$ versus $r_{\mathrm{a}}$ is perfect.

To better compare with their solution, we also embed the BHs into the space (5.32) with $\bar{z} \leq 1$, which is what Figueras-Wiseman did in [20]. Following [20], the freedom of the embedding is fixed by mapping the $r = 0$ ends (i.e., the axis ends) of the horizons to the point $(\bar{r}, \bar{z}) = (0, 1)$ in (5.32), instead of mapping the $z = 1$ ends (i.e., the brane ends) to $\bar{z} = 1$ in (5.32) as what we did in the above sections. i.e. instead of performing the embedding as Fig. 2.3, here we perform the embedding as Fig. 5.13. Please refer to Fig. 5.14 for the results. Fig. 5.14 shows the two results qualitatively agree. However, there are a few differences between these two figures: the largest BH meets the vertical axis at $\bar{r} \approx 0.468$ for our data, but at $\bar{r} \approx 0.457$ for Figueras-Wiseman data; there exist line crossings (can be seen by zooming in) in our data but there is not in Figueras-Wiseman data. Together with the line crossings in Fig. 5.12, it implies that generally apparently stationary states are, close to but distinct from, end states. Here we emphasize that our solution for large BHs were obtained at a resolution that is effectively coarse at the horizon, thus the solution should be less reliable. Furthermore, as one can see from the processes of the settlements to apparently stationary states (Fig. 5.6, Fig. 5.7, Fig. 5.8 and Fig. 5.9), the portion of the apparent horizon that is close to the brane gets settled first, while the remaining portion (the portion that is far from the brane, and close to the symmetric axis) might be still dynamical. This makes the portion that is close to the brane more reliable than the portion that is far from the brane. Therefore, the embedding plot by fixing the brane end of the apparent horizon as done in Fig. 5.12, is more reliable than the embedding plot by fixing the axis end of the apparent horizon as done in Fig. 5.14.

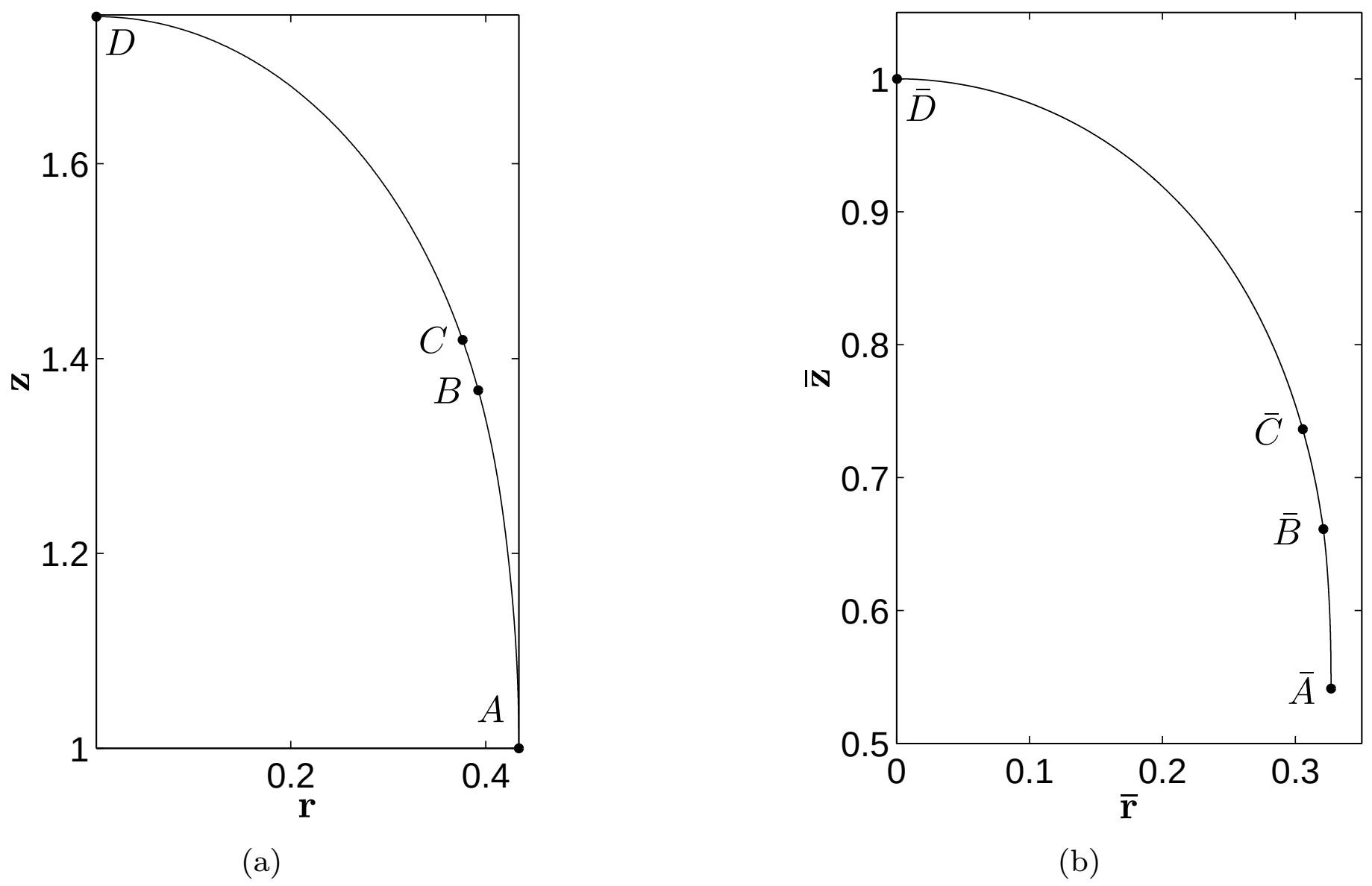

Figure 5.13: The embedding of a closed surface. Fig. (a) is the closed surface in the physical spacetime, and Fig. (b) shows its embedding into the background spacetime (5.32) with $\bar{z} \leq 1$. The freedom of the embedding is fixed by mapping $D$ to $\bar{D}$.

### 5.7.6 Brane Energy

In this section we focus on the physics on the brane, by studying the quantities obtained from the reduced metric ($h_{\mu\nu}$) on the brane. The rationale is that the brane is all one can directly observe (while the bulk is invisible). We compare quantities obtained on the brane, with the same quantities obtained from 4D GR. We will study ADM mass and Hawking mass. As explained in Sec. 2.5, these masses are only defined at asymptotic spatial infinities. However, let us first study the ADM mass and Hawking mass in 4D GR, to obtain some insights.

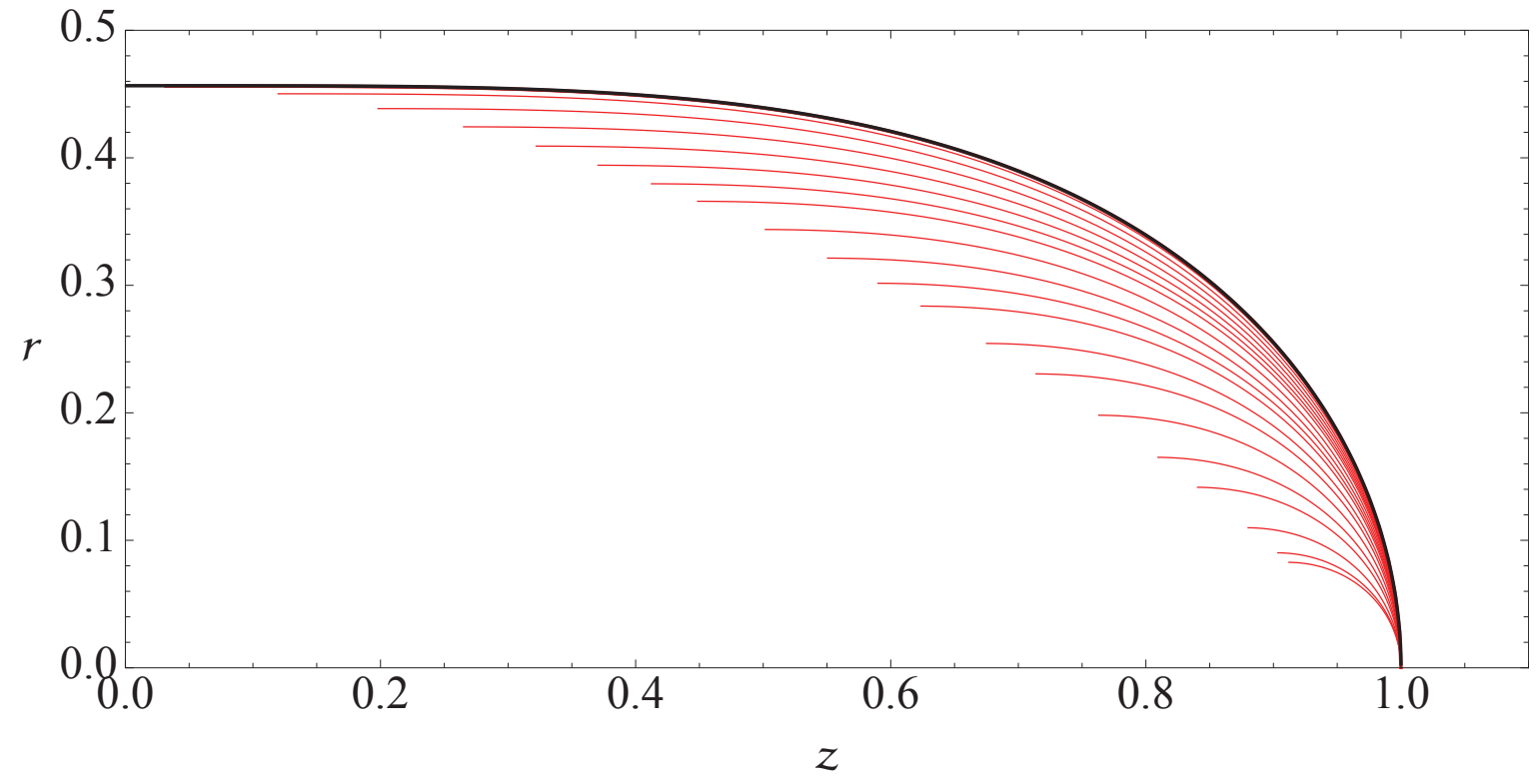

(a) Figueras-Wiseman solution

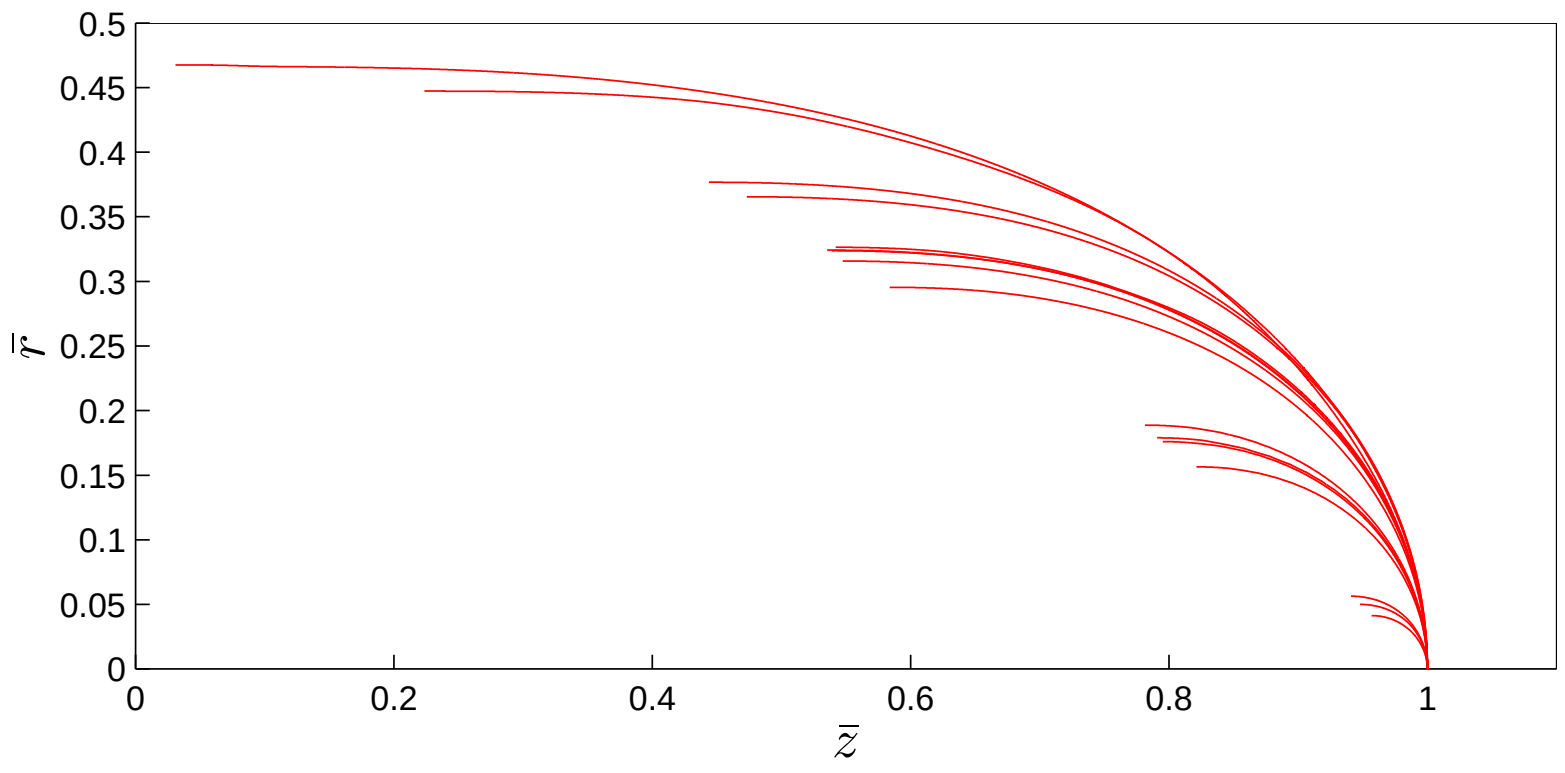

(b) Our solution

Figure 5.14: The comparison with the Figueras-Wiseman solution via embedding into (5.32), with $\bar{z} \leq 1$. The freedom of the embedding is fixed by mapping the $\bar{r} = 0$ ends of the horizons to $\bar{z} = 1$ in (5.32). Note, figure (a) is from [20], and the labels $(r, z)$ should be understood as $(\bar{r}, \bar{z})$. The large BHs have a limiting shape which is the $AdS_5/CFT_4$ solution [20, 21]. In (b), the largest BH is the one with $r_a = 15.1$.

To compare with the braneworld where a spherical symmetry is present on the brane, we first consider gravitational collapse under spherically symmetric configuration in 4D GR. The initial data is time symmetric Gaussian (massless scalar field), centred at $r = 1$. The pulse splits into ingoing and outgoing branches. For data that is not strong enough to produce BH, the ingoing branch is bounced back and then travel outwards. For the ADM mass and Hawking mass in 4D GR, please refer to Fig. 5.15. The figure shows that:

(1) the energy is conserved (remains to be constant at spatial infinity);

(2) the energy is monotonic with radius $r$, which means the energy "density" (not well-defined in general) is positive;

(3) there is a "stair" between two separate matter pulses—this is due to Birkhoff's theorem: vacuum solutions in spherical symmetry are unique (which are Schwarzschild's solutions). Therefore, in the vacuum gap between the two pulses, the solution is should be Schwarzschild, where the ADM/Hawking energy is well-defined as well. Thus Hawking mass is constant within this gap, while ADM mass is approximately constant.

For the ADM mass and Hawking mass on the brane, please refer to Fig. 5.16, where one can see that the above mentioned features (2) and (3) are lost. Feature (1) still holds—this is because the masses are defined as the limit at the spatial infinity utilizing the time translational symmetry. Or in another word, this is because the dynamics happening locally can not propagate to the spatial infinity within finite time. However, it is expected that feature (1) does not to hold in the braneworld. This is because, there is energy exchange between the bulk and the brane, and the simulations show that the "wiggles" are moving from the brane into the bulk. Therefore we need a quantity that can describe this phenomena.

It turns out the energy defined by equation (2.69) can indeed fulfill this task. Please refer to Fig. 5.17. This energy has feature (2) and feature (3) mentioned above. The energy at spatial infinity, on the other hand, changes with time. Fig. 5.17(b) shows the change with time: the brane loses energy in majority of time (which agrees with the phenomena we observed from the simulations we performed), but it gains some energy when the incoming pulse gets bounced back at the origin.

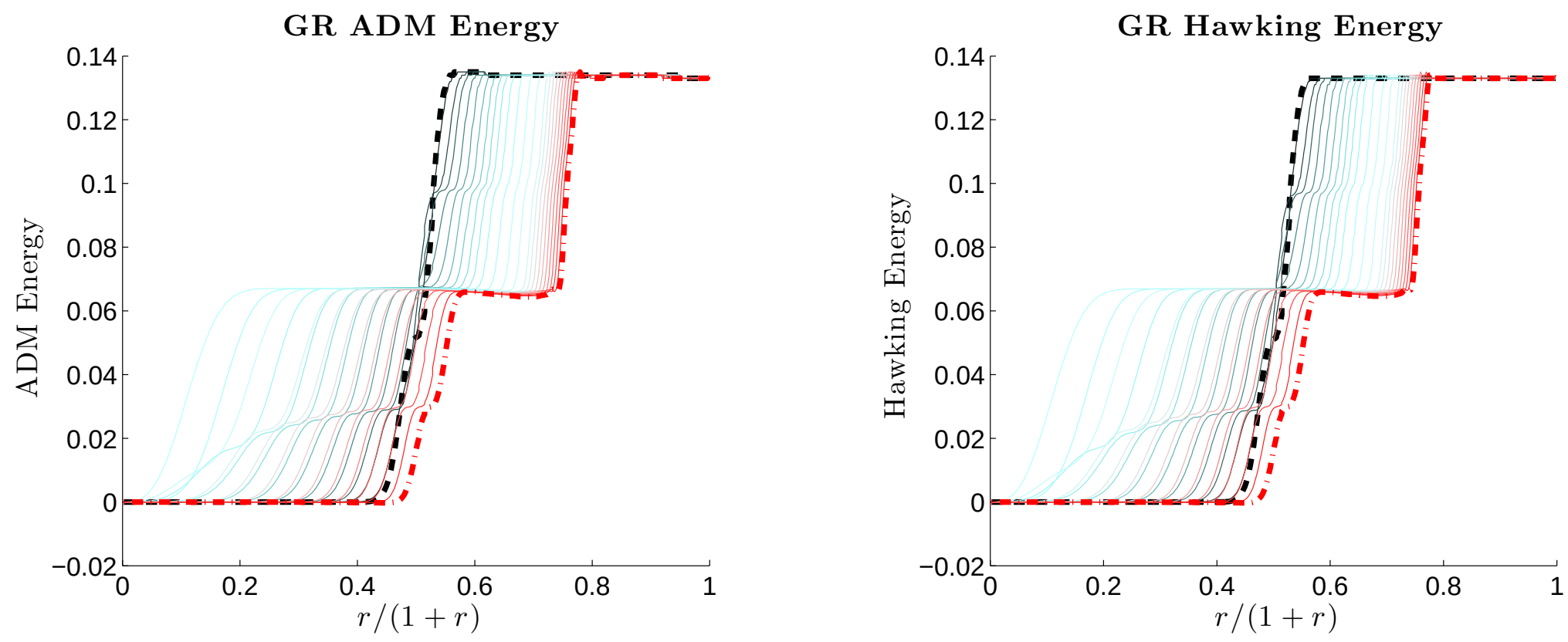

Figure 5.15: Energies in 4D GR. Each line is plotted at an instant (i.e., constant coordinate time). The black dashed line is at the earliest instant, and the red "-." line is at the last instant. Other lines' colour changes gradually from black to red, with respect to coordinate time. Each line only has one colour since it stands for one instant.

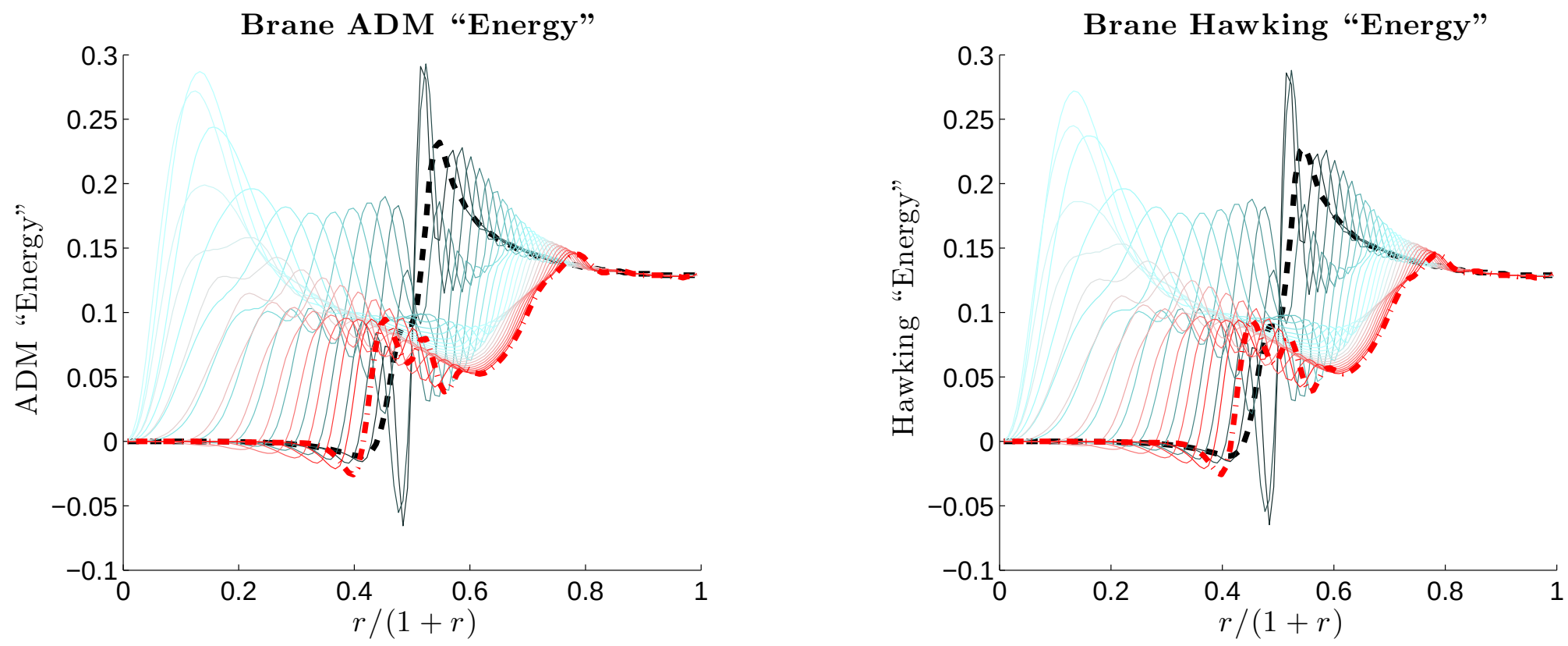

Figure 5.16: "Energies" on Brane. Each line is plotted at an instant (i.e., constant coordinate time). The black dashed line is at the earliest instant, and the red "-." line is at the last instant. Other lines' colour changes gradually from black to red, with respect to coordinate time. Each line only has one colour since it stands for one instant.

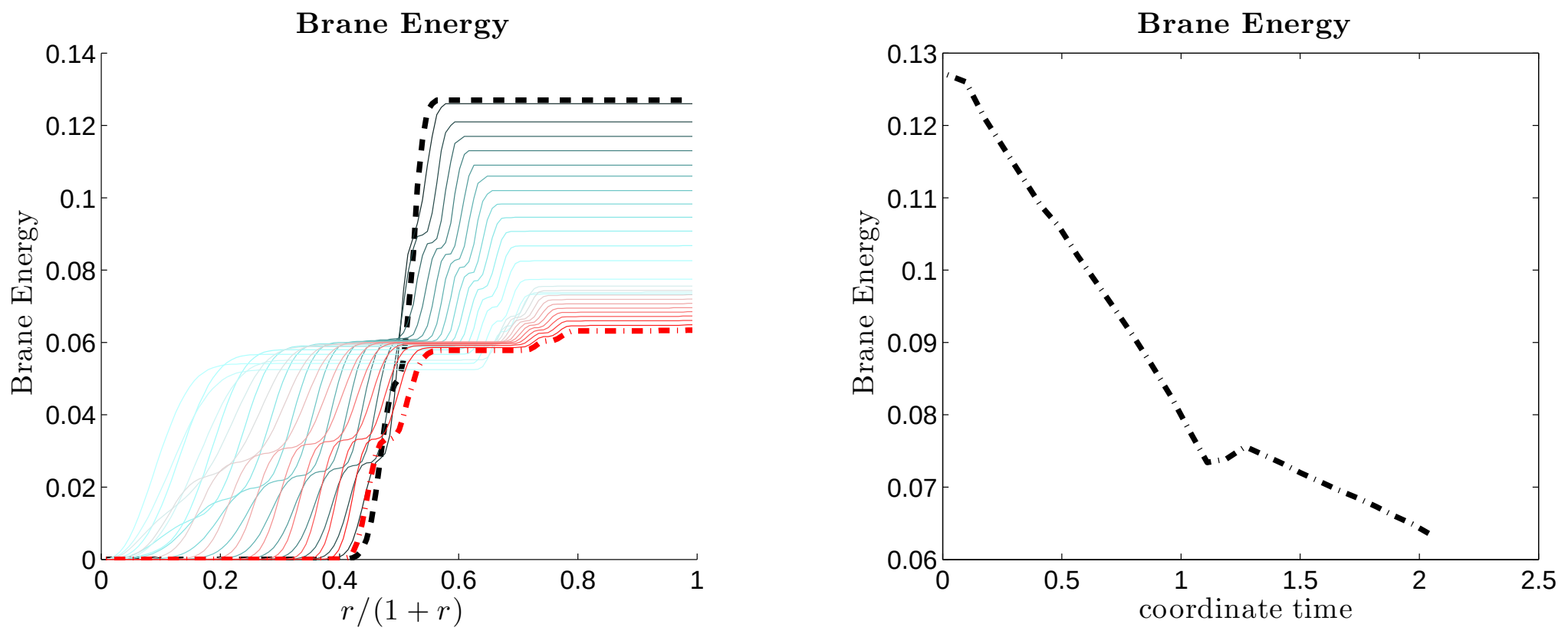

Figure 5.17: This figure shows the brane energy defined by equation (2.69). In (a), each line is plotted at an instant (i.e., constant coordinate time). The black dashed line is at the earliest instant, and the red "-." line is at the last instant. Other lines' colour changes gradually from black to red, with respect to coordinate time. Each line only has one colour since it stands for one instant. (b) shows the energy at spatial infinity changing with coordinate time.

# Chapter 6

# Conclusion and Future Work

The basic idea of braneworld scenarios is that our observable universe could be a 3+1 dimensional surface (the "brane") embedded in a higher dimensional spacetime (the "bulk"). The single brane scenario constructed by Randall and Sundrum (also known as RS-braneworld II), is what we study in this thesis project. The basic setup is as follows: the single brane (the observable universe) is embedded in the bulk with one extra dimension of infinite size. The matter is strictly trapped on the brane while gravity is free to access the bulk. The bulk is therefore "empty", but has an assumed negative cosmological constant. The bulk has a $Z_2$ symmetry with respect to the brane, and, the brane has a tension which enable fine-tuning to any equivalent cosmological constant on the brane. General relativity (GR) is recovered on brane at low energies, but the brane dynamics can be rather different from GR at high energies. This latter regime is the focus of our research. Specifically, we study the dynamical process of black hole formation as a result of gravitational collapse of massless scalar field.

## 6.1 Conclusion

We have achieved the following: in terms of developing the machinery, we discover/develop/invent several novel facts, formalisms and techniques regarding NR and braneworld. The regularity problem in previous NR simulations in axisymmetric (and spherically symmetric) spacetime, is actually associated with neither coordinate systems nor the machine precision. The numerical calculation is regular in any coordinates, provided the fundamental variables (used in simulation) are regular, and their asymptotic behaviour at the vicinity of the axis (or origin) is compatible with the finite difference scheme. Generalized harmonic (GH) and BSSN formalisms for general relativity are developed to make them suitable for simulations in non-Cartesian coordinate under non-flat background. A conformal function of the metric is included into the formalism to simulate the braneworld. The usual constraint damping term used in GH, can not control the severe constraint violation in braneworld. The violation is cured by imposing the constraints properly. In solving elliptic equations (Hamiltonian constraint, for instance), using functions to carry the asymptotic behaviour at spatial infinities could be crucial. The delta-function matter can be simulated by "integrating out" the delta function and the brane content can be encoded in Israel's junction condition.

On the physics side, we perform the first numerical study of gravitational collapse in braneworld within the framework of the single brane model by Randall-Sundrum (RSII). The scheme is capable of obtaining apparently stationary states as the results of gravitational collapse, which are BHs localized on the brane, with finite extension into the bulk. The extension changes from sphere to flattened pancake (or from sphere to cigar, from the embedding point of view) as the size of BH increases. There is strong evidence that the detail of initial data is lost in the resulting BH, therefore no-hair theorem of BH (uniqueness of BH solution) is suggested to hold in RSII spacetimes that are studied in this thesis—these spacetimes are axisymmetric without angular momentum and non-gravitational charges. In particular, the BHs we obtained as the apparently stationary states from the dynamical system, are consistent with the ones previously obtained from a static problem by Figueras and Wiseman. We also obtained some results in closed form without numerical computation: the smoothness of apparent horizon across the brane under perpendicular gauge, the well-definedness of event horizon on the brane, and the equality of ADM mass of the brane with the total mass of braneworld.

## 6.2 Future Work

There are many potential research directions/projects suggested by this project, some of which will become our future work. One such direction is to derive the mass-area relation of the (apparent) horizons in closed form. Another project is to look at the examples related to holographic principle: the equality of bulk energy and the ADM/Hawking energy calculated from the brane geometry which we have proved to hold for two classes of spacetimes. We would like to examine whether the relation holds in general space, and study the asymptotical configurations at spatial infinities.

As for this work per se, it can be expended/upgraded as follows

(1) Apparently stationary states were obtained, but it is not clear whether end states were obtained. The method to identify end state needs to be developed. Some attempts would be to implement the slicing conditions in [118–120].

(2) The current slicing condition and coordinate gauges can perform a wide range of initial data and yield apparently stationary states. However, for certain initial data, the gauges we used (Harmonic gauge, lapse driven gauge, damping wave gauge) can not yield a simulation that last “forever”. Therefore we need to improve the performance by studying coordinate gauges.

(3) There is slight inconsistency in the initial stages of the evolution (which fade out during evolution), which might be due to the truncation error, or inconsistency in initial data. Further study is needed.

(4) The energy of the brane is needed to describe the interaction between the brane and the bulk. The brane energy we obtained is not conceptually satisfying, although the qualitative behaviour is surprisingly good. The energy aspect needs more study.

(5) The current code can run in a parallel environment since it is based on AMRD/PAMR. However, we used a single CPU to do all the simulations since shooting method were employed to search for the apparent horizon. One improvement is to use flow method to search for apparent horizon, which can be used in a parallel environment.

For the same reason, although the code is capable of using the adaptive mesh refinement (AMR) that is built-in in AMRD/PAMR, because of the shooting method used in finding apparent horizon, we performed the simulations under unigrid. Hopefully we can turn on the AMR feature when the shooting method is replaced by flow method. In this way we can also improve the accuracy of the result. Here we emphasize that resolution is important since our calculation of large BHs resulted in apparent horizon at $z \sim 30\ell$, where the effective resolution was quite low and should be improved.

Also, we could try another way (which I called BSSN-like method) to implement the generalized harmonic formalism as discussed in section 5.3.

(6) Generalized harmonic formalism of GR was used in this work. We could also try other formalism such as BSSN, where it is more natural to impose gauges.

# Appendix A

# Generalized BSSN Formalism

The original BSSN [45, 46] formalism is only applicable to 4D spacetime with Cartesian coordinate. It was generalized into non-Cartesian coordinate in [36, 97] for 4D spacetimes with flat backgrounds. We will generalize BSSN into arbitrary dimension, general background, general coordinate and general conformal function (without requiring $\tilde{\gamma} = \bar{\gamma}$, see below), yet free from irregularity issue. This work is essential to make BSSN applicable to higher dimensional spacetime with non-trivial background, such as braneworld simulation. The derivation in [36] is largely borrowed in this section.

Fundamental variables to evolve: $\varphi$, $K$, $\tilde{\gamma}_{ij}$, $\tilde{A}_{ij}$, $\tilde{\Gamma}^i$, which are defined as

$$\tilde{\gamma}_{ij} \equiv e^{-q\varphi}\gamma_{ij}, \tag{A.1}$$

$$A_{ij} \equiv \left(K_{ij} - \frac{1}{d-1}\gamma_{ij}K\right), \tag{A.2}$$

$$\tilde{A}_{ij} \equiv e^{-q\varphi}A_{ij}, \tag{A.3}$$

$$\tilde{\Gamma}^i \equiv \tilde{\gamma}^{jk}\left({}^{(d-1)}\tilde{\Gamma}^i{}_{jk} - {}^{(d-1)}\bar{\Gamma}^i{}_{jk}\right), \tag{A.4}$$

where $\tilde{\gamma}$, the determinant of $\tilde{\gamma}_{ij}$, is set to a time-independent function, such as $\bar{\gamma}$ (the determinant of the background spatial metric). The *bar*-ed quantities are associated with the time-independent background metric.

Let's derive the evolution equations for these quantities.

The starting points are the evolution equations for $K_{\alpha\beta}$ and $\gamma_{\alpha\beta}$ in ADM-York formalism: eq. (1.15) and (1.16). They are written below as

$$\mathcal{L}_{\mathbf{m}}K_{\alpha\beta} = \epsilon D_\alpha D_\beta \alpha + \alpha\left[-\epsilon\ {}^{(d-1)}R_{\alpha\beta} + KK_{\alpha\beta} - 2K_{\alpha\mu}K^{\mu}{}_{\beta} + \epsilon\ k_d\left(S_{\alpha\beta} - \gamma_{\alpha\beta}\frac{S+\epsilon\rho}{d-2}\right)\right], \tag{A.5}$$

$$\mathcal{L}_{\mathbf{m}}\gamma_{\alpha\beta} = -2\alpha K_{\alpha\beta}, \tag{A.6}$$

from which we can derive

$$\mathcal{L}_{\mathbf{m}} \ln \gamma^{1/2} = \frac{1}{2}\gamma^{ij}\mathcal{L}_{\mathbf{m}}\gamma_{ij} = -\alpha K. \tag{A.7}$$

The equation of motion for $K$ is

$$\begin{aligned}\mathcal{L}_{\mathbf{m}}K &= \gamma^{ij}\mathcal{L}_{\mathbf{m}}K_{ij} + K_{ij}\mathcal{L}_{\mathbf{m}}\gamma^{ij} \\ &= \epsilon D^i D_i \alpha + \alpha\left[\tilde{A}_{ij}\tilde{A}^{ij} + \frac{1}{d-1}K^2 + k_d\left(\frac{d-3}{d-2}\rho - \frac{\epsilon S}{d-2}\right)\right],\end{aligned} \tag{A.8}$$

where the Hamiltonian constraint is used. The equations of motion for $\varphi$, $\tilde{\gamma}_{ij}$ are derived to be

$$\mathcal{L}_{\mathbf{m}}\varphi = \frac{2}{q(d-1)}\left[\mathcal{L}_{\mathbf{m}}\ln\gamma^{1/2} - \mathcal{L}_{\mathbf{m}}\ln\tilde{\gamma}^{1/2}\right] = \frac{2}{q(d-1)}\left(-\alpha K - \mathcal{L}_{\mathbf{m}}\ln\tilde{\gamma}^{1/2}\right). \tag{A.9}$$

$$\mathcal{L}_{\mathbf{m}}\tilde{\gamma}_{ij} = e^{-q\varphi}\mathcal{L}_{\mathbf{m}}\gamma_{ij} - qe^{-q\varphi}\gamma_{ij}\mathcal{L}_{\mathbf{m}}\varphi = -2\alpha\tilde{A}_{ij} + \frac{2}{d-1}\tilde{\gamma}_{ij}\mathcal{L}_{\mathbf{m}}\ln\tilde{\gamma}^{1/2}. \tag{A.10}$$

For $\tilde{\Gamma}^i$, we have

$$\tilde{\Gamma}^i = \tilde{\gamma}^{jk}\left({}^{(d-1)}\tilde{\Gamma}^i{}_{jk} - {}^{(d-1)}\bar{\Gamma}^i{}_{jk}\right) = -\bar{D}_j\tilde{\gamma}^{ij} + \tilde{\gamma}^{ij}\partial_j\left[\ln(\bar{\gamma}/\tilde{\gamma})^{1/2}\right]. \tag{A.11}$$

$$\Rightarrow \quad \partial_t\tilde{\Gamma}^i = -\bar{D}_j\left(\partial_t\tilde{\gamma}^{ij}\right) + \partial_t\tilde{\gamma}^{ij}\partial_j\left[\ln(\bar{\gamma}/\tilde{\gamma})^{1/2}\right]. \tag{A.12}$$

where we have applied ${}^{(d-1)}\bar{\Gamma}^j{}_{jk} = \frac{1}{2}\partial_k \ln\bar{\gamma}$ (and ${}^{(d-1)}\tilde{\Gamma}^j{}_{jk} = \frac{1}{2}\partial_k\ln\tilde{\gamma}$), and $\partial_t\bar{D}_j - \bar{D}_j\partial_t = 0$ which is because $\partial_t{}^{(d-1)}\bar{\Gamma}^i{}_{jk} = 0$. $\partial_t\tilde{\gamma}^{ij}$ can be easily evaluated from (A.10) via $\partial_t\tilde{\gamma}^{ij} = -\tilde{\gamma}^{ik}\tilde{\gamma}^{jl}\partial_t\tilde{\gamma}_{kl}$.

The evolution of $\tilde{A}_{ij}$ needs the result of the evolution of $A_{ij}$, which is

$$\begin{aligned}\mathcal{L}_{\mathbf{m}}A_{ij} =& \mathcal{L}_{\mathbf{m}}K_{ij} - \frac{1}{d-1}\left(K\mathcal{L}_{\mathbf{m}}\gamma_{ij} + \gamma_{ij}\mathcal{L}_{\mathbf{m}}K\right) \\ &= \left(\mathcal{L}_{\mathbf{m}}K_{ij}\right)^{\mathrm{TF}} + \frac{2\alpha}{d-1}KK_{ij} - \frac{2\alpha}{d-1}\gamma_{ij}K_{lm}K^{lm},\end{aligned} \tag{A.13}$$

where TF means trace free. Therefore

$$\begin{aligned}\mathcal{L}_{\mathbf{m}}\tilde{A}_{ij} &= e^{-q\varphi}\mathcal{L}_{\mathbf{m}}A_{ij} - qe^{-q\varphi}A_{ij}\mathcal{L}_{\mathbf{m}}\varphi \\ &= e^{-q\varphi}\left[\epsilon D_iD_j\alpha - \alpha\epsilon^{(d-1)}R_{ij} + \epsilon\alpha k_d\left(S_{ij} - \gamma_{ij}\frac{S+\epsilon\rho}{d-2}\right)\right]^{\mathrm{TF}} \\ &\quad + \alpha K\tilde{A}_{ij} - 2\alpha\tilde{A}_{il}\tilde{A}^l{}_j + \frac{2}{d-1}\tilde{A}_{ij}\mathcal{L}_{\mathbf{m}}\ln\tilde{\gamma}^{1/2}.\end{aligned} \tag{A.14}$$

${}^{(d-1)}R_{ij}$ needs to be rewritten in terms of the $\tilde{\Gamma}^i$ defined in (A.11). By the same procedure for the

conformal transformation carried out in [32, 36], we have

$$ {}^{(d-1)}R_{ij} = {}^{(d-1)}\tilde{R}_{ij} - \frac{q(d-3)}{2}\tilde{D}_i\tilde{D}_j\varphi - \frac{q}{2}\tilde{\gamma}_{ij}\,\tilde{D}_k\tilde{D}^k\varphi + \frac{q^2(d-3)}{4}\left(\tilde{D}_i\varphi\,\tilde{D}_j\varphi - \tilde{\gamma}_{ij}\,\tilde{D}_k\varphi\,\tilde{D}^k\varphi\right). \tag{A.15} $$

Therefore

$$ {}^{(d-1)}R = e^{-q\varphi}\left({}^{(d-1)}\tilde{R} - q(d-2)\tilde{D}_k\tilde{D}^k\varphi - \frac{q^2(d-3)(d-2)}{4}\tilde{D}_k\varphi\tilde{D}^k\varphi\right). \tag{A.16} $$

Expressing ${}^{(d-1)}\tilde{R}_{ij}$ and ${}^{(d-1)}\tilde{R}$ in terms of $\tilde{\Gamma}^i$, has been done in section 3.3 (or [32, 36])

$$ {}^{(d-1)}\tilde{R}_{ij} = {}^{(d-1)}\bar{R}_{ij} - \frac{1}{2}\tilde{\gamma}^{kl}\bar{D}_k\bar{D}_l\tilde{\gamma}_{ij} - \bar{D}_k\tilde{\gamma}_{l(i}\bar{D}_{j)}\tilde{\gamma}^{kl} - \bar{D}_{(i}\tilde{\Gamma}_{j)} + \tilde{\Gamma}_k\tilde{C}^k{}_{ij} - \tilde{C}^k{}_{il}\tilde{C}^l{}_{kj}. \tag{A.17} $$

Here we have defined $\tilde{C}^i{}_{jk} \equiv {}^{(d-1)}\tilde{\Gamma}^i{}_{jk} - {}^{(d-1)}\bar{\Gamma}^i{}_{jk}$, and $\tilde{\Gamma}_i \equiv \tilde{\gamma}_{ij}\tilde{\Gamma}^j$. Contracting (A.17) with $\tilde{\gamma}^{ij}$, we can obtain the expression of $\tilde{R}$ in terms of $\tilde{\Gamma}^i$.

In (A.14), $D_iD_j\alpha$ needs to be expressed by its counter parts

$$ \begin{aligned} D_iD_j\alpha &= D_i\tilde{D}_j\alpha = \tilde{D}_i\tilde{D}_j\alpha - C^k{}_{ij}\tilde{D}_k\alpha \\ &= \tilde{D}_i\tilde{D}_j\alpha - \frac{q}{2}\left(\tilde{D}_i\alpha\tilde{D}_j\varphi + \tilde{D}_j\alpha\tilde{D}_i\varphi - \tilde{\gamma}_{ij}\,\tilde{D}^k\varphi\tilde{D}_k\alpha\right), \end{aligned} \tag{A.18} $$

where $C^k{}_{ij} \equiv {}^{(d-1)}\Gamma^k{}_{ij} - {}^{(d-1)}\tilde{\Gamma}^k{}_{ij}$, and $C^k{}_{ij} = \frac{q}{2}\left(\delta^k{}_i\tilde{D}_j\varphi + \delta^k{}_j\tilde{D}_i\varphi - \tilde{\gamma}_{ij}\tilde{D}^k\varphi\right)$ are used, and the latter can be obtained by repeating the relevant derivations in [36].

To perform numerical relativity, all the equations above need to be rewritten as partial derivatives with respect to $t$, which are related to $\mathcal{L}_{\mathbf{m}}$ by $\mathcal{L}_{\mathbf{m}} = \mathcal{L}_{\boldsymbol{\partial}_t} - \mathcal{L}_{\boldsymbol{\beta}}$. We must be careful with the Lie derivatives of tensor densities with respect to $\boldsymbol{\beta}$. An object $X$ is a tensor density of weight $w$, if $X = \text{tensor} \times \gamma^{w/2}$. Its Lie derivative is

$$ \mathcal{L}_{\boldsymbol{\beta}}X = \Big[\mathcal{L}_{\boldsymbol{\beta}}X\Big]_{w=0} + wX\partial_i\beta^i. \tag{A.19} $$

Let's now figure out the weight of fundamental variables. Because of (A.1), we have $e^{\varphi} = (\gamma/\tilde{\gamma})^{\frac{1}{q(d-1)}}$, therefore the weight of $e^{\varphi}$ is $\frac{2-\tilde{w}}{q(d-1)}$, where $\tilde{w}$ is the weight of $\tilde{\gamma}$. Similarly, the weight of $\tilde{\gamma}_{ij}$ and $\tilde{A}_{ij}$ is $\frac{\tilde{w}-2}{d-1}$, the weight of upstairs $\tilde{\gamma}^{ij}$ and $\tilde{A}^{ij}$ is $\frac{2-\tilde{w}}{d-1}$. The value of $\tilde{w}$, can actually be determined as follows. $\tilde{\gamma}/\gamma$ is a scalar, because its value does not change under coordinate transformation. Therefore, $\tilde{\gamma} = (\gamma \cdot \text{tensor})$, which implies $\tilde{w} = 2$. Therefore, $e^{\varphi}, \tilde{\gamma}_{ij}$ and $\tilde{A}_{ij}$ are all tensors[17]. Knowing $\tilde{w} = 2$, it is then easy to derive

$$\mathcal{L}_{\boldsymbol{\beta}} \ln \tilde{\gamma}^{1/2} = \frac{1}{2\tilde{\gamma}} \mathcal{L}_{\boldsymbol{\beta}} \tilde{\gamma} = \cdots = \beta^i \partial_i \ln \tilde{\gamma}^{1/2} + \partial_i \beta^i. \tag{A.20}$$

The formulae can then be rewritten in terms of coordinate derivatives by opening $\mathcal{L}_{\boldsymbol{\beta}} \ln \tilde{\gamma}^{1/2}$. Let's take (A.9) as an example. Using $\boldsymbol{m} = \boldsymbol{\partial}_t - \boldsymbol{\beta}$ and $\partial_t \tilde{\gamma} = 0$, we have

$$\partial_t \varphi = \frac{2}{q(d-1)} \left( -\alpha K + \beta^i \partial_i \ln \tilde{\gamma}^{1/2} \right) + \beta^i \partial_i \varphi + \frac{2}{q(d-1)} \partial_i \beta^i, \tag{A.21}$$

where we have applied $\mathcal{L}_{\boldsymbol{\partial}_t} T = \partial_t T$ in a coordinate system where $t$ is a coordinate, for tensor $T$.

The Hamiltonian constraint is now

$$k_d \ \rho \simeq \frac{1}{2} \left( -\epsilon^{(d-1)} R + \frac{d-2}{d-1} K^2 - \tilde{A}_{ij} \tilde{A}^{ij} \right), \tag{A.22}$$

where ${}^{(d-1)}R$ is of course replaced by its expression in equation (A.16). $\simeq$ means the equation is a constraint relation. The momentum constraint reads

$$\begin{aligned} \epsilon k_d S_i \simeq& \tilde{D}_i K - \left( \tilde{D}_j \tilde{A}^j{}_i + \frac{q(d-1)}{2} \tilde{D}_j \varphi \tilde{A}^j{}_i + \frac{1}{d-1} \tilde{D}_i K \right) \\ =& \frac{d-2}{d-1} \tilde{D}_i K - \left( \tilde{D}_j \tilde{A}^j{}_i + \frac{q(d-1)}{2} \tilde{D}_j \varphi \tilde{A}^j{}_i \right). \end{aligned} \tag{A.23}$$

**Summary**

The fundamental variables are $\varphi$, $K$, $\tilde{\gamma}_{ij}$, $\tilde{A}_{ij}$, $\tilde{\Gamma}^i$. The equations of motion are (A.8, A.9, A.10, A.12, A.14). Constraints are the Hamiltonian constraint (A.22), momentum constraints (A.23), and the definition equations (A.1-A.4), which read

$$\tilde{\gamma} \simeq \text{ the pre-set time-independent function such as } \bar{\gamma}, \tag{A.24}$$

$$\tilde{\gamma}^{ij} \tilde{A}_{ij} \simeq 0, \tag{A.25}$$

$$\tilde{\Gamma}^i \simeq \tilde{\gamma}^{jk} \left( \tilde{\Gamma}^i{}_{jk} - \bar{\Gamma}^i{}_{jk} \right). \tag{A.26}$$

[17] Quite a few authors counted the weights incorrectly—especially under Cartesian coordinate with flat background where $\tilde{\gamma}$ was set to 1, where these authors incorrectly assumed $\tilde{w} = 0$. Fortunately, the values of the weights per se do not matter in the final expressions in terms of partial derivatives. It is the *relative* weight that matters. e.g. one can keep $\tilde{w}$ general (without substituting $\tilde{w} = 2$), therefore the weight of $e^{\varphi}$ is $\frac{2-\tilde{w}}{q(d-1)}$. One can still obtain (A.21) correctly.

### Axisymmetry

For simulations in axisymmetry—take 4D as an example—under cylindrical coordinates $(t, \rho, \phi, z)$, as shown in Chap. 3, the such defined $\tilde{\Gamma}^i$ reduces to the results obtained by our Cartesian components method if the background $\bar{\gamma}_{ij}$ is flat, therefore regular. When the background is not flat, the behaviour needs to be analyzed in a case-by-case basis. The conformal metric and conformal traceless extrinsic curvature are

$$\tilde{\gamma}_{ij} = \begin{pmatrix} \tilde{\gamma}_{\rho\rho} & 0 & \tilde{\gamma}_{\rho z} \\ 0 & \rho^2(\tilde{\gamma}_{\rho\rho} + \rho W) & 0 \\ \tilde{\gamma}_{\rho z} & 0 & \tilde{\gamma}_{zz} \end{pmatrix}, \quad \tilde{A}_{ij} = \begin{pmatrix} \tilde{A}_{\rho\rho} & 0 & \tilde{A}_{\rho z} \\ 0 & \rho^2(\tilde{A}_{\rho\rho} + \rho V) & 0 \\ \tilde{A}_{\rho z} & 0 & \tilde{A}_{zz} \end{pmatrix}, \tag{A.27}$$

where local flatness has been applied. All the fundamental variables depend on $(t, \rho, z)$ only.

# Appendix B

# Extrinsic Curvature as Geodesics Deviation: $C \geq 1$ Case

In this section, we consider the case that a $(d - C)$ dimensional object $\Sigma$ embeds in the $d$ dimensional space $M$. Therefore $C$ is the codimension.

Let ${}^{(1)}n^{\mu}$, ... , ${}^{(C)}n^{\mu}$ be $C$ continuous normal vector fields of $\Sigma$, with unit length ${}^{(1)}\epsilon$, ... , ${}^{(C)}\epsilon$, and ${}^{(I)}n^{\mu}$'s are set to be mutually orthogonal, where the index $I = 1, 2, ..., C$. Each ${}^{(I)}\epsilon = \pm 1$, depending on the spacelike/timelike nature of the corresponding dimension.

The procedure to obtain equation (2.23) can be straightforwardly generalized to $C \geq 1$ case. For example, generalize equation (2.17) to

$$\gamma_{\alpha\beta} = g_{\alpha\beta} - {}^{(1)}\epsilon\ {}^{(1)}n_{\alpha}{}^{(1)}n_{\beta} - ... - {}^{(C)}\epsilon\ {}^{(C)}n_{\alpha}{}^{(C)}n_{\beta}. \tag{B.1}$$

Eventually the deviation of two geodesics produced by $T^{\alpha} \in \Sigma$ is derived to be

$$\sum_{I=1}^{C} {}^{(I)}\epsilon\ {}^{(I)}n^{\mu} T^{\alpha} T^{\beta} \nabla_{\alpha} \left( {}^{(I)}n_{\beta} \right) \equiv T^{\alpha} T^{\beta} \sum_{I=1}^{C} {}^{(I)}\epsilon\ {}^{(I)}n^{\mu}\ {}^{(I)}K_{\alpha\beta}. \tag{B.2}$$

i.e. we have defined $C$ tensors

$${}^{(I)}K_{\alpha\beta} = \gamma_{\alpha}{}^{\mu} \gamma_{\beta}{}^{\nu} \nabla_{\mu} \left( {}^{(I)}n_{\nu} \right), \quad \text{where } I = 1, ..., C. \tag{B.3}$$

On the other hand, NVP takes the "change rate" of the direction of unit normal vector along $\Sigma$ to serve as extrinsic curvature. When $C > 1$, the direction of $C$ dimensional orthogonal space is characterized by the wedge form

$${}^{(1)}n^{\mu} \wedge ... \wedge {}^{(C)}n^{\nu}. \tag{B.4}$$

Therefore, under NVP, the extrinsic curvature is[18]

$$K_{\delta\mu\ldots\nu} = \gamma_\delta{}^\alpha \nabla_\alpha \left( {}^{(1)}n_\mu \wedge \ldots \wedge {}^{(C)}n_\nu \right), \tag{B.5}$$

which is a tensor of rank $C+1$. As the form implies, $K_{\delta\mu\ldots\nu}$ does not lie within $\Sigma$. However, we can prove that NVP is still equivalent to GEP.

First, we define the notation

$$d_\delta \equiv \gamma_\delta{}^\alpha \nabla_\alpha. \tag{B.6}$$

Then it is easy to show that

$$\begin{aligned} K_{\delta\mu\beta\ldots\nu} = d_\delta \left( {}^{(1)}n_\mu \wedge {}^{(2)}n_\beta \wedge \ldots \wedge {}^{(C)}n_\nu \right) = & \left( d_{(\delta)} {}^{(1)}n_\mu \right) \wedge {}^{(2)}n_\beta \wedge \ldots \wedge {}^{(C)}n_\nu \\ + {}^{(1)}n_\mu \wedge \left( d_{(\delta)} {}^{(2)}n_\beta \right) \wedge \ldots \wedge {}^{(C)}n_\nu & + {}^{(1)}n_\mu \wedge {}^{(2)}n_\beta \wedge \ldots \wedge \left( d_{(\delta)} {}^{(C)}n_\nu \right). \end{aligned} \tag{B.7}$$

i.e. Leibniz rule. And of course the operation $d_\delta$ should be excluded from the wedge's. We use a bracket on $\delta$ as $d_{(\delta)}$ to express this fact explicitly.

Before we continue, let us introduce some notations

$$N_{\mu\nu} = g_{\mu\nu} - \gamma_{\mu\nu} = \sum_{I=1}^{C} {}^{(I)}\epsilon \; {}^{(I)}n_\mu {}^{(I)}n_\nu. \tag{B.8}$$

$${}^{(I)}N_{\mu\nu} = N_{\mu\nu} - {}^{(I)}\epsilon \; {}^{(I)}n_\mu {}^{(I)}n_\nu = \sum_{J=1, J\neq I}^{C} {}^{(J)}\epsilon \; {}^{(J)}n_\mu {}^{(J)}n_\nu. \tag{B.9}$$

Noticing that ${}^{(I)}n_\nu {}^{(I)}n^\nu = {}^{(I)}\epsilon =$ constant along $\Sigma$, we have

$$\left( d_\delta {}^{(I)}n_\nu \right) {}^{(I)}n^\nu = 0, \tag{B.10}$$

which implies

$$\begin{aligned} d_\delta {}^{(I)}n_\nu = \gamma_\delta{}^\alpha \nabla_\alpha {}^{(I)}n_\nu &= \left( \gamma_\delta{}^\alpha \nabla_\alpha {}^{(I)}n_\mu \right) g_\nu{}^\mu \\ &= \left( \gamma_\delta{}^\alpha \nabla_\alpha {}^{(I)}n_\mu \right) \left( \gamma_\nu{}^\mu + N_\nu{}^\mu \right) = \left( \gamma_\delta{}^\alpha \nabla_\alpha {}^{(I)}n_\mu \right) \left( \gamma_\nu{}^\mu + {}^{(I)}N_\nu{}^\mu \right), \end{aligned} \tag{B.11}$$

[18]Or define the extrinsic curvature via the derivative of the wedge of the $D - C$ tangent vectors of $\Sigma$.

where we have applied equation (B.10) in deriving the last equal sign. The above equation is just

$$d_\delta{}^{(I)}n_\nu = {}^{(I)}K_{\delta\nu} + \sum_{J=1, J\neq I}^{C} {}^{(J)}\epsilon{}^{(J)}n_\nu{}^{(IJ)}B_\delta, \tag{B.12}$$

where we have defined

$${}^{(IJ)}B_\delta \equiv \left({}^{(J)}n^\mu d_\delta{}^{(I)}n_\mu\right). \tag{B.13}$$

Overall, the above procedure is nothing but projecting $d_\delta{}^{(I)}n_\nu$ into the tangent space of $\Sigma$ and the orthogonal space of $\Sigma$.

Now we can use the fact

$$... \wedge {}^{(J)}n_\nu \wedge ... \wedge {}^{(J)}n_\mu \wedge ... = 0, \tag{B.14}$$

and use equation (B.12) to rewrite equation (B.7) as

$$\begin{aligned} K_{\delta\mu\beta...\nu} = &\ {}^{(1)}K_{(\delta)\mu} \wedge {}^{(2)}n_\beta \wedge ... \wedge {}^{(C)}n_\nu \\ &+ {}^{(1)}n_\mu \wedge {}^{(2)}K_{(\delta)\beta} \wedge ... \wedge {}^{(C)}n_\nu + \cdots + {}^{(1)}n_\mu \wedge {}^{(2)}n_\beta \wedge ... \wedge {}^{(C)}K_{(\delta)\nu}. \end{aligned} \tag{B.15}$$

Again, when doing wedge, $\delta$ is not affected. The terms above are linearly independent, therefore $K_{\delta\mu\beta...\nu}$ is a tensor whose coefficients of linearly independent tensors are ${}^{(I)}K_{\alpha\beta}$. In another word, there is a one-to-one correspondence between $K_{\delta\mu\beta...\nu}$ and a set of ${}^{(I)}K_{\alpha\beta}$. Therefore, NVP extrinsic curvature $K_{\delta\mu\beta...\nu}$ is equivalent to $C$ quantities ${}^{(I)}K_{\mu\nu}$, which characterize extrinsic curvature under GEP. i.e. *GEP and NVP are equivalent for any codimension.*